\documentclass[%
 reprint,
 twocolumn
 amsmath,amssymb,
 aps,
]{revtex4-2}

\usepackage{graphicx}% Include figure files
\usepackage{dcolumn}% Align table columns on decimal point
\usepackage{bm}
\usepackage{caption}% bold math
\usepackage{subcaption}
\usepackage{float}
\usepackage{mathtools}% bold math
\usepackage{multirow}
\usepackage{feynmp-auto}
\usepackage{booktabs}
\usepackage{siunitx}
\usepackage{ragged2e}
\DeclarePairedDelimiterX\braket[2]{\langle}{\rangle}{#1 \delimsize\vert #2}

\usepackage{hyperref}  
\hypersetup{
    colorlinks=true,
    linkcolor=blue,
    filecolor=blue,      
    urlcolor=blue,
    citecolor=blue,
    }

\begin{document}

\preprint{APS/123-QED}

\title{The correlated insulators of magic angle twisted bilayer graphene at zero and one quantum of magnetic flux: a tight-binding study}% Force line breaks with \\

\author{ Miguel Sánchez Sánchez}
\email{miguel.sanchez@csic.es}
 % \altaffiliation[Also at ]{Physics Department, XYZ University.}%Lines break automatically or can be forced with \\
\author{Tobias Stauber}%
\email{tobias.stauber@csic.es}
 % \email{Second.Author@institution.edu}
\affiliation{Instituto de Ciencia de Materiales de Madrid CSIC, Madrid (Spain)}%

% \collaboration{MUSO Collaboration}%\noaffiliation

% \author{Charlie Author}
%  \homepage{http://www.Second.institution.edu/~Charlie.Author}
% \affiliation{
%  Second institution and/or address\\
%  This line break forced% with \\
% }%
% \affiliation{
%  Third institution, the second for Charlie Author
% }%
% \author{Delta Author}
% \affiliation{%
%  Authors' institution and/or address\\
%  This line break forced with \textbackslash\textbackslash
% }%

% \collaboration{CLEO Collaboration}%\noaffiliation

% \date{\today}% It is always \today, today,
             %  but any date may be explicitly specified

\begin{abstract}

Magic angle twisted bilayer graphene (MATBG) has become one of the prominent topics in Condensed Matter during the last few years, however, fully atomistic studies of the interacting physics are missing. In this work, we study the correlated insulator states of MATBG in the setting of a tight-binding model, under a perpendicular magnetic field of $0$ and $26.5$ T, corresponding to zero and one quantum of magnetic flux per unit cell. At zero field and for dopings of two holes ($\nu=-2$) or two electrons ($\nu=+2$) per unit cell, the Kramers intervalley coherent (KIVC) order is the ground state at the Hartree-Fock level, although it is stabilized by a different mechanism to that in continuum model. At charge neutrality, the spin polarized state is competitive with the KIVC due to the on-site Hubbard energy. We obtain a strongly electron-hole asymmetric phase diagram with robust insulators for electron filling and metals for negative filling. In the presence of magnetic flux, we predict an insulator with Chern number $-2$ for $\nu=-2$, a spin polarized state at charge neutrality and competing insulators with Chern numbers $+2$ and $0$ at $\nu=+2$. The stability of the $\nu=+2$ insulators is determined by the screening environment, allowing for the possibility of observing a topological phase transition. 
% The analytical understanding of the phases at $B=26.5$ T requires an intermediate coupling approach to the system.
% \begin{description}
% \item[Usage]
% Secondary publications and information retrieval purposes.
% \item[Structure]
% You may use the \texttt{description} environment to structure your abstract;
% use the optional argument of the \verb+\item+ command to give the category of each item. 
% \end{description}
\end{abstract}

%\keywords{Suggested keywords}%Use showkeys class option if keyword
                              %display desired
\maketitle

%\tableofcontents

\section{Introduction}
Magic angle twisted bilayer graphene (MATBG) is a two dimensional quantum material\cite{Polini22} that exhibits a plethora of exotic phases ranging from superconductors\cite{Cao2018sup,Yankowitz19,Lu2019,liu21} to strange metals\cite{Jaoui2022,cao20}, passing through integer\cite{Wu2021,Stepanov21} and fractional\cite{Xie2021} Chern Insulators. It constitutes a remarkable platform for the understanding of the many-body problem in Condensed Matter and the interplay of strong interactions and topology\cite{repellin2019chern,ledwith20,chew23}, and inaugurated the field of moiré materials\cite{Wang2020,Park2021,scheer2023twistronics,crépel2023chiral}.

The correlated insulators arise when the doping level is such that the number of electrons per unit cell is an integer number. They were discovered before the superconductivity and other phases\cite{Cao2018}, and are one of the most studied phenomena in twisted bilayer graphene (TBG)\cite{Po18, Kang19, Bultinck20, Bultinck21, Lu2019, Yankowitz19,Stepanov2020, Stepanov21, Wu2021, liu21, Sharpe21, Sharpe2019, Cao2018, Zhang22, stauber21, gonzalez20, klebl21, xie20, lin23, vafek20, bernevig321, bernevig421, seo19, Kwan23,ledwith21, Pierce2021, parker21, nuckolls2023quantum,dimitru22,jimenopozo2023short,blason22}.

On another hand, crystalline systems under magnetic fields are controlled by the scale given by the magnetic flux quantum $\Phi_0 = h/e$\cite{Hofstadter76,Herzog20}. When the magnetic field is such that the flux per unit cell is comparable to $\Phi_0$, the system is in the 'Hofstadter regime' and the picture of Landau levels is replaced by a reentrant band structure\cite{Biao20,Guan22}. In typical materials such magnetic fields are of the order of $10^4$ T, but in MATBG the large moiré unit cell allows accessible fields of the order of $30$ T. In MATBG the Landau level spectrum of the correlated insulators has been studied\cite{singh2023topological,wang22,Yankowitz19,Lu2019,Wu2021,Stepanov21}. Also, at one magnetic flux quantum reentrant correlated insulators have been predicted and observed\cite{herzog22_2, herzog22_3, efetov22}.

On the theory side, the Bistritzer-McDonald (BM) or continuum model\cite{McDonald11,Peres12} is a low energy theory that has proven very powerful in  understanding the physics of TBG, revealing the emergent symmetries of the system that have led to the picture of the '$U(4)$ ferromagnets' for the correlated insulators\cite{seo19, Po18,Kang19, vafek20, ledwith21,bernevig321,bernevig421}. However, the model, with only a handful of parameters, cannot capture the finer details of the spectrum and the wave functions. These differences at low energy scales are relevant in the competition between states.

In this work we employ a tight-binding model for MATBG. The high computational cost, which makes atomistic studies scarce in this system\cite{gonzalez20,stauber21,klebl21,Goodwin_2020}, is partially bypassed by a projection onto the subspace of the low energy bands (the 'flat bands'). The external magnetic flux is tuned to zero and one magnetic flux quantum per unit cell, we focus on samples without strain and leave electron-phonon coupling for future work.

The explicit breaking of the $U(4)$ symmetry of MATBG is assessed via the particle-hole asymmetry of the flat bands\cite{vafek23}, the intervalley Hund's coupling\cite{Bultinck20} and the on-site Hubbard interaction. 
% At zero field, the particle-hole asymmetry splits the energy of the valley polarized and Kramers intervalley coherent state, supporting the KIVC at even filling. Previous studies in the BM model established that the KIVC is also supported, but the splitting mechanism is kinetic energy superexchange\cite{Kang19,vafek20,bernevig421,Bultinck20,Kwan23}. The Hubbard interaction makes the spin polarized state competitive at charge neutrality, and 
We find that the intervalley Hund's coupling, relevant for the superconductivity, has antiferromagnetic contributions due to the long range Coulomb interaction. On the other hand, the Hubbard term contributes to a ferromagnetic coupling.

To study the spontaneous symmetry breaking in the correlated states, we perform self-consistent Hartree-Fock simulations. For $\Phi=0$ we find agreement with previous results on the nature of the ground state at even filling\cite{Kang19,vafek20,bernevig421,Bultinck20,Kwan23,Bultinck21}, but the selection mechanism of the ground state is different. Instead of kinetic energy superexchange as expected in the BM model, the appearance of 'inter-Chern' order near the $\Gamma$ point reduces the exchange energy and stabilizes the Kramers intervalley coherent state. In addition, the Hubbard interaction makes the spin polarized state competitive at charge neutrality. The self-consistent states are insulating at electron doping and metallic at hole doping, signalling the experimentally reported many-body electron-hole asymmetry\cite{Pierce2021,Lu2019,Yankowitz19}.

For $\Phi = \Phi_0$, the system departs from the strong coupling $U(4)$ picture due to the increased bandwidth of the kinetic energy bands, in the same manner as the strained samples at zero flux\cite{Bultinck21,parker21}. Consistently for different screening environments, we observe an insulator with Chern number $-2$ at $\nu=-2$ and a spin polarized state at $\nu=0$. For $\nu=+2$ we observe a topological phase transition from an insulator with Chern number $+2$ for small screening to an intervalley coherent trivial insulator for large screening. The Chern $+2$ insulator is compatible with the experimental data of Ref. \cite{efetov22}.

The paper is organized as follows. In section II we describe the tight-binding model of TBG and the Peierls' substitution under magnetic field, in section III we introduce the emergent $U(4)$ symmetry at the magic angle and in section IV we discuss the explicit breaking of $U(4)$ in the lattice model. Finally, in section V we report the correlated states obtained in the Hartree-Fock simulations, and in section VI we draw some conclusions.

\section{The model}
In graphene, the primitive vectors are $\boldsymbol{a}_1 = a(1/2,\sqrt{3}/2)$ and $\boldsymbol{a}_2 = a(-1/2,\sqrt{3}/2)$, with $a = \sqrt{3}a_0$ and $a_0=0.142$ nm the carbon-carbon distance.  Atoms at lattice points belong to sublattice $A$, and their nearest neighbours displaced by $(\boldsymbol{a}_1 + \boldsymbol{a}_2)/3$  to sublattice $B$. 

Consider two graphene layers stacked on top of each other, at $z=-d_0/2$ and $z=d_0/2$ respectively, being $d_0 = 0.335$ nm the interlayer distance, such that top and bottom atoms are vertically aligned. The bottom layer is rotated by an angle $-\theta/2$, and the top layer by $\theta/2$, with the center of rotation being the center of one of the graphene hexagons. We choose a value of $\theta$ that makes the twisted structure commensurate\cite{Peres12}. In our case, we parametrize the angle by an integer $n_\theta$ such that $\cos(\theta) = 1 - 1/2(3n_\theta^2+3n_\theta+1)$. The unit vectors of the superlattice are \begin{align}
\boldsymbol{L}_1 &= R_{-\theta/2}\big(n_\theta\boldsymbol{a}_1 + (n_\theta+1)\boldsymbol{a}_2\big) = L_M (0,1), \nonumber \\
\boldsymbol{L}_2 &= R_{\pi/3}\boldsymbol{L}_1
 = R_{-\theta/2}\big((-n_\theta-1)\boldsymbol{a}_1 + \big(2n_\theta+1)\boldsymbol{a}_2 \big),
\end{align} with $R_{\alpha}$ a rotation by  angle $\alpha$ and $L_M$ the lattice constant. The reciprocal vectors are given by 
\begin{align}
    a_0 \boldsymbol{G}_1 &= G_\theta R_{-\theta/2}\big((3n_\theta+1)\boldsymbol{a}_1+\boldsymbol{a}_2\big),  \nonumber \\
   a_0 \boldsymbol{G}_2 &= R_{-2\pi/3}(\boldsymbol{G}_1) \nonumber \\ &= G_\theta R_{-\theta/2}\big(-(3n_\theta+2)\boldsymbol{a}_1+(3n_\theta+1)\boldsymbol{a}_2\big) ,
\end{align}
where $G_\theta = \frac{4\pi}{3a_0}(9n_\theta^2+9n_\theta+3)^{-1}$. The magic angle is approximately given by $n_\theta=31$ ($1.05^\circ$), corresponding to a Moiré lattice constant of $L_M = 13.4$ nm and $11908$ atoms in the unit cell. 

The point group of this structure is the dihedral group $D_6$, generated by six-fold rotations around the $z$ axis, $C_{6z}$, and two-fold rotations around the $y$ axis, $C_{2y}$, leaving the origin fixed. The combined operation $C_{6z}^3=C_{2z}$ amounts to a two-fold rotation around the $z$ axis, and $C_{2x}= C_{2y}C_{2z}=C_{2x}$ to a two-fold rotation around the $x$ axis.
% $C_{2x}$ and $C_{2y}$ act as layer-interchanging mirror operations. 
The spin-orbit coupling being small, spinless time-reversal $\mathcal{T}$ is also a symmetry.

% Another type of structures, in which the twisting is around one of the A atoms has a smaller symmetry group $D_3$, generated by $C_{3z} = C_{6z}^2$ and $C_{2y}$. However, in the low energy states at small angles $C_{2z}$ becomes an emergent symmetry of the $D_3$ structure\cite{Angeli18}, in agreement with the continuum theories that are not sensitive to the relative translation between layers.\\
\begin{figure}
    \includegraphics[width=.5\linewidth]{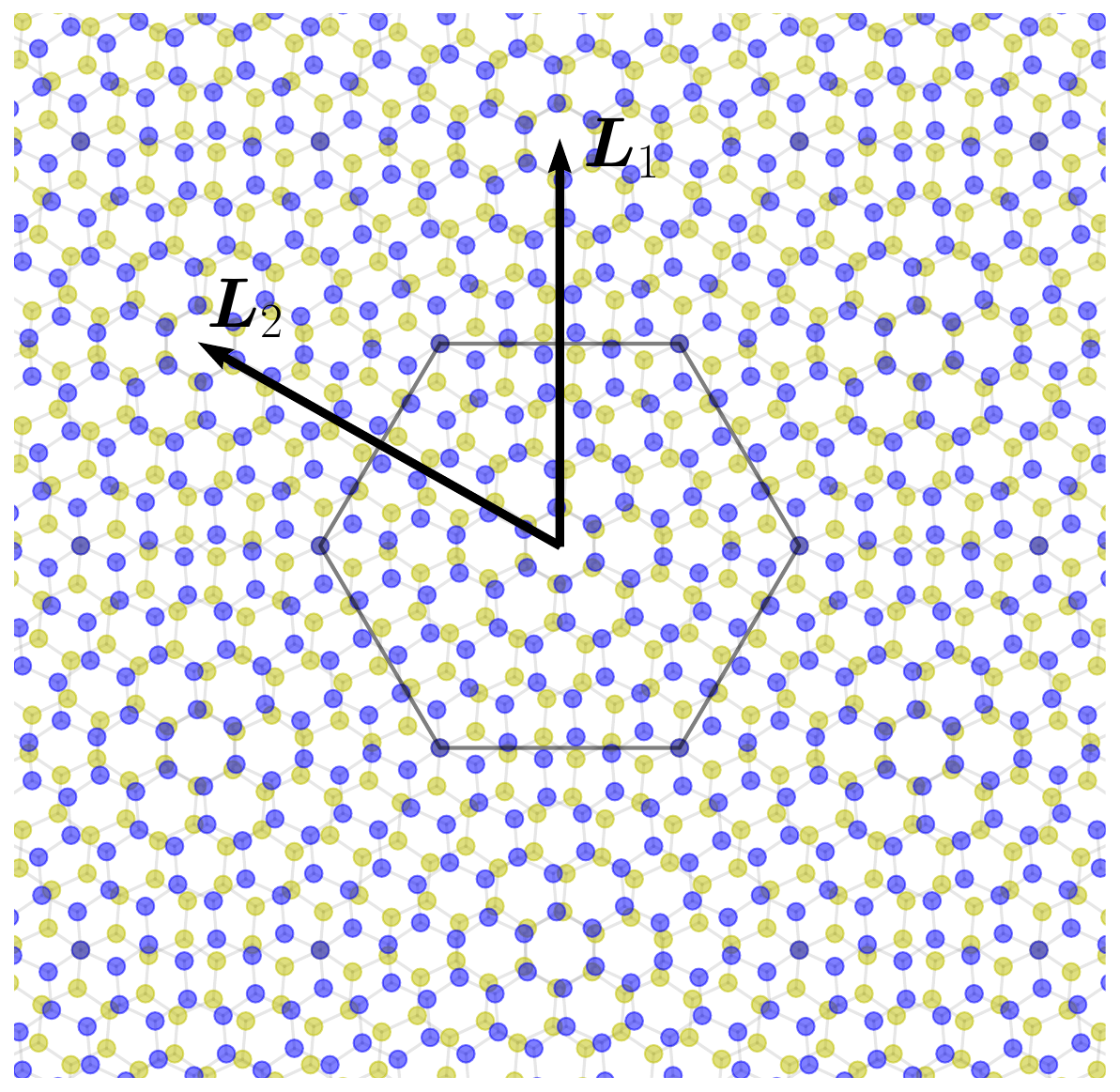}
    \caption{\textbf{Top view of twisted bilayer graphene} for a twist angle of $\theta = 9.43^\circ$. The hexagonal Wigner-Seitz cell is indicated. The center of the unit cell is locally $AA$ stacked (the two layers are on top of each other), and the corners are $AB$, or $BA$ stacked (only $A$ atoms are on top of $B$ atoms, or $\textit{vice versa}$).}
    \label{lattice}
\end{figure}
% An approximate symmetry emergent at small twist angles and low energies is the $U(1)$ valley symmetry. The eigenstates are constructed by monolayer graphene waves of low energy, at the $K$ and $K'$ valleys. The $U(1)_V$ symmetry accounts for the effective decoupling of both valleys, for the interlayer hopping and the screened Coulomb interaction is negligible at momentum transfers of order $||\boldsymbol{K} - \boldsymbol{K'}||$. This symmetry is exact in the continuum models. \\
\begin{figure}
    \centering
    \begin{subfigure}{.58\linewidth}
    \centering
    \includegraphics[width=\linewidth]{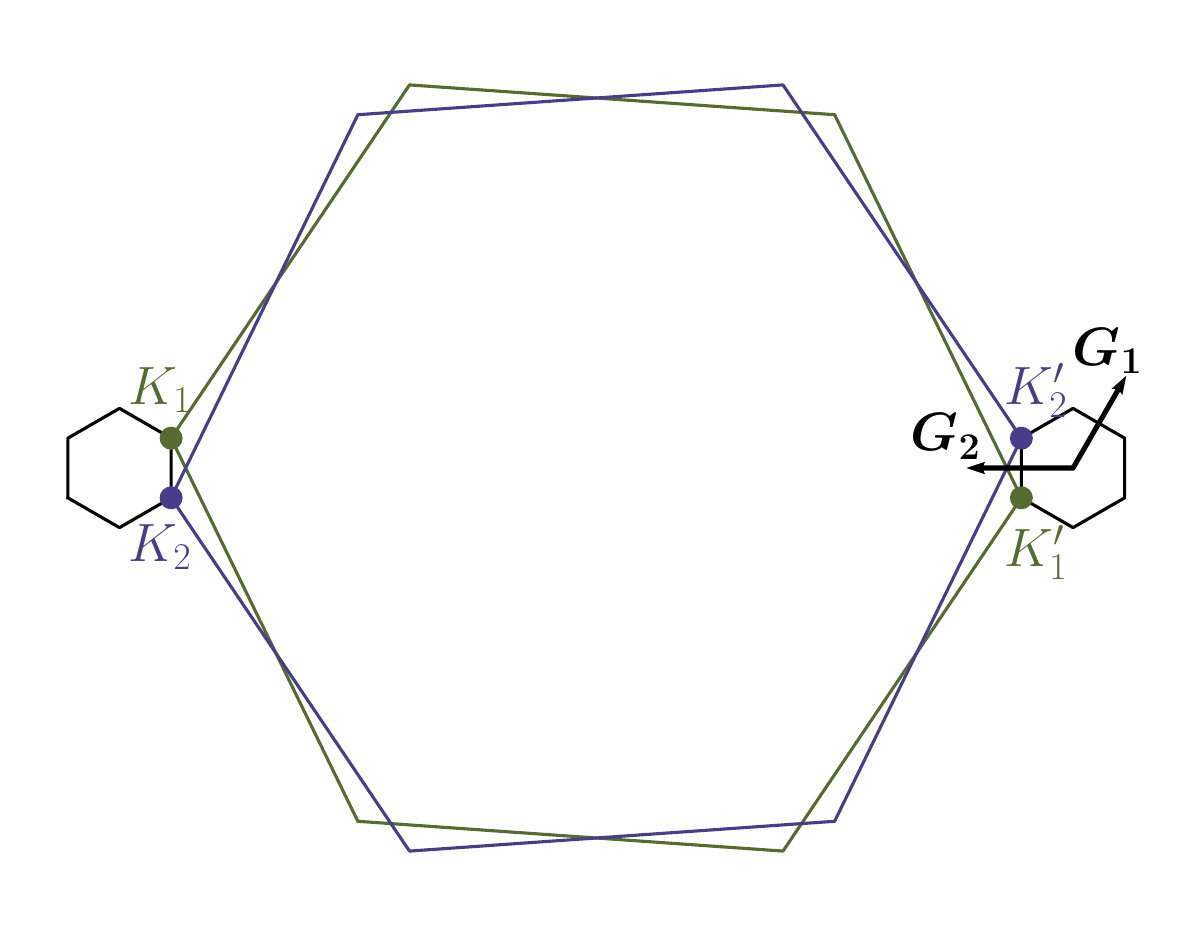}
    \end{subfigure}
    % \hfill
    \begin{subfigure}{.35\linewidth}
    \includegraphics[width=.99\linewidth]{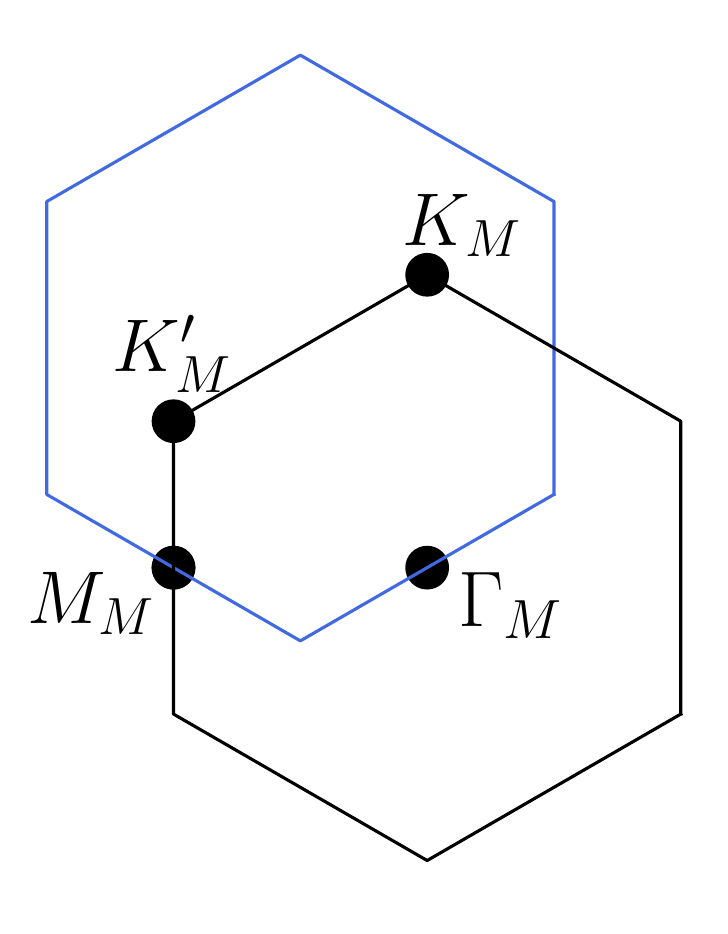}
    \end{subfigure}
    % \hfill
    \caption{\textbf{Brillouin zone of MATBG. Left:} Brillouin zone of the decoupled monolayers in purple and green, and Moiré Brillouin zone of TBG in black. Low energy TBG states belong to valley $K$ or $K'$ of each monolayer. The interlayer tunneling couples both graphene sheets, with negligible mixing between valleys. \textbf{Right:} Redefined high-symmetry momenta at one magnetic flux quantum in the periodic Landau gauge. The BZ gets shifted by $(\boldsymbol{G}_1 + \boldsymbol{G}_2)/2$ (see Appendix \ref{appa}).}
    \label{brzone}
\end{figure}

Lattice relaxation is included via in-plane distortions following the model of Ref.\cite{Koshino17}. The effect of relaxation is to enlarge the AB and BA regions and reduce the AA regions of the Moiré pattern (see Fig. \ref{lattice}), preserving all the crystallographic symmetries.\\ 
We employ the Slater-Koster parametrization of the hopping integral of Ref.\cite{Koshino12}, with a $p_z$ orbital per carbon atom and spin, giving the Hamiltonian 
\begin{align}
    H_0 = \sum_{\boldsymbol{r_i},\boldsymbol{r_j},s} t(\boldsymbol{r_i}-\boldsymbol{r_j}) c^{\dagger}_{\boldsymbol{i}s}c_{\boldsymbol{i}s}, %\nonumber \\
    \end{align}
$c^\dagger_{\boldsymbol{i},s}$ being the creation operator of an electron with spin $s$ at position $\boldsymbol{r_i}$. The hopping integral is decomposed into $\sigma$ and $\pi$-bond hoppings,
\begin{align}
% \end{equation}
% \begin{equation}
    t(\boldsymbol{r}) = - &V_{pp\pi}(r) \Bigg(1 - \bigg(\frac{\boldsymbol{r}\cdot \boldsymbol{\hat{z}}}{ r}\bigg)^2\Bigg) + V_{pp\sigma}(r) \bigg(\frac{\boldsymbol{r}\cdot \boldsymbol{\hat{z}}}{ r}\bigg)^2, \nonumber \\
    &V_{pp\pi}(r) = V_{pp \pi}^0 e^{-(r - a_0)/r_0}, \nonumber \\ 
    &V_{pp\sigma}(r) = V_{pp \sigma}^0 e^{-(r - d_0)/r_0}, 
\end{align}
with the parameters $V^0_{ppp\pi} = 2.7$ eV,  $V^0_{pp\sigma} = 0.48$ eV and $r_0 = 0.0453 $ nm. 

The Coulomb interaction is implemented by the double-gated potential
\begin{align}
    V = \frac{1}{2}\sum_{\boldsymbol{r_i},\boldsymbol{r_j} s_i s_j} V(\boldsymbol{r_i}-\boldsymbol{r_j}) :c^\dagger_{\boldsymbol{i},s_i} c_{\boldsymbol{i}, s_i} c^\dagger_{\boldsymbol{j},s_j} c_{\boldsymbol{j}, s_j}:, \nonumber \\
    V(\boldsymbol{r_i}-\boldsymbol{r_j}) = \frac{e^2}{4\pi \epsilon_0 \epsilon}\sum_{n} \frac{(-1)^n}{||\boldsymbol{r_i} - \boldsymbol{r_j} + n\xi \boldsymbol{\Hat{z}}||},
    \label{potential}
 \end{align}
which applies for the experimental setups where two metallic plates are placed at $z=\pm \xi/2$. Unless stated otherwise, we set $\xi= 10$ nm. The dielectric constant $\epsilon$ accounts for the screening due to the substrate and internal screening due to the electrons. The interaction is normal ordered\cite{giuliani_vignale_2005} with respect to the ground state of two decoupled graphene layers at charge neutrality. This choice of normal ordering is also called graphene subtraction scheme\cite{xie20,lin23}. Under magnetic field, we do not include the Zeeman shift when calculating the graphene state, so that the spin imbalances come entirely from the flat band physics. The on-site Hubbard term is also considered
 \begin{align}
     H_U = U \sum_{\boldsymbol{r_i}} :c^\dagger_{\boldsymbol{i}\uparrow} c_{\boldsymbol{i}\uparrow} c^\dagger_{\boldsymbol{i}\downarrow} c_{\boldsymbol{i}\downarrow}:.
     \label{hubbard}
 \end{align}
 
 The total Hamiltonian is then $H = H_0 + V + H_U$.

\subsection*{Minimal coupling to an external magnetic field}

At nonzero magnetic field, the Peierls' substitution\cite{Luttinger51} adds a phase to the hopping elements,
\begin{align}
    t(\boldsymbol{r_i}-\boldsymbol{r_j}) \to  t(\boldsymbol{r_i}-\boldsymbol{r_j}) e^{i\theta_ {\boldsymbol{i},\boldsymbol{j}}}, \nonumber \\
    \theta_{\boldsymbol{i},\boldsymbol{j}} = \frac{2\pi}{\Phi_0} \int_{\boldsymbol{r_i}\to \boldsymbol{r_j}} \boldsymbol{A}(\boldsymbol{r'}) \cdot d\boldsymbol{r'},
\end{align}
where $\Phi_0 = h/e$ is the quantum of magnetic flux, and the line integral goes from $\boldsymbol{r_i}$ to $\boldsymbol{r_j}$ in a straight line if the orbitals are well localized\cite{Biao20}. \\
In the presence of magnetic flux, the translation operators pick up an Aharonov-Bohm phase. They act on the single-particle states as\cite{Herzog20}
\begin{align}
    \tilde{T}_1 &= \sum_{\boldsymbol{r_i}} e^{-2\pi i \xi_{2\boldsymbol{i}}\phi - i\theta_{\boldsymbol{i},\boldsymbol{i}+\boldsymbol{L_1}}} c^\dagger_{\boldsymbol{i}+\boldsymbol{L}_1}c_{\boldsymbol{i}}, \nonumber \\
    \tilde{T}_2 &= \sum_{\boldsymbol{r_i}} e^{2\pi i \xi_{1\boldsymbol{i}}\phi - i\theta_{\boldsymbol{i},\boldsymbol{i}+\boldsymbol{L_2}}} c^\dagger_{\boldsymbol{i}+\boldsymbol{L_2}}c_{\boldsymbol{i}}.
\end{align}
$\xi_{\boldsymbol{i}1}$ and $\xi_{\boldsymbol{i}2}$ are defined by $\boldsymbol{r_i} = \xi_{\boldsymbol{i}1} \boldsymbol{L_1} + \xi_{\boldsymbol{i}2} \boldsymbol{L_2}$, and  $\phi = \Phi / \Phi_0 = BA_M/\Phi_0$ is the flux per moiré unit cell in units of $\Phi_0$.

It can be shown that $[\mathcal{H}, \tilde{T_1}] = [\mathcal{H}, \tilde{T_2}] = 0$ and $\tilde{T}_1\tilde{T}_2 = e^{-2\pi i \phi }\tilde{T}_2 \tilde{T}_1$, so the translational symmetries are broken in general. However, if $\phi$ is a rational number $p/q$ one can choose the set of commuting operators $\{ \tilde{T}_1, \tilde{T}_2^q\}$, or $\{\tilde{T}_1^q, \tilde{T}_2\}$ and diagonalize them simultaneously with the Hamiltonian. Translational symmetry is then recovered at rational fluxes with a unit cell that is $q$ times larger than at zero flux, and the Bloch waves are generalized to magnetic waves having good $\tilde{T}_1$ and $\tilde{T}_2^q$ quantum numbers.

In the periodic Landau gauge\cite{Cuniberti07}
\begin{align}
    \boldsymbol{A}&(\boldsymbol{r}) = \frac{\Phi}{2\pi}\Bigg(\xi_1 \boldsymbol{G}_2 - 2\pi \boldsymbol{\nabla}\big(\xi_2\left \lfloor{\xi_1 + \epsilon}\right \rfloor\big)  \Bigg) \nonumber \\
    &= \frac{\Phi}{2\pi} \Bigg(-\xi_2 \sum_{n} \delta(\xi_1 - n + \epsilon)\boldsymbol{G}_1 + (\xi_1 - \left \lfloor{\xi_1 + \epsilon}\right \rfloor \boldsymbol{G}_2\Bigg), 
\end{align}
the phases of the translation operators $\tilde{T}_2^q,$ $\tilde{T}_1$ cancel and the Bloch waves have the same form as in zero flux ($\lfloor ... \rfloor$ is the floor function). The infinitesimal $\epsilon$ prevents ambiguities if some atoms lie at integer values of $\xi_1$. The momentum $\boldsymbol{k}$ takes the possible values in the magnetic Brillouin zone of the dual lattice with lattice vectors $\boldsymbol{G}_1$ and $\boldsymbol{G}_2/q$.
% The momentum takes the possible values $\boldsymbol{k} = k_1 \boldsymbol{G}_1 + k_2 \boldsymbol{G}_2/q$, with $k_1 = n_1/N_1$, $k_2=qn_2/N_2$, with $n1$ $(n2) = 0,1,...,N_1-1$ $(N_2/q-1)$ and $N_1$ $(N_2)$ the number of unit cells in the $\boldsymbol{L_1}$ ($\boldsymbol{L_2}$) direction in the sample. \\
Under magnetic flux, time reversal $\mathcal{T}$ and rotations $C_{2y}$, $C_{2x}$ reverse the sign of the external field, but the rotations around the $z$ axis are preserved\cite{Herzog22}.

Besides orbital effects, the Zeeman energy $- g\mu_B B s_z/\hbar$ ($g=2$ is the gyromagnetic ratio of the electron and $\mu_B$ the Bohr magneton) is also taken into account. For $26.5$ T it amounts to $\pm 1.535$ meV.

\begin{figure}[t!]
    \noindent\hspace*{.74cm} $\boldsymbol{\Phi=0}$ \hspace{2.96cm} $\boldsymbol{\Phi = \Phi_0}$  \\
    \centering
    \includegraphics[width=.502\linewidth]{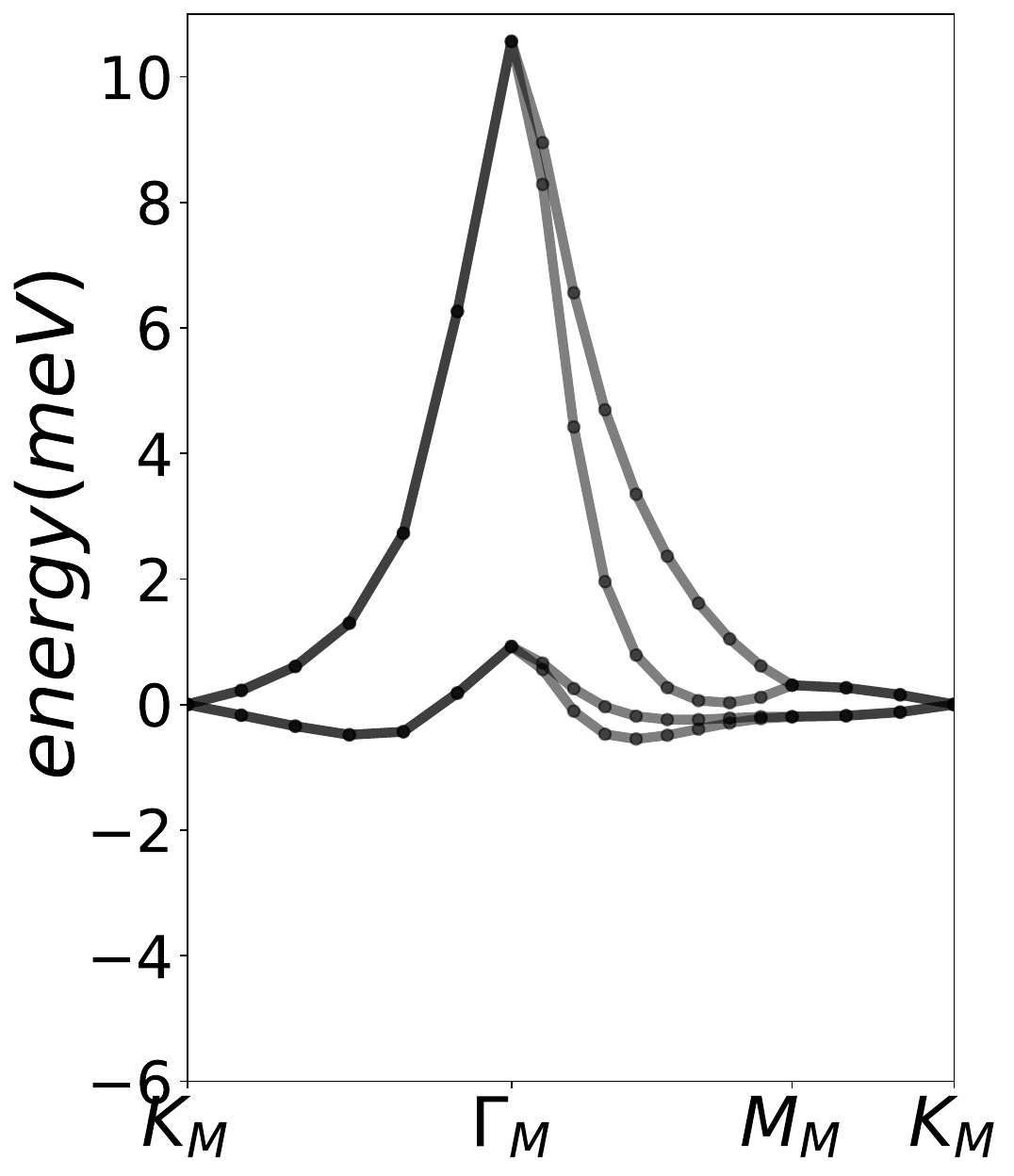}
    \includegraphics[width=.466\linewidth]{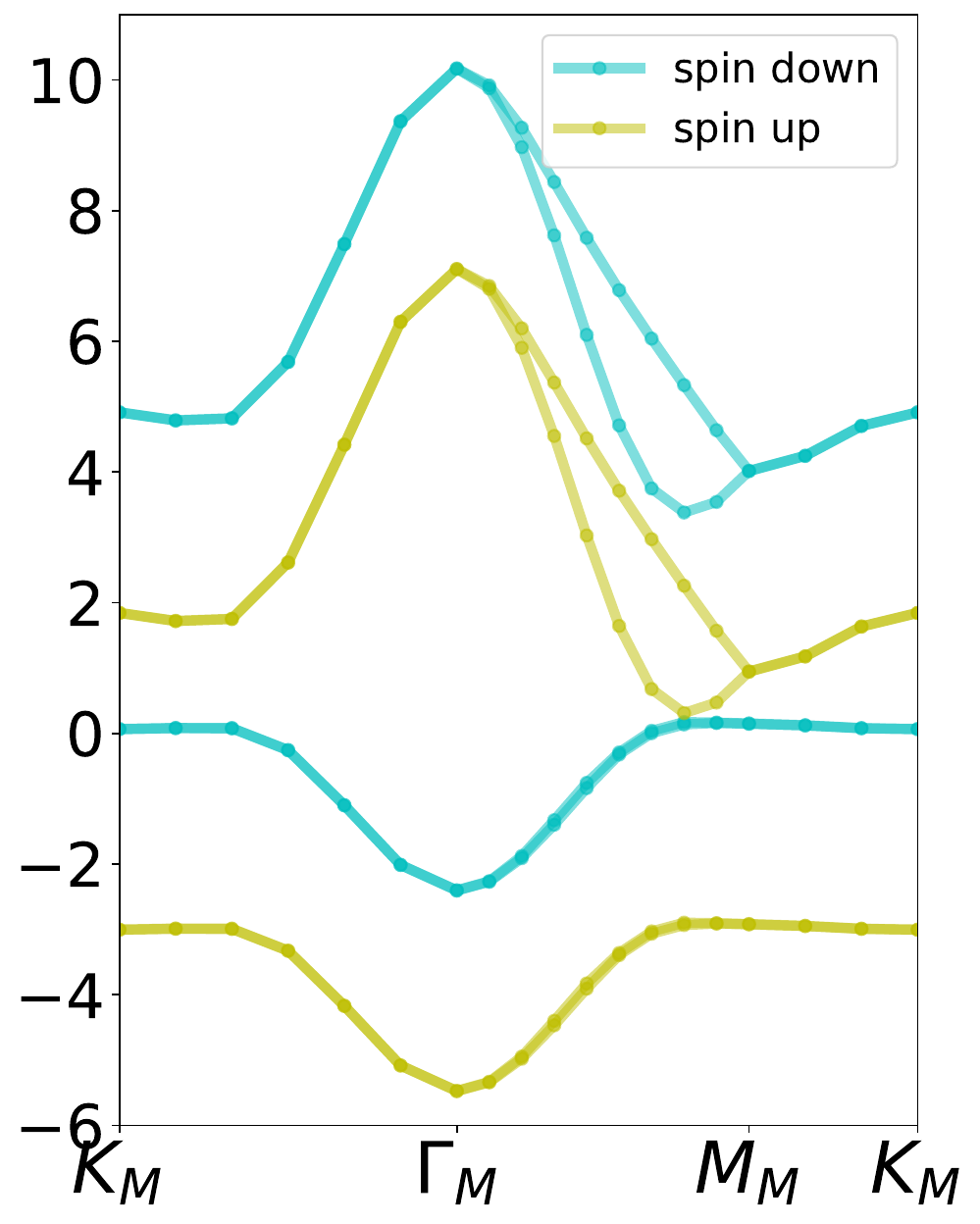}
    % prop 1.077
    \caption{\textbf{Band structure of MATBG ($\boldsymbol{\theta = 1.05^\circ}$)} along the high symmetry $K_M \Gamma_M M_M K_M$ line. \textbf{Left:} flat bands at zero field. \textbf{Right:} flat bands at $B=26.5$ T. The Zeeman energy produces a splitting of $\sim 3$ meV.}
    \label{bands}
\end{figure}
\begin{figure}[t]
    % \noindent\hspace*{.8cm} $\boldsymbol{\Phi=0}$ \hspace{3.0cm} $\boldsymbol{\Phi = \Phi_0}$  \\
    \centering
    \noindent\hspace*{-3.8cm}\textbf{a)} \hspace{3.85cm} \textbf{b)}  \\
    \includegraphics[width=.49\linewidth]{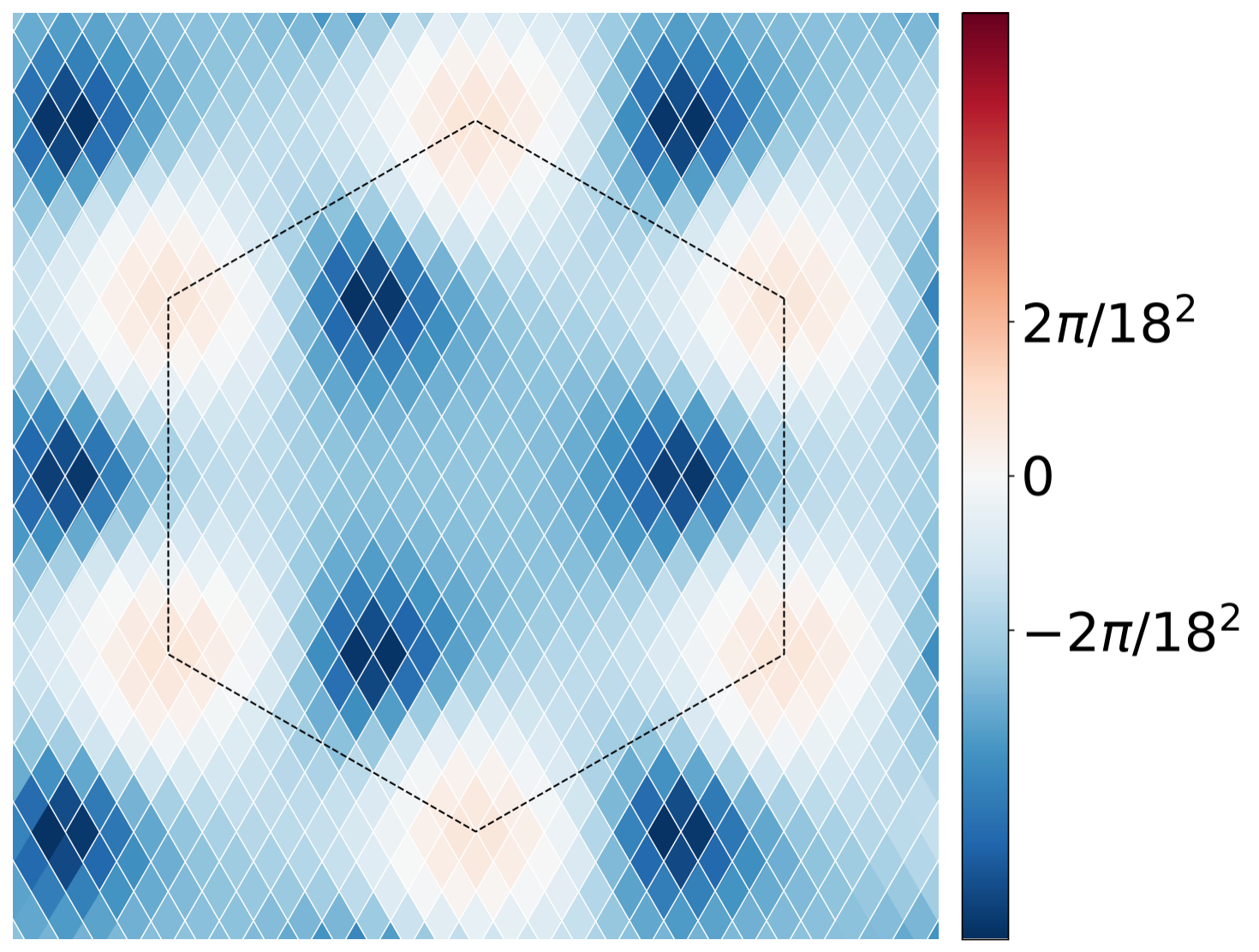}
    \includegraphics[width=.49\linewidth]{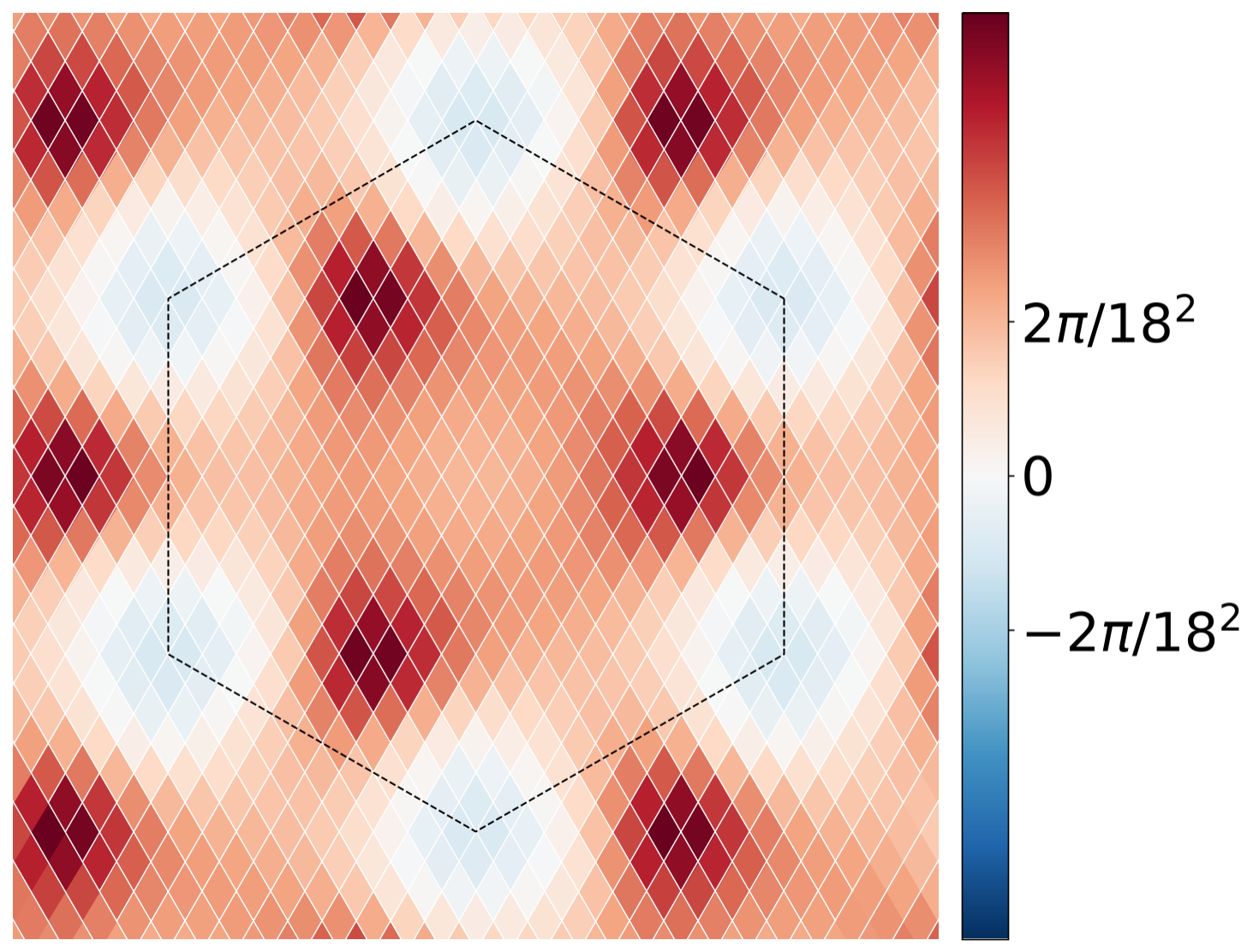}
    \caption{\textbf{Integrated Berry curvature at $\boldsymbol{B = 26.5}$ T.} We calculate the integrated Berry curvature on a $18 \times 18$ grid. \textbf{a)} Valley $K$ valence band (Chern number $-1$). \textbf{b)} Valley $K$ conduction band (Chern number $+1$). $C_{2z}$ symmetry interchanges the valleys of the conduction/valence bands (see the text for the definition of the valley charge).}
    \label{berry}
\end{figure}

\subsection*{Results of the non interacting model}

In Fig. \ref{bands} we plot the spectrum of MATBG for $n_\theta=31$. At the charge neutrality point there are 8 quasi flat bands ($2$ layers $\times$ $2$ valleys $\times$ $2$ spin polarizations) with a bandwidth of around $10$ meV and well separated from the 'remote' bands. The filling is parametrized by $\nu \in (-4,4)$, being $\nu=0$ the neutrality point and $\nu=+4$ $(-4)$ the filled (empty) flat bands. When the external magnetic field is $26.5$ T, we have $\Phi = \Phi_0$ and the flat bands are reentrant. The symmetry $C_{2z}\mathcal{T}$ that preserves the gapless Dirac points in zero flux\cite{junyeong19} is broken, and the Dirac cones are gapped with a Chern number of $-2$\cite{Guan22}. In the flat bands the kinetic energy is small and the interacting physics dominates, giving rise to the rich phase diagram of MATBG. 

In Fig. \ref{berry} we plot the Berry curvatures of the bands at $26.5$ T. In the discretized Brillouin zone, we compute the integrated Berry curvature on the parallelograms defined by the discretization, following Ref. \cite{fukui05}. We see an emergent symmetry relating the Berry curvatures of valence and conduction bands. 

% In the periodic Landau gauge the mirrors and rotations have a non trivial Aharonov-Bohm phase, unlike the translations. The consequence at $p=q=1$ is that the rotations and reflections leave fixed $(\boldsymbol{G}_1 + \boldsymbol{G}_2)/2$ instead of $\Gamma = \boldsymbol{0}$ (see Appendix), thus redefining the high symmetry points in the Brillouin zone, shifting  them by $(\boldsymbol{G}_1 + \boldsymbol{G}_2)/2$.

\section{The U(4) symmetry}
A starting point for the understanding the physics of MATBG is the projected limit, in which the Fermi sea of the remote bands is 'frozen' and the scattering is restricted to states within the flat bands. The correlated insulators  are 'generalized ferromagnets' that spontaneously break a $U(4)$ (or a larger $U(4)\times U(4)$) global symmetry. For a detailed discussion on this symmetry, we refer the reader to Refs. \cite{ledwith20, Bultinck20, bernevig321, herzog22_2}. Here, we describe the most 'physical' subgroups. The full group is generated from these. 
% \begin{widetext}
\begin{itemize}
    \item $U(1)_{v}$. The non interacting eigenstates belong to valley $K$ (valley charge +1) or $K'$ (valley charge -1) of each of the graphene monolayers, see Fig. \ref{brzone}. Charge non conserving terms of the screened Coulomb interaction are suppressed, hence $U(1)_{v}$ is a symmetry of the interacting system. We write $\eta=+1(-1)$ for valley $K$($K'$). The valley charge operator is denoted by $\tau_z$, with eigenvalues $\langle \tau_z \rangle = \eta$. Electric charge $U(1)_{c}$ is also a symmetry.
    \item $SU(2)_{K} \times SU(2)_{K'}$. The exchange integral of pairs of states with different valley charges is suppressed due to the big momentum transfer, and only the Hartree term contributes to the energy. This implies a symmetry of independent spin rotations in each valley, which do not change the total density.
    % Spin rotations are a symmetry both of the kinetic energy and the Coulomb interaction at zero field.
    At nonzero perpendicular magnetic field the Zeeman effect will break the degeneracy and fix the quantization axis to the $z$ axis. 
    \item The particle-hole (p-h) symmetry. In the spirit of the continuum model we can write a generic wave function of valley $\eta$ as
    % \begin{align}
    % \hspace{.6cm}\Psi_{\eta}&(\boldsymbol{r_i}) =e^{i\eta n_\theta \boldsymbol{G_2}\cdot \boldsymbol{r_i}}\nonumber \\ &\times(\Psi_{\eta A b}(\boldsymbol{r_i}), \Psi_{ \eta B b }(\boldsymbol{r_i}), \Psi_{\eta A t}(\boldsymbol{r_i}), \Psi_{\eta B t}(\boldsymbol{r_i}))^T,
    % \label{wf}
    % \end{align}
    \begin{align}
    \hspace{-2.6cm} \Psi_{\eta}(\boldsymbol{r_i}) =e^{i\eta n_\theta \boldsymbol{G_2}\cdot \boldsymbol{r_i}}\Psi_{ \eta \sigma l}(\boldsymbol{r_i})
    \label{wf}
    \end{align}
    where $e^{i \eta n_\theta \boldsymbol{G_2}\cdot \boldsymbol{r_i}}$ is the rapidly oscillating valley phase ($n_{\theta}\boldsymbol{G}_2 \approx \boldsymbol{K}$, the $K$ point of graphene), and $\Psi_{\eta \sigma l}(\boldsymbol{r_i})$ is a smooth envelope that depends on the sublattice $\sigma$ and layer $l$ of the point $\boldsymbol{r_i}$, evaluated at that point. 
    The particle-hole operator $C_{2z}P$ (actually, in the language of Refs. \cite{bernvig2_21,herzog22_2}, the combined operator of $C_{2z}$ and the particle-hole operator $P$) is a hermitian operator that squares to the identity and interchanges the graphene valleys, the sublattice and the layer, and is trivial on the spin. It acts on the wave functions as
    % \begin{align}
    % \hspace{.9cm}C_{2z}&P\big( \Psi_\eta \big)(\boldsymbol{r_i}) 
    % =  \eta e^{-i\eta n_\theta  \boldsymbol{G_2}\cdot \boldsymbol{r_i}}e^{-i \eta \boldsymbol{G_2} \cdot \boldsymbol{r_i}} \nonumber \\ &\times (-\Psi_{\eta B t}(\boldsymbol{r_i}),-\Psi_{\eta A t}(\boldsymbol{r_i}), \Psi_{\eta B b}(\boldsymbol{r_i}), \Psi_{\eta A b}(\boldsymbol{r_i}))^T,
    % \label{c2p}
    % \end{align}
    \begin{align}
    \hspace{0.8cm}C_{2z}P\big( \Psi_\eta \big)(\boldsymbol{r_i}) 
    =  \eta s_l e^{-i\eta n_\theta  \boldsymbol{G_2}\cdot \boldsymbol{r_i}}e^{-i \eta \boldsymbol{G_2} \cdot \boldsymbol{r_i}} \Psi_{\eta \overline{\sigma} \overline{l}}(\boldsymbol{r_i}),
    \label{c2p}
    \end{align}
    where $\overline{\sigma}$ and $\overline{l}$ denote the opposite sublattice and layer to those of $\boldsymbol{r_i}$ and $s_l = +1$ if $l$ is the top layer and $-1$ if $l$ is the bottom layer. In the continuum theory, $C_{2z}P$ commutes with the Coulomb interaction and anticommutes with the kinetic energy, hence the name particle-hole operator. As such, it is the generator of a $U(1)$ subgroup in the flat limit, i.e. when the kinetic energy is negligible compared to the Coulomb energy and set to zero.
\end{itemize}

The $U(1)$ subgroup generated by $C_{z}P$ and $SU(2)_{K} \times SU(2)_{K'}\times U(1)_v \times U(1)_c \simeq U(2)_K \times U(2)_{K'}$ do not commute. They are subgroups of the ubiquitous $U(4)$ symmetry of TBG. This $U(4)$ group can be further enlarged to $U(4)\times U(4)$ if we include another $U(1)$ generated by
\begin{itemize}
    \item the sublattice operator,
    \begin{align}
         C= \sum_{\boldsymbol{i}\in A,s} c^\dagger_{\boldsymbol{i}s} c_{\boldsymbol{i}s}  - \sum_{\boldsymbol{i}\in B,s} c^\dagger_{\boldsymbol{i},s} c_{\boldsymbol{i},s}.
    \end{align} 
    This operator generates a symmetry in the so-called 'chiral' limit\cite{tarn19}, where the projected $C$ operator in the flat band manifold has eigenvalues $\pm 1$ , i.e. there exists a perfectly sublattice polarized basis of the flat bands. In the real system the polarization is around $0.6$-$0.8$\cite{vafek23}, and the symmetry is moderately broken. On the other hand, at one magnetic flux quantum the chiral limit is topologically distinct from the real system\cite{herzog22_2}.
\end{itemize}

The reader might have noticed that both the valley charge and the $C_{2z}P$ operators are emergent in the continuum theory and do not have a direct analogue in the lattice. In Appendix \ref{appb} we describe our implementation of the valley charge and $C_{2z}P$ in the lattice model.
\subsection*{The irrep basis}
% In the following, let us denote by $|u_n(\boldsymbol{k})\rangle$ the Bloch wave functions of the flat bands ($n=1,2,3,4$) at momentum $\boldsymbol{k}$. \\ In order to make manifest the $U(1)_v$ symmetry, we obtain valley-resolved states by diagonalizing the 4$\times$4 matrix $[\tau_{z}(\boldsymbol{k})]_{nn'} = \langle \boldsymbol{k} n |\tau_z| \boldsymbol{k} n' \rangle$ at each value of $\boldsymbol{k}$, with $\tau_z$ the valley charge operator (see Appendix). Of course, the kinetic eigenstates $| \boldsymbol{k} n \rangle$ have definite valley charge, but at the points of degeneracy they are mixed in general and we have to perform the diagonalization.\\
The natural basis of the flat bands in the strong coupling analysis is the so-called 'irrep' basis, with the defining property
\begin{align}
    \langle \boldsymbol{k}\eta \lambda | C_{2z}P | \boldsymbol{k}\eta'\lambda' \rangle = [\tau_y]_{\eta \eta'} [\lambda_z]_{\lambda \lambda'},
\end{align}
% $|\boldsymbol{k}\eta \e_Y \rangle = d^\dagger_{\boldsymbol{k}\eta e_Y} |0\rangle$ 
where $\lambda = \pm 1$ is the irrep number. We write $\tau_{0,x,y,z}$ and  $\lambda_{0,x,y,z}$ for the identity and Pauli matrices in valley and irrep number space, respectively. In this subsection the spin index is omitted, and we construct two identical copies of the irrep basis, one for each spin polarization.

Given the property $\{C,C_{2z}P\} = 0$, the irrep basis is equivalent to the sublattice polarized basis that diagonalizes the projection of $C$ onto the flat bands, which we denote by $\overline{C}(\boldsymbol{k})$ for a given momentum $\boldsymbol{k}$. The sublattice is labeled by $\sigma=A(+1), B(-1)$, and the identity and Pauli matrices in sublattice space by $\sigma_{0,x,y,z}$. $A(B)$ sublattice has eigenvalue $+1(-1)$ under $\sigma_z$. 

However, in the real system the particle-hole symmetry is broken, meaning that $\overline{C_{2z}P}(\boldsymbol{k})$ is not unitary, as we will see. We have to define the irrep basis in a different way.

At zero magnetic field, the sublattice basis is adiabatically connected to the irrep basis of the p-h symmetric limit. This is, if we compute the 'closest' unitary matrix to $\overline{C_{2z}P}(\boldsymbol{k})$ in the sublattice polarized basis, $[\overline{C_{2z}P}(\boldsymbol{k})]_{\eta \sigma \eta' \sigma'} = \langle \boldsymbol{k}\eta \sigma | C_{2z}P | \boldsymbol{k}\eta' \sigma' \rangle$, we get to a very good accuracy (up to a gauge choice) 
\begin{align}
    \overline{C_{2z}P}(\boldsymbol{k})\Big(\overline{C_{2z}P}(\boldsymbol{k})\overline{C_{2z}P}(\boldsymbol{k})^\dagger \Big)^{-1/2} = \sigma_y \tau_x.
\end{align}
Hence, if we identify $\lambda = \eta \sigma$ we conclude that indeed both basis are equivalent also with p-h breaking, and they can be used indistinctly. The sublattice polarized bands $|\boldsymbol{k}\eta \sigma \rangle$ have Chern numbers equal to $\eta \sigma$\cite{Liu19}.
% We also impose the gauge fixing
% \begin{align}
%     \langle \boldsymbol{k} \eta \lambda | C_{2z}\mathcal{T} | \boldsymbol{k} \eta' \lambda' \rangle = [\tau_0]_{\eta \eta'}[\lambda_x]_{\lambda \lambda'}.
%     \label{c2t}
% \end{align}
 
% Although in the real system particle-hole is broken (meaning that $\overline{C_{2z}P}(\boldsymbol{k})$ is not unitary, as we will see), the sublattice polarized basis.  

We impose the following $C_{2z}\mathcal{T}$ gauge fixing,
\begin{align}
    \langle \boldsymbol{k} \rho | C_{2z}\mathcal{T} | \boldsymbol{k} \rho' \rangle = [\sigma_x \tau_0]_{\rho \rho'},
    \label{c2t}
\end{align}
% The broken p-h symmetry in the projected system is generated by the operator
% \begin{align}
%     % S^{y0} =& \sum_{\boldsymbol{k}}\sum_{\{\eta\} \{e_Y\}\{s \}}  d^\dagger_{\boldsymbol{k}\eta e_Y s}[\tau_y]_{\eta \eta'} [\lambda_z]_{e_Y e_Y'} [s_0]_{ss'} d_{\boldsymbol{k}\eta' e_Y' s'} \nonumber \\
%     % =& - \sum_{\boldsymbol{k}}\sum_{\{\eta\} \{\sigma\}\{s\}}  d^\dagger_{\boldsymbol{k}\eta \sigma s}[\sigma_y]_{\sigma \sigma'} [\tau_y]_{\eta \eta'} [s_0]_{ss'}d_{\boldsymbol{k}\eta' \sigma' s'}.
%     % S^{y0} =& \sum_{\boldsymbol{k}}\sum_{\rho\rho'}  d^\dagger_{\boldsymbol{k}\rho } [\tau_y \lambda_z s_0]_{\rho \rho'}  d_{\boldsymbol{k} \rho'} 
%     % \nonumber \\
%     S^{y0} = \sum_{\boldsymbol{k}}\sum_{\rho \rho'}  d^\dagger_{\boldsymbol{k}\rho}[\sigma_y \tau_x s_0]_{\rho\rho'} d_{\boldsymbol{k}\rho'}.
% \end{align}
% Here $\rho$ is the multi-index for valley, sublattice and spin, $s_{0,x,y,z}$ are the Pauli matrices for spin, and $d^\dagger_{\boldsymbol{k}\rho}$ is the creation operator of state $|\boldsymbol{k} \rho \rangle$. \\
with $\rho$ the multi-index for valley and sublattice. We also constrain the representation of $C_{2z}$ and $\mathcal{T}$,
\begin{align}
    \langle [\boldsymbol{-k}] \rho | C_{2z} | \boldsymbol{k} \rho'\rangle=& [\sigma_x\tau_x ]_{\rho \rho'} \ \text{if} \ \boldsymbol{k} \neq [-\boldsymbol{k}] %\nonumber \\
    \ \text{or} \ \boldsymbol{k} = \boldsymbol{\Gamma} \nonumber \\
     =& -[\sigma_x \tau_x]_{\rho \rho'} \ \text{otherwise},\nonumber \\
     \langle [\boldsymbol{-k}] \rho | \mathcal{T} | \boldsymbol{k} \rho'\rangle=& [\sigma_0\tau_x]_{\rho \rho'} \ \text{if} \ \boldsymbol{k} \neq [-\boldsymbol{k}] 
    \ \text{or} \ \boldsymbol{k} = \boldsymbol{\Gamma} \nonumber \\
    =& -[\sigma_0\tau_x]_{\rho \rho'} \ \text{otherwise},
    \label{c2}
\end{align}
 $[\boldsymbol{k}]$ being the momentum  equivalent to $\boldsymbol{k}$ inside the Brillouin zone. The additional signs at parity invariant momenta are due to a topological obstruction\cite{bernevig321}. 
 
Notice that the wave functions are not completely defined, and there exists a phase ambiguity redefining the states as
\begin{align}
    | \boldsymbol{k}\rho \rangle &\longrightarrow [e^{i\alpha(\boldsymbol{k})\sigma_z\tau_z}]_{\rho \rho'} | \boldsymbol{k} \rho' \rangle, \nonumber \\
    \alpha(\boldsymbol{k}) &= \alpha([-\boldsymbol{k}]).
    \label{gaugep0}
\end{align}

At one magnetic flux quantum the irrep basis is defined by
\begin{align}
    % \langle \boldsymbol{k}\eta \lambda | C_{2z}P | \boldsymbol{k}\eta'\lambda' \rangle = [\tau_x]_{\eta \eta'} [\lambda_0]_{\lambda  \lambda'}.
    \overline{C_{2z}P}(\boldsymbol{k})\Big(\overline{C_{2z}P}(\boldsymbol{k})\overline{C_{2z}P}(\boldsymbol{k})^\dagger \Big)^{-1/2} = \tau_x \lambda_0.
    \label{c2pp1}
\end{align}

In the continuum model, the chiral limit is topologically distinct from the real system at one flux quantum\cite{herzog22_2}, a phenomenon that is reflected in the tight-binding model. In this case, the irrep basis is not maximally polarized, hence the sublattice polarization and irrep character cannot be simultaneously manifested.

We further fix the representation of $C_{2z}$ to
\begin{align}
\langle [\boldsymbol{-k}] \rho | C_{2z} | \boldsymbol{k} \rho'\rangle= -[\tau_y \lambda_y]_{\rho \rho'},
\end{align}
with $\rho$ now the multi-index for valley and irrep. 

Because $C_{2z}\mathcal{T}$ is broken due to the magnetic field, there does not exit a remaining local symmetry to further constrain the basis. As a consequence, in flux the irrep basis is only defined up to arbitrary transformations $V(\boldsymbol{k})$ in both valleys
\begin{align}
    &| \boldsymbol{k}\eta \lambda \rangle \longrightarrow [V(\boldsymbol{k})]_{\lambda \lambda'} | \boldsymbol{k} \eta \lambda' \rangle, \nonumber \\
    &V^\dagger([-\boldsymbol{k}])\lambda_y V(\boldsymbol{k}) = \lambda_y.
    \label{irrepv}
\end{align}

\section{EXPLICIT BREAKING OF $\boldsymbol{U(4)}$}
The $U(4)$ (or $U(4) \times U(4)$) symmetry is only approximate, and is broken in the atomistic model down to the physical $SU(2)$ of spin rotations. In this section we study the strength of such symmetry breaking.
\begin{figure}[t]
    \centering \Large{$\Phi=0$} \\
    % \begin{subfigure}{.49\linewidth}
    \centering
    \includegraphics[width=.45\linewidth]{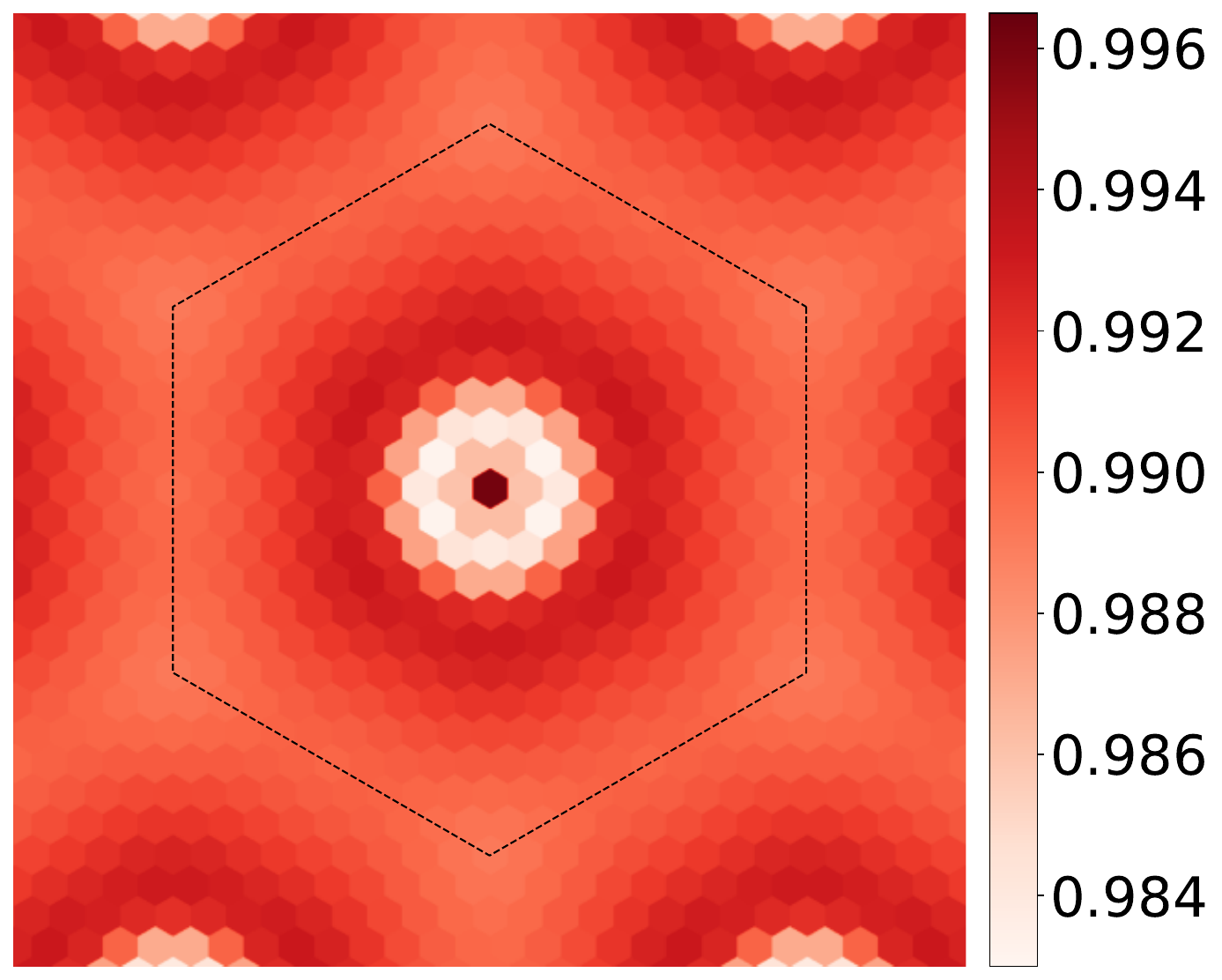}
    % \end{subfigure}
    % \hfill
    % \begin{subfigure}{.49\linewidth}
    \centering
    \includegraphics[width=.45\linewidth]{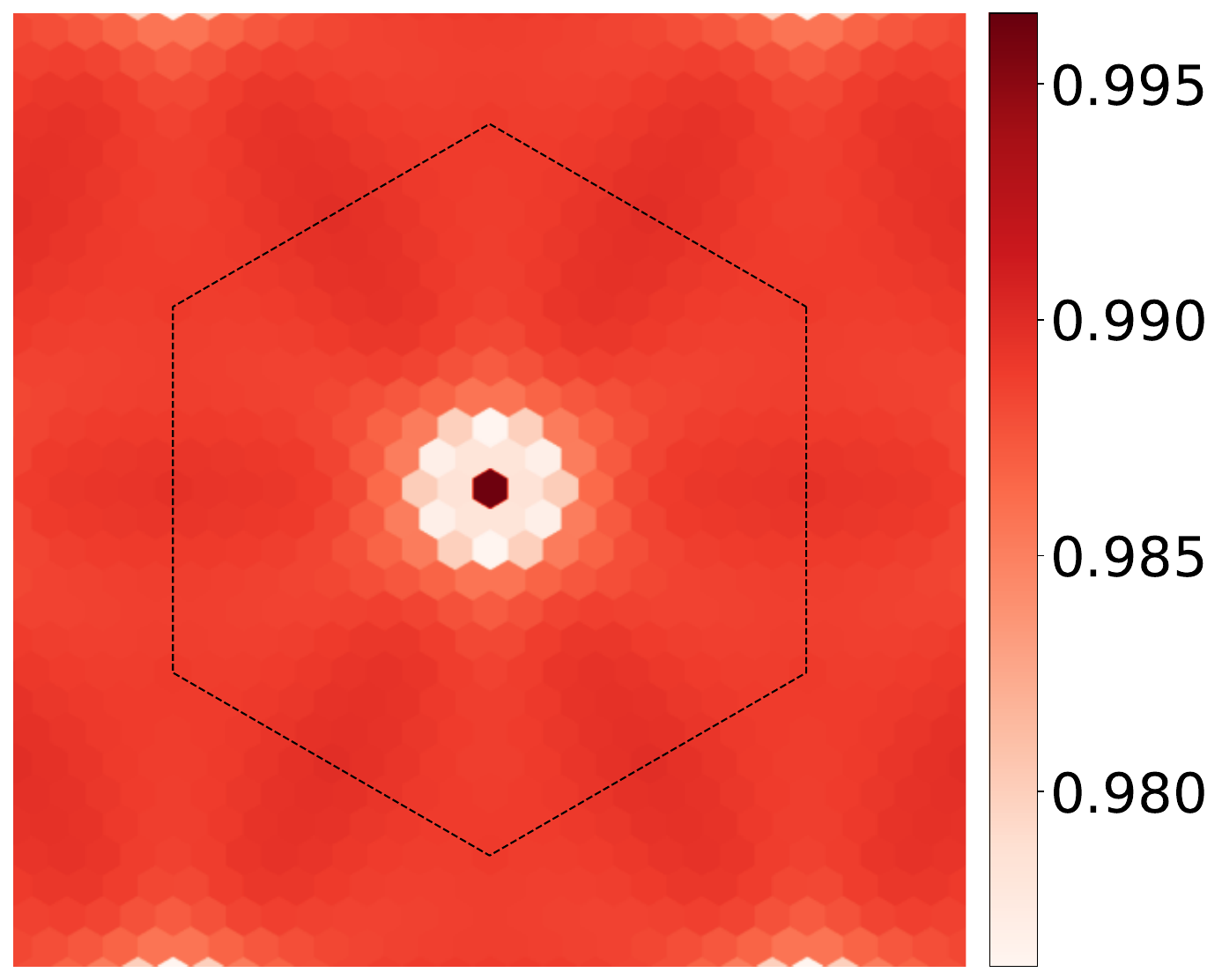}
    % \end{subfigure}
    % \caption{\textbf{Singular values of} $\boldsymbol{\overline{C_{2z}P}}$ \textbf{at} $\boldsymbol{\Phi = 0}$.}
    % \label{c2pphi0}
    % $\Large{\boldsymbol{\Phi} \textbf{=} \boldsymbol{\Phi_0}}$ \\
    \Large{$\Phi = \Phi_0$}\\
    % \begin{subfigure}{.49\linewidth}
    \centering
    \includegraphics[width=.45\linewidth]{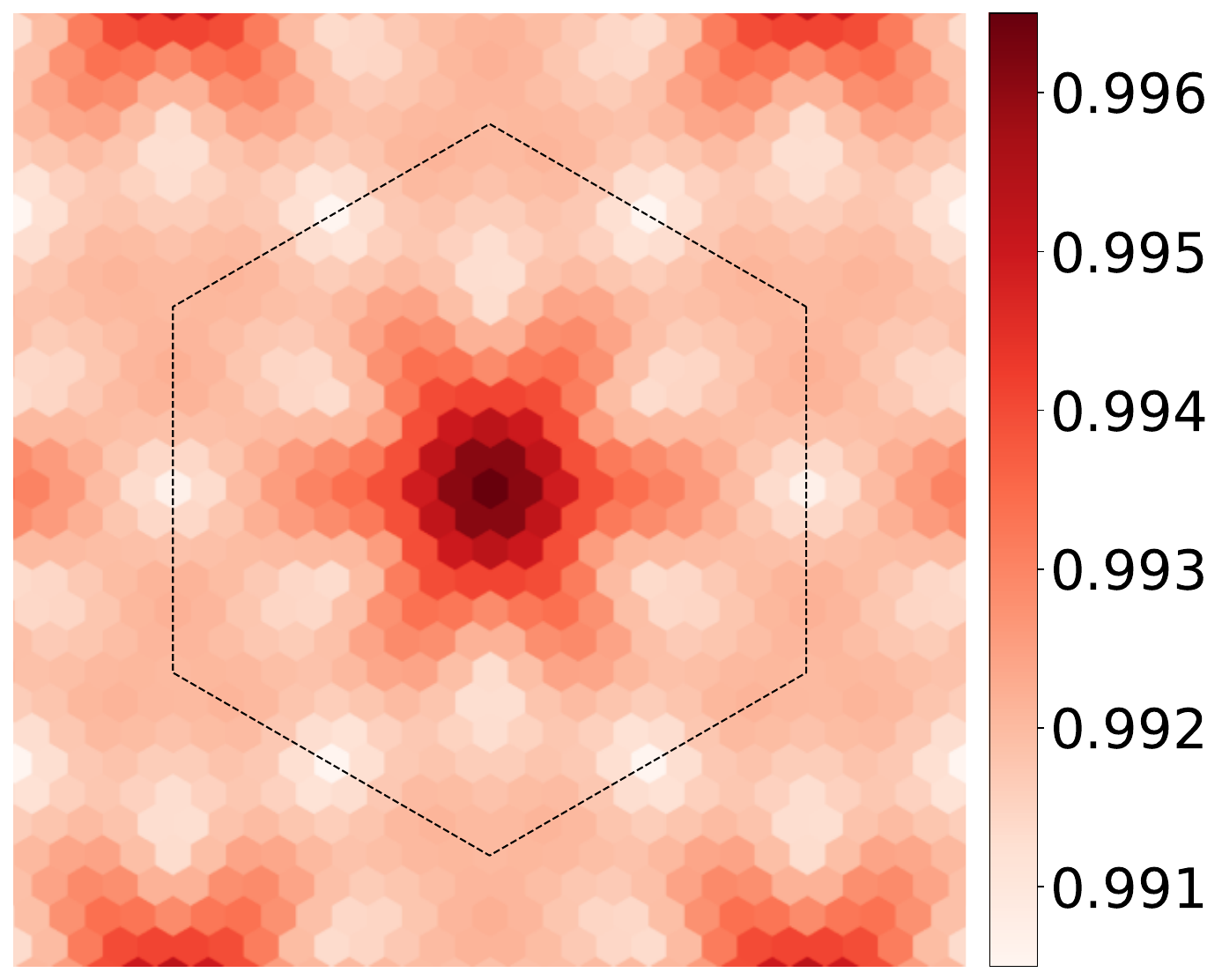}
    % \end{subfigure}
    % \hfill
    % \begin{subfigure}{.49\linewidth}
    \centering
    \includegraphics[width=.45\linewidth]{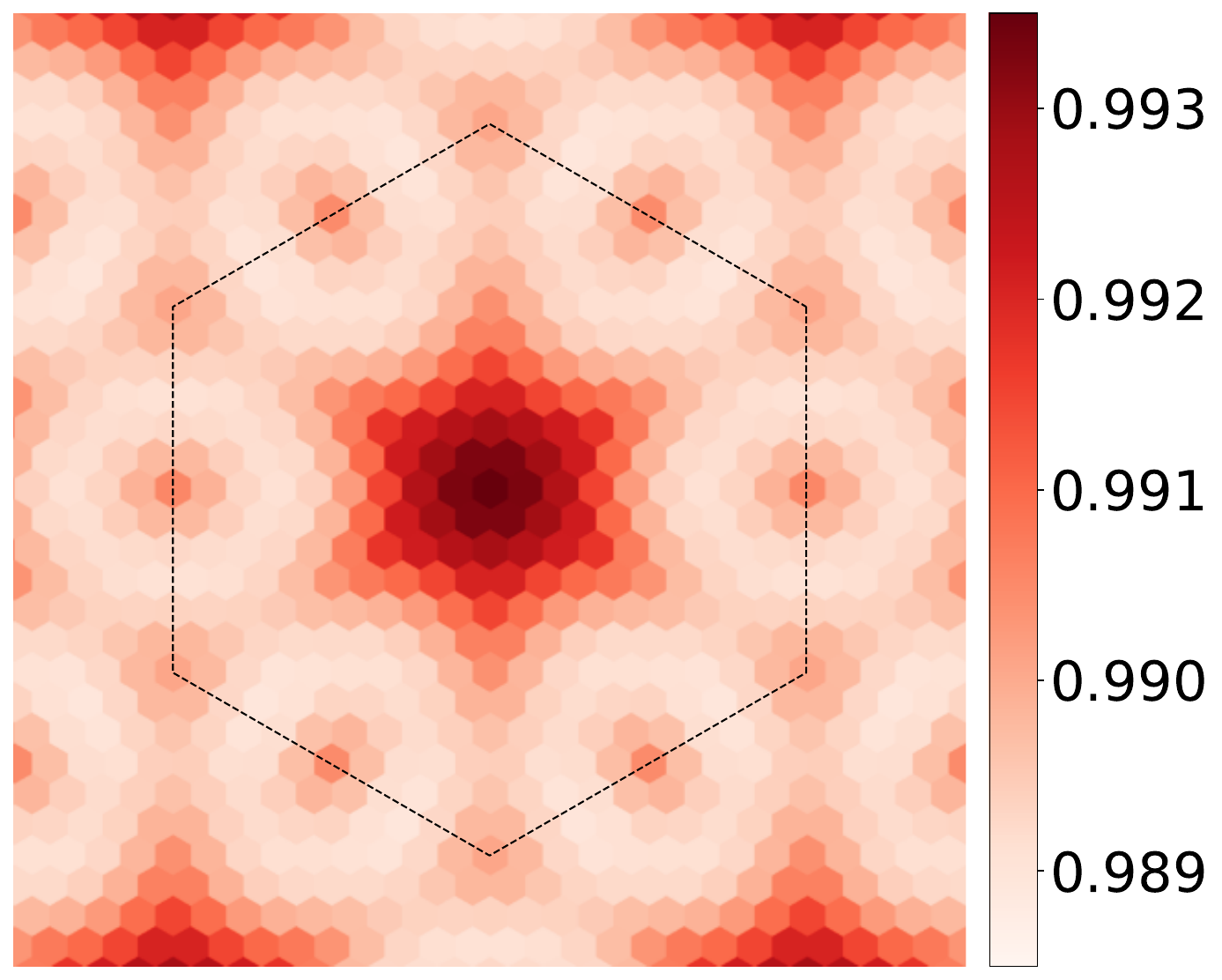}
    % \end{subfigure}
    \caption{\textbf{Singular values of} $\boldsymbol{\overline{C_{2z}P}}$ \textbf{at} $\boldsymbol{\Phi = 0}$ \textbf{and} $\boldsymbol{\Phi_0}$. They are close to $1$, signalling a small particle-hole breaking in the flat bands. For $\Phi=0$ they were previously computed in Ref. \cite{vafek23}}
    \label{c2psv}
\end{figure}

\subsection*{$\boldsymbol{SU(2)_K \times SU(2)_{K'}}$ and p-h breaking}

Assuming that deviations from $[V,C_{2z}P] = 0$ are negligible also in the lattice model, the  projected flat limit enjoys the symmetry generated by the p-h operator if the projected $C_{2z}P$ matrix is unitary. Hence, p-h breaking can be quantified by the singular values (s.v.) of $\overline{C_{2z}P}(\boldsymbol{k})$. Deviations from 1 of the s.v. measure the particle-hole asymmetry of the Hilbert space.

The properties $\overline{C_{2z}P}(\boldsymbol{k})^\dagger = \overline{C_{2z}P}(\boldsymbol{k})$ and $\{\tau_z, C_{2z}P\}=0$ force the s.v. to be degenerate in pairs. We plot the largest and smallest s.v. in Fig. \ref{c2psv} at zero and one flux quantum. The deviations are small and similar in both cases, with mean values of around $0.99$ and a minimal value of about $0.98$ at zero field.
\begin{figure}[b]
\includegraphics[width=.7\linewidth]{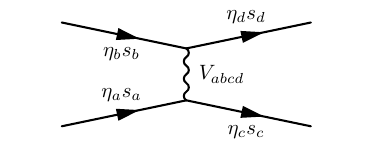}
% \begin{fmffile}{complex-a}
% \begin{fmfgraph*}(150,50)
%     \fmfleft{i1,12}
%     \fmfright{o1,o2}
%     \fmf{fermion,label=$\eta_a s_a$}{i1,w1}
%     \fmf{fermion,label=$\eta_b s_b$}{12,w2}
%     \fmf{fermion,label=$\eta_c s_c$}{w1,o1}
%     \fmf{fermion,label=$\eta_d s_d$}{w2,o2}
%     \fmf{photon,label=$V_{abcd}$}{w1,w2}
% \end{fmfgraph*}
% \end{fmffile}
\caption{\textbf{Coulomb scattering} between electrons with valley charges $\eta_{a,b,c,d}$ and spin projections $s_{a,b,c,d}$. In the continuum approximation $V_{abcd} \sim V(\boldsymbol{q})$, with $\boldsymbol{q} \sim (\eta_{a}-\eta_{c})\boldsymbol{K} \sim  (\eta_{d}-\eta_{b})\boldsymbol{K}$. Lattice-scale effects distort this approximation.}
\label{feynd}
\end{figure}

Regarding $U(1)_v$ and $SU(2)_{K} \times SU(2)_{K'}$, consider the matrix elements of the Coulomb interaction depicted in Fig. \ref{feynd},
\begin{align}
    V_{abcd} = \langle a b | V | c d \rangle,
\end{align}
where the states $|i\rangle$ have valley charge $\eta_{i}$ and spin $s_{i}$, $i=a,b,c,d$. In the continuum theory, the matrix element is to a first approximation equal to the Fourier transform of $V(\boldsymbol{r})$ at momentum $\boldsymbol{q} \sim (\eta_{a}-\eta_{c})\boldsymbol{K} \sim  (\eta_{d}-\eta_{b})\boldsymbol{K}$, with $\boldsymbol{K}$ the corner of the graphene BZ, the midpoint of $\boldsymbol{K_1}$ and $\boldsymbol{K_2}$ in Fig. \ref{brzone}. In turn, $V_{abcd}$ is nonzero only if  $\eta_a = \eta_c$ and $\eta_b = \eta_d$ given that $|| \boldsymbol{K} || \sim a^{-1} \gg \xi^{-1}$, where $\xi$ is the gate distance in Eq. \ref{potential}. The valley charge is conserved, $\eta_a + \eta_b = \eta_c + \eta_d$, so $U(1)_v$ is a symmetry. Moreover, the structure of the matrix elements $V_{abcd} \propto \delta_{\eta_a \eta_c} \delta_{\eta_b \eta_d} \delta_{s_a s_c} \delta_{s_b s_d}$ exhibits the symmetry $SU(2)_K \times SU(2)_{K'}$ consisting of independent spin rotations for each valley sector.
% Different valleys only interact via the Hartree potential, so the group $SU(2)_K \times SU(2)_{K'}$ of independent spin rotations in each valley is also a symmetry.

For the atomistic model, we show in Appendix \ref{appc} that $U(1)_v$ is preserved but $SU(2)_K \times SU(2)_{K'}$ is broken. Furthermore, it is shown that the exchange energy $-\langle a b | V | b a \rangle$ when $\eta_a = -\eta_b$ is always positive, contributing to an antiferomagnetic Hund's coupling $J < 0$ in the language of Ref. \cite{Bultinck20}.

Moreover, the on-site Hubbard Hamiltonian $U \sum_{\boldsymbol{i}} :n_{\boldsymbol{i}\uparrow} n_{\boldsymbol{i}\downarrow}:$ also incorporates symmetry breaking, favouring magnetically ordered phases.

In what follows we quantify the symmetry breaking in the manifold of ground state candidates and discuss its importance.
% \end{widetext}
% At even integer fillings ($\nu=0,\pm 2$), the strong coupling analysis predicts gapped insulators obtained by filling full spin-valley flavors.
% \begin{align}
%     | \nu = -2 \rangle = U \bigg(\prod_{j=1}\prod_{\boldsymbol{k}} d^\dagger_{\boldsymbol{k}\eta_jvs_j} d_{\boldsymbol{k}\eta_jcs_j}\bigg) |0\rangle \nonumber \\
%     | \nu = 0 \rangle = U \bigg( \prod_{j=1}^2 \prod_{\boldsymbol{k}}d^\dagger_{\boldsymbol{k}\eta_jvs_j} d_{\boldsymbol{k}\eta_jcs_j}\bigg) |0\rangle \nonumber \\
%     | \nu = +2 \rangle = U \bigg(\prod_{j=1}^3 \prod_{\boldsymbol{k}} d^\dagger_{\boldsymbol{k}\eta_jvs_j} d_{\boldsymbol{k}\eta_jcs_j}\bigg) |0\rangle \\
% \end{align}
% Here, $d^\dagger_{\boldsymbol{k}\eta_jvs_j}$ ($d^\dagger_{\boldsymbol{k}\eta_jcs_j}$) creates an electron with momentum $\boldsymbol{k}$, valley $\eta_j$ and spin $s_j$ in the valence (conduction) band. $U$ is an arbitrary element of the $U(4)$ symmetry group. Perturbations to the projected-flat limit via explicit $C_{2z}P$ or $SU(2)_{K}\times SU(2)_{K'}$ breaking, kinetic superexchange, strain, electron-phonon coupling or the Zeeman effect select the true ground state out of the $U(4)$ manifold.
\subsection*{The manifold of possible ground states}
For even values of $\nu$, $\nu=0,\pm2$, the ground states in strong coupling are '$U(4)$ ferromagnets'. Any $U(4)$ rotation $U$ of the valley-spin polarized states is a possible ground state\cite{bernevig421,Kang19,Bultinck20,ledwith21,herzog22_3}
\begin{align}
    |\text{GS}\rangle = U \bigg(\prod_{\boldsymbol{k}} \prod_{j=1,\nu/2+2} d^{\dagger}_{\boldsymbol{k}\eta_j+1 s_j } d^\dagger_{\boldsymbol{k}\eta_j -1s_j } | 0 \rangle \bigg),
    % |\nu = 0 \rangle = U \bigg(\prod_{\boldsymbol{k}} \prod_{j=1,2} d^{\dagger}_{\boldsymbol{k}\eta_j +1s_j} d^\dagger_{\boldsymbol{k}\eta_j -1 s_j } \bigg) | 0 \rangle \nonumber \\
    %  |\nu = + 2 \rangle = U \bigg(\prod_{\boldsymbol{k}} \prod_{j=1,2,3}  d^{\dagger}_{\boldsymbol{k}\eta_j+1 s_j} d^\dagger_{\boldsymbol{k}\eta_j -1 s_j} \bigg) | 0 \rangle,
\end{align}
$d^\dagger_{\boldsymbol{k}\eta \lambda s}$ denoting the creation operator of state $|\boldsymbol{k}\eta \lambda s \rangle$, $|0\rangle$ the state with the filled remote bands, and $s=\uparrow, \downarrow$ the spin index. The identity and Pauli matrices in spin will be denoted by $s_{0,x,y,z}$. The valley-spin flavors $\eta_j$, $s_j$ can be chosen arbitrarily, as different choices are related by a $U(4)$ transformation.

% If the symmetry is exact, the (second order) kinetic energy perturbation will select the ground state. In many samples heterostrain is present\cite{Choi2019,Kerelsky2019}, and states with broken translational symmetry can be comptetitive\cite{Bultinck21}. Furthermore, recent experiments\cite{chen2023strong, nuckolls2023quantum} suggest that electron-phonon coupling plays an important role and promotes states in multiplets of the broken $U(4)\times U(4)$ \cite{Kwan23,blason22}.
% The treatment of strain and electron-phonon coupling is beyond the scope of this work, and we focus on the breaking of $C_{2z}P$ and $SU(2)_K \times SU(2)_{K'}$ in unstrained samples, that lead yet to another source for the competition of the different orders. 

% For $\Phi=0$, there is the valley polarized state
% \begin{align}
%     |\text{VP}\rangle =& \prod_{\boldsymbol{k}} \prod_{j=1,\nu/2+2} d^\dagger_{\boldsymbol{k}K A s_j} d^\dagger_{\boldsymbol{k}K B s_j} |0\rangle \nonumber \\
%     =& \prod_{\boldsymbol{k}} \prod_{j=1,\nu/2+2} d^\dagger_{\boldsymbol{k}K +1 s_j} d^\dagger_{\boldsymbol{k}K -1 s_j} |0\rangle.
%     \label{vp}
% \end{align}
% We have written the state in the subalttice basis in the first line and in the irrep basis in the second, which is a valid expression also for $\Phi=\Phi_0$. The spins are chosen to be $s_1 = \uparrow$, $s_2 = \downarrow$, $s_3= \uparrow$.
The valley polarized states $|\text{VP}\rangle$ correspond to choosing $U=1$ and $(\eta_1$, $s_1)=(K$, $\uparrow)$, $(\eta_2$, $s_2)=(K$, $\downarrow)$, $(\eta_3$, $s_3)=(K'$, $\uparrow)$ above, such that the total valley charge is maximized.

For $\Phi=0$, the Kramers intervalley coherent (KIVC) state belongs to the $U(4)$ manifold, and is related to the valley polarized (VP) state by a $C_{2z}P$ angle of $\pi/4$
\begin{align}
    |\text{KIVC}\rangle &=  \text{exp}\bigg(i\frac{\pi}{4} \mathcal{S}\bigg)|\text{VP}\rangle,
\end{align}
with $\mathcal{S}$ the generator of the particle-hole $U(1)$ in the projected system,
\begin{align}
    \mathcal{S} =& \sum_{\boldsymbol{k}}\sum_{\rho \rho'} c^\dagger_{\boldsymbol{k}\rho} [\tau_y \lambda_z s_0]_{\rho \rho'} c_{\boldsymbol{k}\rho'}, 
    % \nonumber \\
    % =& \sum_{\boldsymbol{k}}\sum_{\rho \rho'} c^\dagger_{\boldsymbol{k}\rho} [\sigma_y \tau_x s_0]_{\rho \rho'} c_{\boldsymbol{k}\rho'},
\end{align}
again with the multi-index $\rho$ denoting valley, irrep and spin.
% , or valley, sublattice and spin, depending on the context.
The unitary $\text{exp}\big(i\phi \mathcal{S}\big)$ transforms the basis as 
\begin{align}
     e^{i\phi \mathcal{S}} d^\dagger_{\boldsymbol{k}\rho} e^{-i\phi \mathcal{S}} = \bigg[\text{exp}\Big(i\phi (\tau_y \lambda_z s_0)^T \Big)\bigg]_{\rho \rho'} d^\dagger_{\boldsymbol{k}\rho'}.
    %  \nonumber \\
    % e^{i\phi \mathcal{S}} d^\dagger_{\boldsymbol{k}\rho} e^{-i\phi \mathcal{S}} = \bigg[\text{exp}\Big(i\phi (\sigma_y \tau_x s_0)^T \Big)\bigg]_{\rho \rho'} d^\dagger_{\boldsymbol{k}\rho'}.
\end{align}

On another hand, he chiral $U(4)\times U(4)$ group contain operations that rotate each Chern sector independently\cite{dimitru22}. The time reversal intervalley coherent (TIVC) order relates to VP via a $U(4)\times U(4)$ rotation with $C_{2z}P$ angles of $+ \pi/4$ and $-\pi/4$,
\begin{align}
    | \text{TIVC} \rangle = \text{exp}\bigg(i\frac{\pi}{4} \mathcal{S}_{+1} - i\frac{\pi}{4}\mathcal{S}_{-1}\bigg) |\text{VP} \rangle.
\end{align}
Here $\mathcal{S}_\lambda = P_\lambda \mathcal{S} P_\lambda$,
% is $\mathcal{S}$ restricted to irrep $\lambda$, 
with $P_\lambda$ the projector onto irrep $\lambda$.

In addition, particularizing to $\nu=0$ we consider the fully spin polarized (SP) state,
\begin{align}
       | \text{SP} \rangle &=
       % \prod_{\boldsymbol{k}}  d^\dagger_{\boldsymbol{k}KA\uparrow}d^\dagger_{\boldsymbol{k}KB\uparrow}d^\dagger_{\boldsymbol{k}K'A\uparrow}d^\dagger_{\boldsymbol{k}K'B\uparrow}|0\rangle \nonumber \\ 
       % &=
       \prod_{\boldsymbol{k}}  d^\dagger_{\boldsymbol{k}K+1\uparrow}d^\dagger_{\boldsymbol{k}K-1\uparrow}d^\dagger_{\boldsymbol{k}K'+1\uparrow}d^\dagger_{\boldsymbol{k}K'-1\uparrow}|0\rangle,
       \label{sp}
\end{align}
that is derived from the valley-spin polarized (VSP) state,
\begin{align}
    | \text{VSP} \rangle &=
    % \prod_{\boldsymbol{k}}  d^\dagger_{\boldsymbol{k}KA\uparrow}d^\dagger_{\boldsymbol{k}KB\uparrow}d^\dagger_{\boldsymbol{k}K'A\downarrow}d^\dagger_{\boldsymbol{k}K'B\downarrow}|0\rangle \nonumber \\
    % &=
    \prod_{\boldsymbol{k}}  d^\dagger_{\boldsymbol{k}K+1\uparrow}d^\dagger_{\boldsymbol{k}K-1\uparrow}d^\dagger_{\boldsymbol{k}K'+1\downarrow}d^\dagger_{\boldsymbol{k}K'-1\downarrow}|0\rangle,
    \label{vsp}
\end{align} 
after a spin rotation in valley $K'$, belonging to $SU(2)_K \times SU(2)_{K'}$.
\begin{align}
    | \text{SP} \rangle = \text{exp} \bigg(i \frac{\pi}{2}  P_{K'}s_y P_{K'}\bigg) |\text{VSP}\rangle,
\end{align}
with $P_{K'}$ the projector onto valley $K'$. At zero field, by $\mathcal{T}$ symmetry the VP and the VSP states have the same energy.

At $\Phi=\Phi_0$, the intervalley coherent state corresponds to a $C_{2z}P$ rotation of the VP, of angle $\pi/4$,
\begin{align}
    |\text{IVC}\rangle =&  \text{exp}\bigg(i\frac{\pi}{4} \mathcal{S}\bigg)|\text{VP}\rangle,
\end{align}
with $\mathcal{S}$ taking a different form in accordance with our gauge choice of Eq. \ref{c2pp1},
\begin{align}
    \mathcal{S} = \sum_{\boldsymbol{k}}\sum_{\rho \rho'} c^\dagger_{\boldsymbol{k}\rho} [\lambda_0 \tau_x s_0]_{\rho \rho'} c_{\boldsymbol{k}\rho'}.
\end{align}
% at $\nu=0\pm 2$. Also, the SP and VSP states at charge neutrality are written in the second line in Eqs. \ref{sp} and \ref{vsp}.
\subsection*{Explicit symmetry breaking in the ground state manifold}
\begin{figure}[t!]
    \centering
    \includegraphics[width=.75\linewidth]{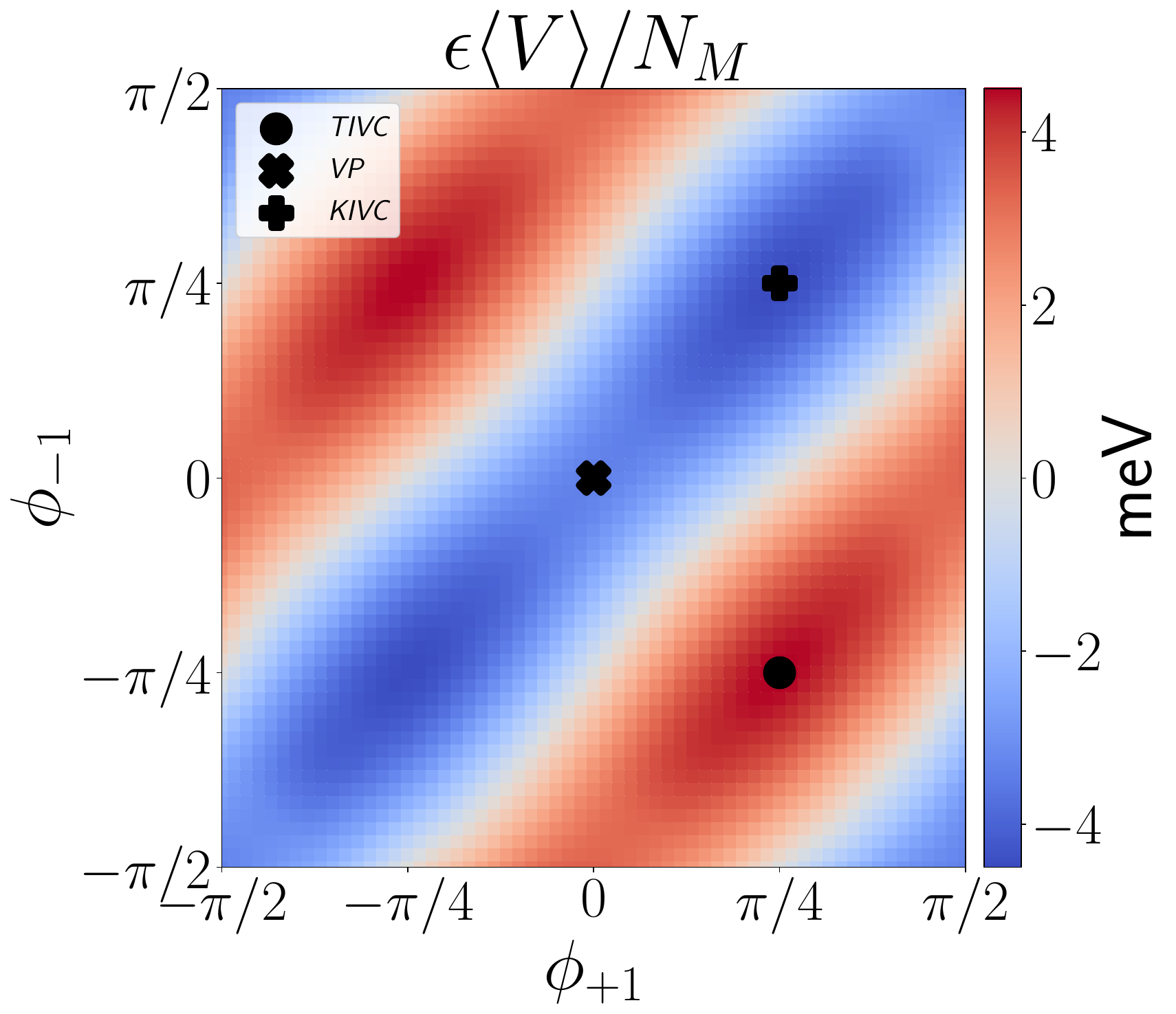}
    \caption{\textbf{Breaking of the chiral $\boldsymbol{U(4)\times U(4)}$ symmetry.} Coulomb energy ($\epsilon$ is set to $1$) per unit cell of candidates states in the $U(4)\times U(4)$ manifold at filling $\nu=-2$, which is explicitly broken to $U(4)$ with $\phi_{+1}=\phi_{-1}$. The zero point of the energy is arbitrary. VP, KIVC and TIVC are the valley polarized, Kramers intervalley coherent and time reversal intervalley coherent states, respectively. Sending $\phi_{\pm 1} \to -\phi_{\pm 1}$ amounts to a $U(1)_v$ transformation of angle $\pi$. $U(1)_v$ is visibly preserved.}
    \label{chbreak}
\end{figure}

The strength of the explicit symmetry breaking processes can be measured by the difference in Coulomb energy of the different ground state candidates. In Tables \ref{energiesp0} and \ref{energiesp1} of Appendix \ref{appf} we tabulate the different contributions to the total energy (Hartree, Fock, kinetic and Hubbard energy) of the states discussed above. We note that, remarkably, the kinetic energy is degenerate for all the states considered at a a given filling. Here we focus on the states at charge neutrality.

At $\Phi=0$, p-h breaking corresponds to the difference between VP and KIVC energies,
\begin{align}
    \epsilon\big(\langle V \rangle_{\text{VP}} -  \langle V \rangle_{\text{KIVC}}\big)/N_M =  2.26\ \text{meV},
    \label{vpkivc}
\end{align}
the breaking of the chiral symmetry by the difference between VP and TIVC states,
\begin{align}
       \epsilon \big(\langle V \rangle_{\text{TIVC}} -  \langle V \rangle_{\text{VP}}\big)/N_M =  14.01 \ \text{meV},
\end{align}
and the breaking of $SU(2)_K \times SU(2)_{K'}$ by the energies of VSP (same energy as VP) and SP,
\begin{align}
    \epsilon\big(\langle V \rangle_{\text{SP}} -  \langle V \rangle_{\text{VP}}\big)/N_M =  10.32 \ \text{meV}.
\end{align}
Here $N_M$ is the number of unit cells, and we have multiplied by $\epsilon$ so that we are comparing energies per unit cell for $\epsilon=1$.

Very similarly for $\Phi=\Phi_0$, p-h breaking corresponds to the difference between VP and IVC energies,
\begin{align}
     \epsilon\big(\langle V \rangle_{\text{IVC}} -  \langle V \rangle_{\text{VP}}\big)/N_M =  0.74\ \text{meV},
\end{align}
and the breaking of $SU(2)_K \times SU(2)_{K'}$ by the VSP and SP states,
\begin{align}
    \epsilon\big(\langle V \rangle_{\text{SP}} -  \langle V \rangle_{\text{VSP}}\big)/N_M = 5.23 \ \text{meV}.
\end{align}
% Interestingly the energy differences are reduced by approximately a factor of $2$ under magnetic flux.

Additionally, in Fig. \ref{chbreak} we plot the Coulomb energy of states of the form $\text{exp}\big(i\phi_{+1}\mathcal{S}_{+1} + i\phi_{-1}\mathcal{S}_{-1}\big) |\text{VP}\rangle$ for $\nu=-2$ and $B=0$ T. The breaking of the chiral $U(4)\times U(4)$ symmetry down to $U(4)$ is evident, and the inversion symmetry of the plot shows the conservation of $U(1)_v$.

In light of the results, we conclude that $C_{2z}P$ breaking is smaller than the breaking of the chiral symmetry at $\Phi=0$ and of $SU(2)_{K}\times SU(2)_{K'}$.
% (the kinetic energy spectrum is very p-h asymmetric at $\Phi=0$ and might be the reason for the p-h asymmetry seen in the experiments).

The $SU(2)_K \times SU(2)_{K'}$ breaking is stronger and comparable to the breaking of the chiral symmetry at zero field. This effect was called intervalley Hund's interaction in Ref. \cite{Bultinck20}, where it was argued that it is the smallest energy scale in the hierarchy of symmetry breakings. 
% In Ref. \cite{chatt20} the splitting between the SP and VSP states was found to be $\sim -0.4$ meV for $\epsilon = 6.6$ in a computation in the continuum theory. We find a value of $+1.56$ meV, four times larger in magnitude and of antiferromagnetic character. 
Our calculations in the tight-binding model involve lattice-scale interactions, providing reliable values for the splitting of the $SU(2)_{K} \times SU(2)_{K'}$ multiplets\cite{chatt20}.

The Hubbard interaction clearly breaks $SU(2)_{K}\times SU(2)_{K'}$, and favours the states with a net spin polarization. On the other hand, we showed that the long ranged Coulomb energy is Hund antiferromagnetic, supporting states with opposite spins in different valleys. Although in the real system we expect their values to be correlated, the interplay between $U$ and $\epsilon$ (and possibly other effects due to phonons\cite{chatt20}) determines the sign of the Hund's coupling $J$. For instance, the splitting between the VSP and SP states changes sign when $\epsilon U  = 8.82$ eV($9.02$ eV) at zero field(one flux quantum). Also, if the Hubbard interaction is strong enough, it can go beyond selecting the state of the $SU(2)_K \times SU(2)_{K'}$ multiplets and stabilize spin polarized phases, as we will see.

Finally, notice that the small $C_{2z}P$ breaking favours the KIVC phase as can be seen in Fig. \ref{chbreak} or Eq. \ref{vpkivc}. However, the gain in energy of the 'dressed' self-consistent states is larger than this small splitting between the VP and KIVC, so 
the p-h breaking is not the decisive factor in the stability of the ground states.

\section{SELF-CONSISTENT HARTREE-FOCK}

We have carried out self-consistent Hartree-Fock simulations in a system of $12 \times 12$ unit cells, focusing on filling factors $\nu=-2,0,+2$. We describe the Hartree-Fock formalism and the flat band projection method in Appendix \ref{appd}.  

Typical values for $\epsilon$ found in the literature range from about $7$ to $12$\cite{Zhang22,Bultinck20}, so we choose $\epsilon=10$ and a realistic value for $U$ of $4$ eV\cite{stauber21,jimenopozo2023short}. However, it has been argued that internal screening is large in these systems and a more appropriate value for $\epsilon$ is several times larger\cite{stauber21,Gonzalez23}. This agrees with the fact that lower values of $\epsilon$ overestimate the gap of the insulators, which in transport are found to be $\lesssim$ 1 meV\cite{Lu2019,Yankowitz19,liu21,efetov22,Pierce2021}. We account for both scenarios and report results also for $\epsilon=50$ and $0.5$ eV.

The self-consistent states are characterized by the $Q$ matrix, defined by 
\begin{align}
    [Q(\boldsymbol{k})]_{\rho \rho'} &= 2[P(\boldsymbol{k})]_{\rho \rho'} - \delta_{\rho\rho'}, \nonumber \\
    [P(\boldsymbol{k})]_{\rho \rho'} &= \langle d^\dagger_{\boldsymbol{k}\rho} d_{\boldsymbol{k}\rho'}\rangle,
\end{align}
with the properties $Q(\boldsymbol{k}) = Q(\boldsymbol{k})^\dagger$, $Q(\boldsymbol{k})^2 = 1$ and $\text{tr}(Q(\boldsymbol{k})) = 2\nu$. In most cases, as we discuss below, $Q$  will be diagonal in the spin, $Q(\boldsymbol{k}) = Q_{\uparrow} (\boldsymbol{k}) P_{\uparrow} + Q_\downarrow (\boldsymbol{k}) P_{\downarrow}$ ($P_{\uparrow(\downarrow)}$ is the projector onto spin $\uparrow(\downarrow)$), with each spin polarization either completely empty ($Q_s(\boldsymbol{k}) = -1$), completely full ($Q_s(\boldsymbol{k}) = +1$) or half filled. If half filled, $Q_s$ can be expressed as a linear combination of products of Pauli matrices,
\begin{align}
    Q_s(\boldsymbol{k}) = \mathop{\sum_{\alpha, \beta = 0,x,y,z}}_{(\alpha, \beta) \neq (0,0)} A^s_{\alpha \beta}(\boldsymbol{k}) \sigma_\alpha \tau_\beta 
    =  \mathop{\sum_{\alpha, \beta = 0,x,y,z}}_{(\alpha, \beta) \neq (0,0)} B^s_{\alpha \beta}(\boldsymbol{k})  \lambda_\alpha \tau_\beta,
\end{align}
with $A^s_{\alpha,\beta}(\boldsymbol{k})$, $B^s_{\alpha, \beta}(\boldsymbol{k})$ real coefficients and $\sum_{\alpha \beta} (A^s_{\alpha \beta}(\boldsymbol{k}))^2 = \sum_{\alpha \beta} (B^s_{\alpha \beta}(\boldsymbol{k}))^2 = 1 $. In the following we will write $\langle \sigma_\alpha \tau_\beta \rangle$, $\langle \lambda_\alpha \tau_\beta \rangle$ to denote the coefficients $A^s_{\alpha \beta}(\boldsymbol{k})$, $B^s_{\alpha \beta}(\boldsymbol{k})$. The momentum dependence is left implicit, and the spin can be deduced depending on the context.

The preferred state of the analytical approaches and numerical studies at zero magnetic field is the KIVC\cite{Bultinck20,bernevig421,Bultinck21,Kwan23}. At $\nu=0$ it can be either spin singlet, $Q(\boldsymbol{k}) = \sigma_y \tau_y$, or 'spin triplet', $Q(\boldsymbol{k}) = \sigma_y \tau_y \boldsymbol{n}\cdot \boldsymbol{s}$, with $\boldsymbol{n}$ denoting an spontaneous direction. An antiferromagnetic Hund's coupling, $J < 0$, favours the KIVC singlet whereas $J > 0$ prefers the 'triplet'\cite{Bultinck20}. However, in our Hartree Fock numerics we restrict the state to be a direct product of spin up and spin down wave functions, so $Q$ will be diagonal in the spin index and we cannot access the 'triplet' state. Setting $\boldsymbol{n} =  \boldsymbol{\Hat{z}}$ amounts to a valley rotation which does not change the kinetic or Coulomb energy. By a similar argument to the one given in Appendix \ref{appd}, the Hubbard energy does not change either.

\begin{figure}[b!]
    \noindent \ \ \textbf{a)} \hfill  \textbf{b)}  \hfill \phantom{phantom}\\
    \centering
    \includegraphics[width=.4\linewidth]{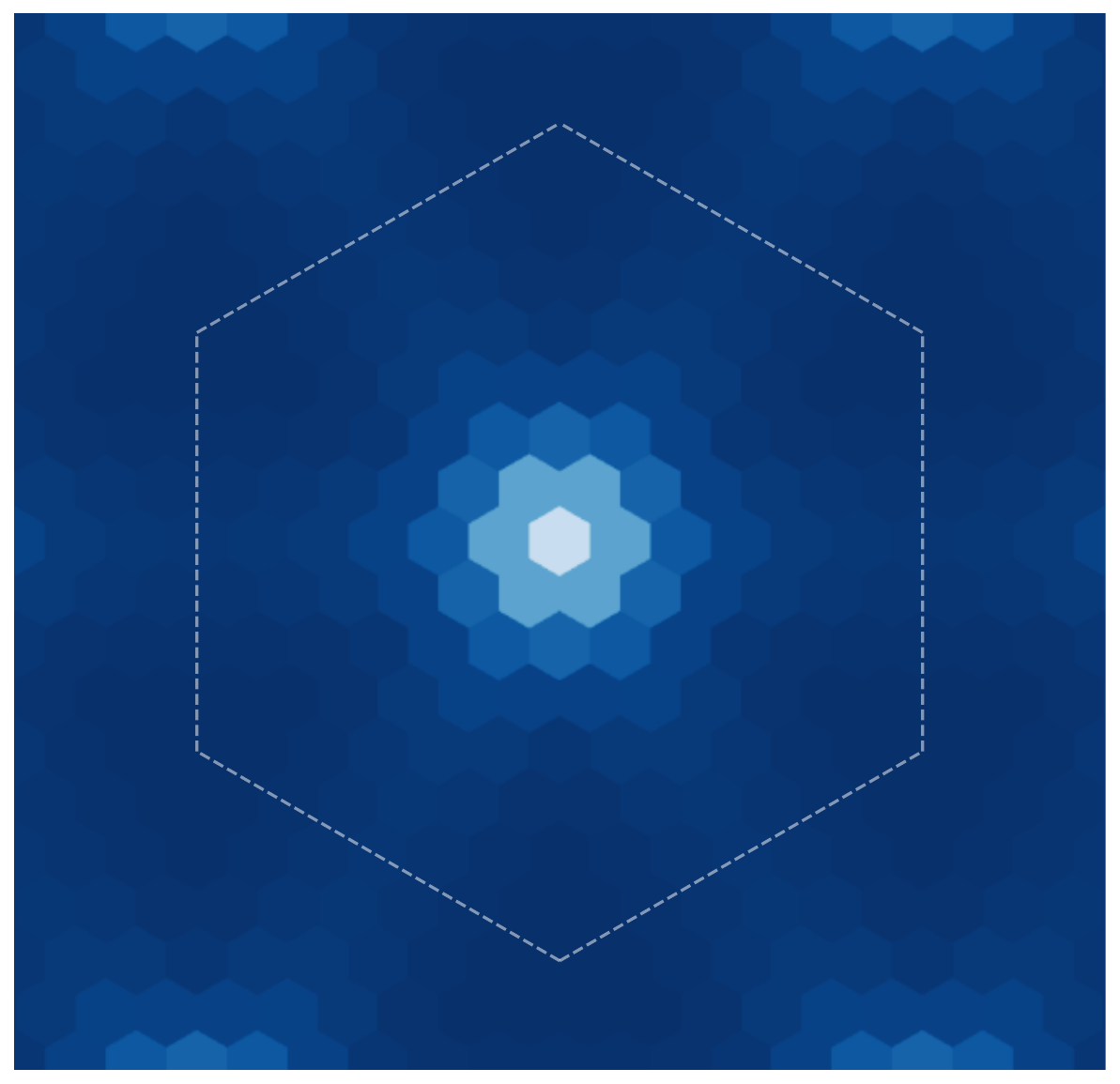}
    \includegraphics[width=.4\linewidth]{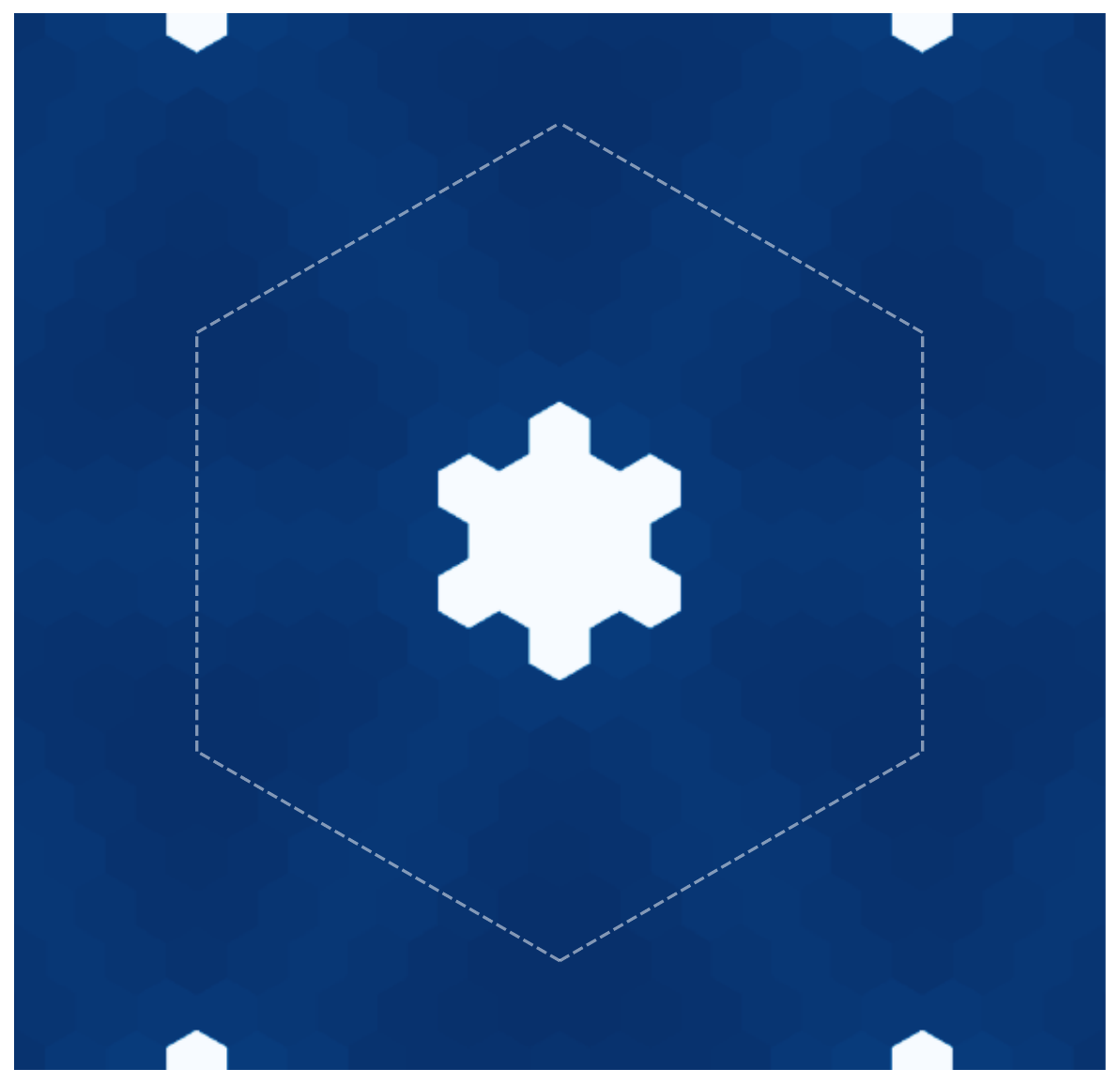}
    \includegraphics[width=.14\linewidth]{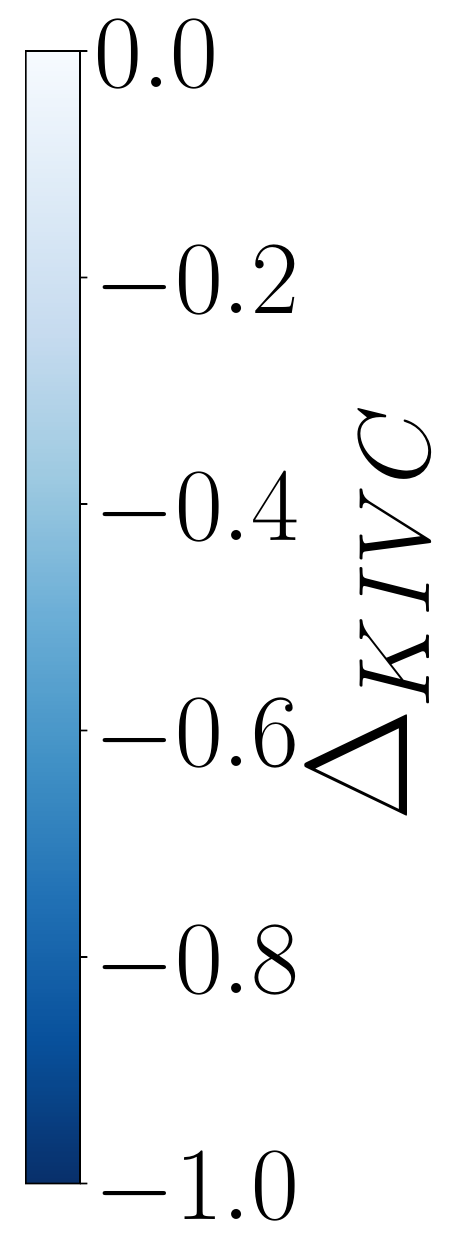}
    \caption{\textbf{KIVC order at $\boldsymbol{\Phi=0}$. a)} $\Delta_{\text{KIVC}}$ at $\nu=+2$ and $\epsilon=10, U=4$ eV. The inter-Chern parameter completes the total weight, $\Delta_{\text{IC}} = 1 - (\Delta_{\text{KIVC}})^2$, reaching a value of $0.95$ at $\Gamma_M$. In the case of $\epsilon= 50, U=0.5$ eV the only difference is that $\Delta_{\text{IC}}$ is smaller with a maximal value of $0.46$ at $\Gamma_M$. For $\nu=0$ the order parameter is almost identical. 
    % to the $\nu=-2$ case, for both values of $\epsilon$.
    %, with  maximal $\Delta_{\text{IC}}$ of $0.95$ and $0.48$. 
    \textbf{b)} KIVC order parameter at $\nu=-2$ and $\epsilon=50,\ U=0.5$ eV. Around $\Gamma_M$ there are extra holes creating a Fermi pocket, and the order parameter is ill-defined in that region.}
    \label{kivcorder}
\end{figure}

For $\nu = \pm 2$, a ferromagnetic Hund's coupling leads to $Q(\boldsymbol{k}) = \sigma_y \tau_y P_\uparrow \pm P_\downarrow$ where the spins in both valleys are aligned. Antiferromagnetic coupling on the other hand promotes the state where the spins of the two valleys are anti-aligned. Again, we can only detect the spin-diagonal order.

Notice that the $U(1)_v$ symmetry allows for an arbitrary global rotation in the order parameter, $\tau_y \to \cos(\theta) \tau_y + \sin(\theta)\tau_x$. However, the difference in the intervalley angle for different values of $\boldsymbol{k}$ cannot be removed and is physical. In any case, we do not observe textures or windings in the IVC angle.

Contrary to the the BM theory\cite{jimenopozo2023short}, in the tight-binding model the on-site Hubbard interaction is implemented naturally. At charge neutrality the SP state is not competitive if we consider only the Coulomb energy, but for sufficiently large $U$ it will be the lowest energy state. For electron and hole dopings, the candidate states are spin polarized, see Table \ref{energiesp0} in Apppendix \ref{appf}, and the preferred states (KIVC, VP) are degenerate in $H_U$, so we do not expect different orders for different $U$.

At flux $\Phi_0$, analytical studies\cite{herzog22_3} suggest that the Zeeman energy drives the system to maximize the spin polarization in the $U(4)$ manifold. At charge neutrality the SP state with $Q(\boldsymbol{k}) = s_z$ is the ground state and at $\nu=-2(+2)$ there are two possible orders, IVC, with $Q(\boldsymbol{k}) = \lambda_0 \tau_y P_\uparrow - P_\downarrow (\lambda_0 \tau_y P_\downarrow + P_\uparrow)$, or valley polarization, $Q(\boldsymbol{k}) = \lambda_0\tau_z P_\uparrow - P_\downarrow (\lambda_0\tau_z P_\downarrow + P_\uparrow)$.

\onecolumngrid

\begin{figure}[H]
    \centering \large{$\epsilon=10, \ U=4 $ eV}\\
    \includegraphics[width=.287\linewidth]{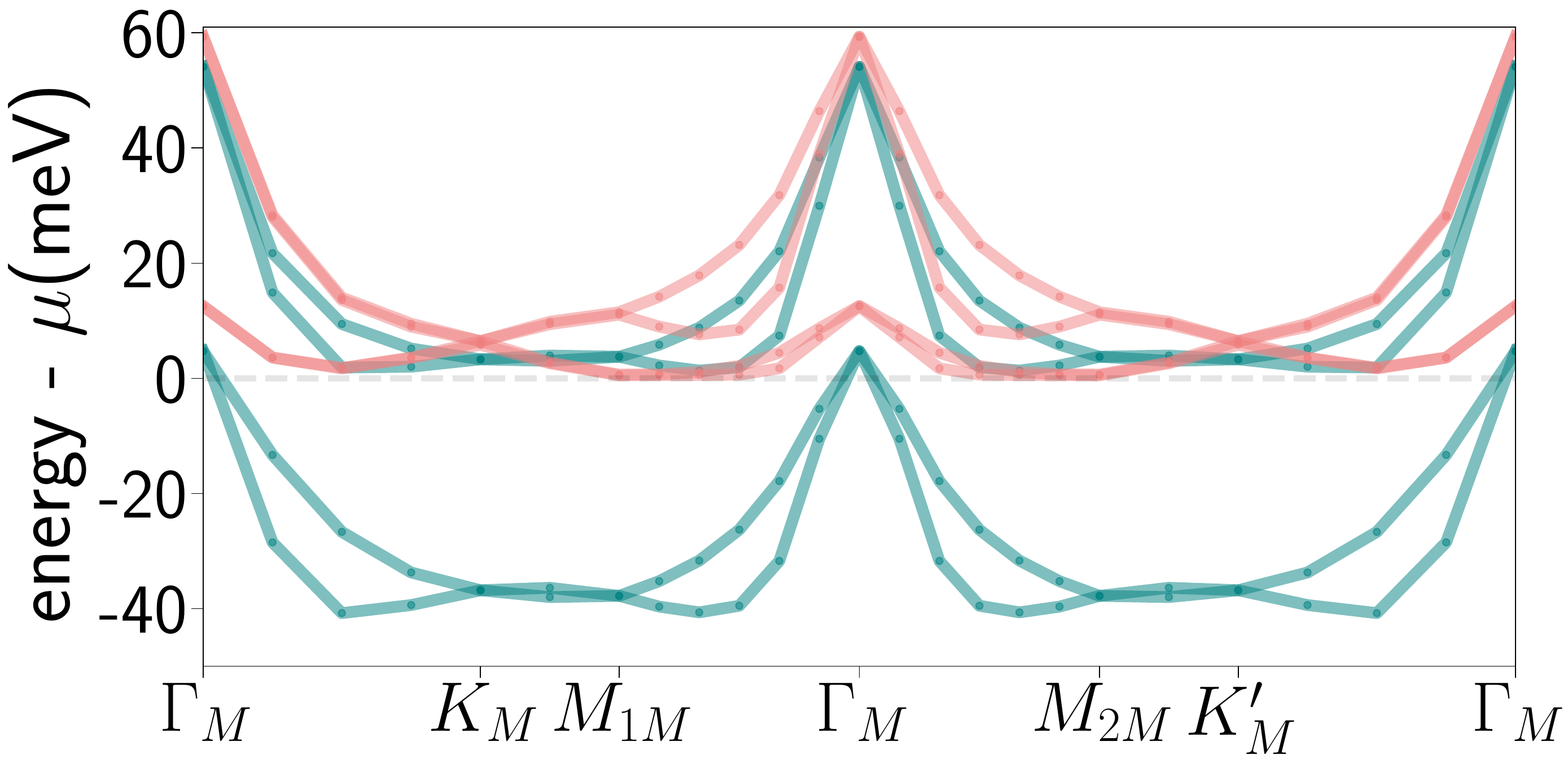}
    \includegraphics[width=.27\linewidth]{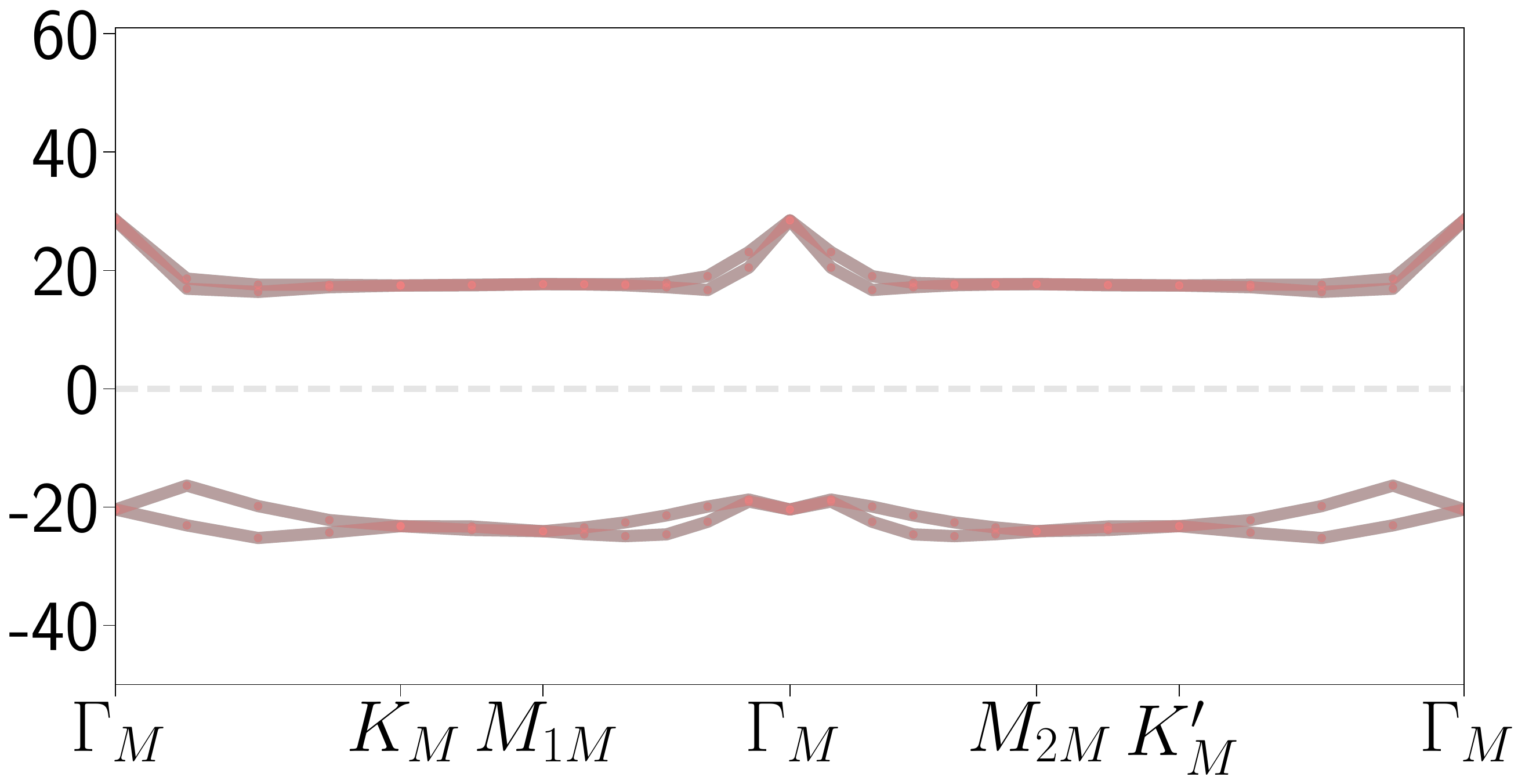}
    \includegraphics[width=.27\linewidth]{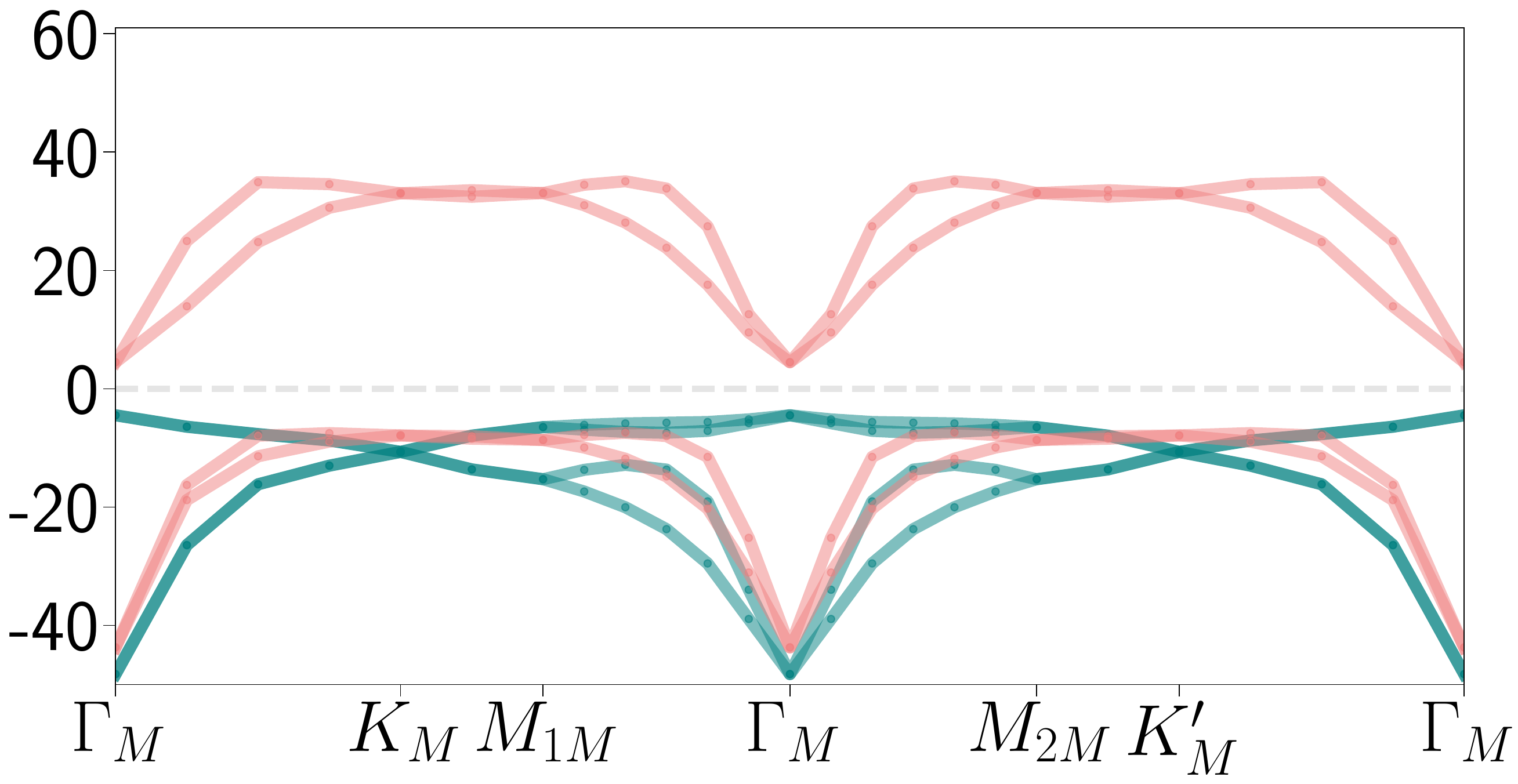}\\
    % proportion 0.942
    \centering \large{$\epsilon=50, \ U=0.5$ eV} \\
    \includegraphics[width=.289\linewidth]{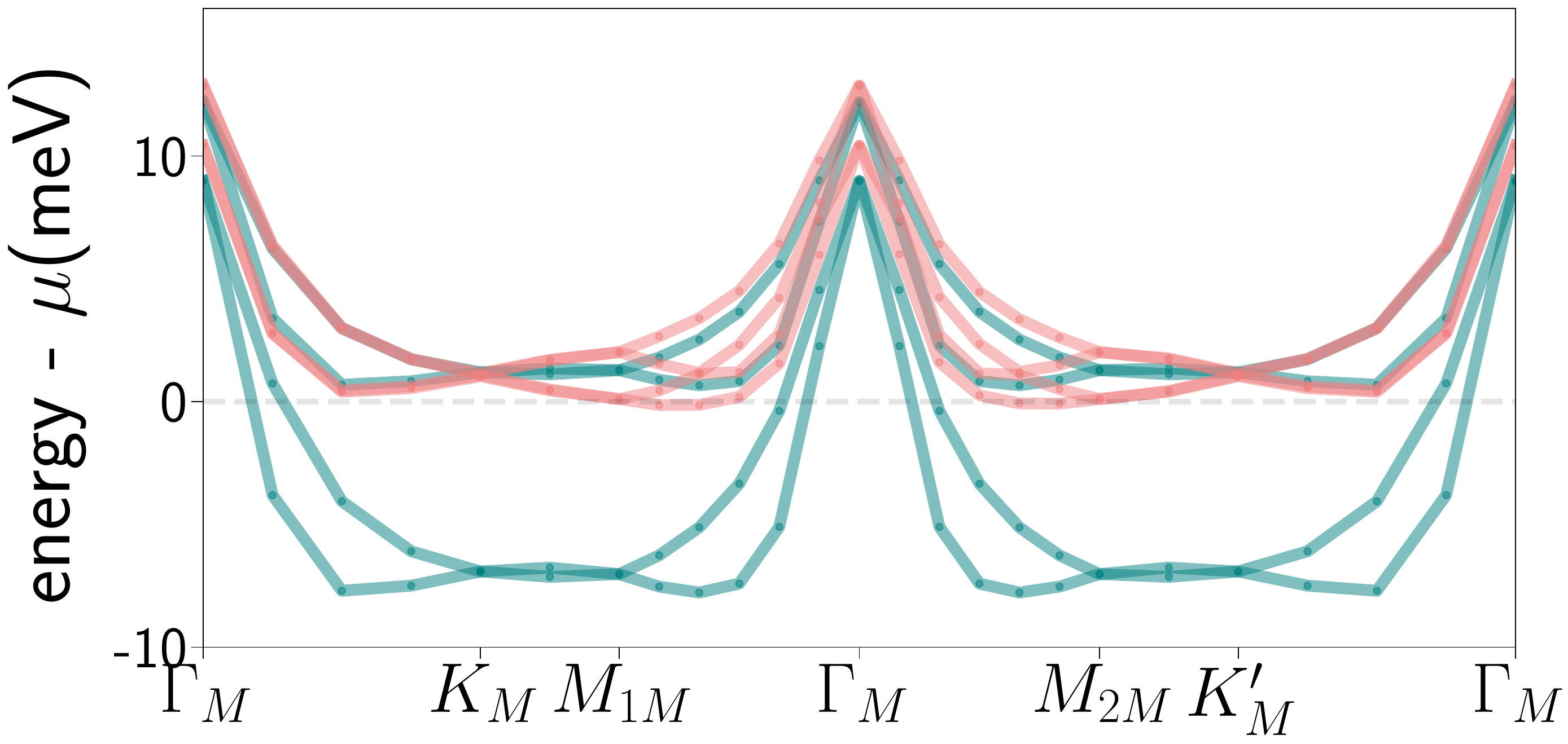}
    \includegraphics[width=.27\linewidth]{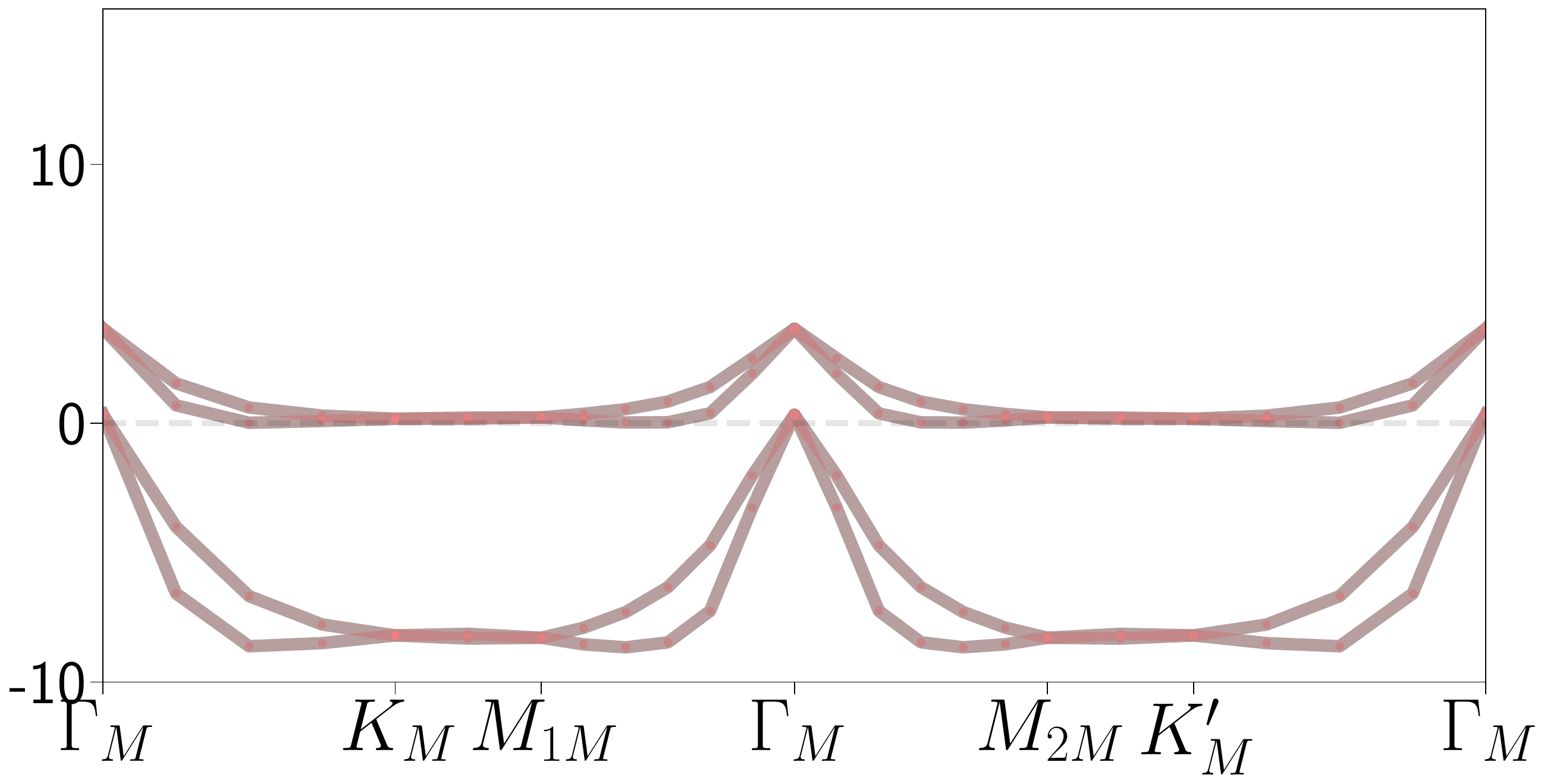}
    \includegraphics[width=.27\linewidth]{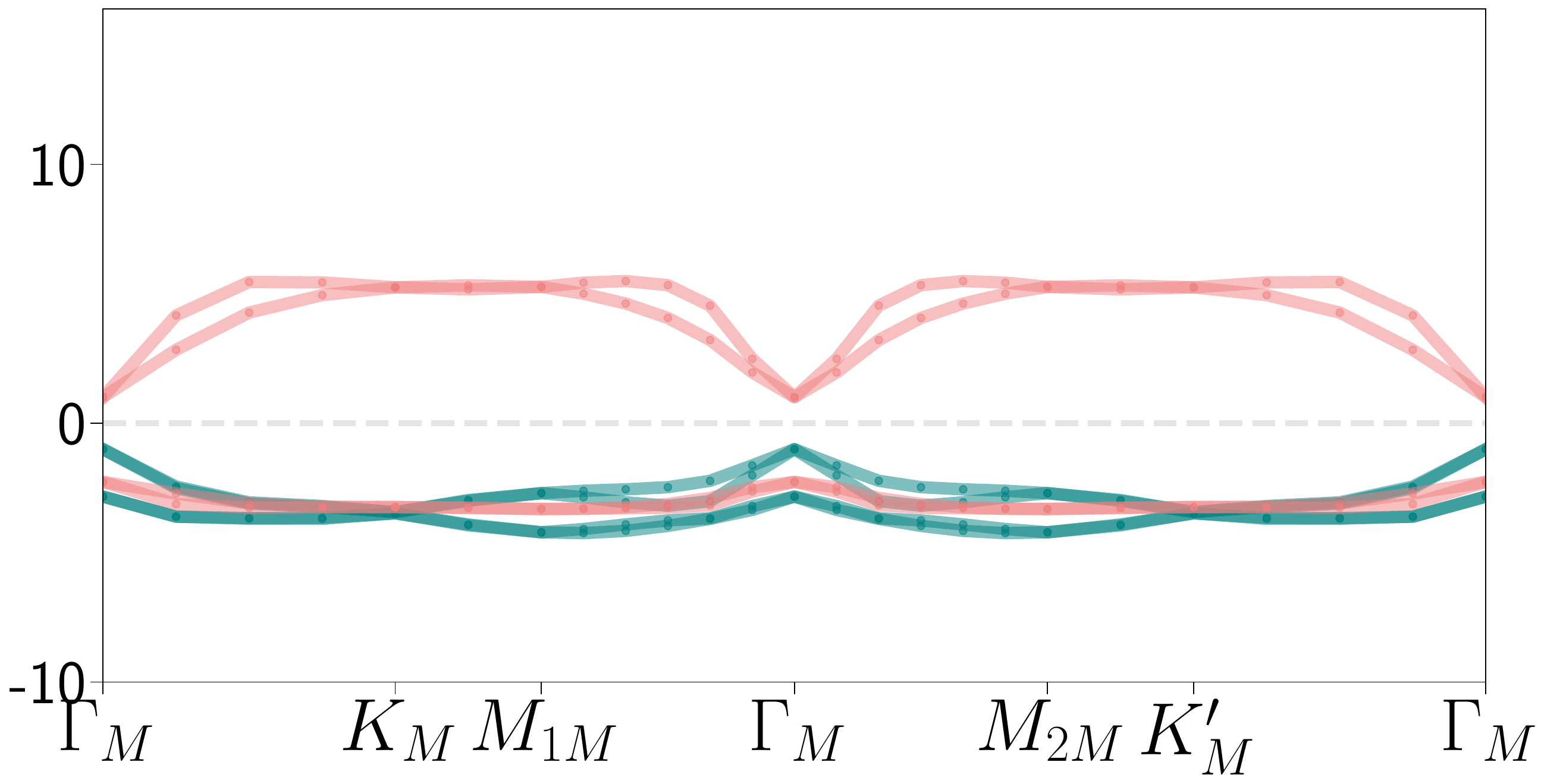}
    \caption{\textbf{KIVC band structures of MATBG at $\boldsymbol{B=0}$ T.} Different spins are shown in different colors. From left to right, the filling is $\nu=-2,0,+2$. At $\nu=\pm2$ these are the true ground states, whereas at charge neutrality there is competition between the KIVC and the spin polarized state. The many-body electron-hole asymmetry is apparent.} 
    \label{bandshfp0}
\end{figure}
\twocolumngrid

\subsection*{Results for $\boldsymbol{\Phi=0}$}
% \textbf{Results at zero field.}

Examining Table \ref{energiesp0} we deduce that the possible orders for $B=0$ T are the VP or KIVC, or the SP at charge neutrality. We compute self-consistent states with initial guesses for the KIVC, VP and SP orders, whose energies are tabulated in Table \ref{energieshfp0} of Appendix \ref{appf}. We plot the band structures of the KIVC states in Fig. \ref{bandshfp0}, of the SP states in Fig. \ref{splnu0} and of the VP states in Fig. \ref{vplhf} of Appendix \ref{appe}. The KIVC, with order parameter
\begin{align}
\Delta_{\text{KIVC}} = \langle \sigma_y \tau_y \rangle,    
\end{align}
is the ground state for $\nu=\pm 2$, and the KIVC and SP are competitive at charge neutrality. 
\begin{figure}[h]
    \large{$\epsilon=10, \ U=4$ eV}\\
    \includegraphics[width=.57\linewidth]{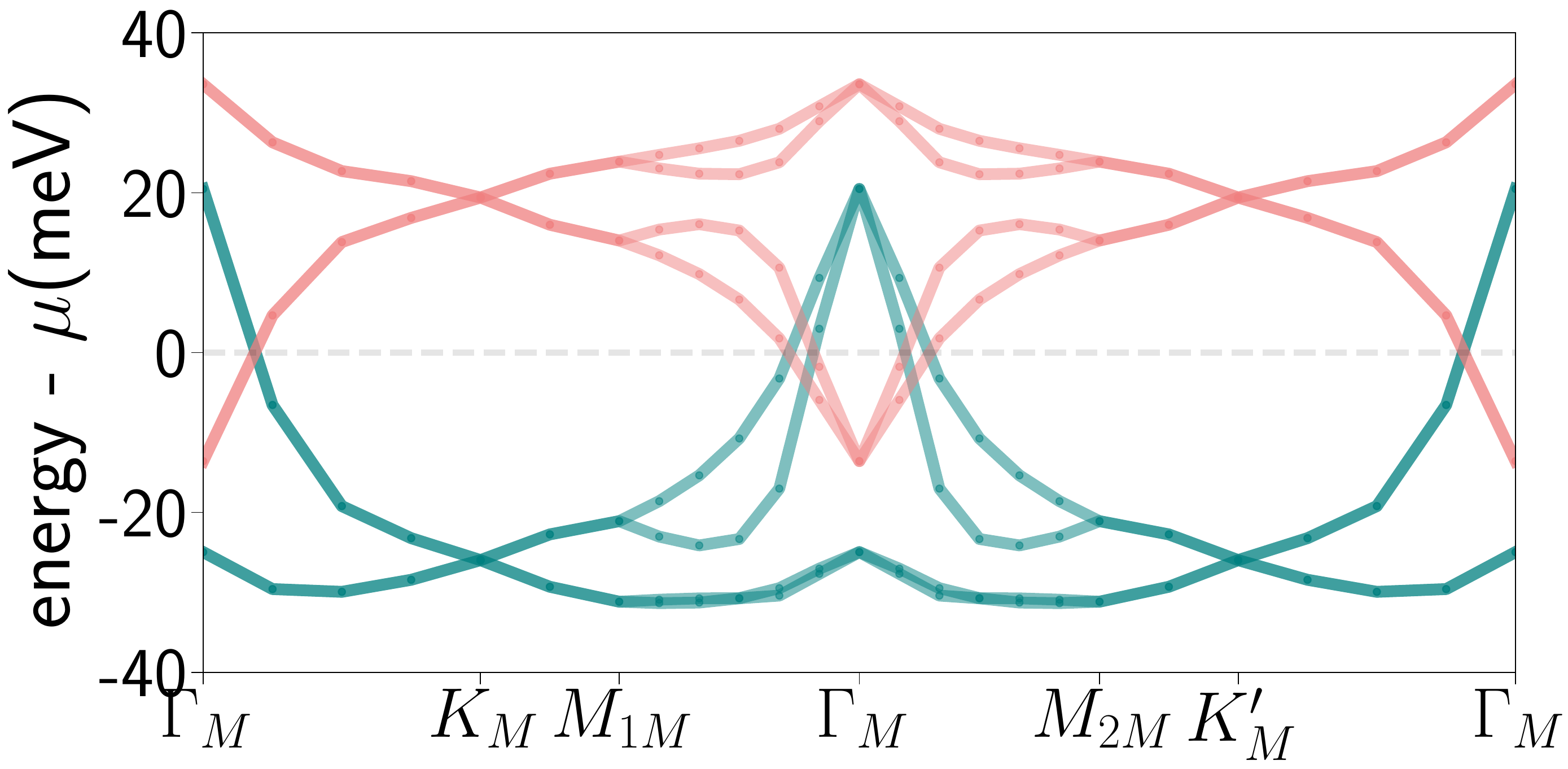}\\
    \large{$\epsilon=50, \ U=0.5$ eV} \\
    \includegraphics[width=.57\linewidth]{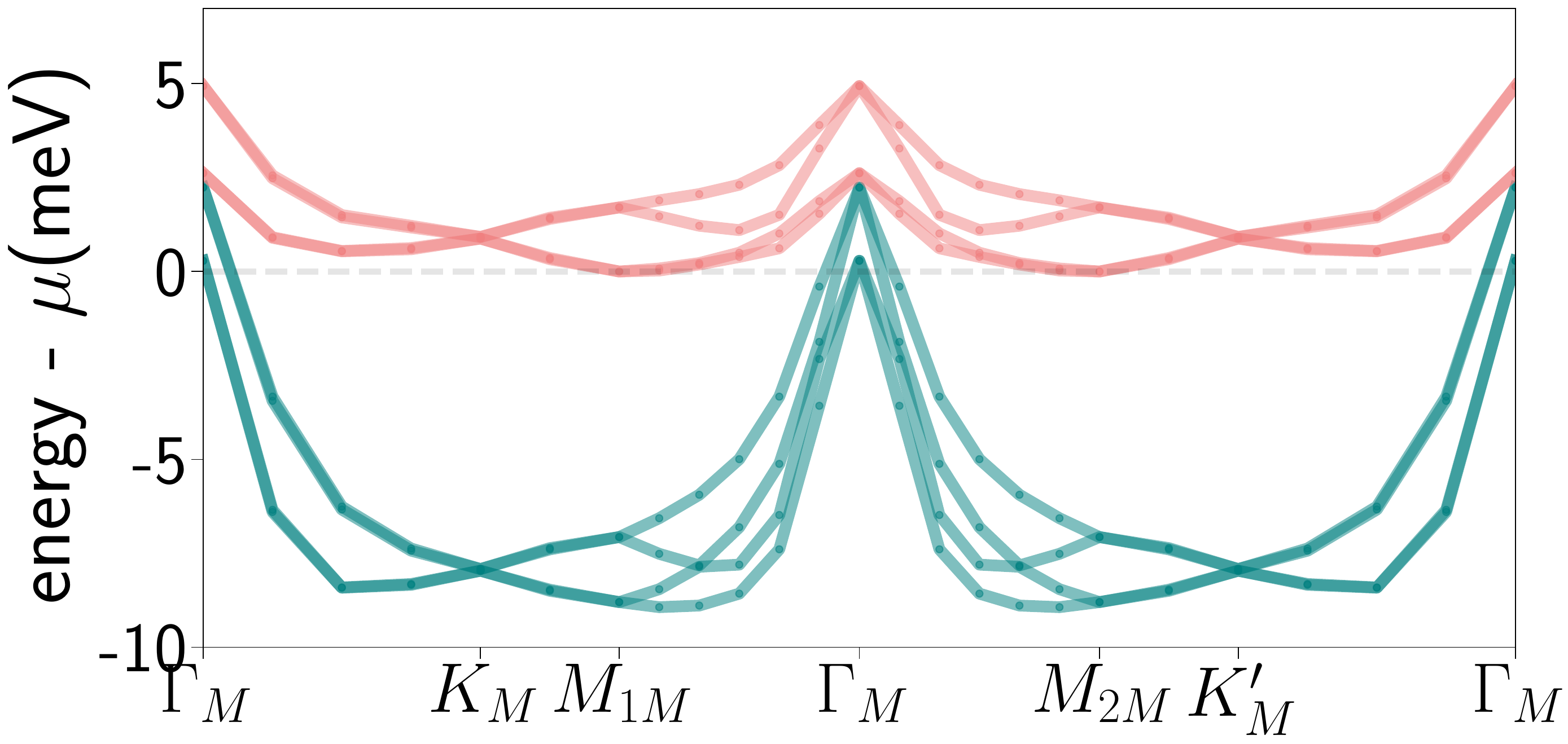}
    \caption{\textbf{Band structures of the spin polarized phase at charge neutrality.} Different spins are shown with different colors. The spin polarization is not full, and the system is a metal.}
    \label{splnu0}
\end{figure}

A general feature of the self-consistent states is that deviations from the predominant order parameter occur near the $\Gamma_M$ point. 

In the gapped phases they involve the inter-Chern parameter,
\begin{align}
    \Delta_{\text{IC}} = \langle \sigma_x\tau_0 \rangle^2 + \langle \sigma_y \tau_z \rangle^2,
\end{align}
signalling coherent superpositions of bands with different Chern numbers. Notice how the expression for $\Delta_{\text{IC}}$ is invariant with respect to the gauge ambiguity of Eq. \ref{gaugep0}. The effect of $\Delta_{\text{IC}}$ near $\Gamma_M$ is to reduce the exchange energy at the expense of kinetic (and Hartree for $\nu=\pm2$) energy. In consequence, $\Delta_{\text{IC}}$ becomes larger for lower $\epsilon$, where the kinetic energy penalty is comparatively less important. 

For $\nu=-2$ the system is metallic with a hole pocket around $\Gamma_M$.
% The wave functions at the edges of the Brillouin zone ($K_M$ and $M_M$ points) are well localized at the center of the unit cell, and near $\Gamma_M$ the density is more homogeneously distributed\cite{Cea22}. 
Because the states near $\Gamma_M$ show a homogeneous density profile, the Hartree energy is optimized if the excess density relative to charge neutrality originates from this region of the Brillouin zone. This is manifested by the characteristic dips at electron doping, or rises at hole doping in the band structures\cite{Cea22}. At $\nu=-2$ the holes are populated, optimizing the Hartree energy.

In Fig. \ref{kivcorder} we plot the KIVC order parameter for two different cases, where both phenomena are illustrated. Additionally, as depicted in Fig. \ref{splnu0} the bands of the spin polarized states show also the Fermi pockets at $\Gamma_M$, and the additional electrons and holes appear in inter-Chern coherence.

The different behaviour at $\nu=+2$ and $\nu=-2$ is a sign of the many-body electron-hole asymmetry, which is a robust experimental feature\cite{Pierce2021,Lu2019,Yankowitz19}. We argue that the flat band limit shows electron-hole symmetry but the kinetic energy disturbs it at weaker couplings. Indeed, the $H_0$ spectrum is very asymmetric at zero flux, see Fig. \ref{bands}. At $\Phi_0$ flux the spectrum is visibly more p-h symmetric, and the electron-hole relation between $\nu=+2$ and $-2$ is more apparent for $\epsilon=10$, $U=4$ eV (Fig. \ref{bandshfp1}). 
% Here it is important to realize that not only the spectrum needs to be p-h symmetric, but the operator $\mathcal{O}$ must satisfy $\{\mathcal{O},H_0\} = 0$

Let us now discuss the energies of the different self-consistent phases at charge neutrality. In the tables below we write the energies of the KIVC, SP and VP self-consistent states at relative to the pure KIVC state, in meV per unit cell (the value of $U$ is in units of eV).
\begin{table}[H]
\centering
$\epsilon=10$, $U=4$ eV\\
\begin{tabular}{SSSSSSSS} \toprule
    {} &{$\epsilon \times$Hartree}& {$\epsilon \times$Fock}& {Hubb./$U$} 
    % & {Hartree} & {Fock} 
    &  {kinetic} & {total}\\ \midrule
    {KIVC}  & {$-0.49$} & {$-64.60$} & {$0.01$} %& {$-15.52$} & {$-0.34$} 
     &  {$1.28$} & {$-5.19$}\\
     {SP}  & {$-0.13$} & {$-8.19$} & {$-1.10$} %& {$-15.52$} & {$-0.34$} 
     &  {$0.61$} & {$-4.62$}\\
    {VP} & {$-0.13$}  & {$-17.97$} & {$0.00$} %&{$-0.03$} & {$-0.19$} 
     & {$0.61$} & {$-1.20$}\\ \bottomrule
\end{tabular}
\end{table}

\begin{table}[H]
\centering
$\epsilon=50$, $U=0.5$ eV\\
\begin{tabular}{SSSSSSSS} \toprule
    {} &{$\epsilon \times$Hartree}& {$\epsilon \times$Fock}& {Hubb./$U$}  
    % & {Hartree} & {Fock} 
    &  {kinetic} & {total}\\ \midrule
    {KIVC}  & {$-0.91$} & {$-31.14$} & {$0.00$} %& {$-15.52$} & {$-0.34$} 
     &  {$0.40$} & {$-0.24$}\\
     {SP}  & {$-0.67$} & {$18.61$} & {$-1.13$} %& {$-15.52$} & {$-0.34$} 
     &  {$-0.16$} & {$-0.37$}\\
    {VP} & {$-0.67$}  & {$8.48$} & {$0.00$} %&{$-0.03$} & {$-0.19$} 
     & {$-0.16$} & {$-0.00$}\\ \bottomrule
\end{tabular}
\end{table}

The KIVC always has lower energy than the VP state, but the competition between KIVC and SP is more complex. For $\epsilon=10$, $U=4$ eV the KIVC state is $0.57$ meV per unit cell lower than the SP state, whereas for $\epsilon=50$, $U=0.5$ eV the SP state is the ground state with $0.13$ meV/cell of difference. Assuming that the self-consistent state does not change much with $U$ (as a matter of fact, we find that the states for $\epsilon=10, U=4$ are almost identical to those with $\epsilon=10, U=0.5$ eV), we expect a transition to the SP state for $U$ grater that a critical value of $U_c=4.51$ eV when $\epsilon=10$, and to the KIVC for $U$ smaller than $U_c=0.39$ when $\epsilon=50$.

% Finally, we comment that for the metallic phases at $\nu=-2$ and $\epsilon=50$, $U=0.5$ eV the energy difference between the KIVC and VP states is only of $0.07$ meV/cell. In this regime, the convergence of the KIVC state is very slow, what we attriibute to the existence of several metastable states nearby. One should take the value of the KIVC state as an upper bound, and we expect a greater difference in the fully self consistent state.
\subsection*{Results for $\boldsymbol{\Phi=\Phi_0}$}
\begin{figure}[H]
    \centering \large{$\epsilon=10, \ U=4$ eV}\\
    \centering
    \includegraphics[width=.265\linewidth]{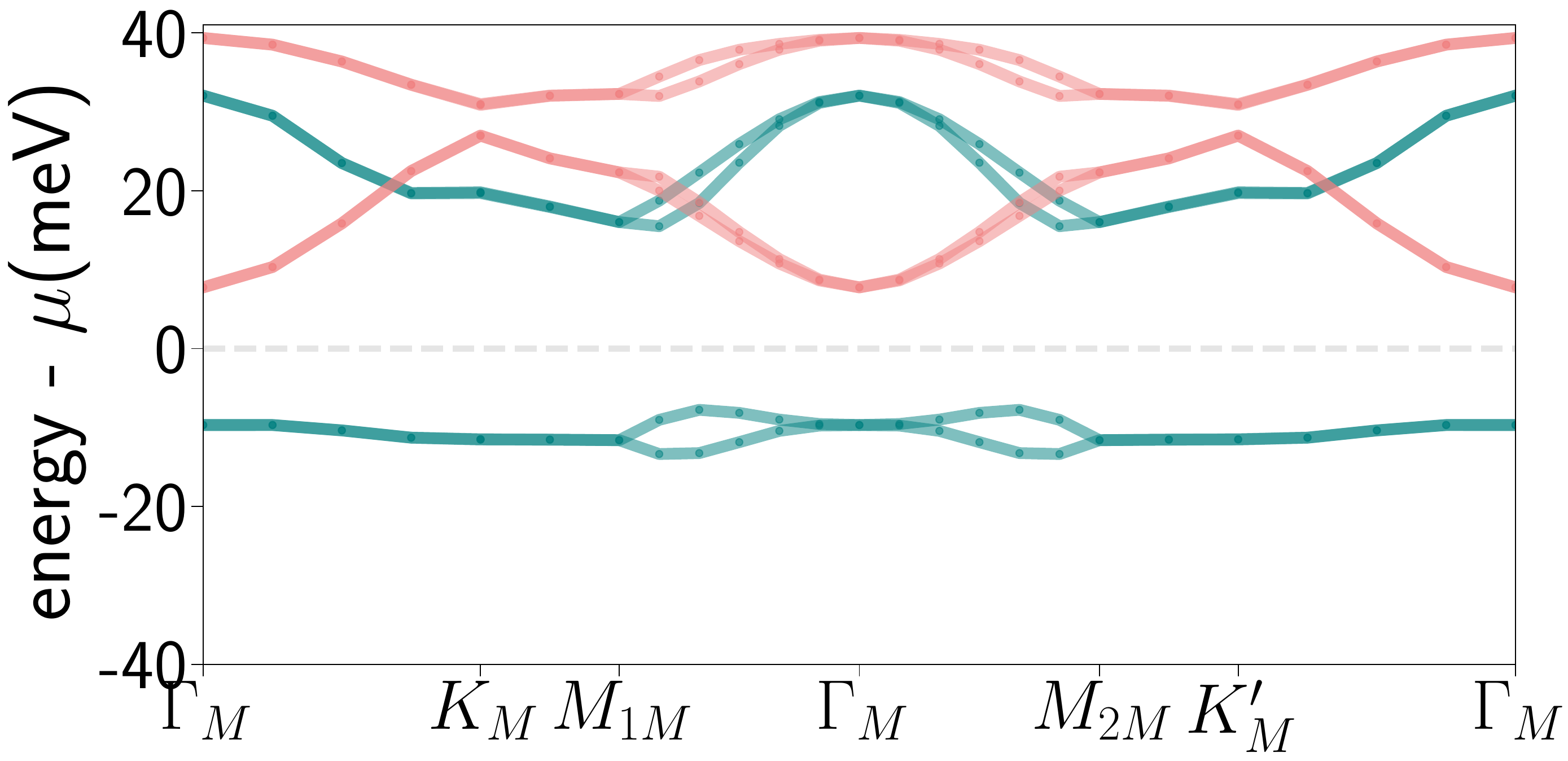}
    \includegraphics[width=.25\linewidth]{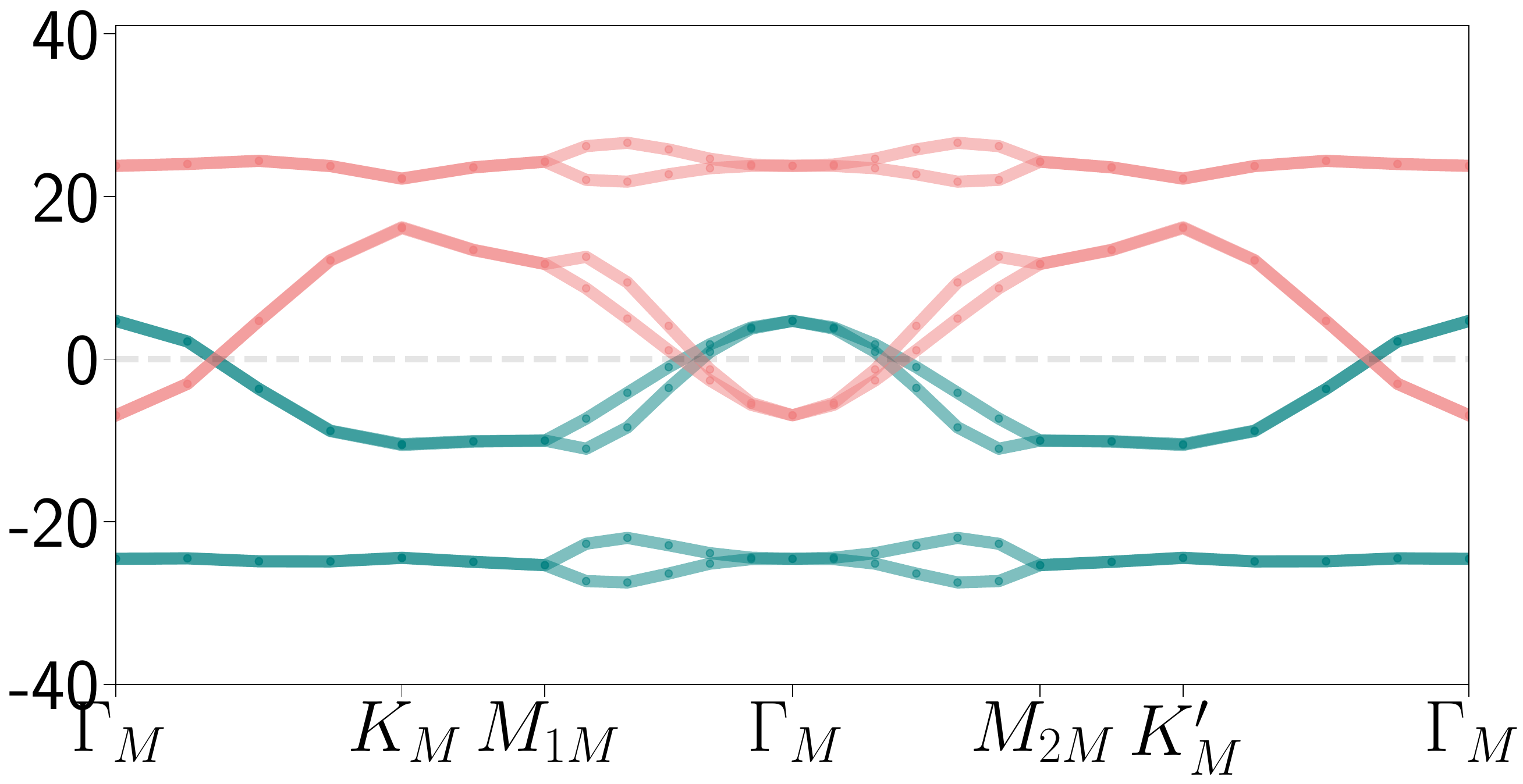}
    \includegraphics[width=.25\linewidth]{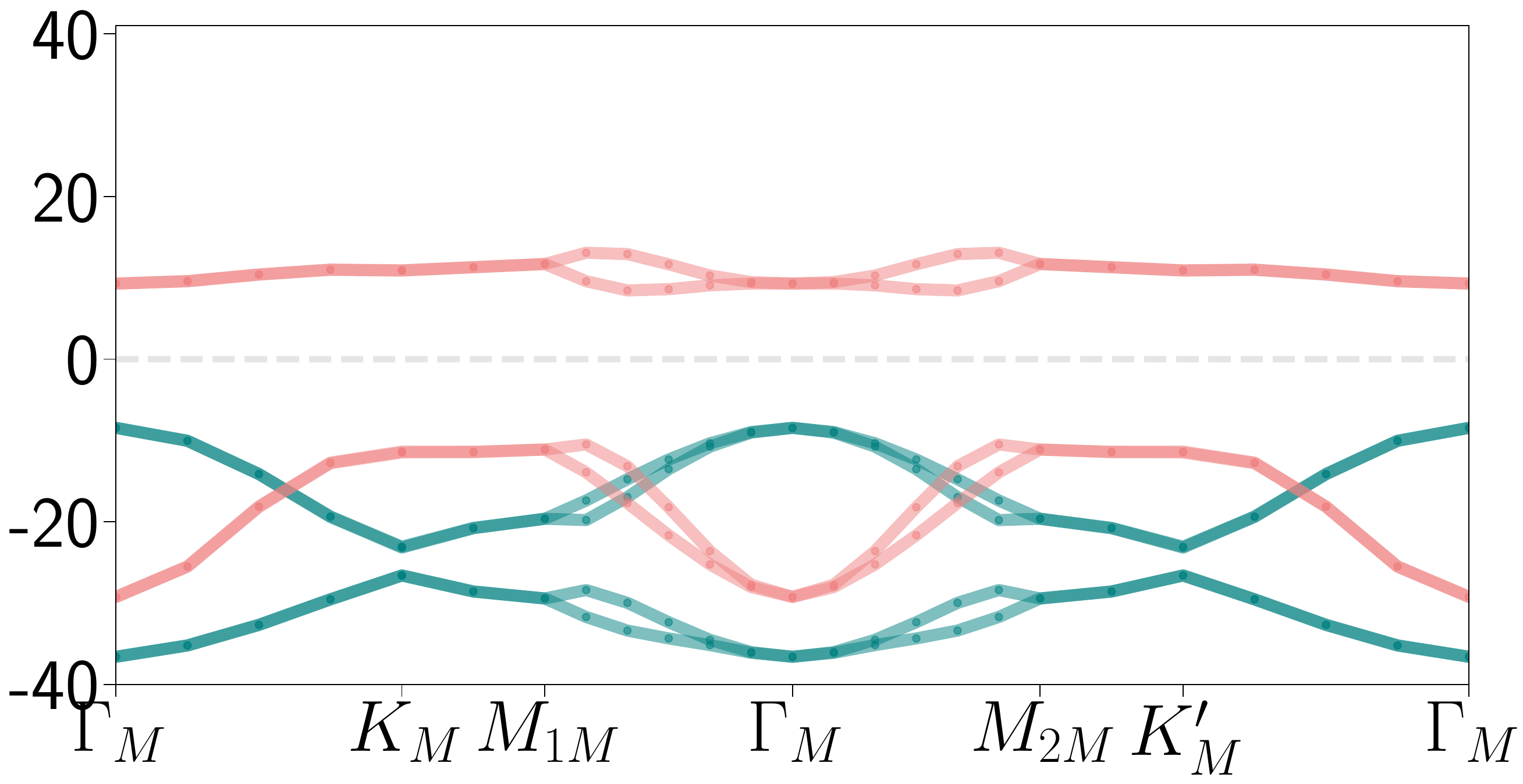}\\
    \centering \large{$\epsilon=50, \ U=0.5$ eV}\\
    \centering
    % \begin{subfigure}{.3\linewidth}
    % \centering
    \includegraphics[width=.265\linewidth]{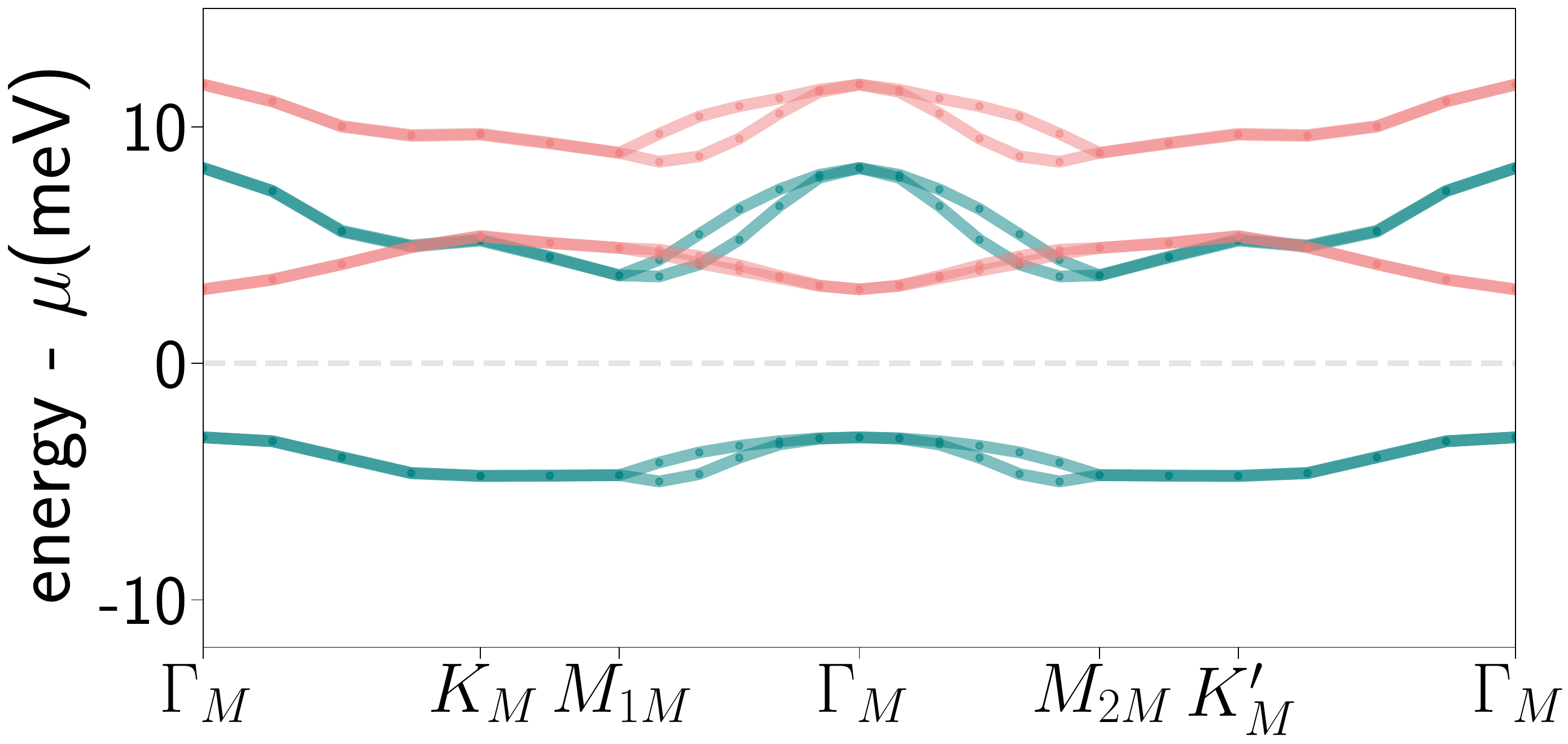}
    \includegraphics[width=.25\linewidth]{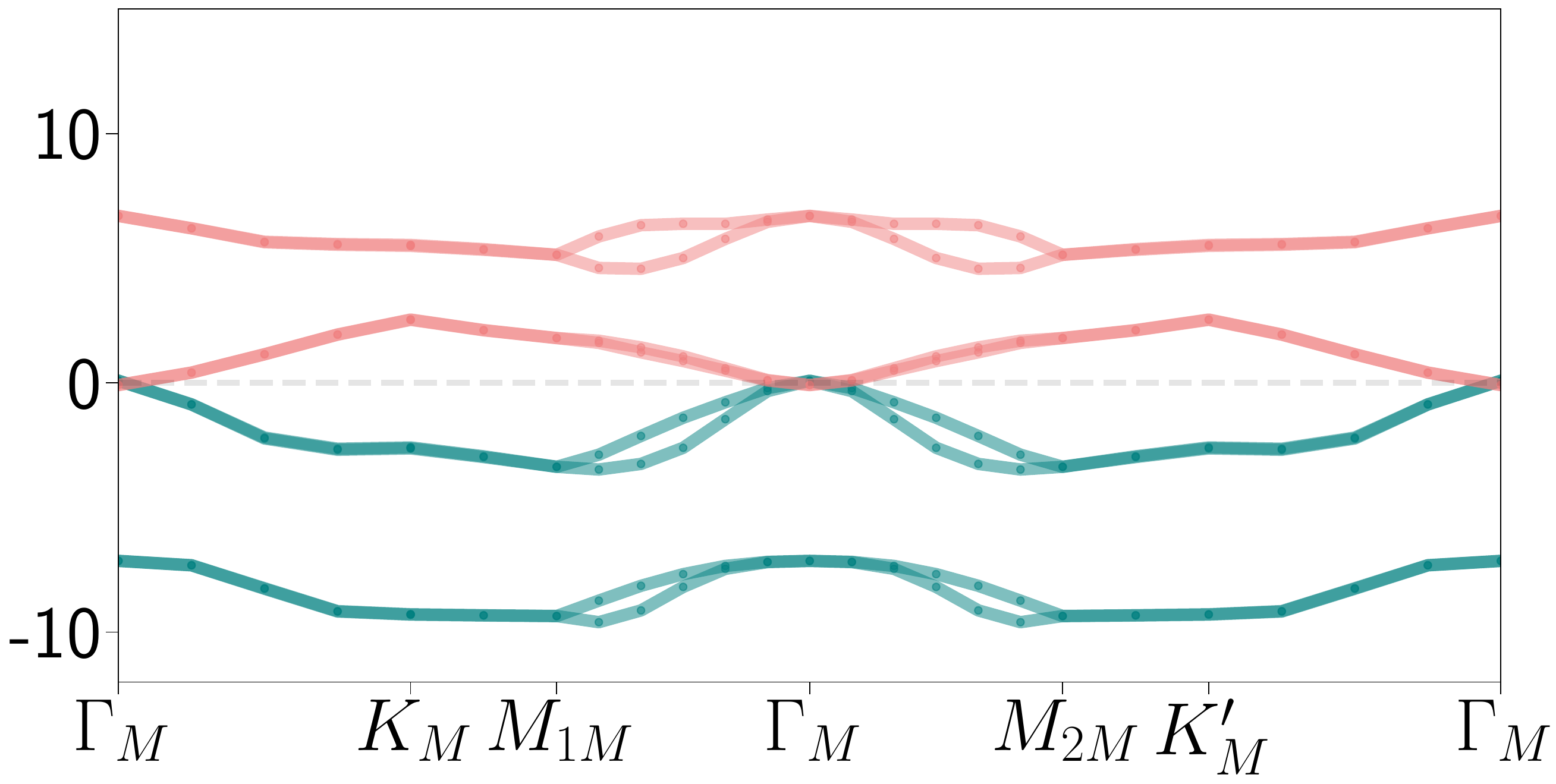}
    \includegraphics[width=.25\linewidth]{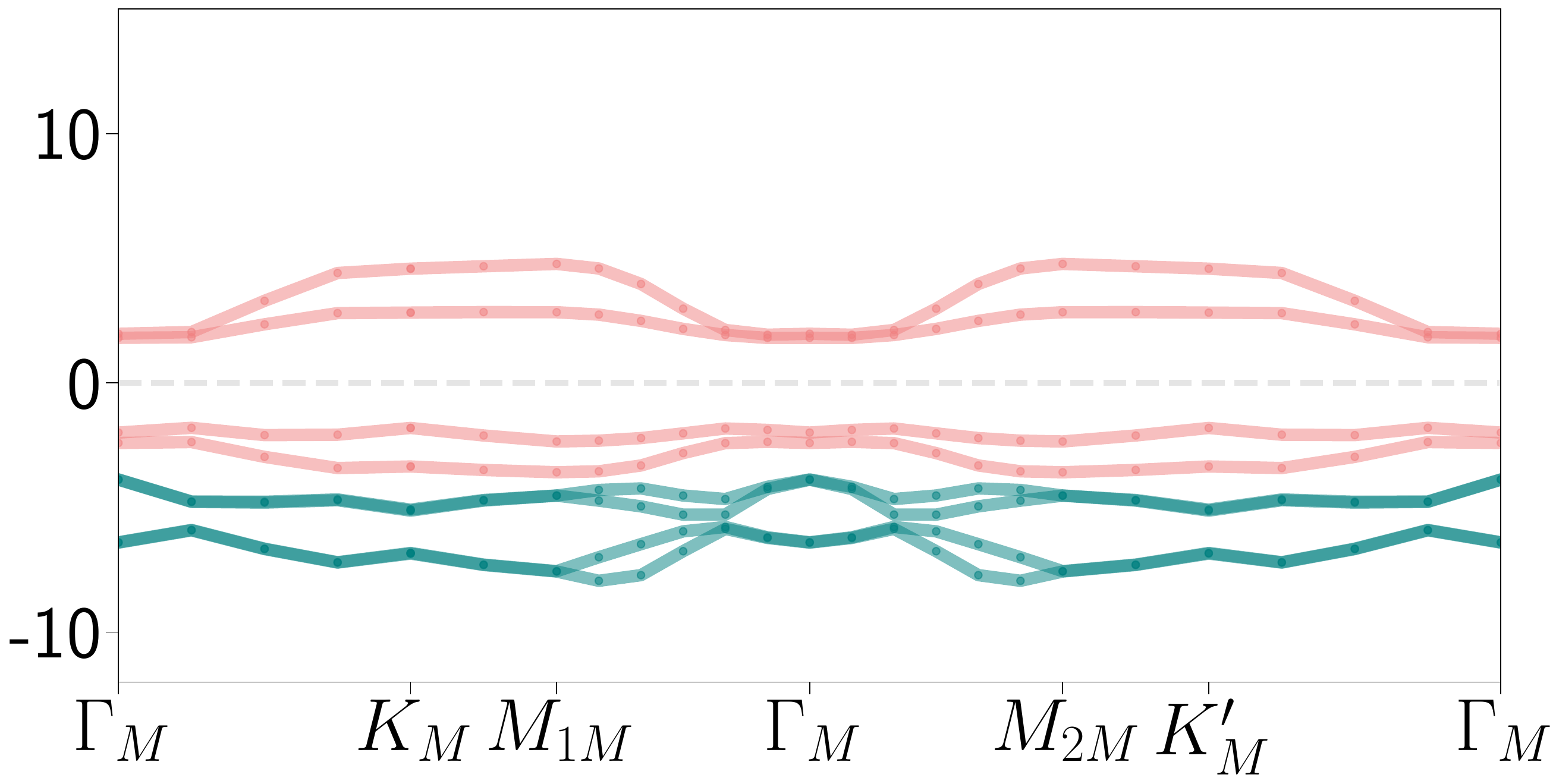}
    \caption{\textbf{Quasiparticle band structures of MATBG at $\boldsymbol{B=26.5}$ T.} Spin up bands are shown in blue and spin down bands in red. From left to right, the filling factor is $\nu=-2,0,+2$. The states are metallic at charge neutrality and insulating for electron and hole doping. When $\epsilon=10, U=4$ eV the $\nu=\pm 2$ states are related by a particle-hole transformation.} 
    \label{bandshfp1}
\end{figure}

% \textbf{Results for $\boldsymbol{\Phi=\Phi_0}$.}
% \centering \textbf{Results for $\boldsymbol{\Phi=\Phi_0}$}\\
The energies of self-consistent states for $\Phi=\Phi_0$ are tabulated in Table \ref{energieshfp1} of Appendix \ref{appf}, and their band structures plotted in Fig. \ref{bandshfp1}. The spin polarization is maximal for all fillings, however, we observe gapped phases with a completely different order compared to the previously predicted for $\nu= \pm 2$.

The order parameter
\begin{align}
    \Delta_{xyz,0} =& \langle \lambda_x \tau_0 \rangle^2 +  \langle \lambda_y \tau_0 \rangle^2 +  \langle \lambda_z \tau_0  \rangle^2, 
    % \nonumber \\
    % \Delta_{xyz,z} =& \langle \lambda_x \tau_z \rangle^2 +  \langle \lambda_y \tau_z \rangle^2 +  \langle \lambda_z \tau_z \rangle^2,
    \label{xyz0}
\end{align}
which is invariant under the gauge ambiguity of Eq. \ref{irrepv}, is predominant at hole doping.
% The ambiguity in the basis choice of Eq. \ref{irrepv} mixes $\lambda_x$, $\lambda_y$ and $\lambda_z$, and the physical order parameters involve the sum of squares above.
$\Delta_{xyz,0}$ corresponds to choosing a particular irrep basis allowed by Eq. \ref{irrepv}, and filling the bands with the same irrep number of the two valleys. 

At electron doping, the state for $\epsilon=10$, $U=4$ eV is very similar to the many-body particle-hole partner of the $\nu=-2$ insulator, with the same order parameter. On the other hand, when $\epsilon$ is larger than some critical value between $30$ and $50$, see Appendix \ref{appe}, there is a phase transition that promotes the IVC state. For both $\epsilon=10$, $U=4$ eV and $\epsilon = 50$, $U=0.5$ eV the second largest order parameter is found to be $\Delta_{xyz,z}$, which is obtained after replacing $\tau_0$ by $\tau_z$ in Eq. \ref{xyz0}. The total 'spectral weight' is essentially saturated by the two terms.

\begin{figure}[t!] 
    % \noindent \ a) \hfill  \hspace{1cm} b)  \hfill \phantom{phantom}\\
    % \noindent \hspace{1.35cm} \large{$\epsilon=10$} \hfill  \hspace{1.4cm} \large{$\epsilon=50$}  \hfill \phantom{phantom}\\
    \centering
    \includegraphics[width=.4\linewidth]{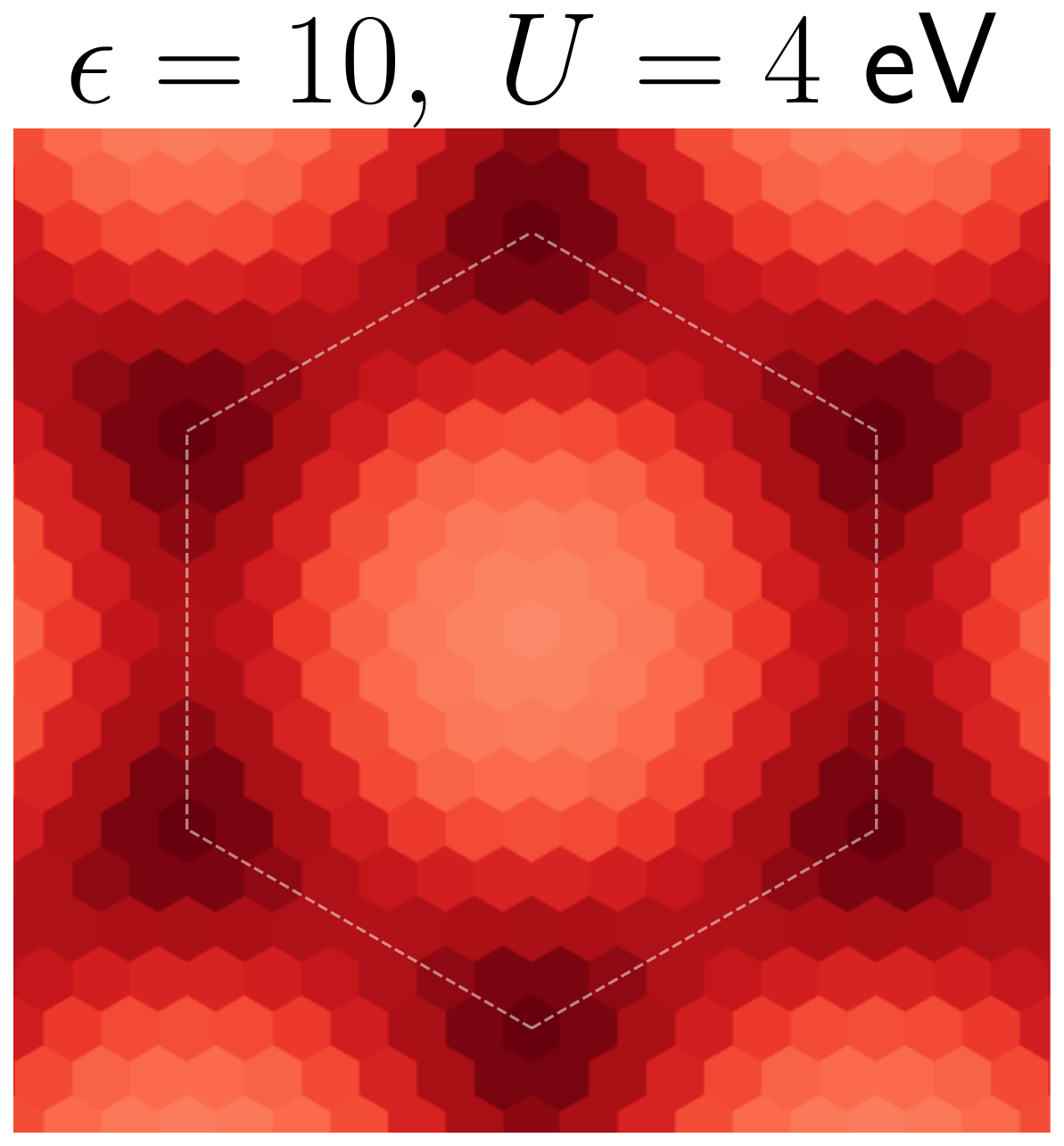}
    \includegraphics[width=.4\linewidth]{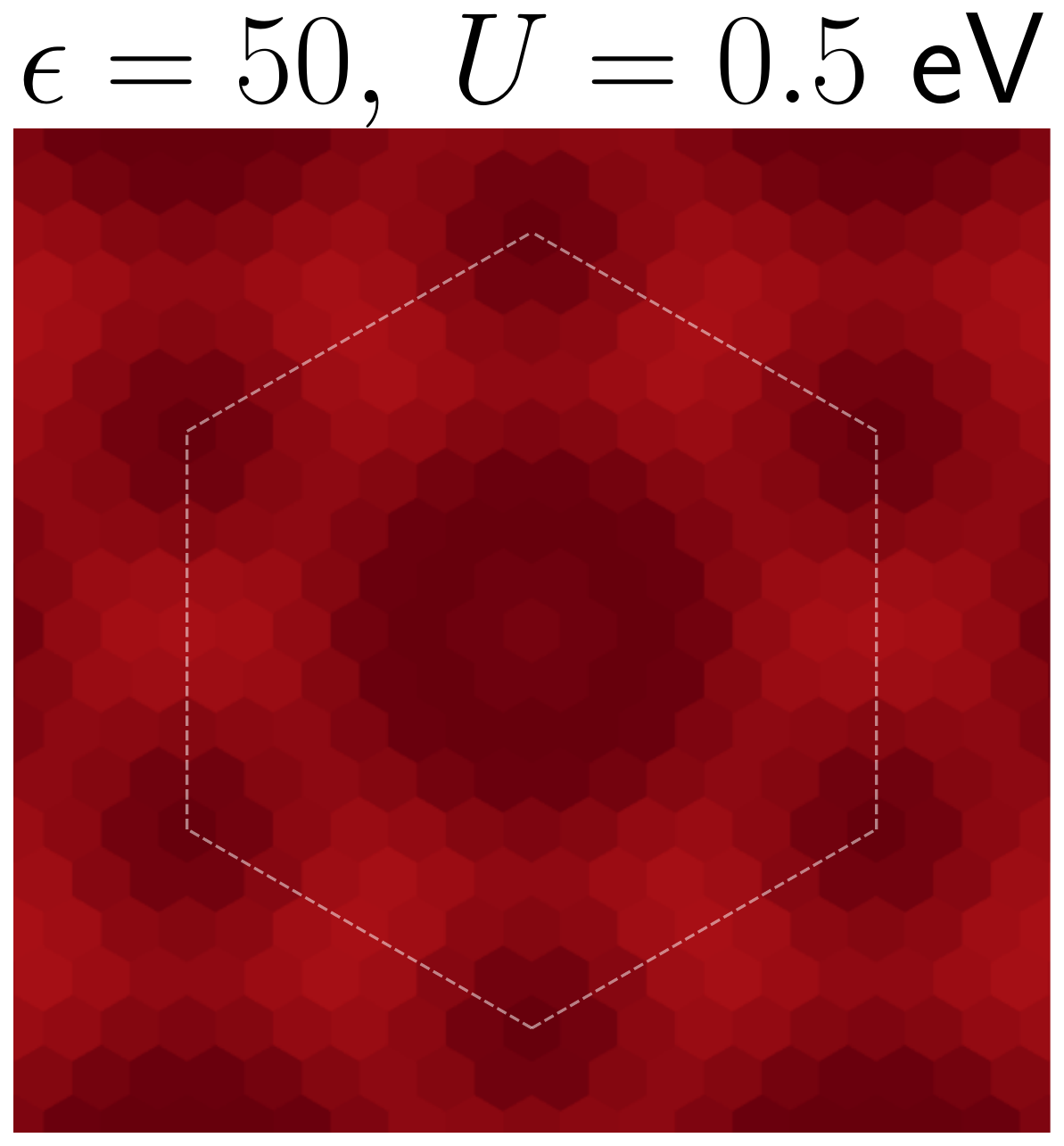}
    \includegraphics[width=.126\linewidth]{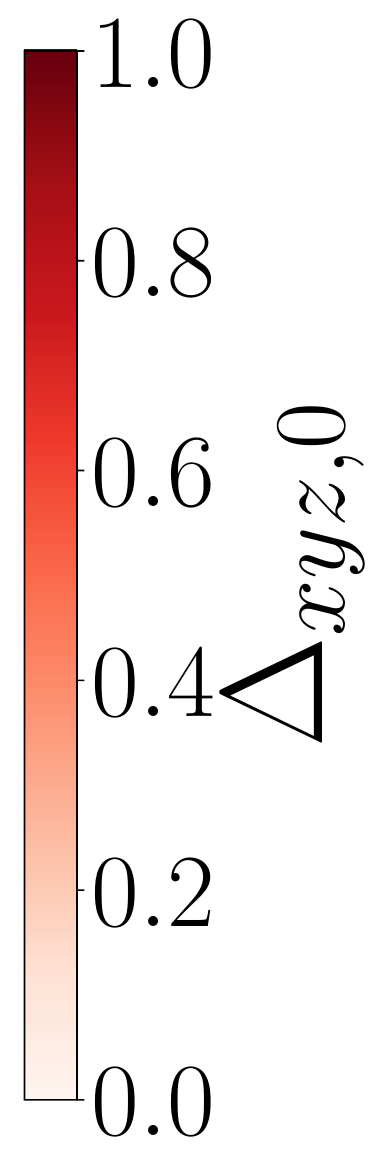}
    \caption{\textbf{The order parameter for $\boldsymbol{\nu=-2}$ at $\boldsymbol{\Phi=\Phi_0}$.} The dominant order parameter of the self-consistent states, $\Delta_{xyz,0}$, is shown. For $\epsilon=10$ it is very similar to its many-body partner at $\nu=+2$ in Fig. \ref{ordernu+2p1}. The $\Delta_{xyz,0}$ character is enhanced for weaker couplings.}
    \label{ordernu-2p1}
\end{figure}

\begin{figure}[t!] 
    % \noindent \ a) \hfill  \hspace{1cm} b)  \hfill \phantom{phantom}\\
    \centering
    \includegraphics[width=.47\linewidth]{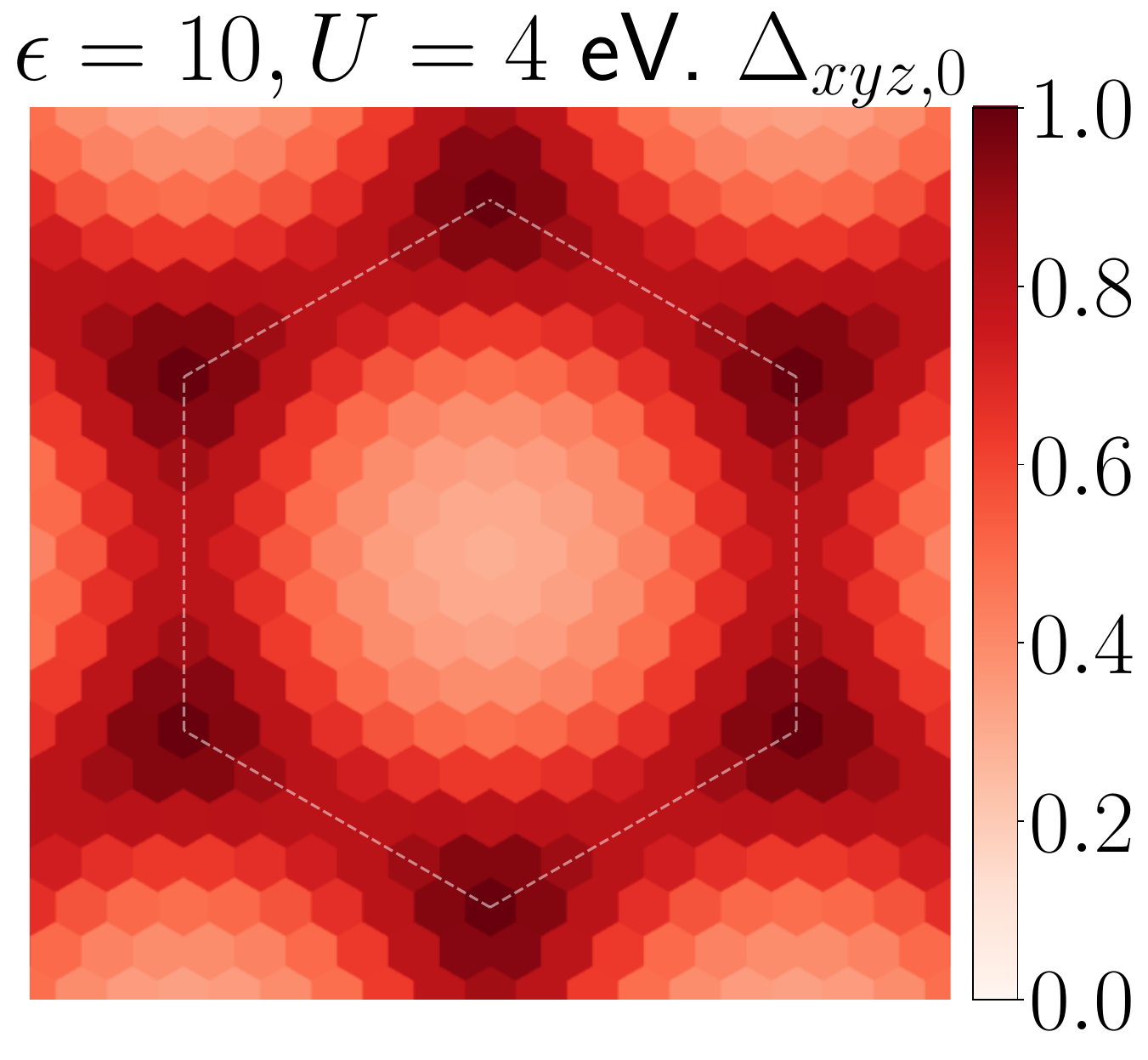}
    \includegraphics[width=.5\linewidth]{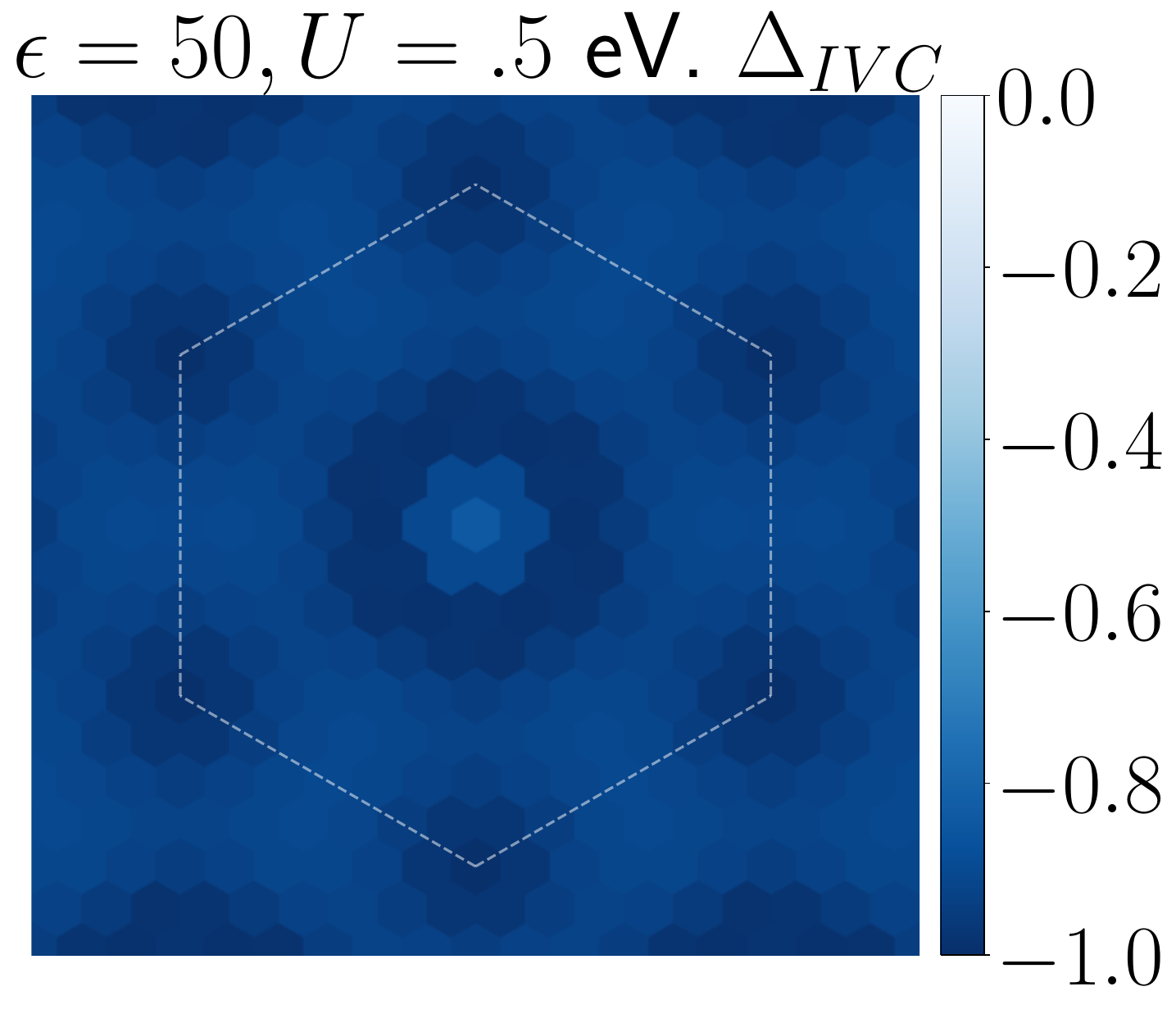}
    \caption{\textbf{Phase transition at $\boldsymbol{\nu=+2}$ and $\boldsymbol{\Phi=\Phi_0}$.} The order parameters of the self-consistent states are shown for $\epsilon=10$, $U=4$ eV and $\epsilon=50$, $U=0.5$ eV. For some value of $\epsilon$ between $50$ and $30$ there is a transition from the IVC to the $\Delta_{xyz,0}$ state.}
    \label{ordernu+2p1}
\end{figure}

Let us discuss the discrepancies between our numerical results and the results of Ref. \cite{herzog22_3}. Firstly, the kinetic spectrum at $B=26.5$ T is much broader than at zero field, with a gap between the valence and conduction bands of about $4$ meV. The system is in an intermediate coupling regime as the kinetic energy is comparable to the interaction energy. Secondly, at $B=0$ T the order is exchange-driven (at least at fillings $\nu=\pm2$ when the Hubbard term is not relevant), and the Fock energy is optimized in the ground state. For $B=26.5$ T we find that the direct term is more influential. In Ref. \cite{herzog22_3} the 'flat metric condition' assumes that the Hartree energy is trivial, and the strong coupling analysis suggests that the kinetic energy is a subdominant scale. Both premises are at odds with our numerical results in the tight-binding model.

The ground state at hole doping has Chern number -2 and is adiabatically connected to the ground state of the non interacting Hamiltonian. The predominant order parameter, $\Delta_{xyz,0}$, is plotted in Fig. \ref{ordernu-2p1}. In the following table we write the energy gain of the $\nu=-2$ ground states with respect to the candidate IVC state for $\epsilon=10$, $U=4$ eV and $\epsilon=50$, $U=0.5$ eV, in meV per unit cell ($U$ is expressed in eV).
\begin{table}[H]
\centering
$\nu=-2$\\
\begin{tabular}{SSSSSSSS} \toprule
    {$\epsilon,\ U$} &{$\epsilon \times$Hartree}& {$\epsilon \times$Fock} & {Hubb./$U$} 
    % & {Hartree} & {Fock} 
    & {kinetic} & {total}\\ \midrule
     {$10,\ 4$}  & {$-100.27$} & {$-41.38$} & {$-0.13$} 
     %&{$-10.03$} & {$-4.14$} 
     & {$-2.89$} & {$-17.58$}\\
     {$50,\ 0.5$} & {$-105.80$}  & {$19.14$}  &{$-0.15$} 
     % &{$-2.12$} & {$0.38$} 
     & {$-4.94$} & {$-6.75$}\\ \bottomrule
     % \\ \bottomrule
\end{tabular}
\end{table}
Clearly, the Hartree and kinetic energies drive the system towards a different order to the previously proposed.

On the other hand, we write in the table below the energy differences of the self-consistent states for $\nu=+2$ and the pure IVC state. 
\begin{table}[H]
\centering
$\nu=+2$\\
\begin{tabular}{SSSSSSSS} \toprule
    {$\epsilon,\ U$} &{$\epsilon \times$Hartree}& {$\epsilon \times$Fock}& {Hubb./$U$}  
    % & {Hartree} & {Fock} 
    &  {kinetic} & {total}\\ \midrule
    {$10,\ 4$}  & {$-155.20$} & {$-3.40$} & {$-0.26$} 
    %& {$-15.52$} & {$-0.34$} 
     &  {$5.21$} & {$-11.69$}\\
     {$50,\ 0.5$} & {$-1.65$}  & {$-9.29$} & {$0.00$} 
     % &{$-0.03$} & {$-0.19$} 
     & {$-0.19$} & {$-0.41$}\\ \bottomrule
\end{tabular}
\end{table}
In this case, we conclude that the interplay between the Hartree and kinetic energies determines the transition between the two competing orders, with their order parameters plotted in Fig. \ref{ordernu+2p1}. The phase transition at intermediate values of $\epsilon$ is topological, since the state for $\epsilon=10$, $U=4$ eV is found to have Chern number $+2$, and the IVC state is trivial with Chern number $0$. 

\onecolumngrid

\twocolumngrid

\section{CONCLUSIONS}

In this work, we have studied magic angle twisted bilayer graphene under external magnetic fluxes of zero and one flux quantum per unit cell, and dopings of $-2$, $0$ and $+2$ electrons per unit cell. In the atomistic tight-binding model we capture details of the physics that are not available in continuum theories.

For instance, we have established numerically and analytically  that the long ranged Coulomb interaction contributes to an antiferromagnetic intervalley Hund's coupling. The on-site Hubbard term, which is generally not included in the continuum model\cite{jimenopozo2023short}, contributes to a ferromagnetic coupling. The character of these valley exchanging interactions is important for the superconducting order parameter in Moiré systems\cite{scheurer20,Chatterjee2022,Lee2019}.

In addition, we studied the particle-hole asymmetry of the Hilbert space of the flat bands, and concluded that p-h breaking effects represent a small energy scale in the problem. In particular, the effects of p-h breaking cannot stabilize the time reversal intervalley coherent (TIVC) state at $\nu=-2$ observed recently\cite{nuckolls2023quantum}, and it must rely on other mechanisms like electron-phonon coupling\cite{Kwan23,chen2023strong}.

In order to elucidate the nature of the ground state we have performed self-consistent Hartree-Fock simulations. 
For $\Phi=0$, the Kramers intervalley coherent state is the ground state at electron and hole doping, and is competitive with the spin polarized state, supported by the Hubbard energy, at charge neutrality. In contrast to the BM model, where the self-consistent KIVC is favoured by kinetic energy superexchange\cite{Kang19,vafek20,bernevig421,Bultinck20,Kwan23,Bultinck21}, in the tight-binding model the inter-Chern order at $\Gamma_M$ lowers the exchange at the expense of kinetic and Hartree energies. Also, in the BM model the on-site Hubbard interaction is generally not included\cite{jimenopozo2023short}, and the SP state is never a candidate. We predict a phase transition from the KIVC to the SP at a critical value of $U_c=4.51$ eV when $\epsilon=10$. Finally, we also detect the many-body electron-hole asymmetry found in the experiments\cite{Lu2019,Pierce2021,Yankowitz19}. %another effect that is not properly included in the BM model\cite{Kwan23,vafek23}.

For $\Phi_0$ flux, the self-consistent states are spin polarized due to the Zeeman effect. Also, the Dirac cones are gapped and the bandwidth of the flat bands is increased, driving the system to an intermediate coupling regime. There are correlated insulators for $\nu=\pm 2$, and at charge neutrality the state is (almost) fully spin polarized. For electron doping and small screening, we observe an insulator with Chern number $+2$, consistent with experiments\cite{efetov22}, and a trivial insulator for large screening. We envision the possibility of observing a topological phase transition by changing the screening environment in the experimental setup.

Strain\cite{Kwan23,parker21} and electron-phonon coupling\cite{Bultinck21,blason22,chen2023strong} are relevant to the physics and their treatment is left for future work.  
Also, an accurate understanding of internal screening is needed, and it has been suggested that it is large in MATBG\cite{goodwin19,stauber21}. Improvements of the Hartree-Fock method such as the inclusion of more bands in the projection, and specially allowing for coherence between spins, like in the KIVC 'triplet' state, will further refine the outcome.

Finally, our results shed light into the correlated states in magic angle twisted bilayer graphene with precise atomic-scale modelling, both at $B=0$ and $26.5$ T. Furthermore, our calculations constitute the first self-consistent study of the correlated phases in MATBG under one magnetic flux quantum.

\section{Acknowledgements}
This work has been supported by MICINN (Spain) under Grant No. PID2020-113164GBI00, as well as by the CSIC Research Platform on Quantum Technologies PTI-001. The access to computational resources of CESGA (Centro de
Supercomputación de Galicia) is also gratefully acknowledged.

% \clearpage
% \nocite{*}
\bibliography{apssamp}% Produces the bibliography via BibTeX.

%apsrev4-2.bst 2019-01-14 (MD) hand-edited version of apsrev4-1.bst
%Control: key (0)
%Control: author (8) initials jnrlst
%Control: editor formatted (1) identically to author
%Control: production of article title (0) allowed
%Control: page (0) single
%Control: year (1) truncated
%Control: production of eprint (0) enabled
\providecommand{\noopsort}[1]{}\providecommand{\singleletter}[1]{#1}%
\begin{thebibliography}{76}%
\makeatletter
\providecommand \@ifxundefined [1]{%
 \@ifx{#1\undefined}
}%
\providecommand \@ifnum [1]{%
 \ifnum #1\expandafter \@firstoftwo
 \else \expandafter \@secondoftwo
 \fi
}%
\providecommand \@ifx [1]{%
 \ifx #1\expandafter \@firstoftwo
 \else \expandafter \@secondoftwo
 \fi
}%
\providecommand \natexlab [1]{#1}%
\providecommand \enquote  [1]{``#1''}%
\providecommand \bibnamefont  [1]{#1}%
\providecommand \bibfnamefont [1]{#1}%
\providecommand \citenamefont [1]{#1}%
\providecommand \href@noop [0]{\@secondoftwo}%
\providecommand \href [0]{\begingroup \@sanitize@url \@href}%
\providecommand \@href[1]{\@@startlink{#1}\@@href}%
\providecommand \@@href[1]{\endgroup#1\@@endlink}%
\providecommand \@sanitize@url [0]{\catcode `\\12\catcode `\$12\catcode
  `\&12\catcode `\#12\catcode `\^12\catcode `\_12\catcode `\%12\relax}%
\providecommand \@@startlink[1]{}%
\providecommand \@@endlink[0]{}%
\providecommand \url  [0]{\begingroup\@sanitize@url \@url }%
\providecommand \@url [1]{\endgroup\@href {#1}{\urlprefix }}%
\providecommand \urlprefix  [0]{URL }%
\providecommand \Eprint [0]{\href }%
\providecommand \doibase [0]{https://doi.org/}%
\providecommand \selectlanguage [0]{\@gobble}%
\providecommand \bibinfo  [0]{\@secondoftwo}%
\providecommand \bibfield  [0]{\@secondoftwo}%
\providecommand \translation [1]{[#1]}%
\providecommand \BibitemOpen [0]{}%
\providecommand \bibitemStop [0]{}%
\providecommand \bibitemNoStop [0]{.\EOS\space}%
\providecommand \EOS [0]{\spacefactor3000\relax}%
\providecommand \BibitemShut  [1]{\csname bibitem#1\endcsname}%
\let\auto@bib@innerbib\@empty
%</preamble>
\bibitem [{\citenamefont {Polini}\ \emph {et~al.}(2022)\citenamefont {Polini},
  \citenamefont {Giazotto}, \citenamefont {Fong}, \citenamefont {Pop},
  \citenamefont {Schuck}, \citenamefont {Boccali}, \citenamefont {Signorelli},
  \citenamefont {D'Elia}, \citenamefont {Hadfield}, \citenamefont
  {Giovannetti}, \citenamefont {Rossini}, \citenamefont {Tredicucci},
  \citenamefont {Efetov}, \citenamefont {Koppens}, \citenamefont
  {Jarillo-Herrero}, \citenamefont {Grassellino},\ and\ \citenamefont
  {Pisignano}}]{Polini22}%
  \BibitemOpen
  \bibfield  {author} {\bibinfo {author} {\bibfnamefont {M.}~\bibnamefont
  {Polini}}, \bibinfo {author} {\bibfnamefont {F.}~\bibnamefont {Giazotto}},
  \bibinfo {author} {\bibfnamefont {K.~C.}\ \bibnamefont {Fong}}, \bibinfo
  {author} {\bibfnamefont {I.~M.}\ \bibnamefont {Pop}}, \bibinfo {author}
  {\bibfnamefont {C.}~\bibnamefont {Schuck}}, \bibinfo {author} {\bibfnamefont
  {T.}~\bibnamefont {Boccali}}, \bibinfo {author} {\bibfnamefont
  {G.}~\bibnamefont {Signorelli}}, \bibinfo {author} {\bibfnamefont
  {M.}~\bibnamefont {D'Elia}}, \bibinfo {author} {\bibfnamefont {R.~H.}\
  \bibnamefont {Hadfield}}, \bibinfo {author} {\bibfnamefont {V.}~\bibnamefont
  {Giovannetti}}, \bibinfo {author} {\bibfnamefont {D.}~\bibnamefont
  {Rossini}}, \bibinfo {author} {\bibfnamefont {A.}~\bibnamefont {Tredicucci}},
  \bibinfo {author} {\bibfnamefont {D.~K.}\ \bibnamefont {Efetov}}, \bibinfo
  {author} {\bibfnamefont {F.~H.~L.}\ \bibnamefont {Koppens}}, \bibinfo
  {author} {\bibfnamefont {P.}~\bibnamefont {Jarillo-Herrero}}, \bibinfo
  {author} {\bibfnamefont {A.}~\bibnamefont {Grassellino}},\ and\ \bibinfo
  {author} {\bibfnamefont {D.}~\bibnamefont {Pisignano}},\ }\href
  {https://doi.org/10.48550/ARXIV.2201.09260} {\bibinfo {title} {Materials and
  devices for fundamental quantum science and quantum technologies}} (\bibinfo
  {year} {2022})\BibitemShut {NoStop}%
\bibitem [{\citenamefont {Cao}\ \emph {et~al.}(2018{\natexlab{a}})\citenamefont
  {Cao}, \citenamefont {Fatemi}, \citenamefont {Fang}, \citenamefont
  {Watanabe}, \citenamefont {Taniguchi}, \citenamefont {Kaxiras},\ and\
  \citenamefont {Jarillo-Herrero}}]{Cao2018sup}%
  \BibitemOpen
  \bibfield  {author} {\bibinfo {author} {\bibfnamefont {Y.}~\bibnamefont
  {Cao}}, \bibinfo {author} {\bibfnamefont {V.}~\bibnamefont {Fatemi}},
  \bibinfo {author} {\bibfnamefont {S.}~\bibnamefont {Fang}}, \bibinfo {author}
  {\bibfnamefont {K.}~\bibnamefont {Watanabe}}, \bibinfo {author}
  {\bibfnamefont {T.}~\bibnamefont {Taniguchi}}, \bibinfo {author}
  {\bibfnamefont {E.}~\bibnamefont {Kaxiras}},\ and\ \bibinfo {author}
  {\bibfnamefont {P.}~\bibnamefont {Jarillo-Herrero}},\ }\bibfield  {title}
  {\bibinfo {title} {Unconventional superconductivity in magic-angle graphene
  superlattices},\ }\href {https://doi.org/10.1038/nature26160} {\bibfield
  {journal} {\bibinfo  {journal} {Nature}\ }\textbf {\bibinfo {volume} {556}},\
  \bibinfo {pages} {43} (\bibinfo {year} {2018}{\natexlab{a}})}\BibitemShut
  {NoStop}%
\bibitem [{\citenamefont {Yankowitz}\ \emph {et~al.}(2019)\citenamefont
  {Yankowitz}, \citenamefont {Chen}, \citenamefont {Polshyn}, \citenamefont
  {Zhang}, \citenamefont {Watanabe}, \citenamefont {Taniguchi}, \citenamefont
  {Graf}, \citenamefont {Young},\ and\ \citenamefont {Dean}}]{Yankowitz19}%
  \BibitemOpen
  \bibfield  {author} {\bibinfo {author} {\bibfnamefont {M.}~\bibnamefont
  {Yankowitz}}, \bibinfo {author} {\bibfnamefont {S.}~\bibnamefont {Chen}},
  \bibinfo {author} {\bibfnamefont {H.}~\bibnamefont {Polshyn}}, \bibinfo
  {author} {\bibfnamefont {Y.}~\bibnamefont {Zhang}}, \bibinfo {author}
  {\bibfnamefont {K.}~\bibnamefont {Watanabe}}, \bibinfo {author}
  {\bibfnamefont {T.}~\bibnamefont {Taniguchi}}, \bibinfo {author}
  {\bibfnamefont {D.}~\bibnamefont {Graf}}, \bibinfo {author} {\bibfnamefont
  {A.~F.}\ \bibnamefont {Young}},\ and\ \bibinfo {author} {\bibfnamefont
  {C.~R.}\ \bibnamefont {Dean}},\ }\bibfield  {title} {\bibinfo {title} {Tuning
  superconductivity in twisted bilayer graphene},\ }\href
  {https://doi.org/10.1126/science.aav1910} {\bibfield  {journal} {\bibinfo
  {journal} {Science}\ }\textbf {\bibinfo {volume} {363}},\ \bibinfo {pages}
  {1059} (\bibinfo {year} {2019})},\ \Eprint
  {https://arxiv.org/abs/https://www.science.org/doi/pdf/10.1126/science.aav1910}
  {https://www.science.org/doi/pdf/10.1126/science.aav1910} \BibitemShut
  {NoStop}%
\bibitem [{\citenamefont {Lu}\ \emph {et~al.}(2019)\citenamefont {Lu},
  \citenamefont {Stepanov}, \citenamefont {Yang}, \citenamefont {Xie},
  \citenamefont {Aamir}, \citenamefont {Das}, \citenamefont {Urgell},
  \citenamefont {Watanabe}, \citenamefont {Taniguchi}, \citenamefont {Zhang},
  \citenamefont {Bachtold}, \citenamefont {MacDonald},\ and\ \citenamefont
  {Efetov}}]{Lu2019}%
  \BibitemOpen
  \bibfield  {author} {\bibinfo {author} {\bibfnamefont {X.}~\bibnamefont
  {Lu}}, \bibinfo {author} {\bibfnamefont {P.}~\bibnamefont {Stepanov}},
  \bibinfo {author} {\bibfnamefont {W.}~\bibnamefont {Yang}}, \bibinfo {author}
  {\bibfnamefont {M.}~\bibnamefont {Xie}}, \bibinfo {author} {\bibfnamefont
  {M.~A.}\ \bibnamefont {Aamir}}, \bibinfo {author} {\bibfnamefont
  {I.}~\bibnamefont {Das}}, \bibinfo {author} {\bibfnamefont {C.}~\bibnamefont
  {Urgell}}, \bibinfo {author} {\bibfnamefont {K.}~\bibnamefont {Watanabe}},
  \bibinfo {author} {\bibfnamefont {T.}~\bibnamefont {Taniguchi}}, \bibinfo
  {author} {\bibfnamefont {G.}~\bibnamefont {Zhang}}, \bibinfo {author}
  {\bibfnamefont {A.}~\bibnamefont {Bachtold}}, \bibinfo {author}
  {\bibfnamefont {A.~H.}\ \bibnamefont {MacDonald}},\ and\ \bibinfo {author}
  {\bibfnamefont {D.~K.}\ \bibnamefont {Efetov}},\ }\bibfield  {title}
  {\bibinfo {title} {Superconductors, orbital magnets and correlated states in
  magic-angle bilayer graphene},\ }\href
  {https://doi.org/10.1038/s41586-019-1695-0} {\bibfield  {journal} {\bibinfo
  {journal} {Nature}\ }\textbf {\bibinfo {volume} {574}},\ \bibinfo {pages}
  {653} (\bibinfo {year} {2019})}\BibitemShut {NoStop}%
\bibitem [{\citenamefont {Liu}\ \emph {et~al.}(2021)\citenamefont {Liu},
  \citenamefont {Wang}, \citenamefont {Watanabe}, \citenamefont {Taniguchi},
  \citenamefont {Vafek},\ and\ \citenamefont {Li}}]{liu21}%
  \BibitemOpen
  \bibfield  {author} {\bibinfo {author} {\bibfnamefont {X.}~\bibnamefont
  {Liu}}, \bibinfo {author} {\bibfnamefont {Z.}~\bibnamefont {Wang}}, \bibinfo
  {author} {\bibfnamefont {K.}~\bibnamefont {Watanabe}}, \bibinfo {author}
  {\bibfnamefont {T.}~\bibnamefont {Taniguchi}}, \bibinfo {author}
  {\bibfnamefont {O.}~\bibnamefont {Vafek}},\ and\ \bibinfo {author}
  {\bibfnamefont {J.~I.~A.}\ \bibnamefont {Li}},\ }\bibfield  {title} {\bibinfo
  {title} {Tuning electron correlation in magic-angle twisted bilayer graphene
  using coulomb screening},\ }\href {https://doi.org/10.1126/science.abb8754}
  {\bibfield  {journal} {\bibinfo  {journal} {Science}\ }\textbf {\bibinfo
  {volume} {371}},\ \bibinfo {pages} {1261} (\bibinfo {year} {2021})},\ \Eprint
  {https://arxiv.org/abs/https://www.science.org/doi/pdf/10.1126/science.abb8754}
  {https://www.science.org/doi/pdf/10.1126/science.abb8754} \BibitemShut
  {NoStop}%
\bibitem [{\citenamefont {Jaoui}\ \emph {et~al.}(2022)\citenamefont {Jaoui},
  \citenamefont {Das}, \citenamefont {Di~Battista}, \citenamefont
  {D{\'i}ez-M{\'e}rida}, \citenamefont {Lu}, \citenamefont {Watanabe},
  \citenamefont {Taniguchi}, \citenamefont {Ishizuka}, \citenamefont
  {Levitov},\ and\ \citenamefont {Efetov}}]{Jaoui2022}%
  \BibitemOpen
  \bibfield  {author} {\bibinfo {author} {\bibfnamefont {A.}~\bibnamefont
  {Jaoui}}, \bibinfo {author} {\bibfnamefont {I.}~\bibnamefont {Das}}, \bibinfo
  {author} {\bibfnamefont {G.}~\bibnamefont {Di~Battista}}, \bibinfo {author}
  {\bibfnamefont {J.}~\bibnamefont {D{\'i}ez-M{\'e}rida}}, \bibinfo {author}
  {\bibfnamefont {X.}~\bibnamefont {Lu}}, \bibinfo {author} {\bibfnamefont
  {K.}~\bibnamefont {Watanabe}}, \bibinfo {author} {\bibfnamefont
  {T.}~\bibnamefont {Taniguchi}}, \bibinfo {author} {\bibfnamefont
  {H.}~\bibnamefont {Ishizuka}}, \bibinfo {author} {\bibfnamefont
  {L.}~\bibnamefont {Levitov}},\ and\ \bibinfo {author} {\bibfnamefont {D.~K.}\
  \bibnamefont {Efetov}},\ }\bibfield  {title} {\bibinfo {title} {Quantum
  critical behaviour in magic-angle twisted bilayer graphene},\ }\href
  {https://doi.org/10.1038/s41567-022-01556-5} {\bibfield  {journal} {\bibinfo
  {journal} {Nature Physics}\ }\textbf {\bibinfo {volume} {18}},\ \bibinfo
  {pages} {633} (\bibinfo {year} {2022})}\BibitemShut {NoStop}%
\bibitem [{\citenamefont {Cao}\ \emph {et~al.}(2020)\citenamefont {Cao},
  \citenamefont {Chowdhury}, \citenamefont {Rodan-Legrain}, \citenamefont
  {Rubies-Bigorda}, \citenamefont {Watanabe}, \citenamefont {Taniguchi},
  \citenamefont {Senthil},\ and\ \citenamefont {Jarillo-Herrero}}]{cao20}%
  \BibitemOpen
  \bibfield  {author} {\bibinfo {author} {\bibfnamefont {Y.}~\bibnamefont
  {Cao}}, \bibinfo {author} {\bibfnamefont {D.}~\bibnamefont {Chowdhury}},
  \bibinfo {author} {\bibfnamefont {D.}~\bibnamefont {Rodan-Legrain}}, \bibinfo
  {author} {\bibfnamefont {O.}~\bibnamefont {Rubies-Bigorda}}, \bibinfo
  {author} {\bibfnamefont {K.}~\bibnamefont {Watanabe}}, \bibinfo {author}
  {\bibfnamefont {T.}~\bibnamefont {Taniguchi}}, \bibinfo {author}
  {\bibfnamefont {T.}~\bibnamefont {Senthil}},\ and\ \bibinfo {author}
  {\bibfnamefont {P.}~\bibnamefont {Jarillo-Herrero}},\ }\bibfield  {title}
  {\bibinfo {title} {Strange metal in magic-angle graphene with near planckian
  dissipation},\ }\href {https://doi.org/10.1103/PhysRevLett.124.076801}
  {\bibfield  {journal} {\bibinfo  {journal} {Phys. Rev. Lett.}\ }\textbf
  {\bibinfo {volume} {124}},\ \bibinfo {pages} {076801} (\bibinfo {year}
  {2020})}\BibitemShut {NoStop}%
\bibitem [{\citenamefont {Wu}\ \emph {et~al.}(2021)\citenamefont {Wu},
  \citenamefont {Zhang}, \citenamefont {Watanabe}, \citenamefont {Taniguchi},\
  and\ \citenamefont {Andrei}}]{Wu2021}%
  \BibitemOpen
  \bibfield  {author} {\bibinfo {author} {\bibfnamefont {S.}~\bibnamefont
  {Wu}}, \bibinfo {author} {\bibfnamefont {Z.}~\bibnamefont {Zhang}}, \bibinfo
  {author} {\bibfnamefont {K.}~\bibnamefont {Watanabe}}, \bibinfo {author}
  {\bibfnamefont {T.}~\bibnamefont {Taniguchi}},\ and\ \bibinfo {author}
  {\bibfnamefont {E.~Y.}\ \bibnamefont {Andrei}},\ }\bibfield  {title}
  {\bibinfo {title} {Chern insulators, van hove singularities and topological
  flat bands in magic-angle twisted bilayer graphene},\ }\href
  {https://doi.org/10.1038/s41563-020-00911-2} {\bibfield  {journal} {\bibinfo
  {journal} {Nature Materials}\ }\textbf {\bibinfo {volume} {20}},\ \bibinfo
  {pages} {488} (\bibinfo {year} {2021})}\BibitemShut {NoStop}%
\bibitem [{\citenamefont {Stepanov}\ \emph {et~al.}(2021)\citenamefont
  {Stepanov}, \citenamefont {Xie}, \citenamefont {Taniguchi}, \citenamefont
  {Watanabe}, \citenamefont {Lu}, \citenamefont {MacDonald}, \citenamefont
  {Bernevig},\ and\ \citenamefont {Efetov}}]{Stepanov21}%
  \BibitemOpen
  \bibfield  {author} {\bibinfo {author} {\bibfnamefont {P.}~\bibnamefont
  {Stepanov}}, \bibinfo {author} {\bibfnamefont {M.}~\bibnamefont {Xie}},
  \bibinfo {author} {\bibfnamefont {T.}~\bibnamefont {Taniguchi}}, \bibinfo
  {author} {\bibfnamefont {K.}~\bibnamefont {Watanabe}}, \bibinfo {author}
  {\bibfnamefont {X.}~\bibnamefont {Lu}}, \bibinfo {author} {\bibfnamefont
  {A.~H.}\ \bibnamefont {MacDonald}}, \bibinfo {author} {\bibfnamefont {B.~A.}\
  \bibnamefont {Bernevig}},\ and\ \bibinfo {author} {\bibfnamefont {D.~K.}\
  \bibnamefont {Efetov}},\ }\bibfield  {title} {\bibinfo {title} {Competing
  zero-field chern insulators in superconducting twisted bilayer graphene},\
  }\href {https://doi.org/10.1103/PhysRevLett.127.197701} {\bibfield  {journal}
  {\bibinfo  {journal} {Phys. Rev. Lett.}\ }\textbf {\bibinfo {volume} {127}},\
  \bibinfo {pages} {197701} (\bibinfo {year} {2021})}\BibitemShut {NoStop}%
\bibitem [{\citenamefont {Xie}\ \emph {et~al.}(2021)\citenamefont {Xie},
  \citenamefont {Pierce}, \citenamefont {Park}, \citenamefont {Parker},
  \citenamefont {Khalaf}, \citenamefont {Ledwith}, \citenamefont {Cao},
  \citenamefont {Lee}, \citenamefont {Chen}, \citenamefont {Forrester},
  \citenamefont {Watanabe}, \citenamefont {Taniguchi}, \citenamefont
  {Vishwanath}, \citenamefont {Jarillo-Herrero},\ and\ \citenamefont
  {Yacoby}}]{Xie2021}%
  \BibitemOpen
  \bibfield  {author} {\bibinfo {author} {\bibfnamefont {Y.}~\bibnamefont
  {Xie}}, \bibinfo {author} {\bibfnamefont {A.~T.}\ \bibnamefont {Pierce}},
  \bibinfo {author} {\bibfnamefont {J.~M.}\ \bibnamefont {Park}}, \bibinfo
  {author} {\bibfnamefont {D.~E.}\ \bibnamefont {Parker}}, \bibinfo {author}
  {\bibfnamefont {E.}~\bibnamefont {Khalaf}}, \bibinfo {author} {\bibfnamefont
  {P.}~\bibnamefont {Ledwith}}, \bibinfo {author} {\bibfnamefont
  {Y.}~\bibnamefont {Cao}}, \bibinfo {author} {\bibfnamefont {S.~H.}\
  \bibnamefont {Lee}}, \bibinfo {author} {\bibfnamefont {S.}~\bibnamefont
  {Chen}}, \bibinfo {author} {\bibfnamefont {P.~R.}\ \bibnamefont {Forrester}},
  \bibinfo {author} {\bibfnamefont {K.}~\bibnamefont {Watanabe}}, \bibinfo
  {author} {\bibfnamefont {T.}~\bibnamefont {Taniguchi}}, \bibinfo {author}
  {\bibfnamefont {A.}~\bibnamefont {Vishwanath}}, \bibinfo {author}
  {\bibfnamefont {P.}~\bibnamefont {Jarillo-Herrero}},\ and\ \bibinfo {author}
  {\bibfnamefont {A.}~\bibnamefont {Yacoby}},\ }\bibfield  {title} {\bibinfo
  {title} {Fractional chern insulators in magic-angle twisted bilayer
  graphene},\ }\href {https://doi.org/10.1038/s41586-021-04002-3} {\bibfield
  {journal} {\bibinfo  {journal} {Nature}\ }\textbf {\bibinfo {volume} {600}},\
  \bibinfo {pages} {439} (\bibinfo {year} {2021})}\BibitemShut {NoStop}%
\bibitem [{\citenamefont {Repellin}\ and\ \citenamefont
  {Senthil}(2019)}]{repellin2019chern}%
  \BibitemOpen
  \bibfield  {author} {\bibinfo {author} {\bibfnamefont {C.}~\bibnamefont
  {Repellin}}\ and\ \bibinfo {author} {\bibfnamefont {T.}~\bibnamefont
  {Senthil}},\ }\href@noop {} {\bibinfo {title} {Chern bands of twisted bilayer
  graphene: fractional chern insulators and spin phase transition}} (\bibinfo
  {year} {2019}),\ \Eprint {https://arxiv.org/abs/1912.11469} {arXiv:1912.11469
  [cond-mat.str-el]} \BibitemShut {NoStop}%
\bibitem [{\citenamefont {Ledwith}\ \emph {et~al.}(2020)\citenamefont
  {Ledwith}, \citenamefont {Tarnopolsky}, \citenamefont {Khalaf},\ and\
  \citenamefont {Vishwanath}}]{ledwith20}%
  \BibitemOpen
  \bibfield  {author} {\bibinfo {author} {\bibfnamefont {P.~J.}\ \bibnamefont
  {Ledwith}}, \bibinfo {author} {\bibfnamefont {G.}~\bibnamefont
  {Tarnopolsky}}, \bibinfo {author} {\bibfnamefont {E.}~\bibnamefont
  {Khalaf}},\ and\ \bibinfo {author} {\bibfnamefont {A.}~\bibnamefont
  {Vishwanath}},\ }\bibfield  {title} {\bibinfo {title} {Fractional chern
  insulator states in twisted bilayer graphene: An analytical approach},\
  }\href {https://doi.org/10.1103/PhysRevResearch.2.023237} {\bibfield
  {journal} {\bibinfo  {journal} {Phys. Rev. Res.}\ }\textbf {\bibinfo {volume}
  {2}},\ \bibinfo {pages} {023237} (\bibinfo {year} {2020})}\BibitemShut
  {NoStop}%
\bibitem [{\citenamefont {Chew}\ \emph {et~al.}(2023)\citenamefont {Chew},
  \citenamefont {Wang}, \citenamefont {Bernevig},\ and\ \citenamefont
  {Song}}]{chew23}%
  \BibitemOpen
  \bibfield  {author} {\bibinfo {author} {\bibfnamefont {A.}~\bibnamefont
  {Chew}}, \bibinfo {author} {\bibfnamefont {Y.}~\bibnamefont {Wang}}, \bibinfo
  {author} {\bibfnamefont {B.~A.}\ \bibnamefont {Bernevig}},\ and\ \bibinfo
  {author} {\bibfnamefont {Z.-D.}\ \bibnamefont {Song}},\ }\bibfield  {title}
  {\bibinfo {title} {Higher-order topological superconductivity in twisted
  bilayer graphene},\ }\href {https://doi.org/10.1103/PhysRevB.107.094512}
  {\bibfield  {journal} {\bibinfo  {journal} {Phys. Rev. B}\ }\textbf {\bibinfo
  {volume} {107}},\ \bibinfo {pages} {094512} (\bibinfo {year}
  {2023})}\BibitemShut {NoStop}%
\bibitem [{\citenamefont {Wang}\ \emph {et~al.}(2020)\citenamefont {Wang},
  \citenamefont {Shih}, \citenamefont {Ghiotto}, \citenamefont {Xian},
  \citenamefont {Rhodes}, \citenamefont {Tan}, \citenamefont {Claassen},
  \citenamefont {Kennes}, \citenamefont {Bai}, \citenamefont {Kim},
  \citenamefont {Watanabe}, \citenamefont {Taniguchi}, \citenamefont {Zhu},
  \citenamefont {Hone}, \citenamefont {Rubio}, \citenamefont {Pasupathy},\ and\
  \citenamefont {Dean}}]{Wang2020}%
  \BibitemOpen
  \bibfield  {author} {\bibinfo {author} {\bibfnamefont {L.}~\bibnamefont
  {Wang}}, \bibinfo {author} {\bibfnamefont {E.-M.}\ \bibnamefont {Shih}},
  \bibinfo {author} {\bibfnamefont {A.}~\bibnamefont {Ghiotto}}, \bibinfo
  {author} {\bibfnamefont {L.}~\bibnamefont {Xian}}, \bibinfo {author}
  {\bibfnamefont {D.~A.}\ \bibnamefont {Rhodes}}, \bibinfo {author}
  {\bibfnamefont {C.}~\bibnamefont {Tan}}, \bibinfo {author} {\bibfnamefont
  {M.}~\bibnamefont {Claassen}}, \bibinfo {author} {\bibfnamefont {D.~M.}\
  \bibnamefont {Kennes}}, \bibinfo {author} {\bibfnamefont {Y.}~\bibnamefont
  {Bai}}, \bibinfo {author} {\bibfnamefont {B.}~\bibnamefont {Kim}}, \bibinfo
  {author} {\bibfnamefont {K.}~\bibnamefont {Watanabe}}, \bibinfo {author}
  {\bibfnamefont {T.}~\bibnamefont {Taniguchi}}, \bibinfo {author}
  {\bibfnamefont {X.}~\bibnamefont {Zhu}}, \bibinfo {author} {\bibfnamefont
  {J.}~\bibnamefont {Hone}}, \bibinfo {author} {\bibfnamefont {A.}~\bibnamefont
  {Rubio}}, \bibinfo {author} {\bibfnamefont {A.~N.}\ \bibnamefont
  {Pasupathy}},\ and\ \bibinfo {author} {\bibfnamefont {C.~R.}\ \bibnamefont
  {Dean}},\ }\bibfield  {title} {\bibinfo {title} {Correlated electronic phases
  in twisted bilayer transition metal dichalcogenides},\ }\href
  {https://doi.org/10.1038/s41563-020-0708-6} {\bibfield  {journal} {\bibinfo
  {journal} {Nature Materials}\ }\textbf {\bibinfo {volume} {19}},\ \bibinfo
  {pages} {861} (\bibinfo {year} {2020})}\BibitemShut {NoStop}%
\bibitem [{\citenamefont {Park}\ \emph {et~al.}(2021)\citenamefont {Park},
  \citenamefont {Cao}, \citenamefont {Watanabe}, \citenamefont {Taniguchi},\
  and\ \citenamefont {Jarillo-Herrero}}]{Park2021}%
  \BibitemOpen
  \bibfield  {author} {\bibinfo {author} {\bibfnamefont {J.~M.}\ \bibnamefont
  {Park}}, \bibinfo {author} {\bibfnamefont {Y.}~\bibnamefont {Cao}}, \bibinfo
  {author} {\bibfnamefont {K.}~\bibnamefont {Watanabe}}, \bibinfo {author}
  {\bibfnamefont {T.}~\bibnamefont {Taniguchi}},\ and\ \bibinfo {author}
  {\bibfnamefont {P.}~\bibnamefont {Jarillo-Herrero}},\ }\bibfield  {title}
  {\bibinfo {title} {Tunable strongly coupled superconductivity in magic-angle
  twisted trilayer graphene},\ }\href
  {https://doi.org/10.1038/s41586-021-03192-0} {\bibfield  {journal} {\bibinfo
  {journal} {Nature}\ }\textbf {\bibinfo {volume} {590}},\ \bibinfo {pages}
  {249} (\bibinfo {year} {2021})}\BibitemShut {NoStop}%
\bibitem [{\citenamefont {Scheer}\ and\ \citenamefont
  {Lian}(2023)}]{scheer2023twistronics}%
  \BibitemOpen
  \bibfield  {author} {\bibinfo {author} {\bibfnamefont {M.~G.}\ \bibnamefont
  {Scheer}}\ and\ \bibinfo {author} {\bibfnamefont {B.}~\bibnamefont {Lian}},\
  }\href@noop {} {\bibinfo {title} {Twistronics of kekul\'e graphene: Honeycomb
  and kagome flat bands}} (\bibinfo {year} {2023}),\ \Eprint
  {https://arxiv.org/abs/2305.19927} {arXiv:2305.19927 [cond-mat.mes-hall]}
  \BibitemShut {NoStop}%
\bibitem [{\citenamefont {Crépel}\ \emph {et~al.}(2023)\citenamefont
  {Crépel}, \citenamefont {Dunbrack}, \citenamefont {Guerci}, \citenamefont
  {Bonini},\ and\ \citenamefont {Cano}}]{crépel2023chiral}%
  \BibitemOpen
  \bibfield  {author} {\bibinfo {author} {\bibfnamefont {V.}~\bibnamefont
  {Crépel}}, \bibinfo {author} {\bibfnamefont {A.}~\bibnamefont {Dunbrack}},
  \bibinfo {author} {\bibfnamefont {D.}~\bibnamefont {Guerci}}, \bibinfo
  {author} {\bibfnamefont {J.}~\bibnamefont {Bonini}},\ and\ \bibinfo {author}
  {\bibfnamefont {J.}~\bibnamefont {Cano}},\ }\href@noop {} {\bibinfo {title}
  {Chiral model of twisted bilayer graphene realized in a monolayer}} (\bibinfo
  {year} {2023}),\ \Eprint {https://arxiv.org/abs/2305.14423} {arXiv:2305.14423
  [cond-mat.mes-hall]} \BibitemShut {NoStop}%
\bibitem [{\citenamefont {Cao}\ \emph {et~al.}(2018{\natexlab{b}})\citenamefont
  {Cao}, \citenamefont {Fatemi}, \citenamefont {Demir}, \citenamefont {Fang},
  \citenamefont {Tomarken}, \citenamefont {Luo}, \citenamefont
  {Sanchez-Yamagishi}, \citenamefont {Watanabe}, \citenamefont {Taniguchi},
  \citenamefont {Kaxiras}, \citenamefont {Ashoori},\ and\ \citenamefont
  {Jarillo-Herrero}}]{Cao2018}%
  \BibitemOpen
  \bibfield  {author} {\bibinfo {author} {\bibfnamefont {Y.}~\bibnamefont
  {Cao}}, \bibinfo {author} {\bibfnamefont {V.}~\bibnamefont {Fatemi}},
  \bibinfo {author} {\bibfnamefont {A.}~\bibnamefont {Demir}}, \bibinfo
  {author} {\bibfnamefont {S.}~\bibnamefont {Fang}}, \bibinfo {author}
  {\bibfnamefont {S.~L.}\ \bibnamefont {Tomarken}}, \bibinfo {author}
  {\bibfnamefont {J.~Y.}\ \bibnamefont {Luo}}, \bibinfo {author} {\bibfnamefont
  {J.~D.}\ \bibnamefont {Sanchez-Yamagishi}}, \bibinfo {author} {\bibfnamefont
  {K.}~\bibnamefont {Watanabe}}, \bibinfo {author} {\bibfnamefont
  {T.}~\bibnamefont {Taniguchi}}, \bibinfo {author} {\bibfnamefont
  {E.}~\bibnamefont {Kaxiras}}, \bibinfo {author} {\bibfnamefont {R.~C.}\
  \bibnamefont {Ashoori}},\ and\ \bibinfo {author} {\bibfnamefont
  {P.}~\bibnamefont {Jarillo-Herrero}},\ }\bibfield  {title} {\bibinfo {title}
  {Correlated insulator behaviour at half-filling in magic-angle graphene
  superlattices},\ }\href {https://doi.org/10.1038/nature26154} {\bibfield
  {journal} {\bibinfo  {journal} {Nature}\ }\textbf {\bibinfo {volume} {556}},\
  \bibinfo {pages} {80} (\bibinfo {year} {2018}{\natexlab{b}})}\BibitemShut
  {NoStop}%
\bibitem [{\citenamefont {Po}\ \emph {et~al.}(2018)\citenamefont {Po},
  \citenamefont {Zou}, \citenamefont {Vishwanath},\ and\ \citenamefont
  {Senthil}}]{Po18}%
  \BibitemOpen
  \bibfield  {author} {\bibinfo {author} {\bibfnamefont {H.~C.}\ \bibnamefont
  {Po}}, \bibinfo {author} {\bibfnamefont {L.}~\bibnamefont {Zou}}, \bibinfo
  {author} {\bibfnamefont {A.}~\bibnamefont {Vishwanath}},\ and\ \bibinfo
  {author} {\bibfnamefont {T.}~\bibnamefont {Senthil}},\ }\bibfield  {title}
  {\bibinfo {title} {Origin of mott insulating behavior and superconductivity
  in twisted bilayer graphene},\ }\href
  {https://doi.org/10.1103/PhysRevX.8.031089} {\bibfield  {journal} {\bibinfo
  {journal} {Phys. Rev. X}\ }\textbf {\bibinfo {volume} {8}},\ \bibinfo {pages}
  {031089} (\bibinfo {year} {2018})}\BibitemShut {NoStop}%
\bibitem [{\citenamefont {Kang}\ and\ \citenamefont {Vafek}(2019)}]{Kang19}%
  \BibitemOpen
  \bibfield  {author} {\bibinfo {author} {\bibfnamefont {J.}~\bibnamefont
  {Kang}}\ and\ \bibinfo {author} {\bibfnamefont {O.}~\bibnamefont {Vafek}},\
  }\bibfield  {title} {\bibinfo {title} {Strong coupling phases of partially
  filled twisted bilayer graphene narrow bands},\ }\href
  {https://doi.org/10.1103/PhysRevLett.122.246401} {\bibfield  {journal}
  {\bibinfo  {journal} {Phys. Rev. Lett.}\ }\textbf {\bibinfo {volume} {122}},\
  \bibinfo {pages} {246401} (\bibinfo {year} {2019})}\BibitemShut {NoStop}%
\bibitem [{\citenamefont {Bultinck}\ \emph {et~al.}(2020)\citenamefont
  {Bultinck}, \citenamefont {Khalaf}, \citenamefont {Liu}, \citenamefont
  {Chatterjee}, \citenamefont {Vishwanath},\ and\ \citenamefont
  {Zaletel}}]{Bultinck20}%
  \BibitemOpen
  \bibfield  {author} {\bibinfo {author} {\bibfnamefont {N.}~\bibnamefont
  {Bultinck}}, \bibinfo {author} {\bibfnamefont {E.}~\bibnamefont {Khalaf}},
  \bibinfo {author} {\bibfnamefont {S.}~\bibnamefont {Liu}}, \bibinfo {author}
  {\bibfnamefont {S.}~\bibnamefont {Chatterjee}}, \bibinfo {author}
  {\bibfnamefont {A.}~\bibnamefont {Vishwanath}},\ and\ \bibinfo {author}
  {\bibfnamefont {M.~P.}\ \bibnamefont {Zaletel}},\ }\bibfield  {title}
  {\bibinfo {title} {Ground state and hidden symmetry of magic-angle graphene
  at even integer filling},\ }\href
  {https://doi.org/10.1103/PhysRevX.10.031034} {\bibfield  {journal} {\bibinfo
  {journal} {Phys. Rev. X}\ }\textbf {\bibinfo {volume} {10}},\ \bibinfo
  {pages} {031034} (\bibinfo {year} {2020})}\BibitemShut {NoStop}%
\bibitem [{\citenamefont {Kwan}\ \emph {et~al.}(2021)\citenamefont {Kwan},
  \citenamefont {Wagner}, \citenamefont {Soejima}, \citenamefont {Zaletel},
  \citenamefont {Simon}, \citenamefont {Parameswaran},\ and\ \citenamefont
  {Bultinck}}]{Bultinck21}%
  \BibitemOpen
  \bibfield  {author} {\bibinfo {author} {\bibfnamefont {Y.~H.}\ \bibnamefont
  {Kwan}}, \bibinfo {author} {\bibfnamefont {G.}~\bibnamefont {Wagner}},
  \bibinfo {author} {\bibfnamefont {T.}~\bibnamefont {Soejima}}, \bibinfo
  {author} {\bibfnamefont {M.~P.}\ \bibnamefont {Zaletel}}, \bibinfo {author}
  {\bibfnamefont {S.~H.}\ \bibnamefont {Simon}}, \bibinfo {author}
  {\bibfnamefont {S.~A.}\ \bibnamefont {Parameswaran}},\ and\ \bibinfo {author}
  {\bibfnamefont {N.}~\bibnamefont {Bultinck}},\ }\bibfield  {title} {\bibinfo
  {title} {Kekul\'e spiral order at all nonzero integer fillings in twisted
  bilayer graphene},\ }\href {https://doi.org/10.1103/PhysRevX.11.041063}
  {\bibfield  {journal} {\bibinfo  {journal} {Phys. Rev. X}\ }\textbf {\bibinfo
  {volume} {11}},\ \bibinfo {pages} {041063} (\bibinfo {year}
  {2021})}\BibitemShut {NoStop}%
\bibitem [{\citenamefont {Stepanov}\ \emph {et~al.}(2020)\citenamefont
  {Stepanov}, \citenamefont {Das}, \citenamefont {Lu}, \citenamefont
  {Fahimniya}, \citenamefont {Watanabe}, \citenamefont {Taniguchi},
  \citenamefont {Koppens}, \citenamefont {Lischner}, \citenamefont {Levitov},\
  and\ \citenamefont {Efetov}}]{Stepanov2020}%
  \BibitemOpen
  \bibfield  {author} {\bibinfo {author} {\bibfnamefont {P.}~\bibnamefont
  {Stepanov}}, \bibinfo {author} {\bibfnamefont {I.}~\bibnamefont {Das}},
  \bibinfo {author} {\bibfnamefont {X.}~\bibnamefont {Lu}}, \bibinfo {author}
  {\bibfnamefont {A.}~\bibnamefont {Fahimniya}}, \bibinfo {author}
  {\bibfnamefont {K.}~\bibnamefont {Watanabe}}, \bibinfo {author}
  {\bibfnamefont {T.}~\bibnamefont {Taniguchi}}, \bibinfo {author}
  {\bibfnamefont {F.~H.~L.}\ \bibnamefont {Koppens}}, \bibinfo {author}
  {\bibfnamefont {J.}~\bibnamefont {Lischner}}, \bibinfo {author}
  {\bibfnamefont {L.}~\bibnamefont {Levitov}},\ and\ \bibinfo {author}
  {\bibfnamefont {D.~K.}\ \bibnamefont {Efetov}},\ }\bibfield  {title}
  {\bibinfo {title} {Untying the insulating and superconducting orders in
  magic-angle graphene},\ }\href {https://doi.org/10.1038/s41586-020-2459-6}
  {\bibfield  {journal} {\bibinfo  {journal} {Nature}\ }\textbf {\bibinfo
  {volume} {583}},\ \bibinfo {pages} {375} (\bibinfo {year}
  {2020})}\BibitemShut {NoStop}%
\bibitem [{\citenamefont {Sharpe}\ \emph {et~al.}(2021)\citenamefont {Sharpe},
  \citenamefont {Fox}, \citenamefont {Barnard}, \citenamefont {Finney},
  \citenamefont {Watanabe}, \citenamefont {Taniguchi}, \citenamefont
  {Kastner},\ and\ \citenamefont {Goldhaber-Gordon}}]{Sharpe21}%
  \BibitemOpen
  \bibfield  {author} {\bibinfo {author} {\bibfnamefont {A.~L.}\ \bibnamefont
  {Sharpe}}, \bibinfo {author} {\bibfnamefont {E.~J.}\ \bibnamefont {Fox}},
  \bibinfo {author} {\bibfnamefont {A.~W.}\ \bibnamefont {Barnard}}, \bibinfo
  {author} {\bibfnamefont {J.}~\bibnamefont {Finney}}, \bibinfo {author}
  {\bibfnamefont {K.}~\bibnamefont {Watanabe}}, \bibinfo {author}
  {\bibfnamefont {T.}~\bibnamefont {Taniguchi}}, \bibinfo {author}
  {\bibfnamefont {M.~A.}\ \bibnamefont {Kastner}},\ and\ \bibinfo {author}
  {\bibfnamefont {D.}~\bibnamefont {Goldhaber-Gordon}},\ }\bibfield  {title}
  {\bibinfo {title} {Evidence of orbital ferromagnetism in twisted bilayer
  graphene aligned to hexagonal boron nitride},\ }\href
  {https://doi.org/10.1021/acs.nanolett.1c00696} {\bibfield  {journal}
  {\bibinfo  {journal} {Nano Letters}\ }\textbf {\bibinfo {volume} {21}},\
  \bibinfo {pages} {4299} (\bibinfo {year} {2021})},\ \bibinfo {note} {pMID:
  33970644},\ \Eprint
  {https://arxiv.org/abs/https://doi.org/10.1021/acs.nanolett.1c00696}
  {https://doi.org/10.1021/acs.nanolett.1c00696} \BibitemShut {NoStop}%
\bibitem [{\citenamefont {Sharpe}\ \emph {et~al.}(2019)\citenamefont {Sharpe},
  \citenamefont {Fox}, \citenamefont {Barnard}, \citenamefont {Finney},
  \citenamefont {Watanabe}, \citenamefont {Taniguchi}, \citenamefont
  {Kastner},\ and\ \citenamefont {Goldhaber-Gordon}}]{Sharpe2019}%
  \BibitemOpen
  \bibfield  {author} {\bibinfo {author} {\bibfnamefont {A.~L.}\ \bibnamefont
  {Sharpe}}, \bibinfo {author} {\bibfnamefont {E.~J.}\ \bibnamefont {Fox}},
  \bibinfo {author} {\bibfnamefont {A.~W.}\ \bibnamefont {Barnard}}, \bibinfo
  {author} {\bibfnamefont {J.}~\bibnamefont {Finney}}, \bibinfo {author}
  {\bibfnamefont {K.}~\bibnamefont {Watanabe}}, \bibinfo {author}
  {\bibfnamefont {T.}~\bibnamefont {Taniguchi}}, \bibinfo {author}
  {\bibfnamefont {M.~A.}\ \bibnamefont {Kastner}},\ and\ \bibinfo {author}
  {\bibfnamefont {D.}~\bibnamefont {Goldhaber-Gordon}},\ }\bibfield  {title}
  {\bibinfo {title} {Emergent ferromagnetism near three-quarters filling in
  twisted bilayer graphene},\ }\href {https://doi.org/10.1126/science.aaw3780}
  {\bibfield  {journal} {\bibinfo  {journal} {Science}\ }\textbf {\bibinfo
  {volume} {365}},\ \bibinfo {pages} {605} (\bibinfo {year}
  {2019})}\BibitemShut {NoStop}%
\bibitem [{\citenamefont {Zhang}\ \emph {et~al.}(2022)\citenamefont {Zhang},
  \citenamefont {Lu},\ and\ \citenamefont {Liu}}]{Zhang22}%
  \BibitemOpen
  \bibfield  {author} {\bibinfo {author} {\bibfnamefont {S.}~\bibnamefont
  {Zhang}}, \bibinfo {author} {\bibfnamefont {X.}~\bibnamefont {Lu}},\ and\
  \bibinfo {author} {\bibfnamefont {J.}~\bibnamefont {Liu}},\ }\bibfield
  {title} {\bibinfo {title} {Correlated insulators, density wave states, and
  their nonlinear optical response in magic-angle twisted bilayer graphene},\
  }\href {https://doi.org/10.1103/PhysRevLett.128.247402} {\bibfield  {journal}
  {\bibinfo  {journal} {Phys. Rev. Lett.}\ }\textbf {\bibinfo {volume} {128}},\
  \bibinfo {pages} {247402} (\bibinfo {year} {2022})}\BibitemShut {NoStop}%
\bibitem [{\citenamefont {Gonz\'alez}\ and\ \citenamefont
  {Stauber}(2021)}]{stauber21}%
  \BibitemOpen
  \bibfield  {author} {\bibinfo {author} {\bibfnamefont {J.}~\bibnamefont
  {Gonz\'alez}}\ and\ \bibinfo {author} {\bibfnamefont {T.}~\bibnamefont
  {Stauber}},\ }\bibfield  {title} {\bibinfo {title} {Magnetic phases from
  competing hubbard and extended coulomb interactions in twisted bilayer
  graphene},\ }\href {https://doi.org/10.1103/PhysRevB.104.115110} {\bibfield
  {journal} {\bibinfo  {journal} {Phys. Rev. B}\ }\textbf {\bibinfo {volume}
  {104}},\ \bibinfo {pages} {115110} (\bibinfo {year} {2021})}\BibitemShut
  {NoStop}%
\bibitem [{\citenamefont {Gonz\'alez}\ and\ \citenamefont
  {Stauber}(2020)}]{gonzalez20}%
  \BibitemOpen
  \bibfield  {author} {\bibinfo {author} {\bibfnamefont {J.}~\bibnamefont
  {Gonz\'alez}}\ and\ \bibinfo {author} {\bibfnamefont {T.}~\bibnamefont
  {Stauber}},\ }\bibfield  {title} {\bibinfo {title} {Time-reversal symmetry
  breaking versus chiral symmetry breaking in twisted bilayer graphene},\
  }\href {https://doi.org/10.1103/PhysRevB.102.081118} {\bibfield  {journal}
  {\bibinfo  {journal} {Phys. Rev. B}\ }\textbf {\bibinfo {volume} {102}},\
  \bibinfo {pages} {081118} (\bibinfo {year} {2020})}\BibitemShut {NoStop}%
\bibitem [{\citenamefont {Klebl}\ \emph {et~al.}(2021)\citenamefont {Klebl},
  \citenamefont {Goodwin}, \citenamefont {Mostofi}, \citenamefont {Kennes},\
  and\ \citenamefont {Lischner}}]{klebl21}%
  \BibitemOpen
  \bibfield  {author} {\bibinfo {author} {\bibfnamefont {L.}~\bibnamefont
  {Klebl}}, \bibinfo {author} {\bibfnamefont {Z.~A.~H.}\ \bibnamefont
  {Goodwin}}, \bibinfo {author} {\bibfnamefont {A.~A.}\ \bibnamefont
  {Mostofi}}, \bibinfo {author} {\bibfnamefont {D.~M.}\ \bibnamefont
  {Kennes}},\ and\ \bibinfo {author} {\bibfnamefont {J.}~\bibnamefont
  {Lischner}},\ }\bibfield  {title} {\bibinfo {title} {Importance of
  long-ranged electron-electron interactions for the magnetic phase diagram of
  twisted bilayer graphene},\ }\href
  {https://doi.org/10.1103/PhysRevB.103.195127} {\bibfield  {journal} {\bibinfo
   {journal} {Phys. Rev. B}\ }\textbf {\bibinfo {volume} {103}},\ \bibinfo
  {pages} {195127} (\bibinfo {year} {2021})}\BibitemShut {NoStop}%
\bibitem [{\citenamefont {Xie}\ and\ \citenamefont {MacDonald}(2020)}]{xie20}%
  \BibitemOpen
  \bibfield  {author} {\bibinfo {author} {\bibfnamefont {M.}~\bibnamefont
  {Xie}}\ and\ \bibinfo {author} {\bibfnamefont {A.~H.}\ \bibnamefont
  {MacDonald}},\ }\bibfield  {title} {\bibinfo {title} {Nature of the
  correlated insulator states in twisted bilayer graphene},\ }\href
  {https://doi.org/10.1103/PhysRevLett.124.097601} {\bibfield  {journal}
  {\bibinfo  {journal} {Phys. Rev. Lett.}\ }\textbf {\bibinfo {volume} {124}},\
  \bibinfo {pages} {097601} (\bibinfo {year} {2020})}\BibitemShut {NoStop}%
\bibitem [{\citenamefont {Faulstich}\ \emph {et~al.}(2023)\citenamefont
  {Faulstich}, \citenamefont {Stubbs}, \citenamefont {Zhu}, \citenamefont
  {Soejima}, \citenamefont {Dilip}, \citenamefont {Zhai}, \citenamefont {Kim},
  \citenamefont {Zaletel}, \citenamefont {Chan},\ and\ \citenamefont
  {Lin}}]{lin23}%
  \BibitemOpen
  \bibfield  {author} {\bibinfo {author} {\bibfnamefont {F.~M.}\ \bibnamefont
  {Faulstich}}, \bibinfo {author} {\bibfnamefont {K.~D.}\ \bibnamefont
  {Stubbs}}, \bibinfo {author} {\bibfnamefont {Q.}~\bibnamefont {Zhu}},
  \bibinfo {author} {\bibfnamefont {T.}~\bibnamefont {Soejima}}, \bibinfo
  {author} {\bibfnamefont {R.}~\bibnamefont {Dilip}}, \bibinfo {author}
  {\bibfnamefont {H.}~\bibnamefont {Zhai}}, \bibinfo {author} {\bibfnamefont
  {R.}~\bibnamefont {Kim}}, \bibinfo {author} {\bibfnamefont {M.~P.}\
  \bibnamefont {Zaletel}}, \bibinfo {author} {\bibfnamefont {G.~K.-L.}\
  \bibnamefont {Chan}},\ and\ \bibinfo {author} {\bibfnamefont
  {L.}~\bibnamefont {Lin}},\ }\bibfield  {title} {\bibinfo {title} {Interacting
  models for twisted bilayer graphene: A quantum chemistry approach},\ }\href
  {https://doi.org/10.1103/PhysRevB.107.235123} {\bibfield  {journal} {\bibinfo
   {journal} {Phys. Rev. B}\ }\textbf {\bibinfo {volume} {107}},\ \bibinfo
  {pages} {235123} (\bibinfo {year} {2023})}\BibitemShut {NoStop}%
\bibitem [{\citenamefont {Vafek}\ and\ \citenamefont {Kang}(2020)}]{vafek20}%
  \BibitemOpen
  \bibfield  {author} {\bibinfo {author} {\bibfnamefont {O.}~\bibnamefont
  {Vafek}}\ and\ \bibinfo {author} {\bibfnamefont {J.}~\bibnamefont {Kang}},\
  }\bibfield  {title} {\bibinfo {title} {Renormalization group study of hidden
  symmetry in twisted bilayer graphene with coulomb interactions},\ }\href
  {https://doi.org/10.1103/PhysRevLett.125.257602} {\bibfield  {journal}
  {\bibinfo  {journal} {Phys. Rev. Lett.}\ }\textbf {\bibinfo {volume} {125}},\
  \bibinfo {pages} {257602} (\bibinfo {year} {2020})}\BibitemShut {NoStop}%
\bibitem [{\citenamefont {Bernevig}\ \emph {et~al.}(2021)\citenamefont
  {Bernevig}, \citenamefont {Song}, \citenamefont {Regnault},\ and\
  \citenamefont {Lian}}]{bernevig321}%
  \BibitemOpen
  \bibfield  {author} {\bibinfo {author} {\bibfnamefont {B.~A.}\ \bibnamefont
  {Bernevig}}, \bibinfo {author} {\bibfnamefont {Z.-D.}\ \bibnamefont {Song}},
  \bibinfo {author} {\bibfnamefont {N.}~\bibnamefont {Regnault}},\ and\
  \bibinfo {author} {\bibfnamefont {B.}~\bibnamefont {Lian}},\ }\bibfield
  {title} {\bibinfo {title} {Twisted bilayer graphene. iii. interacting
  hamiltonian and exact symmetries},\ }\href
  {https://doi.org/10.1103/PhysRevB.103.205413} {\bibfield  {journal} {\bibinfo
   {journal} {Phys. Rev. B}\ }\textbf {\bibinfo {volume} {103}},\ \bibinfo
  {pages} {205413} (\bibinfo {year} {2021})}\BibitemShut {NoStop}%
\bibitem [{\citenamefont {Lian}\ \emph {et~al.}(2021)\citenamefont {Lian},
  \citenamefont {Song}, \citenamefont {Regnault}, \citenamefont {Efetov},
  \citenamefont {Yazdani},\ and\ \citenamefont {Bernevig}}]{bernevig421}%
  \BibitemOpen
  \bibfield  {author} {\bibinfo {author} {\bibfnamefont {B.}~\bibnamefont
  {Lian}}, \bibinfo {author} {\bibfnamefont {Z.-D.}\ \bibnamefont {Song}},
  \bibinfo {author} {\bibfnamefont {N.}~\bibnamefont {Regnault}}, \bibinfo
  {author} {\bibfnamefont {D.~K.}\ \bibnamefont {Efetov}}, \bibinfo {author}
  {\bibfnamefont {A.}~\bibnamefont {Yazdani}},\ and\ \bibinfo {author}
  {\bibfnamefont {B.~A.}\ \bibnamefont {Bernevig}},\ }\bibfield  {title}
  {\bibinfo {title} {Twisted bilayer graphene. iv. exact insulator ground
  states and phase diagram},\ }\href
  {https://doi.org/10.1103/PhysRevB.103.205414} {\bibfield  {journal} {\bibinfo
   {journal} {Phys. Rev. B}\ }\textbf {\bibinfo {volume} {103}},\ \bibinfo
  {pages} {205414} (\bibinfo {year} {2021})}\BibitemShut {NoStop}%
\bibitem [{\citenamefont {Seo}\ \emph {et~al.}(2019)\citenamefont {Seo},
  \citenamefont {Kotov},\ and\ \citenamefont {Uchoa}}]{seo19}%
  \BibitemOpen
  \bibfield  {author} {\bibinfo {author} {\bibfnamefont {K.}~\bibnamefont
  {Seo}}, \bibinfo {author} {\bibfnamefont {V.~N.}\ \bibnamefont {Kotov}},\
  and\ \bibinfo {author} {\bibfnamefont {B.}~\bibnamefont {Uchoa}},\ }\bibfield
   {title} {\bibinfo {title} {Ferromagnetic mott state in twisted graphene
  bilayers at the magic angle},\ }\href
  {https://doi.org/10.1103/PhysRevLett.122.246402} {\bibfield  {journal}
  {\bibinfo  {journal} {Phys. Rev. Lett.}\ }\textbf {\bibinfo {volume} {122}},\
  \bibinfo {pages} {246402} (\bibinfo {year} {2019})}\BibitemShut {NoStop}%
\bibitem [{\citenamefont {Kwan}\ \emph {et~al.}(2023)\citenamefont {Kwan},
  \citenamefont {Wagner}, \citenamefont {Bultinck}, \citenamefont {Simon},
  \citenamefont {Berg},\ and\ \citenamefont {Parameswaran}}]{Kwan23}%
  \BibitemOpen
  \bibfield  {author} {\bibinfo {author} {\bibfnamefont {Y.~H.}\ \bibnamefont
  {Kwan}}, \bibinfo {author} {\bibfnamefont {G.}~\bibnamefont {Wagner}},
  \bibinfo {author} {\bibfnamefont {N.}~\bibnamefont {Bultinck}}, \bibinfo
  {author} {\bibfnamefont {S.~H.}\ \bibnamefont {Simon}}, \bibinfo {author}
  {\bibfnamefont {E.}~\bibnamefont {Berg}},\ and\ \bibinfo {author}
  {\bibfnamefont {S.~A.}\ \bibnamefont {Parameswaran}},\ }\href@noop {}
  {\bibinfo {title} {Electron-phonon coupling and competing kekul\'e orders in
  twisted bilayer graphene}} (\bibinfo {year} {2023}),\ \Eprint
  {https://arxiv.org/abs/2303.13602} {arXiv:2303.13602 [cond-mat.str-el]}
  \BibitemShut {NoStop}%
\bibitem [{\citenamefont {Ledwith}\ \emph {et~al.}(2021)\citenamefont
  {Ledwith}, \citenamefont {Khalaf},\ and\ \citenamefont
  {Vishwanath}}]{ledwith21}%
  \BibitemOpen
  \bibfield  {author} {\bibinfo {author} {\bibfnamefont {P.~J.}\ \bibnamefont
  {Ledwith}}, \bibinfo {author} {\bibfnamefont {E.}~\bibnamefont {Khalaf}},\
  and\ \bibinfo {author} {\bibfnamefont {A.}~\bibnamefont {Vishwanath}},\
  }\bibfield  {title} {\bibinfo {title} {Strong coupling theory of magic-angle
  graphene: A pedagogical introduction},\ }\href
  {https://doi.org/https://doi.org/10.1016/j.aop.2021.168646} {\bibfield
  {journal} {\bibinfo  {journal} {Annals of Physics}\ }\textbf {\bibinfo
  {volume} {435}},\ \bibinfo {pages} {168646} (\bibinfo {year} {2021})},\
  \bibinfo {note} {special issue on Philip W. Anderson}\BibitemShut {NoStop}%
\bibitem [{\citenamefont {Pierce}\ \emph {et~al.}(2021)\citenamefont {Pierce},
  \citenamefont {Xie}, \citenamefont {Park}, \citenamefont {Khalaf},
  \citenamefont {Lee}, \citenamefont {Cao}, \citenamefont {Parker},
  \citenamefont {Forrester}, \citenamefont {Chen}, \citenamefont {Watanabe},
  \citenamefont {Taniguchi}, \citenamefont {Vishwanath}, \citenamefont
  {Jarillo-Herrero},\ and\ \citenamefont {Yacoby}}]{Pierce2021}%
  \BibitemOpen
  \bibfield  {author} {\bibinfo {author} {\bibfnamefont {A.~T.}\ \bibnamefont
  {Pierce}}, \bibinfo {author} {\bibfnamefont {Y.}~\bibnamefont {Xie}},
  \bibinfo {author} {\bibfnamefont {J.~M.}\ \bibnamefont {Park}}, \bibinfo
  {author} {\bibfnamefont {E.}~\bibnamefont {Khalaf}}, \bibinfo {author}
  {\bibfnamefont {S.~H.}\ \bibnamefont {Lee}}, \bibinfo {author} {\bibfnamefont
  {Y.}~\bibnamefont {Cao}}, \bibinfo {author} {\bibfnamefont {D.~E.}\
  \bibnamefont {Parker}}, \bibinfo {author} {\bibfnamefont {P.~R.}\
  \bibnamefont {Forrester}}, \bibinfo {author} {\bibfnamefont {S.}~\bibnamefont
  {Chen}}, \bibinfo {author} {\bibfnamefont {K.}~\bibnamefont {Watanabe}},
  \bibinfo {author} {\bibfnamefont {T.}~\bibnamefont {Taniguchi}}, \bibinfo
  {author} {\bibfnamefont {A.}~\bibnamefont {Vishwanath}}, \bibinfo {author}
  {\bibfnamefont {P.}~\bibnamefont {Jarillo-Herrero}},\ and\ \bibinfo {author}
  {\bibfnamefont {A.}~\bibnamefont {Yacoby}},\ }\bibfield  {title} {\bibinfo
  {title} {Unconventional sequence of correlated chern insulators in
  magic-angle twisted bilayer graphene},\ }\href
  {https://doi.org/10.1038/s41567-021-01347-4} {\bibfield  {journal} {\bibinfo
  {journal} {Nature Physics}\ }\textbf {\bibinfo {volume} {17}},\ \bibinfo
  {pages} {1210} (\bibinfo {year} {2021})}\BibitemShut {NoStop}%
\bibitem [{\citenamefont {Parker}\ \emph {et~al.}(2021)\citenamefont {Parker},
  \citenamefont {Soejima}, \citenamefont {Hauschild}, \citenamefont {Zaletel},\
  and\ \citenamefont {Bultinck}}]{parker21}%
  \BibitemOpen
  \bibfield  {author} {\bibinfo {author} {\bibfnamefont {D.~E.}\ \bibnamefont
  {Parker}}, \bibinfo {author} {\bibfnamefont {T.}~\bibnamefont {Soejima}},
  \bibinfo {author} {\bibfnamefont {J.}~\bibnamefont {Hauschild}}, \bibinfo
  {author} {\bibfnamefont {M.~P.}\ \bibnamefont {Zaletel}},\ and\ \bibinfo
  {author} {\bibfnamefont {N.}~\bibnamefont {Bultinck}},\ }\bibfield  {title}
  {\bibinfo {title} {Strain-induced quantum phase transitions in magic-angle
  graphene},\ }\href {https://doi.org/10.1103/PhysRevLett.127.027601}
  {\bibfield  {journal} {\bibinfo  {journal} {Phys. Rev. Lett.}\ }\textbf
  {\bibinfo {volume} {127}},\ \bibinfo {pages} {027601} (\bibinfo {year}
  {2021})}\BibitemShut {NoStop}%
\bibitem [{\citenamefont {Nuckolls}\ \emph {et~al.}(2023)\citenamefont
  {Nuckolls}, \citenamefont {Lee}, \citenamefont {Oh}, \citenamefont {Wong},
  \citenamefont {Soejima}, \citenamefont {Hong}, \citenamefont {Călugăru},
  \citenamefont {Herzog-Arbeitman}, \citenamefont {Bernevig}, \citenamefont
  {Watanabe}, \citenamefont {Taniguchi}, \citenamefont {Regnault},
  \citenamefont {Zaletel},\ and\ \citenamefont
  {Yazdani}}]{nuckolls2023quantum}%
  \BibitemOpen
  \bibfield  {author} {\bibinfo {author} {\bibfnamefont {K.~P.}\ \bibnamefont
  {Nuckolls}}, \bibinfo {author} {\bibfnamefont {R.~L.}\ \bibnamefont {Lee}},
  \bibinfo {author} {\bibfnamefont {M.}~\bibnamefont {Oh}}, \bibinfo {author}
  {\bibfnamefont {D.}~\bibnamefont {Wong}}, \bibinfo {author} {\bibfnamefont
  {T.}~\bibnamefont {Soejima}}, \bibinfo {author} {\bibfnamefont {J.~P.}\
  \bibnamefont {Hong}}, \bibinfo {author} {\bibfnamefont {D.}~\bibnamefont
  {Călugăru}}, \bibinfo {author} {\bibfnamefont {J.}~\bibnamefont
  {Herzog-Arbeitman}}, \bibinfo {author} {\bibfnamefont {B.~A.}\ \bibnamefont
  {Bernevig}}, \bibinfo {author} {\bibfnamefont {K.}~\bibnamefont {Watanabe}},
  \bibinfo {author} {\bibfnamefont {T.}~\bibnamefont {Taniguchi}}, \bibinfo
  {author} {\bibfnamefont {N.}~\bibnamefont {Regnault}}, \bibinfo {author}
  {\bibfnamefont {M.~P.}\ \bibnamefont {Zaletel}},\ and\ \bibinfo {author}
  {\bibfnamefont {A.}~\bibnamefont {Yazdani}},\ }\href@noop {} {\bibinfo
  {title} {Quantum textures of the many-body wavefunctions in magic-angle
  graphene}} (\bibinfo {year} {2023}),\ \Eprint
  {https://arxiv.org/abs/2303.00024} {arXiv:2303.00024 [cond-mat.mes-hall]}
  \BibitemShut {NoStop}%
\bibitem [{\citenamefont {C\ifmmode \u{a}\else \u{a}\fi{}lug\ifmmode~\u{a}\else
  \u{a}\fi{}ru}\ \emph {et~al.}(2022)\citenamefont {C\ifmmode \u{a}\else
  \u{a}\fi{}lug\ifmmode~\u{a}\else \u{a}\fi{}ru}, \citenamefont {Regnault},
  \citenamefont {Oh}, \citenamefont {Nuckolls}, \citenamefont {Wong},
  \citenamefont {Lee}, \citenamefont {Yazdani}, \citenamefont {Vafek},\ and\
  \citenamefont {Bernevig}}]{dimitru22}%
  \BibitemOpen
  \bibfield  {author} {\bibinfo {author} {\bibfnamefont {D.}~\bibnamefont
  {C\ifmmode \u{a}\else \u{a}\fi{}lug\ifmmode~\u{a}\else \u{a}\fi{}ru}},
  \bibinfo {author} {\bibfnamefont {N.}~\bibnamefont {Regnault}}, \bibinfo
  {author} {\bibfnamefont {M.}~\bibnamefont {Oh}}, \bibinfo {author}
  {\bibfnamefont {K.~P.}\ \bibnamefont {Nuckolls}}, \bibinfo {author}
  {\bibfnamefont {D.}~\bibnamefont {Wong}}, \bibinfo {author} {\bibfnamefont
  {R.~L.}\ \bibnamefont {Lee}}, \bibinfo {author} {\bibfnamefont
  {A.}~\bibnamefont {Yazdani}}, \bibinfo {author} {\bibfnamefont
  {O.}~\bibnamefont {Vafek}},\ and\ \bibinfo {author} {\bibfnamefont {B.~A.}\
  \bibnamefont {Bernevig}},\ }\bibfield  {title} {\bibinfo {title}
  {Spectroscopy of twisted bilayer graphene correlated insulators},\ }\href
  {https://doi.org/10.1103/PhysRevLett.129.117602} {\bibfield  {journal}
  {\bibinfo  {journal} {Phys. Rev. Lett.}\ }\textbf {\bibinfo {volume} {129}},\
  \bibinfo {pages} {117602} (\bibinfo {year} {2022})}\BibitemShut {NoStop}%
\bibitem [{\citenamefont {Jimeno-Pozo}\ \emph {et~al.}(2023)\citenamefont
  {Jimeno-Pozo}, \citenamefont {Goodwin}, \citenamefont {Pantaleón},
  \citenamefont {Vitale}, \citenamefont {Klebl}, \citenamefont {Kennes},
  \citenamefont {Mostofi}, \citenamefont {Lischner},\ and\ \citenamefont
  {Guinea}}]{jimenopozo2023short}%
  \BibitemOpen
  \bibfield  {author} {\bibinfo {author} {\bibfnamefont {A.}~\bibnamefont
  {Jimeno-Pozo}}, \bibinfo {author} {\bibfnamefont {Z.~A.~H.}\ \bibnamefont
  {Goodwin}}, \bibinfo {author} {\bibfnamefont {P.~A.}\ \bibnamefont
  {Pantaleón}}, \bibinfo {author} {\bibfnamefont {V.}~\bibnamefont {Vitale}},
  \bibinfo {author} {\bibfnamefont {L.}~\bibnamefont {Klebl}}, \bibinfo
  {author} {\bibfnamefont {D.~M.}\ \bibnamefont {Kennes}}, \bibinfo {author}
  {\bibfnamefont {A.}~\bibnamefont {Mostofi}}, \bibinfo {author} {\bibfnamefont
  {J.}~\bibnamefont {Lischner}},\ and\ \bibinfo {author} {\bibfnamefont
  {F.}~\bibnamefont {Guinea}},\ }\href@noop {} {\bibinfo {title} {Short vs.
  long range exchange interactions in twisted bilayer graphene}} (\bibinfo
  {year} {2023}),\ \Eprint {https://arxiv.org/abs/2303.18025} {arXiv:2303.18025
  [cond-mat.mes-hall]} \BibitemShut {NoStop}%
\bibitem [{\citenamefont {Blason}\ and\ \citenamefont
  {Fabrizio}(2022)}]{blason22}%
  \BibitemOpen
  \bibfield  {author} {\bibinfo {author} {\bibfnamefont {A.}~\bibnamefont
  {Blason}}\ and\ \bibinfo {author} {\bibfnamefont {M.}~\bibnamefont
  {Fabrizio}},\ }\bibfield  {title} {\bibinfo {title} {Local kekul\'e
  distortion turns twisted bilayer graphene into topological mott insulators
  and superconductors},\ }\href {https://doi.org/10.1103/PhysRevB.106.235112}
  {\bibfield  {journal} {\bibinfo  {journal} {Phys. Rev. B}\ }\textbf {\bibinfo
  {volume} {106}},\ \bibinfo {pages} {235112} (\bibinfo {year}
  {2022})}\BibitemShut {NoStop}%
\bibitem [{\citenamefont {Hofstadter}(1976)}]{Hofstadter76}%
  \BibitemOpen
  \bibfield  {author} {\bibinfo {author} {\bibfnamefont {D.~R.}\ \bibnamefont
  {Hofstadter}},\ }\bibfield  {title} {\bibinfo {title} {Energy levels and wave
  functions of bloch electrons in rational and irrational magnetic fields},\
  }\href {https://doi.org/10.1103/PhysRevB.14.2239} {\bibfield  {journal}
  {\bibinfo  {journal} {Phys. Rev. B}\ }\textbf {\bibinfo {volume} {14}},\
  \bibinfo {pages} {2239} (\bibinfo {year} {1976})}\BibitemShut {NoStop}%
\bibitem [{\citenamefont {Herzog-Arbeitman}\ \emph {et~al.}(2020)\citenamefont
  {Herzog-Arbeitman}, \citenamefont {Song}, \citenamefont {Regnault},\ and\
  \citenamefont {Bernevig}}]{Herzog20}%
  \BibitemOpen
  \bibfield  {author} {\bibinfo {author} {\bibfnamefont {J.}~\bibnamefont
  {Herzog-Arbeitman}}, \bibinfo {author} {\bibfnamefont {Z.-D.}\ \bibnamefont
  {Song}}, \bibinfo {author} {\bibfnamefont {N.}~\bibnamefont {Regnault}},\
  and\ \bibinfo {author} {\bibfnamefont {B.~A.}\ \bibnamefont {Bernevig}},\
  }\bibfield  {title} {\bibinfo {title} {Hofstadter topology: Noncrystalline
  topological materials at high flux},\ }\href
  {https://doi.org/10.1103/PhysRevLett.125.236804} {\bibfield  {journal}
  {\bibinfo  {journal} {Phys. Rev. Lett.}\ }\textbf {\bibinfo {volume} {125}},\
  \bibinfo {pages} {236804} (\bibinfo {year} {2020})}\BibitemShut {NoStop}%
\bibitem [{\citenamefont {Lian}\ \emph {et~al.}(2020)\citenamefont {Lian},
  \citenamefont {Xie},\ and\ \citenamefont {Bernevig}}]{Biao20}%
  \BibitemOpen
  \bibfield  {author} {\bibinfo {author} {\bibfnamefont {B.}~\bibnamefont
  {Lian}}, \bibinfo {author} {\bibfnamefont {F.}~\bibnamefont {Xie}},\ and\
  \bibinfo {author} {\bibfnamefont {B.~A.}\ \bibnamefont {Bernevig}},\
  }\bibfield  {title} {\bibinfo {title} {Landau level of fragile topology},\
  }\href {https://doi.org/10.1103/PhysRevB.102.041402} {\bibfield  {journal}
  {\bibinfo  {journal} {Phys. Rev. B}\ }\textbf {\bibinfo {volume} {102}},\
  \bibinfo {pages} {041402} (\bibinfo {year} {2020})}\BibitemShut {NoStop}%
\bibitem [{\citenamefont {Guan}\ \emph {et~al.}(2022)\citenamefont {Guan},
  \citenamefont {Yazyev},\ and\ \citenamefont {Kruchkov}}]{Guan22}%
  \BibitemOpen
  \bibfield  {author} {\bibinfo {author} {\bibfnamefont {Y.}~\bibnamefont
  {Guan}}, \bibinfo {author} {\bibfnamefont {O.~V.}\ \bibnamefont {Yazyev}},\
  and\ \bibinfo {author} {\bibfnamefont {A.}~\bibnamefont {Kruchkov}},\
  }\bibfield  {title} {\bibinfo {title} {Reentrant magic-angle phenomena in
  twisted bilayer graphene in integer magnetic fluxes},\ }\href
  {https://doi.org/10.1103/PhysRevB.106.L121115} {\bibfield  {journal}
  {\bibinfo  {journal} {Phys. Rev. B}\ }\textbf {\bibinfo {volume} {106}},\
  \bibinfo {pages} {L121115} (\bibinfo {year} {2022})}\BibitemShut {NoStop}%
\bibitem [{\citenamefont {Singh}\ \emph {et~al.}(2023)\citenamefont {Singh},
  \citenamefont {Chew}, \citenamefont {Herzog-Arbeitman}, \citenamefont
  {Bernevig},\ and\ \citenamefont {Vafek}}]{singh2023topological}%
  \BibitemOpen
  \bibfield  {author} {\bibinfo {author} {\bibfnamefont {K.}~\bibnamefont
  {Singh}}, \bibinfo {author} {\bibfnamefont {A.}~\bibnamefont {Chew}},
  \bibinfo {author} {\bibfnamefont {J.}~\bibnamefont {Herzog-Arbeitman}},
  \bibinfo {author} {\bibfnamefont {B.~A.}\ \bibnamefont {Bernevig}},\ and\
  \bibinfo {author} {\bibfnamefont {O.}~\bibnamefont {Vafek}},\ }\href@noop {}
  {\bibinfo {title} {Topological heavy fermions in magnetic field}} (\bibinfo
  {year} {2023}),\ \Eprint {https://arxiv.org/abs/2305.08171} {arXiv:2305.08171
  [cond-mat.str-el]} \BibitemShut {NoStop}%
\bibitem [{\citenamefont {Wang}\ and\ \citenamefont {Vafek}(2022)}]{wang22}%
  \BibitemOpen
  \bibfield  {author} {\bibinfo {author} {\bibfnamefont {X.}~\bibnamefont
  {Wang}}\ and\ \bibinfo {author} {\bibfnamefont {O.}~\bibnamefont {Vafek}},\
  }\bibfield  {title} {\bibinfo {title} {Narrow bands in magnetic field and
  strong-coupling hofstadter spectra},\ }\href
  {https://doi.org/10.1103/PhysRevB.106.L121111} {\bibfield  {journal}
  {\bibinfo  {journal} {Phys. Rev. B}\ }\textbf {\bibinfo {volume} {106}},\
  \bibinfo {pages} {L121111} (\bibinfo {year} {2022})}\BibitemShut {NoStop}%
\bibitem [{\citenamefont {Herzog-Arbeitman}\ \emph
  {et~al.}(2022{\natexlab{a}})\citenamefont {Herzog-Arbeitman}, \citenamefont
  {Chew},\ and\ \citenamefont {Bernevig}}]{herzog22_2}%
  \BibitemOpen
  \bibfield  {author} {\bibinfo {author} {\bibfnamefont {J.}~\bibnamefont
  {Herzog-Arbeitman}}, \bibinfo {author} {\bibfnamefont {A.}~\bibnamefont
  {Chew}},\ and\ \bibinfo {author} {\bibfnamefont {B.~A.}\ \bibnamefont
  {Bernevig}},\ }\bibfield  {title} {\bibinfo {title} {Magnetic bloch theorem
  and reentrant flat bands in twisted bilayer graphene at $2\ensuremath{\pi}$
  flux},\ }\href {https://doi.org/10.1103/PhysRevB.106.085140} {\bibfield
  {journal} {\bibinfo  {journal} {Phys. Rev. B}\ }\textbf {\bibinfo {volume}
  {106}},\ \bibinfo {pages} {085140} (\bibinfo {year}
  {2022}{\natexlab{a}})}\BibitemShut {NoStop}%
\bibitem [{\citenamefont {Herzog-Arbeitman}\ \emph
  {et~al.}(2022{\natexlab{b}})\citenamefont {Herzog-Arbeitman}, \citenamefont
  {Chew}, \citenamefont {Efetov},\ and\ \citenamefont {Bernevig}}]{herzog22_3}%
  \BibitemOpen
  \bibfield  {author} {\bibinfo {author} {\bibfnamefont {J.}~\bibnamefont
  {Herzog-Arbeitman}}, \bibinfo {author} {\bibfnamefont {A.}~\bibnamefont
  {Chew}}, \bibinfo {author} {\bibfnamefont {D.~K.}\ \bibnamefont {Efetov}},\
  and\ \bibinfo {author} {\bibfnamefont {B.~A.}\ \bibnamefont {Bernevig}},\
  }\bibfield  {title} {\bibinfo {title} {Reentrant correlated insulators in
  twisted bilayer graphene at 25 t ($2\ensuremath{\pi}$ flux)},\ }\href
  {https://doi.org/10.1103/PhysRevLett.129.076401} {\bibfield  {journal}
  {\bibinfo  {journal} {Phys. Rev. Lett.}\ }\textbf {\bibinfo {volume} {129}},\
  \bibinfo {pages} {076401} (\bibinfo {year} {2022}{\natexlab{b}})}\BibitemShut
  {NoStop}%
\bibitem [{\citenamefont {Das}\ \emph {et~al.}(2022)\citenamefont {Das},
  \citenamefont {Shen}, \citenamefont {Jaoui}, \citenamefont
  {Herzog-Arbeitman}, \citenamefont {Chew}, \citenamefont {Cho}, \citenamefont
  {Watanabe}, \citenamefont {Taniguchi}, \citenamefont {Piot}, \citenamefont
  {Bernevig},\ and\ \citenamefont {Efetov}}]{efetov22}%
  \BibitemOpen
  \bibfield  {author} {\bibinfo {author} {\bibfnamefont {I.}~\bibnamefont
  {Das}}, \bibinfo {author} {\bibfnamefont {C.}~\bibnamefont {Shen}}, \bibinfo
  {author} {\bibfnamefont {A.}~\bibnamefont {Jaoui}}, \bibinfo {author}
  {\bibfnamefont {J.}~\bibnamefont {Herzog-Arbeitman}}, \bibinfo {author}
  {\bibfnamefont {A.}~\bibnamefont {Chew}}, \bibinfo {author} {\bibfnamefont
  {C.-W.}\ \bibnamefont {Cho}}, \bibinfo {author} {\bibfnamefont
  {K.}~\bibnamefont {Watanabe}}, \bibinfo {author} {\bibfnamefont
  {T.}~\bibnamefont {Taniguchi}}, \bibinfo {author} {\bibfnamefont {B.~A.}\
  \bibnamefont {Piot}}, \bibinfo {author} {\bibfnamefont {B.~A.}\ \bibnamefont
  {Bernevig}},\ and\ \bibinfo {author} {\bibfnamefont {D.~K.}\ \bibnamefont
  {Efetov}},\ }\bibfield  {title} {\bibinfo {title} {Observation of reentrant
  correlated insulators and interaction-driven fermi-surface reconstructions at
  one magnetic flux quantum per moir\'e unit cell in magic-angle twisted
  bilayer graphene},\ }\href {https://doi.org/10.1103/PhysRevLett.128.217701}
  {\bibfield  {journal} {\bibinfo  {journal} {Phys. Rev. Lett.}\ }\textbf
  {\bibinfo {volume} {128}},\ \bibinfo {pages} {217701} (\bibinfo {year}
  {2022})}\BibitemShut {NoStop}%
\bibitem [{\citenamefont {Bistritzer}\ and\ \citenamefont
  {MacDonald}(2011)}]{McDonald11}%
  \BibitemOpen
  \bibfield  {author} {\bibinfo {author} {\bibfnamefont {R.}~\bibnamefont
  {Bistritzer}}\ and\ \bibinfo {author} {\bibfnamefont {A.~H.}\ \bibnamefont
  {MacDonald}},\ }\bibfield  {title} {\bibinfo {title} {Moiré bands in twisted
  double-layer graphene},\ }\href {https://doi.org/10.1073/pnas.1108174108}
  {\bibfield  {journal} {\bibinfo  {journal} {Proceedings of the National
  Academy of Sciences}\ }\textbf {\bibinfo {volume} {108}},\ \bibinfo {pages}
  {12233} (\bibinfo {year} {2011})},\ \Eprint
  {https://arxiv.org/abs/https://www.pnas.org/doi/pdf/10.1073/pnas.1108174108}
  {https://www.pnas.org/doi/pdf/10.1073/pnas.1108174108} \BibitemShut {NoStop}%
\bibitem [{\citenamefont {Lopes~dos Santos}\ \emph {et~al.}(2012)\citenamefont
  {Lopes~dos Santos}, \citenamefont {Peres},\ and\ \citenamefont
  {Castro~Neto}}]{Peres12}%
  \BibitemOpen
  \bibfield  {author} {\bibinfo {author} {\bibfnamefont {J.~M.~B.}\
  \bibnamefont {Lopes~dos Santos}}, \bibinfo {author} {\bibfnamefont
  {N.~M.~R.}\ \bibnamefont {Peres}},\ and\ \bibinfo {author} {\bibfnamefont
  {A.~H.}\ \bibnamefont {Castro~Neto}},\ }\bibfield  {title} {\bibinfo {title}
  {Continuum model of the twisted graphene bilayer},\ }\href
  {https://doi.org/10.1103/PhysRevB.86.155449} {\bibfield  {journal} {\bibinfo
  {journal} {Phys. Rev. B}\ }\textbf {\bibinfo {volume} {86}},\ \bibinfo
  {pages} {155449} (\bibinfo {year} {2012})}\BibitemShut {NoStop}%
\bibitem [{\citenamefont {Goodwin}\ \emph {et~al.}(2020)\citenamefont
  {Goodwin}, \citenamefont {Vitale}, \citenamefont {Liang}, \citenamefont
  {Mostofi},\ and\ \citenamefont {Lischner}}]{Goodwin_2020}%
  \BibitemOpen
  \bibfield  {author} {\bibinfo {author} {\bibfnamefont {Z.~A.~H.}\
  \bibnamefont {Goodwin}}, \bibinfo {author} {\bibfnamefont {V.}~\bibnamefont
  {Vitale}}, \bibinfo {author} {\bibfnamefont {X.}~\bibnamefont {Liang}},
  \bibinfo {author} {\bibfnamefont {A.~A.}\ \bibnamefont {Mostofi}},\ and\
  \bibinfo {author} {\bibfnamefont {J.}~\bibnamefont {Lischner}},\ }\bibfield
  {title} {\bibinfo {title} {Hartree theory calculations of quasiparticle
  properties in twisted bilayer graphene},\ }\href
  {https://doi.org/10.1088/2516-1075/ab9f94} {\bibfield  {journal} {\bibinfo
  {journal} {Electronic Structure}\ }\textbf {\bibinfo {volume} {2}},\ \bibinfo
  {pages} {034001} (\bibinfo {year} {2020})}\BibitemShut {NoStop}%
\bibitem [{\citenamefont {Kang}\ and\ \citenamefont {Vafek}(2023)}]{vafek23}%
  \BibitemOpen
  \bibfield  {author} {\bibinfo {author} {\bibfnamefont {J.}~\bibnamefont
  {Kang}}\ and\ \bibinfo {author} {\bibfnamefont {O.}~\bibnamefont {Vafek}},\
  }\bibfield  {title} {\bibinfo {title} {Pseudomagnetic fields, particle-hole
  asymmetry, and microscopic effective continuum hamiltonians of twisted
  bilayer graphene},\ }\href {https://doi.org/10.1103/PhysRevB.107.075408}
  {\bibfield  {journal} {\bibinfo  {journal} {Phys. Rev. B}\ }\textbf {\bibinfo
  {volume} {107}},\ \bibinfo {pages} {075408} (\bibinfo {year}
  {2023})}\BibitemShut {NoStop}%
\bibitem [{\citenamefont {Nam}\ and\ \citenamefont
  {Koshino}(2017)}]{Koshino17}%
  \BibitemOpen
  \bibfield  {author} {\bibinfo {author} {\bibfnamefont {N.~N.~T.}\
  \bibnamefont {Nam}}\ and\ \bibinfo {author} {\bibfnamefont {M.}~\bibnamefont
  {Koshino}},\ }\bibfield  {title} {\bibinfo {title} {Lattice relaxation and
  energy band modulation in twisted bilayer graphene},\ }\href
  {https://doi.org/10.1103/PhysRevB.96.075311} {\bibfield  {journal} {\bibinfo
  {journal} {Phys. Rev. B}\ }\textbf {\bibinfo {volume} {96}},\ \bibinfo
  {pages} {075311} (\bibinfo {year} {2017})}\BibitemShut {NoStop}%
\bibitem [{\citenamefont {Moon}\ and\ \citenamefont
  {Koshino}(2012)}]{Koshino12}%
  \BibitemOpen
  \bibfield  {author} {\bibinfo {author} {\bibfnamefont {P.}~\bibnamefont
  {Moon}}\ and\ \bibinfo {author} {\bibfnamefont {M.}~\bibnamefont {Koshino}},\
  }\bibfield  {title} {\bibinfo {title} {Energy spectrum and quantum hall
  effect in twisted bilayer graphene},\ }\href
  {https://doi.org/10.1103/PhysRevB.85.195458} {\bibfield  {journal} {\bibinfo
  {journal} {Phys. Rev. B}\ }\textbf {\bibinfo {volume} {85}},\ \bibinfo
  {pages} {195458} (\bibinfo {year} {2012})}\BibitemShut {NoStop}%
\bibitem [{\citenamefont {Giuliani}\ and\ \citenamefont
  {Vignale}(2005)}]{giuliani_vignale_2005}%
  \BibitemOpen
  \bibfield  {author} {\bibinfo {author} {\bibfnamefont {G.}~\bibnamefont
  {Giuliani}}\ and\ \bibinfo {author} {\bibfnamefont {G.}~\bibnamefont
  {Vignale}},\ }\href {https://doi.org/10.1017/CBO9780511619915} {\emph
  {\bibinfo {title} {Quantum Theory of the Electron Liquid}}}\ (\bibinfo
  {publisher} {Cambridge University Press},\ \bibinfo {year}
  {2005})\BibitemShut {NoStop}%
\bibitem [{\citenamefont {Luttinger}(1951)}]{Luttinger51}%
  \BibitemOpen
  \bibfield  {author} {\bibinfo {author} {\bibfnamefont {J.~M.}\ \bibnamefont
  {Luttinger}},\ }\bibfield  {title} {\bibinfo {title} {The effect of a
  magnetic field on electrons in a periodic potential},\ }\href
  {https://doi.org/10.1103/PhysRev.84.814} {\bibfield  {journal} {\bibinfo
  {journal} {Phys. Rev.}\ }\textbf {\bibinfo {volume} {84}},\ \bibinfo {pages}
  {814} (\bibinfo {year} {1951})}\BibitemShut {NoStop}%
\bibitem [{\citenamefont {Nemec}\ and\ \citenamefont
  {Cuniberti}(2007)}]{Cuniberti07}%
  \BibitemOpen
  \bibfield  {author} {\bibinfo {author} {\bibfnamefont {N.}~\bibnamefont
  {Nemec}}\ and\ \bibinfo {author} {\bibfnamefont {G.}~\bibnamefont
  {Cuniberti}},\ }\bibfield  {title} {\bibinfo {title} {Hofstadter butterflies
  of bilayer graphene},\ }\href {https://doi.org/10.1103/PhysRevB.75.201404}
  {\bibfield  {journal} {\bibinfo  {journal} {Phys. Rev. B}\ }\textbf {\bibinfo
  {volume} {75}},\ \bibinfo {pages} {201404} (\bibinfo {year}
  {2007})}\BibitemShut {NoStop}%
\bibitem [{\citenamefont {Herzog-Arbeitman}\ \emph
  {et~al.}(2022{\natexlab{c}})\citenamefont {Herzog-Arbeitman}, \citenamefont
  {Song}, \citenamefont {Elcoro},\ and\ \citenamefont {Bernevig}}]{Herzog22}%
  \BibitemOpen
  \bibfield  {author} {\bibinfo {author} {\bibfnamefont {J.}~\bibnamefont
  {Herzog-Arbeitman}}, \bibinfo {author} {\bibfnamefont {Z.-D.}\ \bibnamefont
  {Song}}, \bibinfo {author} {\bibfnamefont {L.}~\bibnamefont {Elcoro}},\ and\
  \bibinfo {author} {\bibfnamefont {B.~A.}\ \bibnamefont {Bernevig}},\ }\href
  {https://doi.org/10.48550/ARXIV.2209.10559} {\bibinfo {title} {Hofstadter
  topology with real space invariants and reentrant projective symmetries}}
  (\bibinfo {year} {2022}{\natexlab{c}})\BibitemShut {NoStop}%
\bibitem [{\citenamefont {Ahn}\ \emph {et~al.}(2019)\citenamefont {Ahn},
  \citenamefont {Park},\ and\ \citenamefont {Yang}}]{junyeong19}%
  \BibitemOpen
  \bibfield  {author} {\bibinfo {author} {\bibfnamefont {J.}~\bibnamefont
  {Ahn}}, \bibinfo {author} {\bibfnamefont {S.}~\bibnamefont {Park}},\ and\
  \bibinfo {author} {\bibfnamefont {B.-J.}\ \bibnamefont {Yang}},\ }\bibfield
  {title} {\bibinfo {title} {Failure of nielsen-ninomiya theorem and fragile
  topology in two-dimensional systems with space-time inversion symmetry:
  Application to twisted bilayer graphene at magic angle},\ }\href
  {https://doi.org/10.1103/PhysRevX.9.021013} {\bibfield  {journal} {\bibinfo
  {journal} {Phys. Rev. X}\ }\textbf {\bibinfo {volume} {9}},\ \bibinfo {pages}
  {021013} (\bibinfo {year} {2019})}\BibitemShut {NoStop}%
\bibitem [{\citenamefont {Fukui}\ \emph {et~al.}(2005)\citenamefont {Fukui},
  \citenamefont {Hatsugai},\ and\ \citenamefont {Suzuki}}]{fukui05}%
  \BibitemOpen
  \bibfield  {author} {\bibinfo {author} {\bibfnamefont {T.}~\bibnamefont
  {Fukui}}, \bibinfo {author} {\bibfnamefont {Y.}~\bibnamefont {Hatsugai}},\
  and\ \bibinfo {author} {\bibfnamefont {H.}~\bibnamefont {Suzuki}},\
  }\bibfield  {title} {\bibinfo {title} {Chern numbers in discretized brillouin
  zone: Efficient method of computing (spin) hall conductances},\ }\href
  {https://doi.org/10.1143/JPSJ.74.1674} {\bibfield  {journal} {\bibinfo
  {journal} {Journal of the Physical Society of Japan}\ }\textbf {\bibinfo
  {volume} {74}},\ \bibinfo {pages} {1674} (\bibinfo {year} {2005})},\ \Eprint
  {https://arxiv.org/abs/https://doi.org/10.1143/JPSJ.74.1674}
  {https://doi.org/10.1143/JPSJ.74.1674} \BibitemShut {NoStop}%
\bibitem [{\citenamefont {Song}\ \emph {et~al.}(2021)\citenamefont {Song},
  \citenamefont {Lian}, \citenamefont {Regnault},\ and\ \citenamefont
  {Bernevig}}]{bernvig2_21}%
  \BibitemOpen
  \bibfield  {author} {\bibinfo {author} {\bibfnamefont {Z.-D.}\ \bibnamefont
  {Song}}, \bibinfo {author} {\bibfnamefont {B.}~\bibnamefont {Lian}}, \bibinfo
  {author} {\bibfnamefont {N.}~\bibnamefont {Regnault}},\ and\ \bibinfo
  {author} {\bibfnamefont {B.~A.}\ \bibnamefont {Bernevig}},\ }\bibfield
  {title} {\bibinfo {title} {Twisted bilayer graphene. ii. stable symmetry
  anomaly},\ }\href {https://doi.org/10.1103/PhysRevB.103.205412} {\bibfield
  {journal} {\bibinfo  {journal} {Phys. Rev. B}\ }\textbf {\bibinfo {volume}
  {103}},\ \bibinfo {pages} {205412} (\bibinfo {year} {2021})}\BibitemShut
  {NoStop}%
\bibitem [{\citenamefont {Tarnopolsky}\ \emph {et~al.}(2019)\citenamefont
  {Tarnopolsky}, \citenamefont {Kruchkov},\ and\ \citenamefont
  {Vishwanath}}]{tarn19}%
  \BibitemOpen
  \bibfield  {author} {\bibinfo {author} {\bibfnamefont {G.}~\bibnamefont
  {Tarnopolsky}}, \bibinfo {author} {\bibfnamefont {A.~J.}\ \bibnamefont
  {Kruchkov}},\ and\ \bibinfo {author} {\bibfnamefont {A.}~\bibnamefont
  {Vishwanath}},\ }\bibfield  {title} {\bibinfo {title} {Origin of magic angles
  in twisted bilayer graphene},\ }\href
  {https://doi.org/10.1103/PhysRevLett.122.106405} {\bibfield  {journal}
  {\bibinfo  {journal} {Phys. Rev. Lett.}\ }\textbf {\bibinfo {volume} {122}},\
  \bibinfo {pages} {106405} (\bibinfo {year} {2019})}\BibitemShut {NoStop}%
\bibitem [{\citenamefont {Liu}\ \emph {et~al.}(2019)\citenamefont {Liu},
  \citenamefont {Liu},\ and\ \citenamefont {Dai}}]{Liu19}%
  \BibitemOpen
  \bibfield  {author} {\bibinfo {author} {\bibfnamefont {J.}~\bibnamefont
  {Liu}}, \bibinfo {author} {\bibfnamefont {J.}~\bibnamefont {Liu}},\ and\
  \bibinfo {author} {\bibfnamefont {X.}~\bibnamefont {Dai}},\ }\bibfield
  {title} {\bibinfo {title} {Pseudo landau level representation of twisted
  bilayer graphene: Band topology and implications on the correlated insulating
  phase},\ }\href {https://doi.org/10.1103/PhysRevB.99.155415} {\bibfield
  {journal} {\bibinfo  {journal} {Phys. Rev. B}\ }\textbf {\bibinfo {volume}
  {99}},\ \bibinfo {pages} {155415} (\bibinfo {year} {2019})}\BibitemShut
  {NoStop}%
\bibitem [{\citenamefont {Chatterjee}\ \emph {et~al.}(2020)\citenamefont
  {Chatterjee}, \citenamefont {Bultinck},\ and\ \citenamefont
  {Zaletel}}]{chatt20}%
  \BibitemOpen
  \bibfield  {author} {\bibinfo {author} {\bibfnamefont {S.}~\bibnamefont
  {Chatterjee}}, \bibinfo {author} {\bibfnamefont {N.}~\bibnamefont
  {Bultinck}},\ and\ \bibinfo {author} {\bibfnamefont {M.~P.}\ \bibnamefont
  {Zaletel}},\ }\bibfield  {title} {\bibinfo {title} {Symmetry breaking and
  skyrmionic transport in twisted bilayer graphene},\ }\href
  {https://doi.org/10.1103/PhysRevB.101.165141} {\bibfield  {journal} {\bibinfo
   {journal} {Phys. Rev. B}\ }\textbf {\bibinfo {volume} {101}},\ \bibinfo
  {pages} {165141} (\bibinfo {year} {2020})}\BibitemShut {NoStop}%
\bibitem [{\citenamefont {Gonz{\'a}lez}\ and\ \citenamefont
  {Stauber}(2023)}]{Gonzalez23}%
  \BibitemOpen
  \bibfield  {author} {\bibinfo {author} {\bibfnamefont {J.}~\bibnamefont
  {Gonz{\'a}lez}}\ and\ \bibinfo {author} {\bibfnamefont {T.}~\bibnamefont
  {Stauber}},\ }\bibfield  {title} {\bibinfo {title} {Ising superconductivity
  induced from spin-selective valley symmetry breaking in twisted trilayer
  graphene},\ }\href {https://doi.org/10.1038/s41467-023-38250-w} {\bibfield
  {journal} {\bibinfo  {journal} {Nature Communications}\ }\textbf {\bibinfo
  {volume} {14}},\ \bibinfo {pages} {2746} (\bibinfo {year}
  {2023})}\BibitemShut {NoStop}%
\bibitem [{\citenamefont {Cea}\ \emph {et~al.}(2022)\citenamefont {Cea},
  \citenamefont {Pantaleón}, \citenamefont {Walet},\ and\ \citenamefont
  {Guinea}}]{Cea22}%
  \BibitemOpen
  \bibfield  {author} {\bibinfo {author} {\bibfnamefont {T.}~\bibnamefont
  {Cea}}, \bibinfo {author} {\bibfnamefont {P.~A.}\ \bibnamefont {Pantaleón}},
  \bibinfo {author} {\bibfnamefont {N.~R.}\ \bibnamefont {Walet}},\ and\
  \bibinfo {author} {\bibfnamefont {F.}~\bibnamefont {Guinea}},\ }\bibfield
  {title} {\bibinfo {title} {Electrostatic interactions in twisted bilayer
  graphene},\ }\href
  {https://doi.org/https://doi.org/10.1016/j.nanoms.2021.10.001} {\bibfield
  {journal} {\bibinfo  {journal} {Nano Materials Science}\ }\textbf {\bibinfo
  {volume} {4}},\ \bibinfo {pages} {27} (\bibinfo {year} {2022})},\ \bibinfo
  {note} {special issue on Graphene and 2D Alternative Materials}\BibitemShut
  {NoStop}%
\bibitem [{\citenamefont {Scheurer}\ and\ \citenamefont
  {Samajdar}(2020)}]{scheurer20}%
  \BibitemOpen
  \bibfield  {author} {\bibinfo {author} {\bibfnamefont {M.~S.}\ \bibnamefont
  {Scheurer}}\ and\ \bibinfo {author} {\bibfnamefont {R.}~\bibnamefont
  {Samajdar}},\ }\bibfield  {title} {\bibinfo {title} {Pairing in
  graphene-based moir\'e superlattices},\ }\href
  {https://doi.org/10.1103/PhysRevResearch.2.033062} {\bibfield  {journal}
  {\bibinfo  {journal} {Phys. Rev. Res.}\ }\textbf {\bibinfo {volume} {2}},\
  \bibinfo {pages} {033062} (\bibinfo {year} {2020})}\BibitemShut {NoStop}%
\bibitem [{\citenamefont {Chatterjee}\ \emph {et~al.}(2022)\citenamefont
  {Chatterjee}, \citenamefont {Wang}, \citenamefont {Berg},\ and\ \citenamefont
  {Zaletel}}]{Chatterjee2022}%
  \BibitemOpen
  \bibfield  {author} {\bibinfo {author} {\bibfnamefont {S.}~\bibnamefont
  {Chatterjee}}, \bibinfo {author} {\bibfnamefont {T.}~\bibnamefont {Wang}},
  \bibinfo {author} {\bibfnamefont {E.}~\bibnamefont {Berg}},\ and\ \bibinfo
  {author} {\bibfnamefont {M.~P.}\ \bibnamefont {Zaletel}},\ }\bibfield
  {title} {\bibinfo {title} {Inter-valley coherent order and isospin
  fluctuation mediated superconductivity in rhombohedral trilayer graphene},\
  }\href {https://doi.org/10.1038/s41467-022-33561-w} {\bibfield  {journal}
  {\bibinfo  {journal} {Nature Communications}\ }\textbf {\bibinfo {volume}
  {13}},\ \bibinfo {pages} {6013} (\bibinfo {year} {2022})}\BibitemShut
  {NoStop}%
\bibitem [{\citenamefont {Lee}\ \emph {et~al.}(2019)\citenamefont {Lee},
  \citenamefont {Khalaf}, \citenamefont {Liu}, \citenamefont {Liu},
  \citenamefont {Hao}, \citenamefont {Kim},\ and\ \citenamefont
  {Vishwanath}}]{Lee2019}%
  \BibitemOpen
  \bibfield  {author} {\bibinfo {author} {\bibfnamefont {J.~Y.}\ \bibnamefont
  {Lee}}, \bibinfo {author} {\bibfnamefont {E.}~\bibnamefont {Khalaf}},
  \bibinfo {author} {\bibfnamefont {S.}~\bibnamefont {Liu}}, \bibinfo {author}
  {\bibfnamefont {X.}~\bibnamefont {Liu}}, \bibinfo {author} {\bibfnamefont
  {Z.}~\bibnamefont {Hao}}, \bibinfo {author} {\bibfnamefont {P.}~\bibnamefont
  {Kim}},\ and\ \bibinfo {author} {\bibfnamefont {A.}~\bibnamefont
  {Vishwanath}},\ }\bibfield  {title} {\bibinfo {title} {Theory of correlated
  insulating behaviour and spin-triplet superconductivity in twisted double
  bilayer graphene},\ }\href {https://doi.org/10.1038/s41467-019-12981-1}
  {\bibfield  {journal} {\bibinfo  {journal} {Nature Communications}\ }\textbf
  {\bibinfo {volume} {10}},\ \bibinfo {pages} {5333} (\bibinfo {year}
  {2019})}\BibitemShut {NoStop}%
\bibitem [{\citenamefont {Chen}\ \emph {et~al.}(2023)\citenamefont {Chen},
  \citenamefont {Nuckolls}, \citenamefont {Ding}, \citenamefont {Miao},
  \citenamefont {Wong}, \citenamefont {Oh}, \citenamefont {Lee}, \citenamefont
  {He}, \citenamefont {Peng}, \citenamefont {Pei}, \citenamefont {Li},
  \citenamefont {Zhang}, \citenamefont {Liu}, \citenamefont {Liu},
  \citenamefont {Jozwiak}, \citenamefont {Bostwick}, \citenamefont {Rotenberg},
  \citenamefont {Li}, \citenamefont {Han}, \citenamefont {Pan}, \citenamefont
  {Dai}, \citenamefont {Liu}, \citenamefont {Bernevig}, \citenamefont {Wang},
  \citenamefont {Yazdani},\ and\ \citenamefont {Chen}}]{chen2023strong}%
  \BibitemOpen
  \bibfield  {author} {\bibinfo {author} {\bibfnamefont {C.}~\bibnamefont
  {Chen}}, \bibinfo {author} {\bibfnamefont {K.~P.}\ \bibnamefont {Nuckolls}},
  \bibinfo {author} {\bibfnamefont {S.}~\bibnamefont {Ding}}, \bibinfo {author}
  {\bibfnamefont {W.}~\bibnamefont {Miao}}, \bibinfo {author} {\bibfnamefont
  {D.}~\bibnamefont {Wong}}, \bibinfo {author} {\bibfnamefont {M.}~\bibnamefont
  {Oh}}, \bibinfo {author} {\bibfnamefont {R.~L.}\ \bibnamefont {Lee}},
  \bibinfo {author} {\bibfnamefont {S.}~\bibnamefont {He}}, \bibinfo {author}
  {\bibfnamefont {C.}~\bibnamefont {Peng}}, \bibinfo {author} {\bibfnamefont
  {D.}~\bibnamefont {Pei}}, \bibinfo {author} {\bibfnamefont {Y.}~\bibnamefont
  {Li}}, \bibinfo {author} {\bibfnamefont {S.}~\bibnamefont {Zhang}}, \bibinfo
  {author} {\bibfnamefont {J.}~\bibnamefont {Liu}}, \bibinfo {author}
  {\bibfnamefont {Z.}~\bibnamefont {Liu}}, \bibinfo {author} {\bibfnamefont
  {C.}~\bibnamefont {Jozwiak}}, \bibinfo {author} {\bibfnamefont
  {A.}~\bibnamefont {Bostwick}}, \bibinfo {author} {\bibfnamefont
  {E.}~\bibnamefont {Rotenberg}}, \bibinfo {author} {\bibfnamefont
  {C.}~\bibnamefont {Li}}, \bibinfo {author} {\bibfnamefont {X.}~\bibnamefont
  {Han}}, \bibinfo {author} {\bibfnamefont {D.}~\bibnamefont {Pan}}, \bibinfo
  {author} {\bibfnamefont {X.}~\bibnamefont {Dai}}, \bibinfo {author}
  {\bibfnamefont {C.}~\bibnamefont {Liu}}, \bibinfo {author} {\bibfnamefont
  {B.~A.}\ \bibnamefont {Bernevig}}, \bibinfo {author} {\bibfnamefont
  {Y.}~\bibnamefont {Wang}}, \bibinfo {author} {\bibfnamefont {A.}~\bibnamefont
  {Yazdani}},\ and\ \bibinfo {author} {\bibfnamefont {Y.}~\bibnamefont
  {Chen}},\ }\href@noop {} {\bibinfo {title} {Strong inter-valley
  electron-phonon coupling in magic-angle twisted bilayer graphene}} (\bibinfo
  {year} {2023}),\ \Eprint {https://arxiv.org/abs/2303.14903} {arXiv:2303.14903
  [cond-mat.mes-hall]} \BibitemShut {NoStop}%
\bibitem [{\citenamefont {Goodwin}\ \emph {et~al.}(2019)\citenamefont
  {Goodwin}, \citenamefont {Corsetti}, \citenamefont {Mostofi},\ and\
  \citenamefont {Lischner}}]{goodwin19}%
  \BibitemOpen
  \bibfield  {author} {\bibinfo {author} {\bibfnamefont {Z.~A.~H.}\
  \bibnamefont {Goodwin}}, \bibinfo {author} {\bibfnamefont {F.}~\bibnamefont
  {Corsetti}}, \bibinfo {author} {\bibfnamefont {A.~A.}\ \bibnamefont
  {Mostofi}},\ and\ \bibinfo {author} {\bibfnamefont {J.}~\bibnamefont
  {Lischner}},\ }\bibfield  {title} {\bibinfo {title} {Attractive
  electron-electron interactions from internal screening in magic-angle twisted
  bilayer graphene},\ }\href {https://doi.org/10.1103/PhysRevB.100.235424}
  {\bibfield  {journal} {\bibinfo  {journal} {Phys. Rev. B}\ }\textbf {\bibinfo
  {volume} {100}},\ \bibinfo {pages} {235424} (\bibinfo {year}
  {2019})}\BibitemShut {NoStop}%
\bibitem [{\citenamefont {Ramires}\ and\ \citenamefont {Lado}(2019)}]{Lado19}%
  \BibitemOpen
  \bibfield  {author} {\bibinfo {author} {\bibfnamefont {A.}~\bibnamefont
  {Ramires}}\ and\ \bibinfo {author} {\bibfnamefont {J.~L.}\ \bibnamefont
  {Lado}},\ }\bibfield  {title} {\bibinfo {title} {Impurity-induced triple
  point fermions in twisted bilayer graphene},\ }\href
  {https://doi.org/10.1103/PhysRevB.99.245118} {\bibfield  {journal} {\bibinfo
  {journal} {Phys. Rev. B}\ }\textbf {\bibinfo {volume} {99}},\ \bibinfo
  {pages} {245118} (\bibinfo {year} {2019})}\BibitemShut {NoStop}%
\end{thebibliography}%
\onecolumngrid
\appendix
\setcounter{figure}{0}
\renewcommand\thefigure{\thesection.\arabic{figure}}    
\setcounter{table}{0}
\renewcommand{\thetable}{F\arabic{table}}
\section{The symmetry operations under magnetic fields}\label{appa}

We look for unitary operators realizing the $C_{3z}$ and $C_{2z}$ symmetries, acting on the creation operators as
\begin{align}
g c^\dagger_{\boldsymbol{i}} g^{-1} = \exp(i\chi_g(g(\boldsymbol{r_i}))) c^\dagger_{g(\boldsymbol{i})}.
\end{align}
Here we use indistinctly $g$ for the unitary operators and for the linear transformations acting on points of the lattice. These can always be distinguished by the context. As in the main text, $c^\dagger_{\boldsymbol{i}}(c^\dagger_{g(\boldsymbol{i})})$ is the creation operator at position $\boldsymbol{r_i}$ $(g(\boldsymbol{r_i}))$. The action on the Hamiltonian is
% \begin{align}
% \end{align}
\begin{align}
g H_0 g^{-1} = \sum_{\boldsymbol{i},\boldsymbol{j}} t(\boldsymbol{r_i}-\boldsymbol{r_j}) \exp(i\theta_{\boldsymbol{i},\boldsymbol{j}})\exp(i(\chi_g(g(\boldsymbol{r_i}) - \chi_g(g(\boldsymbol{r_j}))) c_{g(\boldsymbol{i})}^\dagger c_{g(\boldsymbol{j})}.
\end{align}
We are dealing with symmetries at zero flux, so $t(g(\boldsymbol{r_i})-g(\boldsymbol{r_j})) = t(\boldsymbol{r_i} - \boldsymbol{r_j})$. Then to realize the symmetry (this is, for $g H_0 g^{-1} = H_0$) $\chi_g(\boldsymbol{r})$ must obey 
\begin{align}
\theta_{g^{-1}(\boldsymbol{i}), g^{-1}(\boldsymbol{j})} + \chi_g(\boldsymbol{r_i}) - \chi_g(\boldsymbol{j}) = \theta_{\boldsymbol{i},\boldsymbol{j}} \nonumber \\
\frac{2\pi}{\Phi_0}\int_{g^{-1}(\boldsymbol{r_i})}^{g^{-1}(\boldsymbol{r_j})} \boldsymbol{A}(\boldsymbol{r'}) \cdot d\boldsymbol{r'} - \frac{2\pi}{\Phi_0}\int_{\boldsymbol{r_i}}^{\boldsymbol{r_j}} \boldsymbol{A}(\boldsymbol{r'}) \cdot d\boldsymbol{r'} = \int_{\boldsymbol{r_i}}^{\boldsymbol{r_j}} \boldsymbol{\nabla} \chi_g(\boldsymbol{r'}) \cdot d\boldsymbol{r'} \nonumber \\
\frac{2\pi}{\Phi_0}\Bigg(g\big(\boldsymbol{A}(g^{-1}(\boldsymbol{r}))\big) - \boldsymbol{A}(\boldsymbol{r})\Bigg) = \boldsymbol{\nabla}\chi_g(\boldsymbol{r}).
\end{align}
In the periodic Landau gauge $\boldsymbol{A}(\boldsymbol{r}) = \frac{\Phi}{2\pi}\Bigg(\xi_1 \boldsymbol{G}_2 - 2\pi\boldsymbol{\nabla}\big(\xi_2\left \lfloor{\xi_1 + \epsilon}\right \rfloor\big) \Bigg) $ where $\xi_{1}$ and $\xi_{2}$ are defined by $\boldsymbol{r} = \xi_{1} \boldsymbol{L}_1 + \xi_{2} \boldsymbol{L}_2$. 

We have for $C_{3z}$
\begin{align}
C_{3z}\big(\boldsymbol{A}(C_{3z}^{-1}(\boldsymbol{r}))\big) = \frac{\Phi}{2\pi}\Bigg(-\xi_2 (\boldsymbol{G}_2 - \boldsymbol{G}_1) + 2\pi \boldsymbol{\nabla}\bigg((\xi_2+ \xi_1)\left \lfloor{\xi_2 + \epsilon}\right \rfloor\bigg) \Bigg), 
\end{align}
and hence
\begin{align}
\chi_{C_{3z}}(\boldsymbol{r}) = \frac{2 \pi p}{q} \Bigg(( \xi_1 + \xi_2)\lfloor \xi_2 + \epsilon \rfloor + \xi_2 \lfloor \xi_1 + \epsilon  \rfloor - \xi_1 \xi_2 - \frac{\xi_2^2}{2}\Bigg).
\end{align}
Similarly for $C_{2z}$ we get
\begin{align}
C_{2z}\Big(\boldsymbol{A}\big(C_{2z}^{-1}(\boldsymbol{r})\big)\Big) &= \frac{\Phi}{2\pi}\Bigg(\xi_1 \boldsymbol{G}_2  + 2\pi \boldsymbol{\nabla}\Big(\xi_2\left \lfloor{-\xi_1 + \epsilon}\right \rfloor\Big) \Bigg), \nonumber \\
\chi_{C_{2z}}(\boldsymbol{r}) &= \frac{2\pi p}{q} \Bigg(\xi_2 \lfloor \xi_1 + \epsilon \rfloor + \xi_2 \lfloor - \xi_1 + \epsilon \rfloor \Bigg). 
\end{align}
Above we have used the facts that for orthogonal transformations $g$ and scalar functions $f(\boldsymbol{r})$ and $h(\boldsymbol{r}) = f(g^{-1}(\boldsymbol{r}))$, we have $\boldsymbol{\nabla} h|_{\boldsymbol{r}}  = g\big(\boldsymbol{\nabla}f|_{g^{-1}(\boldsymbol{r})}\big)$, and that for a function of $\xi_1$ and $\xi_2$ we have $2\pi \boldsymbol{\nabla} f = \frac{\partial f}{\partial \xi_1} \boldsymbol{G}_1+ \frac{\partial f}{\partial \xi_2}  \boldsymbol{G}_2$. The functions $\chi_{C_{3z}}$ and $\chi_{C_{2z}}$ have the following periodicity properties 
\begin{align}
    \chi_{C_{3z}}(\boldsymbol{r} + q \boldsymbol{L}_2) =  \chi_{C_{3z}}(\boldsymbol{r}) + \pi p q \  \text{mod} \ 2\pi,& \quad  \chi_{C_{3z}}(\boldsymbol{r} + \boldsymbol{L}_1) = \chi_{C_{3z}}(\boldsymbol{r}) + \frac{2 \pi p}{q}\lfloor \xi_2 + \epsilon \rfloor \nonumber \\
    \chi_{C_{2z}}(\boldsymbol{r} + q \boldsymbol{L}_2) =  \chi_{C_{2z}}(\boldsymbol{r}) \  \text{mod} \ 2\pi,& \quad  \chi_{C_{2z}}(\boldsymbol{r} + \boldsymbol{L}_1) = \chi_{C_{2z}}(\boldsymbol{r}).
\end{align}
We are interested in $p=q=1$, so we can write
\begin{align}
e^{i \chi_{C_{3z}}(\boldsymbol{r})} =& e^{-i\boldsymbol{G}_2\cdot \boldsymbol{r}/2}e^{i\overline{\chi}_{C_{3z}}(\boldsymbol{r})} \nonumber \\
e^{i \chi_{C_{2
z}}(\boldsymbol{r})} = & e^{i \overline{\chi}_{C_{2z}}(\boldsymbol{r})},
\end{align}
where barred phases are periodic in the Moiré unit cell. As we will see now, the phases $e^{i \chi_{C_{3z}}(\boldsymbol{r})}$ and $e^{i \chi_{C_{2z}}(\boldsymbol{r})}$ modify the transformations of the Bloch waves, redefining the high symmetry points in flux.

The Bloch waves are written
\begin{align}
    c^\dagger_{\boldsymbol{k},\boldsymbol{i}} = \frac{1}{\sqrt{N_M}} \sum_{\boldsymbol{l}} e^{i\boldsymbol{k}\cdot (\boldsymbol{R_l} + \boldsymbol{\delta_i})} c^\dagger_{\boldsymbol{l},\boldsymbol{i}},
\end{align}
with $\boldsymbol{k}$ belonging to the Moiré Brillouin zone, and here $c^\dagger_{\boldsymbol{l},\boldsymbol{i}}$ creates an electron at position $\boldsymbol{R_l} + \boldsymbol{\delta_i}$ where $\boldsymbol{R_l}$ is a lattice vector and $\boldsymbol{\delta_i}$ belongs to the Wigner-Seitz cell. 

Under $C_{3z}$, $  c^\dagger_{\boldsymbol{k},\boldsymbol{i}}$ transforms as
\begin{align}
    C_{3z} c^\dagger_{\boldsymbol{k},\boldsymbol{i}} (C_{3z})^{-1} = \frac{1}{\sqrt{N_M}} \sum_{\boldsymbol{l}} e^{i(C_{3z}(\boldsymbol{k}) - \boldsymbol{G}_2/2)\cdot  C_{3z}(\boldsymbol{R_l} + \boldsymbol{\delta_i})} e^{i\overline{\chi}_{C_{3z}}(C_{3z}(\boldsymbol{\delta_i}))}c^\dagger_{C_{3z}(\boldsymbol{l},\boldsymbol{i})}.
\end{align}
Here, $c^\dagger_{C_{3z}(\boldsymbol{l},\boldsymbol{i})}$ creates an electron at position $C_{3z}(\boldsymbol{R_l} + \boldsymbol{\delta_i})$. We see that $C_{3z}$ sends momentum $\boldsymbol{k}$ to $C_{3z}(\boldsymbol{k}) - \boldsymbol{G}_2/2$. Via the embedding relation $c^\dagger_{\boldsymbol{k} + \boldsymbol{G},\boldsymbol{i}} = e^{i\boldsymbol{G}\cdot \boldsymbol{\delta_i}} c^\dagger_{\boldsymbol{k},\boldsymbol{i}}$ for $\boldsymbol{G}$ a reciprocal lattice vector, the three-fold rotation in flux acts in the momenta as follows,
\begin{align}
    \boldsymbol{k} \xrightarrow{C_{3z}} C_{3z}(\boldsymbol{k}) - \boldsymbol{G}_2/2 \sim C_{3z}\Big(\boldsymbol{k} - (\boldsymbol{G}_1 + \boldsymbol{G}_2)/2\Big) + (\boldsymbol{G}_1 + \boldsymbol{G}_2)/2.
\end{align}
Also, given that $\chi_{C_{2z}}(\boldsymbol{r})$ is periodic mod $2\pi$ on the unit cell, the momentum transforms like in zero flux, 
\begin{align}
    \boldsymbol{k} \xrightarrow{C_{2z}} C_{2z}(\boldsymbol{k}) \sim C_{2z}\Big(\boldsymbol{k} - (\boldsymbol{G}_1 + \boldsymbol{G}_2)/2\Big) + (\boldsymbol{G}_1 + \boldsymbol{G}_2)/2.
\end{align}

The center of rotations has shifted from $\boldsymbol{\Gamma_M}$ to one of the other parity invariant points, $(\boldsymbol{G}_1 + \boldsymbol{G}_2)/2$, at one magnetic flux quantum.

Now we look for the operator realizing $C_{2y}$. The procedure is the same, but in this case $C_{2y}H_0(C_{2y})^{-1}$ should be equal to the Hamiltonian with the sign of the magnetic field reversed. Hence, $\chi_{C_{2y}}(\boldsymbol{r})$ must obey
\begin{align}
    C_{2y}\big(\boldsymbol{A}(C_{2y}^{-1}(\boldsymbol{r}))\big) + \boldsymbol{A}(\boldsymbol{r}) = \boldsymbol{\nabla}\chi_{C_{2y}}(\boldsymbol{r}).
\end{align}
We obtain for $\chi_{C_{2y}}(\boldsymbol{r})$
\begin{align}
    \chi_{C_{2y}}(\boldsymbol{r}) = \frac{2 \pi p}{q} \Bigg(-\xi_2\lfloor \xi_1 + \epsilon \rfloor + \xi_2 \lfloor \xi_1 + \xi_2 + \epsilon  \rfloor - \frac{\xi_2^2}{2}\Bigg),
\end{align} which obeys the properties
\begin{align}
     \chi_{C_{2y}}(\boldsymbol{r} + q \boldsymbol{L}_2) =  \chi_{C_{2y}}(\boldsymbol{r}) - \pi p q \  \text{mod} \ 2\pi,& \quad  \chi_{C_{2y}}(\boldsymbol{r} + \boldsymbol{L}_1) = \chi_{C_{2y}}(\boldsymbol{r}).
\end{align}
The same analysis as before applies, and under $C_{2y}$ the momentum transform as
\begin{align}
    \boldsymbol{k} \xrightarrow{C_{2y}} C_{2y}(\boldsymbol{k}) - \boldsymbol{G}_2/2 \sim C_{2y}\Big(\boldsymbol{k} - (\boldsymbol{G}_1 + \boldsymbol{G}_2)/2\Big) +  (\boldsymbol{G}_1 + \boldsymbol{G}_2)/2.
\end{align}

For the time reversal operator $\mathcal{T}$, the magnetic field should be reversed also, and it is trivial to see that the action is the same as for the zero flux case. $\mathcal{T}$ is an antiunitary operator satisfying $\mathcal{T}c^\dagger_{\boldsymbol{i}}\mathcal{T}^{-1} = c_{\boldsymbol{i}}^\dagger$, and transforming the momentum as 
\begin{align}
    \boldsymbol{k} \xrightarrow{\mathcal{T}} - \boldsymbol{k} \sim -\Big(\boldsymbol{k} - (\boldsymbol{G}_1 + \boldsymbol{G}_2)/2\Big) + (\boldsymbol{G}_1 + \boldsymbol{G}_2)/2.
\end{align}

In conclusion, the action of symmetry operators under one magnetic flux quantum effectively shift the Brillouin zone by $(\boldsymbol{G}_1 + \boldsymbol{G_2})/2$, redefining the high symmetry points.

\section{Valley charge and $C_{2z}P$ operator on the lattice}\label{appb}

We wish to find an operator $\tau_z$ implementing the valley charge on the lattice, such that $\langle \tau_z \rangle = +1$ on states nearby the $K$ point of graphene and $-1$ near the $K'$ point. We adopt a slight generalization of the valley operator of Ref. \cite{Lado19}
% \begin{align}
%     \tau_z = \sum_{\triangle} e^{i \theta_{\boldsymbol{}}}c^\dagger_{\boldsymbol{r}(\triangle(1))}  c^\dagger_{\boldsymbol{r}(\triangle(2))} + c^\dagger_{\boldsymbol{r}(\triangle(2))}  c^\dagger_{\boldsymbol{r}(\triangle(3))} + c^\dagger_{\boldsymbol{r}(\triangle(3))}  c^\dagger_{\boldsymbol{r}(\triangle(1))}
% \end{align}
\begin{align}
    \tau_z = \frac{i}{3\sqrt{3}}\sum_{l}\bigg(&\sum_{\bigtriangledown} e^{-i \theta_{\bigtriangledown(1),\bigtriangledown(2)}} c^\dagger_{\bigtriangledown(1)}  c_{\bigtriangledown(2)} + e^{-i \theta_{\bigtriangledown(2),\bigtriangledown(3)}} c^\dagger_{\bigtriangledown(2)}  c_{\bigtriangledown(3)} + e^{-i \theta_{\bigtriangledown(3),\bigtriangledown(1)}} c^\dagger_{\bigtriangledown(3)}  c_{\bigtriangledown(1)} \nonumber \\ &- \sum_{\triangle} e^{-i \theta_{\triangle(1),\triangle(2)}} c^\dagger_{\triangle(1)}  c_{\triangle(2)} + e^{-i \theta_{\triangle(2),\triangle(3)}} c^\dagger_{\triangle(2)}  c_{\triangle(3)} + e^{-i \theta_{\triangle(3),\triangle(1)}} c^\dagger_{\triangle(3)}  c_{\triangle(1)}  \bigg) + \text{h.c.}.
\end{align}

The sums are over triangles upside down of sublattice $A$ atoms, and triangles of sublattice $B$, and $l$ denotes the sum over the two layers. We draw an example of each kind of triangle in Figure \ref{vop}. The phases are the Peierls' phases defined in the main text. It can be shown that valley $K$ states have $\langle \tau_z \rangle = +1 + O(a/L_M)$ and valley $K'$  states have $\langle \tau_z \rangle = -1 + O(a/L_M)$.

For the particle-hole operator $C_{2z}P$ (Eq.\ref{c2p}) we proceed as follows. Once we have valley  polarized states, we obtain the envelope functions $\Psi_{\eta \sigma l}$ by multiplying  $\Psi_\eta$ with the valley factor $e^{-i\eta n_\theta \boldsymbol{G_2}\cdot \boldsymbol{r}}$. Afterwards, we perform a smooth interpolation of the data $\Psi_{\eta \sigma l}(\boldsymbol{r_i})$ being $\boldsymbol{r_i}$ the positions of the atoms at sublattice $\sigma$ and layer $l$. Finally, the smooth functions are sampled at the points of the opposite sublattice and layer, and the factor $\eta s_l e^{-i\eta n_\theta  \boldsymbol{G_2}\cdot \boldsymbol{r}}e^{-i \eta \boldsymbol{G_2} \cdot \boldsymbol{r}}$ is also added. As a note, When $\Phi = \Phi_0$ the envelopes have a discontinuity at $\xi_1 = $ an integer in the periodic Landau gauge, and special care is needed when performing the interpolation.

The projected operator in the flat bands $[\overline{C_{2z}P}(\boldsymbol{k})]_{\rho \rho'} = \langle \boldsymbol{k} \rho | C_{2z}P | \boldsymbol{k}\rho' \rangle = \langle \boldsymbol{k} \rho | C_{2z}P( \boldsymbol{k}\rho') \rangle  $  is then constructed. We have checked that the particular choice of basis for the interpolation is irrelevant, and the matrix elements of $\overline{C_{2z}P}(\boldsymbol{k})$ in a different basis computed via interpolation or unitary conjugation of the original matrix are essentially identical.

In order to check the validity of our definitions, we have computed several benchmarks. The mean of the absolute value of the matrix elements of $\overline{C_{2z}P}(\boldsymbol{k}) - \overline{C_{2z}P}(\boldsymbol{k})^\dagger$ are always less than $10^{-5}$, so we capture well the hermiticity of $C_{2z}P$. Also, the mean of the absolute value of the valley off-diagonal matrix elements of $\overline{C_{2z}P}(\boldsymbol{k})$ are always less than $10^{-5}$ for $\Phi_0$ flux and $10^{-7}$ at zero field. We conclude that our implementations of the valley charge and the p-h operator are trustworthy.

\begin{figure}[H]
\centering
% \begin{subfigure}{.5\textwidth}
  \centering
  \includegraphics[width=.45\linewidth]{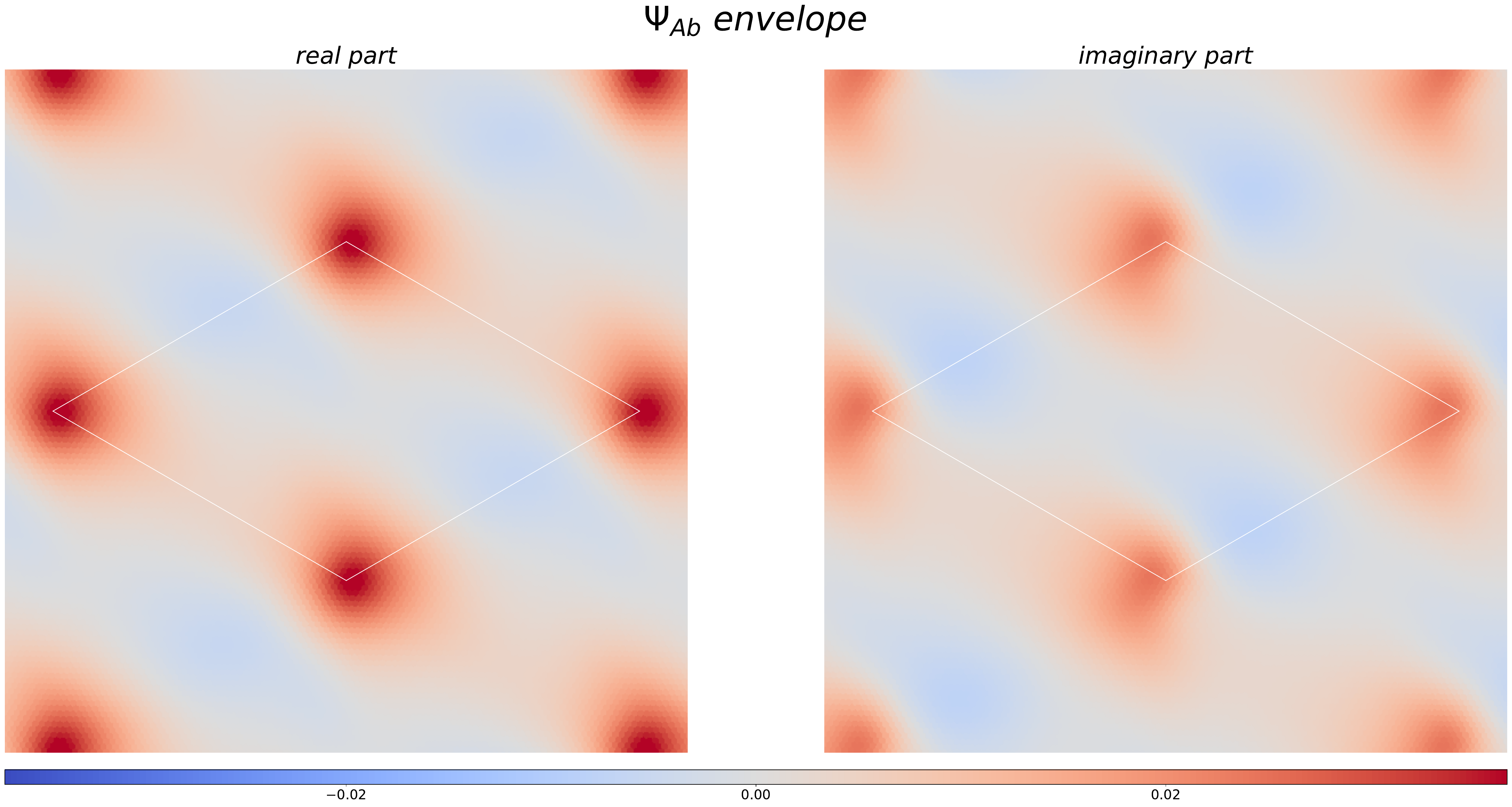}
  \includegraphics[width=.45\linewidth]{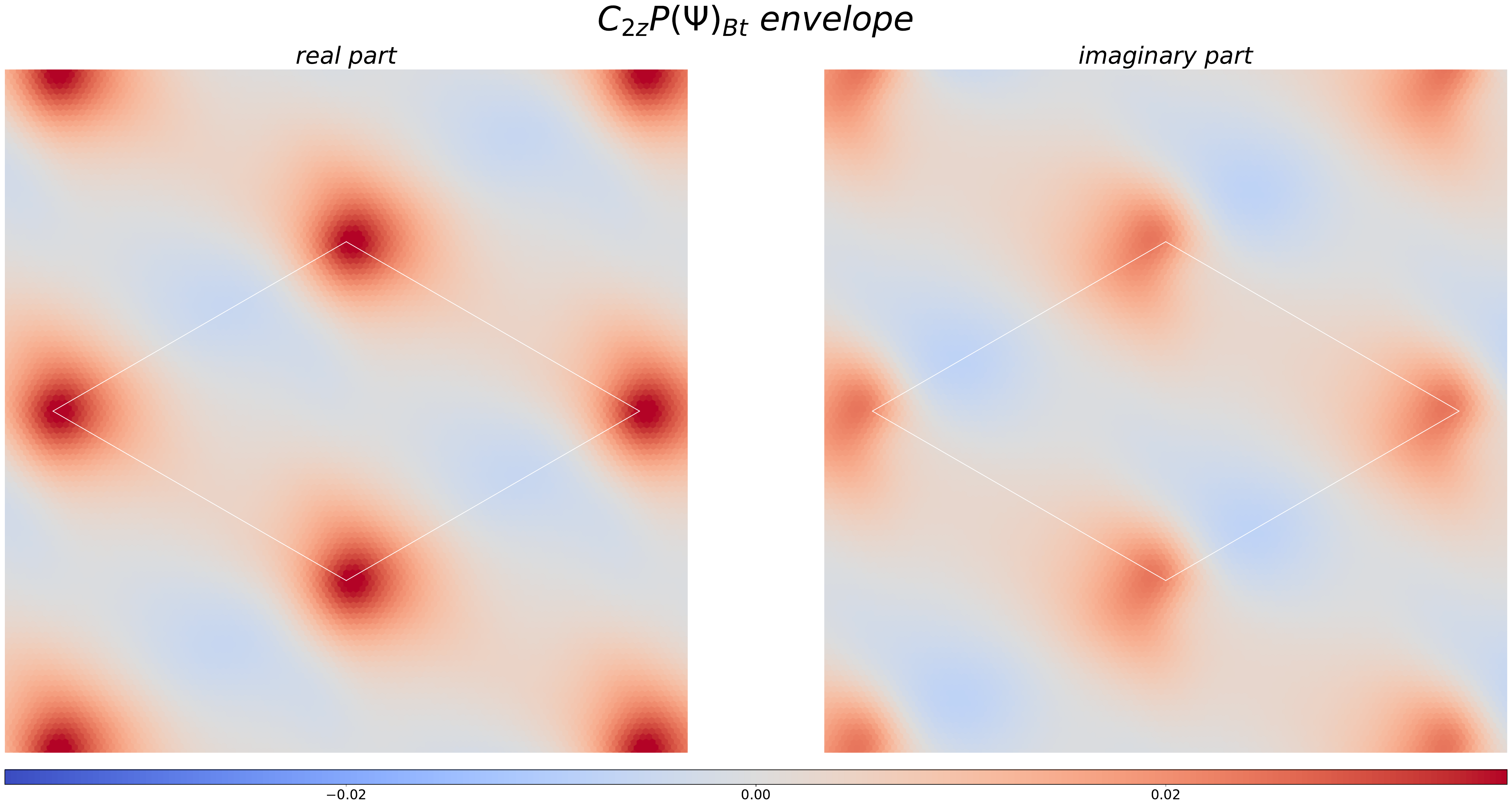}
\caption{\textbf{$\boldsymbol{C_{2z}P}$ operator at zero flux. Left:} for the state $|\boldsymbol{M_M} K A \rangle$ at zero field we plot the smooth envelope at sublattice $A$ bottom layer, $\Psi_{K A b}(\boldsymbol{r})$. \textbf{Right:} for the $C_{2z}P$ transformed state, we plot $e^{in_\theta \boldsymbol{G}_2 \cdot \boldsymbol{r}} e^{i \boldsymbol{G}_2 \cdot \boldsymbol{r}}C_{2z}P(\Psi_K)_{Bt}(\boldsymbol{r})$. }
\label{figc2}
\end{figure}

\begin{figure}[H]
\centering
  \centering
  \includegraphics[width=.45\linewidth]{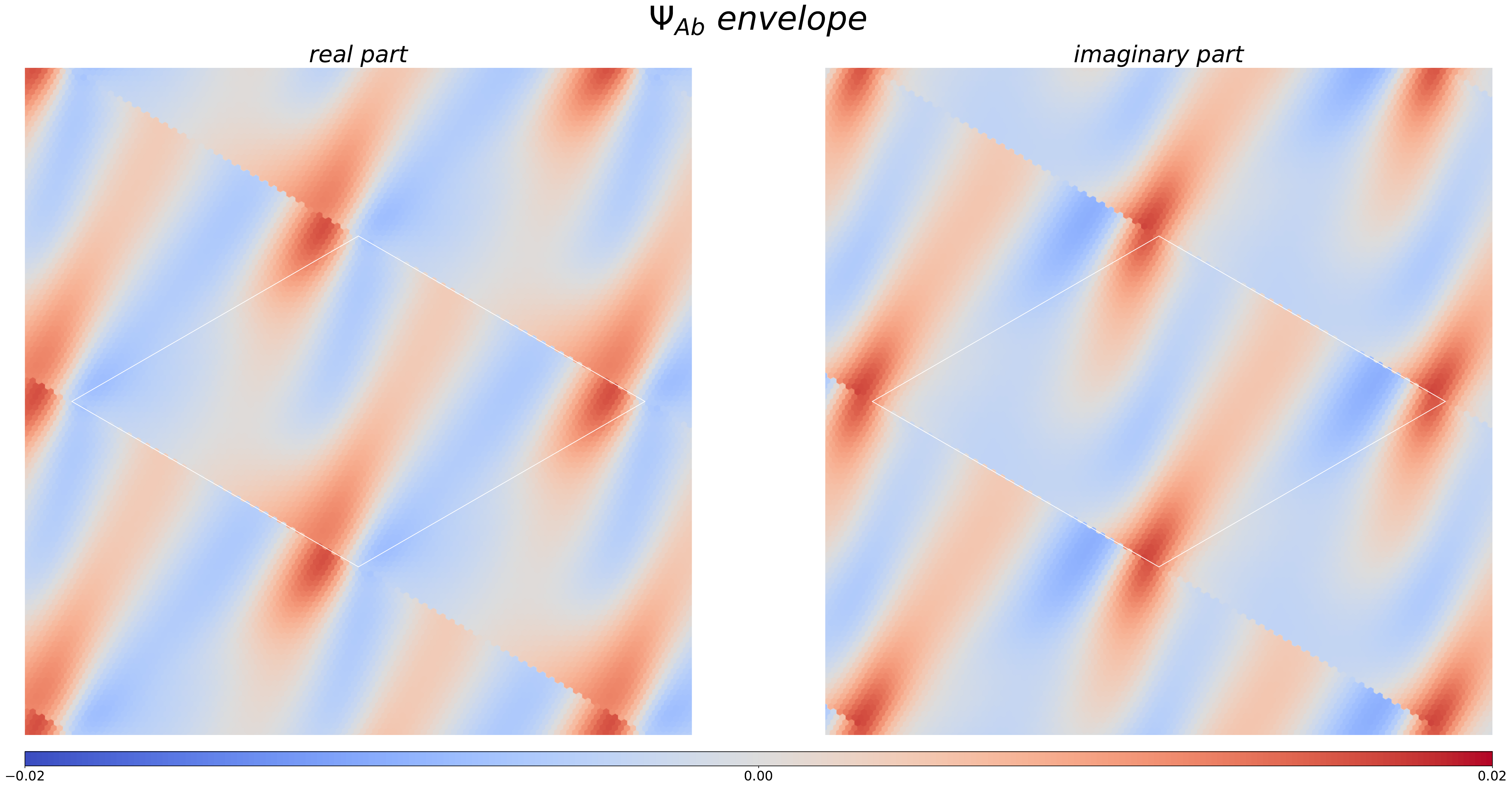}
  \includegraphics[width=.45\linewidth]{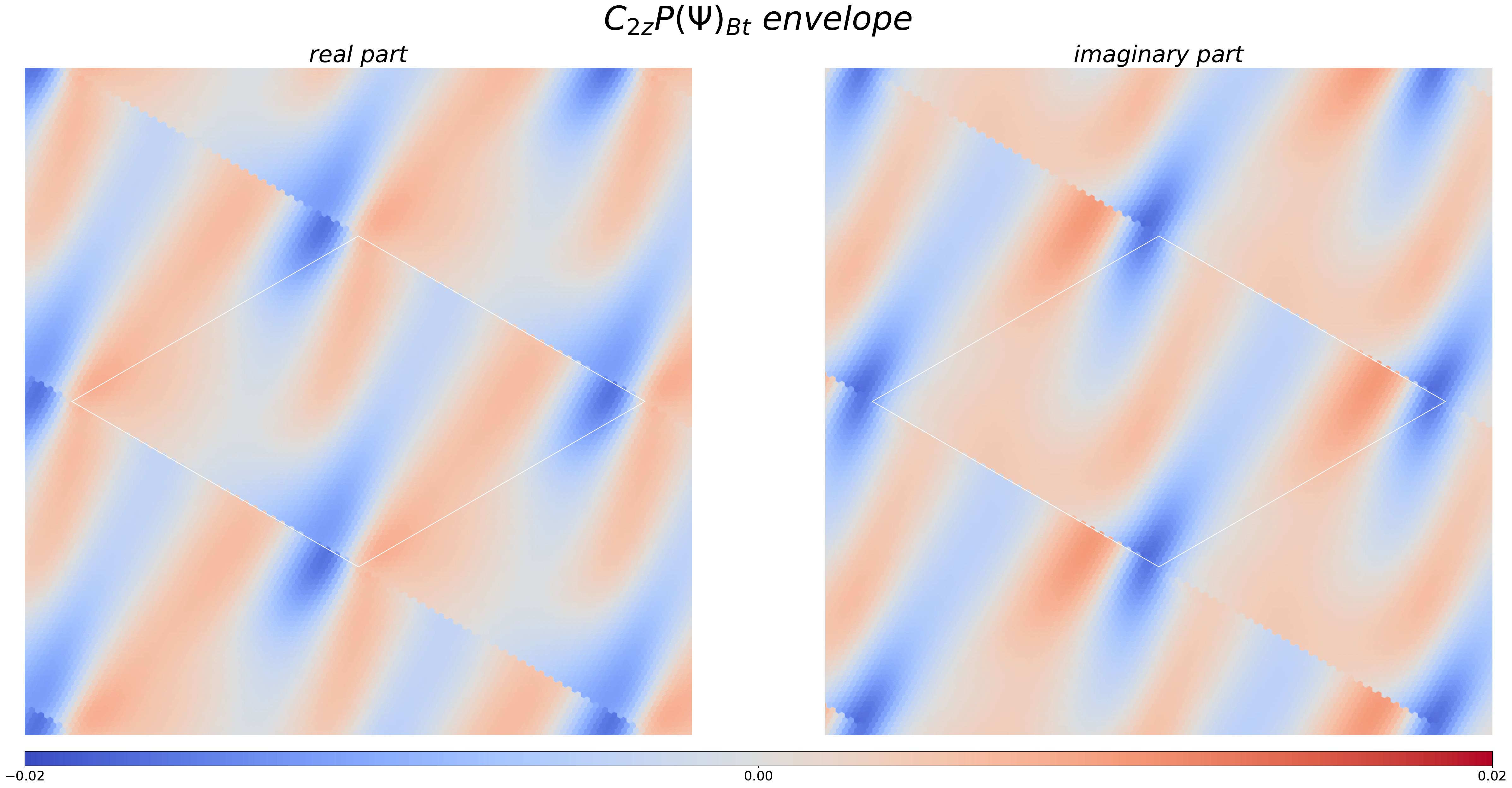}
\label{fig:test}
\caption{\textbf{$\boldsymbol{C_{2z}P}$ operator at one flux quantum.} In this case we plot the envelope of $|\boldsymbol{M_M} K' +1 \rangle$ at $26.5$ T in the right, and $e^{-in_\theta \boldsymbol{G}_2 \cdot \boldsymbol{r}} e^{-i \boldsymbol{G}_2 \cdot \boldsymbol{r}}C_{2z}P(\Psi_K')_{Bt}(\boldsymbol{r})$ in the left plot. Notice the discontinuities of the wave function on the unit cell edge due to the periodic Landau gauge, and the extra minus sign $\eta=-1$ in the transformed wave fucntion, see Eq. \ref{c2p}.}
\end{figure}

\section{$\boldsymbol{U(1)_v}$ AND $\boldsymbol{SU(2)_{K}\times SU(2)_{K'}}$ symmetries in the lattice model}\label{appc}

In the BM model, a general wave function can be written similarly to Eq. \ref{wf}, in a $4$ component notation for sublattice and layer ($t$=top, $b$=bottom),
\begin{align}
    u(\boldsymbol{r}) =e^{i\boldsymbol{K_u}\cdot \boldsymbol{r}} (u_{A t}(\boldsymbol{r}), u_{ B t }(\boldsymbol{r}), u_{A b}(\boldsymbol{r}), u_{B b}(\boldsymbol{r}))^T,
\end{align}
with $\boldsymbol{K_u} =\pm\boldsymbol{K}$, the graphene Dirac points, and the envelope functions are smooth on the graphene scale.
The matrix elements of the Coulomb interaction (actually, the four fermion part of the normal ordered operator) read
\begin{align}
    \langle a b | V | c d \rangle = \frac{1}{2} \sum_{\{\sigma\},\{l\}}
    \int d\boldsymbol{r_i} \int d\boldsymbol{r_j} V(\boldsymbol{r_i} - \boldsymbol{r_j})  e^{i(\boldsymbol{K_c} -\boldsymbol{K_a})\cdot \boldsymbol{r_i}} e^{i(\boldsymbol{K_d} -\boldsymbol{K_b})\cdot \boldsymbol{r_j}} a^*_{\sigma_a l_a}(\boldsymbol{r_i}) b^*_{\sigma_b l_b}(\boldsymbol{r_j}) c_{\sigma_c l_c}(\boldsymbol{r_i}) d_{\sigma_d l_d}(\boldsymbol{r_j}).
\end{align}

This integral computes the Fourier transform of the potential at momenta $\sim \boldsymbol{K_a} - \boldsymbol{K_{c}}$, or equivalently $\sim \boldsymbol{K_d} - \boldsymbol{K_{b}}$. Hence, unless $\boldsymbol{K_a} = \boldsymbol{K_{c}}$ and $\boldsymbol{K_b} = \boldsymbol{K_{d}}$ the form factor is negligible given $||\boldsymbol{K}|| \sim a^{-1} \gg \xi^{-1}$. This is the origin of $U(1)_v$ and $SU(2)_{K} \times SU(2)_{K'}$ in TBG.

\begin{figure}[b]
    \centering
    \includegraphics[width=.3\linewidth]{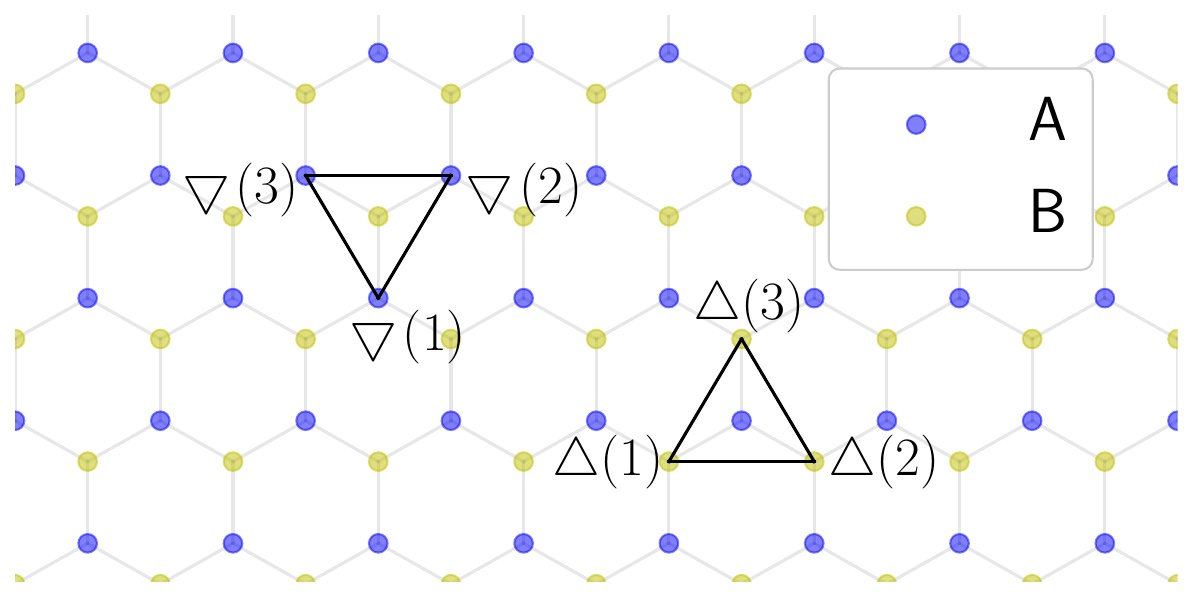}
    \includegraphics[width=.2\linewidth]{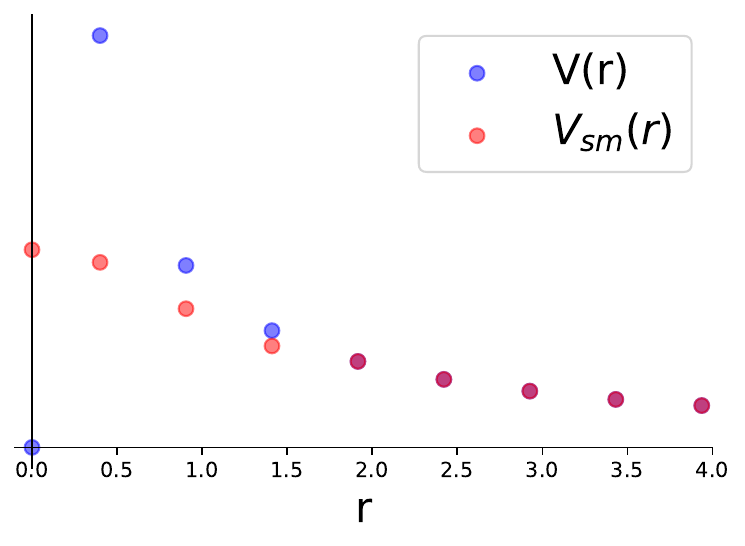}
    \caption{\textbf{Left:} Triangular loops that calculate the valley charge. \textbf{Right:} Decomposition of the potential. $V_{sm}(r)$ is a smoothed version of $V(r)$}
    \label{vop}
\end{figure}

In the tight-binding, we write states $|u\rangle$, with wavefunctions
\begin{align}
    \langle \boldsymbol{r_i} | u \rangle = e^{i \boldsymbol{K_u}\cdot \boldsymbol{r_i}} u(\boldsymbol{r_i}).
     %\\ \nonumber
    % \langle \boldsymbol{r_i} | b \rangle = e^{i \boldsymbol{K_b}\cdot \boldsymbol{r_i}} b(\boldsymbol{r_i}) \\ \nonumber
    % \langle \boldsymbol{r_i} | c \rangle = e^{i \boldsymbol{K_c}\cdot \boldsymbol{r_i}} c(\boldsymbol{r_i}) \\ \nonumber
    % \langle \boldsymbol{r_i} | d \rangle = e^{i \boldsymbol{K_d}\cdot \boldsymbol{r_i}} d(\boldsymbol{r_i}) \\ \nonumber
\end{align}
$u(\boldsymbol{r_i})$ is further decomposed depending on the sublattice and layer of the point $\boldsymbol{r_i}$, $u(\boldsymbol{r_i}) = \sum_{\sigma l }u_{\sigma l}(\boldsymbol{r_i}) \delta_{\boldsymbol{r_i} \in \sigma l}$. The functions $ u_{\sigma l}(\boldsymbol{r_i})$ vary slowly on the graphene scale, such that $u_{\sigma l}(\boldsymbol{r_i} + \boldsymbol{a}_{1,2}) \approx u_{\sigma l}(\boldsymbol{r_i})$. Consider the matrix element of the interaction (actually, the four fermion part of the normal ordered operator)
\begin{align}
    \langle a b | V | c d \rangle = \frac{1}{2}\sum_{\boldsymbol{r_i}, \boldsymbol{r_j}} V(\boldsymbol{r_i} - \boldsymbol{r_j}) e^{-i(\boldsymbol{K_a}-\boldsymbol{K_c})\cdot \boldsymbol{r_i}} e^{-i(\boldsymbol{K_b}-\boldsymbol{K_d})\cdot \boldsymbol{r_j}} a^*(\boldsymbol{r_i}) b^*(\boldsymbol{r_j}) c(\boldsymbol{r_i}) d(\boldsymbol{r_j})
\end{align}

We will see that if the valley charge is not conserved, i.e. $\boldsymbol{K_a} + \boldsymbol{K_b} - \boldsymbol{K_c} - \boldsymbol{K_d} \neq \boldsymbol{0}$ the matrix element vanishes to a first approximation. Let us take for example $\boldsymbol{K_a} = \boldsymbol{K_c} = \boldsymbol{K_d} = -\boldsymbol{K_b} =\boldsymbol{K}$, then
\begin{align}
    \langle a b | V | c d \rangle =  \frac{1}{2}\sum_{\boldsymbol{r_i},\boldsymbol{r_j}} V(\boldsymbol{r_i} - \boldsymbol{r_j}) e^{-2i\boldsymbol{K}\cdot \boldsymbol{r_i}} a^*(\boldsymbol{r_i}) b^*(\boldsymbol{r_j}) c(\boldsymbol{r_i}) d(\boldsymbol{r_j}).
\end{align}
Shifting the integration variables $\boldsymbol{r_{i,j}} \to \boldsymbol{r_{i,j}} + \boldsymbol{a}_1$ and using the property of the envelope functions $u(\boldsymbol{r_i} + \boldsymbol{a}_{1,2}) \approx u(\boldsymbol{r_i})$ we get 
\begin{align}
        \langle a b | V | c d \rangle = e^{-2i \boldsymbol{K}\cdot \boldsymbol{a}_1}   \langle a b | V | c d \rangle =  e^{-2 \pi i/3}   \langle a b | V | c d \rangle = 0.
\end{align}
The argument for other charge non conserving combinations of $\boldsymbol{K_{a,b,c,d}}$ is the same and we conclude that $U(1)_v$ is a symmetry in the atomistic model also. 

If $\boldsymbol{K_a} = \boldsymbol{K_d} = - \boldsymbol{K_b} = - \boldsymbol{K_c}$, say $\boldsymbol{K_a} = \boldsymbol{K}$, we have
\begin{align}
    \langle a b | V | c d \rangle =   \frac{1}{2} \sum_{\boldsymbol{r_i},\boldsymbol{r_j}} V(\boldsymbol{r_i} - \boldsymbol{r_j}) e^{-2i\boldsymbol{K}\cdot (\boldsymbol{r_i} - \boldsymbol{r_j})}  a^*(\boldsymbol{r_i}) b^*(\boldsymbol{r_j}) c(\boldsymbol{r_i}) d(\boldsymbol{r_j}),
\end{align}
and the shifting argument does not work because the phases for $\boldsymbol{r_i}$ and $\boldsymbol{r_j}$ cancel. In general these matrix elements are non zero and break the $SU(2)_K \times SU(2)_{K'}$ symmetry of independent spin rotations in each valley. 

Furthermore, the exchange energy of states with different valley charge, say $|a\rangle = |d\rangle$  and  $|b\rangle = |c\rangle$, is always positive. For that, decompose the potential into a smooth part $V_{sm}(\boldsymbol{r})$ and a short-range part  $V_{sr}(\boldsymbol{r})$. $V_{sm}(\boldsymbol{r})$ is slowly varying on the graphene scale, is equal to $V(\boldsymbol{r})$ at long distances (say $||\boldsymbol{r}|| > 2a$) and $V_{sm}(\boldsymbol{0})$ equals a positive constant. On the other hand, $V_{sr}(\boldsymbol{r}) = V(\boldsymbol{r}) - V_{sm}(\boldsymbol{r})$ is equal to $-V_{sm}(\boldsymbol{0})$ at $\boldsymbol{r} = 0$ (remember the on-site interaction is treated separately by the Hubbard term, so $V(\boldsymbol{0})=0$), exhibits the steep Coulomb repulsion at short distances and vanishes at long distances. In Fig. \ref{vop} we sketch this decompositon.

Given that $V_{sm}(\boldsymbol{r}+\boldsymbol{a}_{1,2})\approx V_{sm}(\boldsymbol{r})$, one can apply the shifting only to the fisrt variable $\boldsymbol{r_i}$, and get
\begin{align}
        \langle a b | V_{sm} | b a \rangle = e^{-2i \boldsymbol{K}\cdot \boldsymbol{a}_1}   \langle a b | V_{sm} | b a \rangle =  e^{-2 \pi i/3}   \langle a b | V_{sm} | b a \rangle = 0.
\end{align}

For $V_{sr}(\boldsymbol{r})$ let us consider only the dominant contributions of intralayer terms with $||\boldsymbol{r_i}-\boldsymbol{r_j}|| = 0 , a_0$ and $a$.
% \begin{widetext}
\begin{align}
        % % &\langle a b | V_{sr} | b a \rangle = \sum_{\boldsymbol{r_i},\boldsymbol{r_j}}  V_{sr}(\boldsymbol{r_i} - \boldsymbol{r_j}) e^{-2i\boldsymbol{K}\cdot (\boldsymbol{r_i} - \boldsymbol{r_j})} \nonumber \\ 
        % & \hspace{2.7cm} \times a^*(\boldsymbol{r_i}) b^*(\boldsymbol{r_j}) b(\boldsymbol{r_i}) a(\boldsymbol{r_j}) = \nonumber \\&
        \langle a b | V_{sr} | b a \rangle  = \frac{1}{2}&\sum_{\boldsymbol{r_j}} V_0 |a_{\sigma_j l_j}(\boldsymbol{r_j})|^2 |b_{\sigma_j l_j}(\boldsymbol{r_j})|^2 + 3 V_a(e^{2 \pi i /3} + e^{-2 \pi i /3})|a_{\sigma_j l_j}(\boldsymbol{r_j})|^2 |b_{\sigma_j l_j}(\boldsymbol{r_j})|^2 
        \nonumber \\ 
        &\hspace{.5cm} + V_{a_0}(1 + e^{2\pi i /3} + e^{-2 \pi i /3}) a^*_{\Bar{\sigma}_j l_j}(\boldsymbol{r_j})a_{\sigma_j l_j}(\boldsymbol{r_j}) b^*_{\sigma_j l_j}(\boldsymbol{r_j})b_{\Bar{\sigma}_j l_j}(\boldsymbol{r_j})  %\nonumber \\  
        % V_a(3e^{-2 \pi i /3} + 3e^{-2 \pi i /3})|a(\boldsymbol{r_j})|^2 |b(\boldsymbol{r_j})|^2 
        \nonumber \\
        =& \frac{1}{2} \sum_{\boldsymbol{r_j}} (V_0 - 3V_{a}) |a(\boldsymbol{r_j})|^2 |b(\boldsymbol{r_j})|^2 < 0,
\end{align}
$\sigma_j$ and $l_j$ denoting the sublattice and layer of the point $\boldsymbol{r_j}$ and $\Bar{\sigma}_j$ the opposite sublattice to $\sigma_j$. $V_r$ encodes the value of $V_{sr}(\boldsymbol{r})$ when $||\boldsymbol{r}|| = r$. Notice that $V_0 < 0$ and $V_a > 0$.
In turn, the exchange energy $- \langle a b | V | b a \rangle$ is always positive
\section{The Hartree-Fock method and flat band projection}\label{appd}
Consider the normal ordered interaction of Eqs. \ref{potential} and \ref{hubbard},
\begin{align}
     V + H_U = \frac{1}{2}\sum_{\boldsymbol{r_i},\boldsymbol{r_j} s_i s_j} V(\boldsymbol{r_i}-\boldsymbol{r_j}) :c^\dagger_{\boldsymbol{i},s_i} c_{\boldsymbol{i}, s_i} c^\dagger_{\boldsymbol{j},s_j} c_{\boldsymbol{j}, s_j}: + U \sum_{\boldsymbol{i}} :c_{\boldsymbol{i}\uparrow}^\dagger c_{\boldsymbol{r_i}\uparrow} c_{\boldsymbol{i}\downarrow}^\dagger c_{\boldsymbol{i}\downarrow}:
\end{align}

The choice of the normal ordering with respect to the ground state of graphene at charge neutrality is necessary to avoid double counting the interaction\cite{xie20,Bultinck20}. This is, we assume that the hopping integrals $t(\boldsymbol{r})$ are already renormalized by the interactions with the deep Fermi sea of graphene. After expanding the normal ordered product\cite{giuliani_vignale_2005} and performing the Hartree-Fock decoupling, the Hamiltonian reads
\begin{align}
    V_{\text{HF}} + H_{U\text{HF}} =& \sum_{\boldsymbol{r_i},\boldsymbol{r_j},s_i,s_j} V(\boldsymbol{r_i}-\boldsymbol{r_j}) c_{\boldsymbol{i}s_i}^\dagger c_{\boldsymbol{i}s_i} \Big(\langle c_{\boldsymbol{j}s_j}^\dagger c_{\boldsymbol{j}s_j}\rangle - \langle c^\dagger_{\boldsymbol{j}s_j}c_{\boldsymbol{j}s_j}\rangle_0 \Big) -  \sum_{\boldsymbol{r_i},\boldsymbol{r_j},s} V(\boldsymbol{r_i}-\boldsymbol{r_j}) c_{\boldsymbol{i}s}^\dagger c_{\boldsymbol{j}s} \Big(\langle c_{\boldsymbol{i}s}^\dagger c_{\boldsymbol{j}s}\rangle - \langle c^\dagger_{\boldsymbol{i}s}c_{\boldsymbol{j}s}\rangle_0 \Big)^* \nonumber \\
    &+ U \sum_{\boldsymbol{r_i}} c_{\boldsymbol{i}\uparrow}^\dagger c_{\boldsymbol{i}\uparrow} \Big( \langle c_{\boldsymbol{i}\downarrow}^\dagger c_{\boldsymbol{i}\downarrow} \rangle - \langle c_{\boldsymbol{i}\downarrow}^\dagger c_{\boldsymbol{i}\downarrow} \rangle_0 \Big) + U \sum_{\boldsymbol{r_i}} c_{\boldsymbol{i}\downarrow}^\dagger c_{\boldsymbol{i}\downarrow} \Big( \langle c_{\boldsymbol{i}\uparrow}^\dagger c_{\boldsymbol{i}\uparrow} \rangle - \langle c_{\boldsymbol{i}\uparrow}^\dagger c_{\boldsymbol{i}\uparrow} \rangle_0 \Big) + \text{constant},
\end{align}
with $\langle ... \rangle_0$ denoting the expectation value in the ground state of graphene at charge neutrality, and $\langle ... \rangle$ the expectation value in the particular state of our Hartree-Fock decoupling. In our implementation we restrict the wave function to be a direct product of spin up and down electrons, such that $\langle c^\dagger_{\boldsymbol{i}\uparrow} c_{\boldsymbol{j}\downarrow} \rangle = 0$ for all $\boldsymbol{r_i},\boldsymbol{r_j}$.

In the projected limit we assume that the remote bands are filled and the relevant physics takes place in the flat bands. In this spirit we compute mean field interaction restricted to the subspace of the flat bands,
\begin{align}
    [V_{\text{HF,p}}(\boldsymbol{k},\boldsymbol{k'}) + H_{U\text{HF,p}}(\boldsymbol{k},\boldsymbol{k'})]_{\rho \rho'} = \Big(\langle \text{FS}| \otimes \langle \boldsymbol{k}\rho | \Big) \Big( V_{\text{HF}} + H_{U\text{HF}}\Big) \Big(| \text{FS} \rangle \otimes | \boldsymbol{k'} \rho' \rangle \Big),
\end{align}
with $ |\text{FS}\rangle \otimes |\boldsymbol{k}\rho \rangle$ denoting the direct product of the state with the filled remote bands and the state with momentum $\boldsymbol{k}$ and multi-index $\rho$. We further assume translational symmetry that makes the mean field Hamiltonian block-diagonal in momentum space, $V_{\text{HF,p}}(\boldsymbol{k},\boldsymbol{k'}) + H_{U\text{HF,p}}(\boldsymbol{k},\boldsymbol{k'}) = \Big(V_{\text{HF,p}}(\boldsymbol{k}) + H_{U\text{HF,p}}(\boldsymbol{k})\Big)\delta_{\boldsymbol{k},\boldsymbol{k'}}$.

The self-consistent method starts by proposing an ansatz for the ground state at any given filling, computing the mean field Hamiltonian and performing the flat band projection. Next, we solve the projected mean filed Hamiltonian 
\begin{align}
H_{\text{HF,p}}(\boldsymbol{k},\boldsymbol{k'}) = \Big(H_{0\text{,p}}(\boldsymbol{k}) + V_{\text{HF,p}}(\boldsymbol{k})+H_{U\text{HF,p}}(\boldsymbol{k})\Big) \delta_{\boldsymbol{k},\boldsymbol{k'}},    
\end{align}
with $H_{0\text{,p}}(\boldsymbol{k})\delta_{\boldsymbol{k},\boldsymbol{k'}}$ the projected kinetic energy operator. The ground state of this Hamiltonian is then a new ansatz for the self-consistent ground state and the process is repeated until convergence is reached.

The energy of the self-consistent state is
\begin{align}
    \langle H \rangle =& \langle V \rangle + \langle H_U \rangle + \langle H_0 \rangle \nonumber \\
    =& \frac{1}{2} \sum_{\boldsymbol{r_i},\boldsymbol{r_j},s_i,s_j} V(\boldsymbol{r_i}-\boldsymbol{r_j}) \Big(\langle c_{\boldsymbol{i}s_i}^\dagger c_{\boldsymbol{i}s_i}\rangle - \langle c^\dagger_{\boldsymbol{i}s_i}c_{\boldsymbol{i}s_i}\rangle_0 \Big) \Big(\langle c_{\boldsymbol{j}s_j}^\dagger c_{\boldsymbol{j}s_j}\rangle - \langle c^\dagger_{\boldsymbol{j}s_j}c_{\boldsymbol{j}s_j}\rangle_0 \Big) - \frac{1}{2} \sum_{\boldsymbol{r_i},\boldsymbol{r_j},s} V(\boldsymbol{r_i}-\boldsymbol{r_j}) \Big\lvert \Big\lvert\langle c_{\boldsymbol{i}s}^\dagger c_{\boldsymbol{j}s}\rangle - \langle c^\dagger_{\boldsymbol{i}s}c_{\boldsymbol{j}s}\rangle_0 \Big\rvert \Big\rvert^2 \nonumber \\ &+ U \sum_{\boldsymbol{r_i}} \Big(\langle c_{\boldsymbol{i}\uparrow}^\dagger c_{\boldsymbol{i}\uparrow}\rangle - \langle c^\dagger_{\boldsymbol{i}\uparrow}c_{\boldsymbol{i}\uparrow}\rangle_0 \Big) \Big(\langle c_{\boldsymbol{i}\downarrow}^\dagger c_{\boldsymbol{i}\downarrow}\rangle - \langle c^\dagger_{\boldsymbol{i}\downarrow}c_{\boldsymbol{i}\downarrow}\rangle_0 \Big) + \sum_{\boldsymbol{r_i},\boldsymbol{r_j}s} t(\boldsymbol{r_i}-\boldsymbol{r_j})e^{i\theta_{\boldsymbol{i},\boldsymbol{j}}} \langle c_{\boldsymbol{i}s}^\dagger c_{\boldsymbol{j}s} \rangle.
\end{align}
The Coulomb interaction is split into the Hartree or direct and Fock or exchange terms, with the plus and minus signs in front respectively. In our algorithm, we always work with the Fock matrix $\langle c_{\boldsymbol{i}s}^\dagger c_{\boldsymbol{i}s}\rangle - \langle c^\dagger_{\boldsymbol{i}s}c_{\boldsymbol{i}s}\rangle_0$, so the values reported for the kinetic energy have a constant offset of $\sum_{\boldsymbol{r_i},\boldsymbol{r_j}s} t(\boldsymbol{r_i}-\boldsymbol{r_j})e^{i\theta_{\boldsymbol{i},\boldsymbol{j}}} \langle c_{\boldsymbol{i}s}^\dagger c_{\boldsymbol{j}s} \rangle_0$.
% It is important to notice that both the Hartee and Fock energies, as they are written above, have contributions which are not physical coming from self interactions of the electron. These contributions couple the density of one of the orbitals in the Slater determinant with itself. They have opposite signs in the Hartree and Fock parts and cancel each other in the total energy.
\clearpage
\section{Additional Hartree-Fock results}\label{appe}

\begin{figure}[H]
    \noindent\hspace{-6.5cm} \textbf{a)}\\
    \centering \large{$\Phi = 0 $}\\
    \includegraphics[width=.287\linewidth]{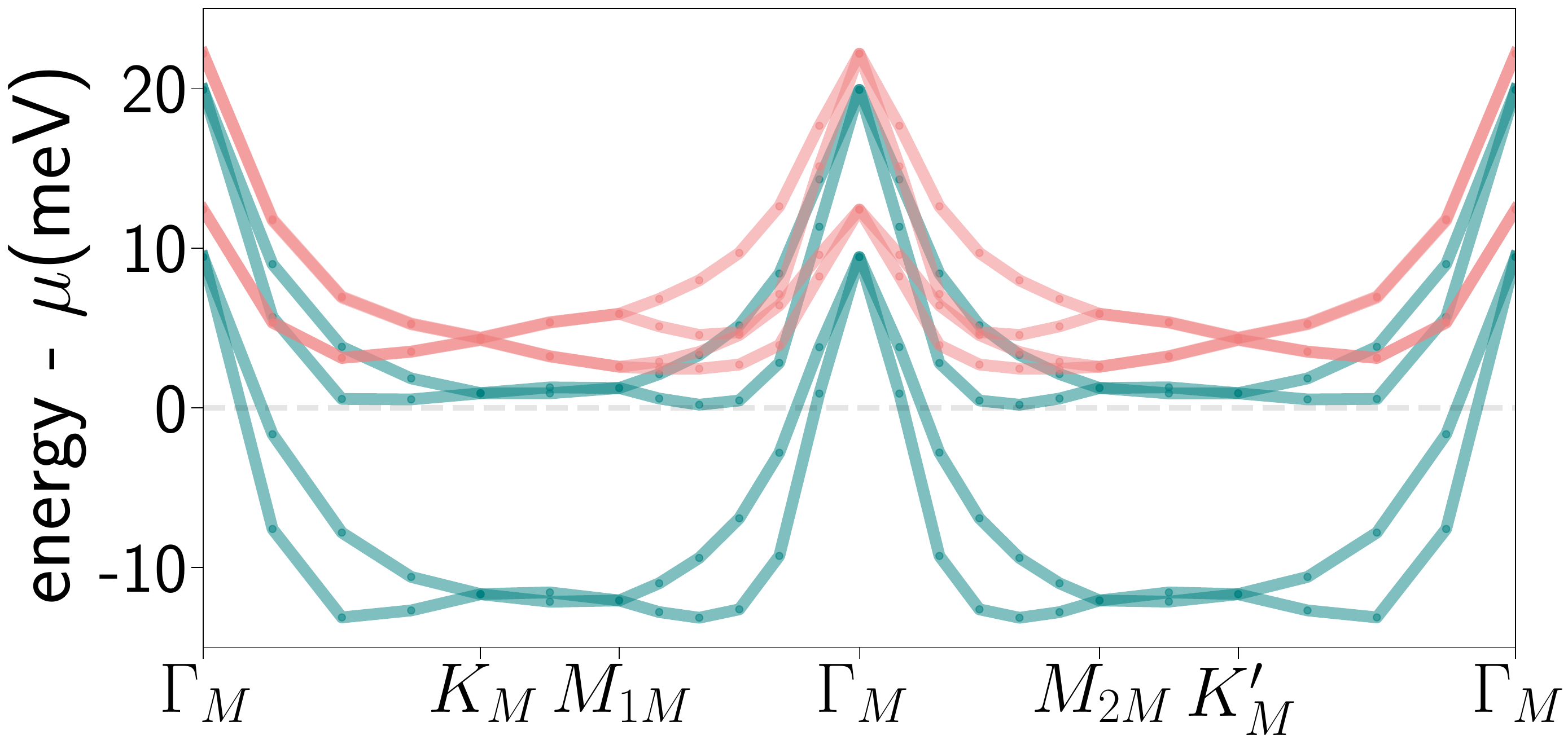}
    \includegraphics[width=.27\linewidth]{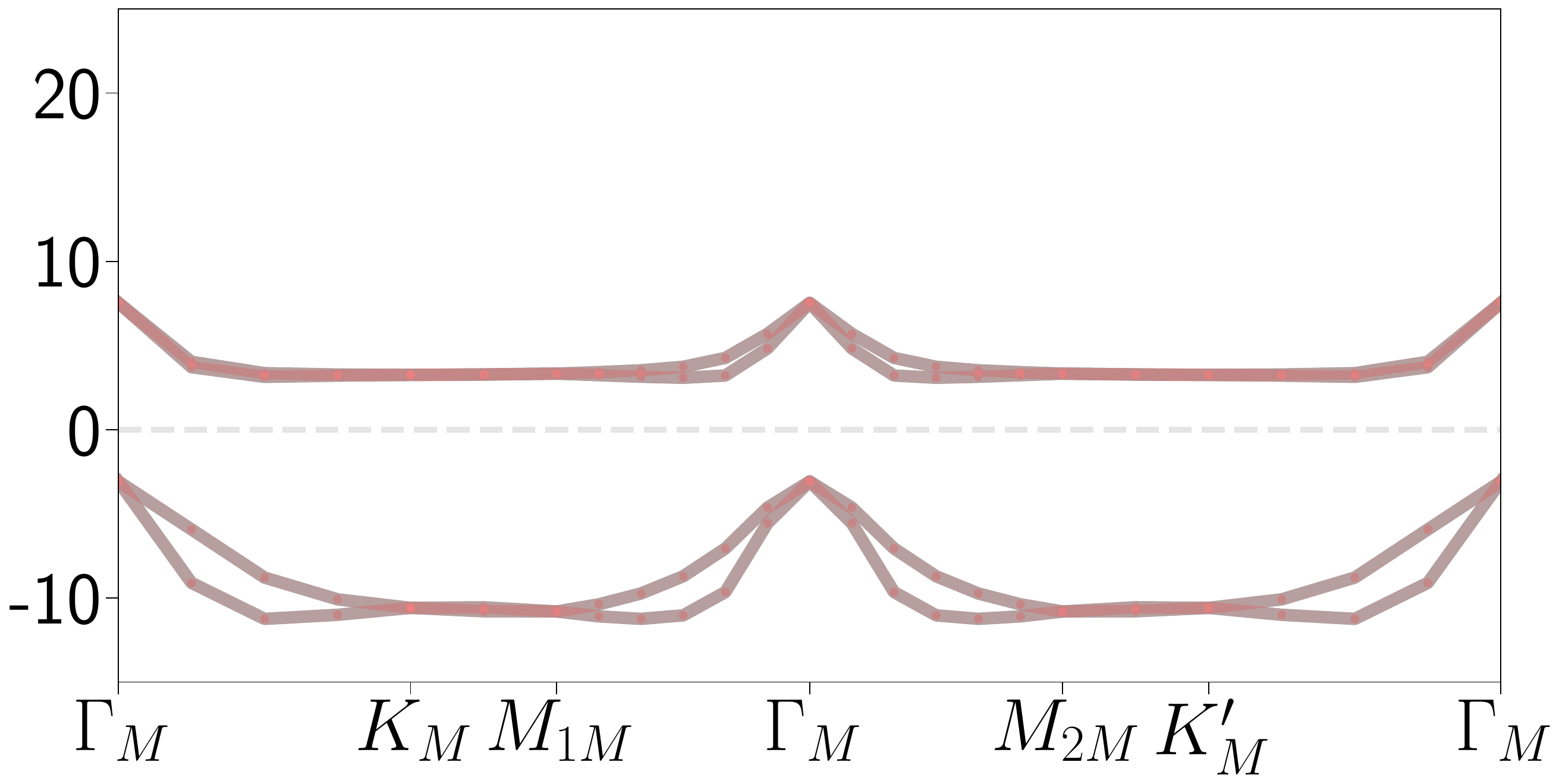}
    \includegraphics[width=.27\linewidth]{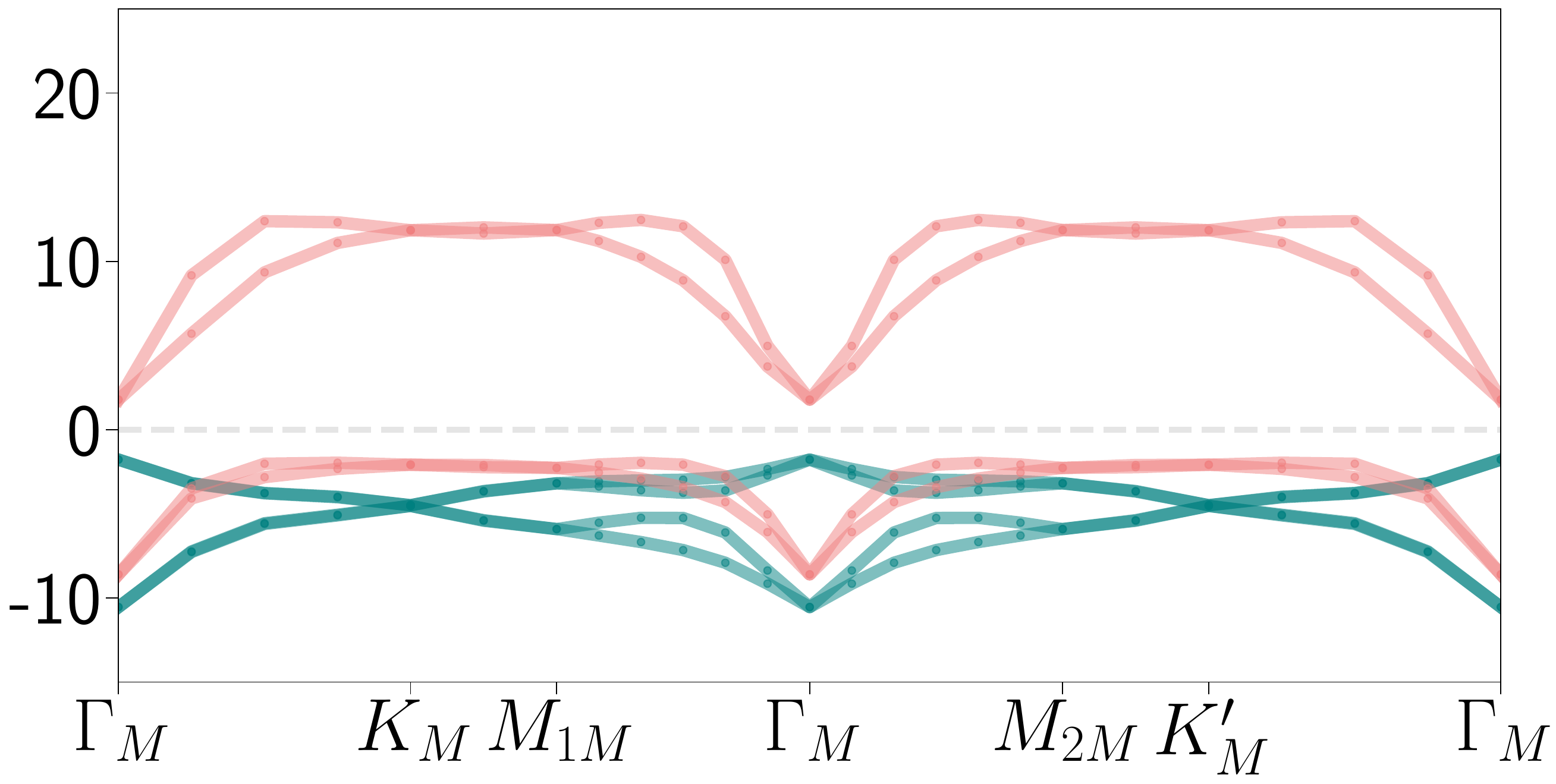}\\
    \centering \large{$\Phi = \Phi_0$}\\ 
    \includegraphics[width=.287\linewidth]{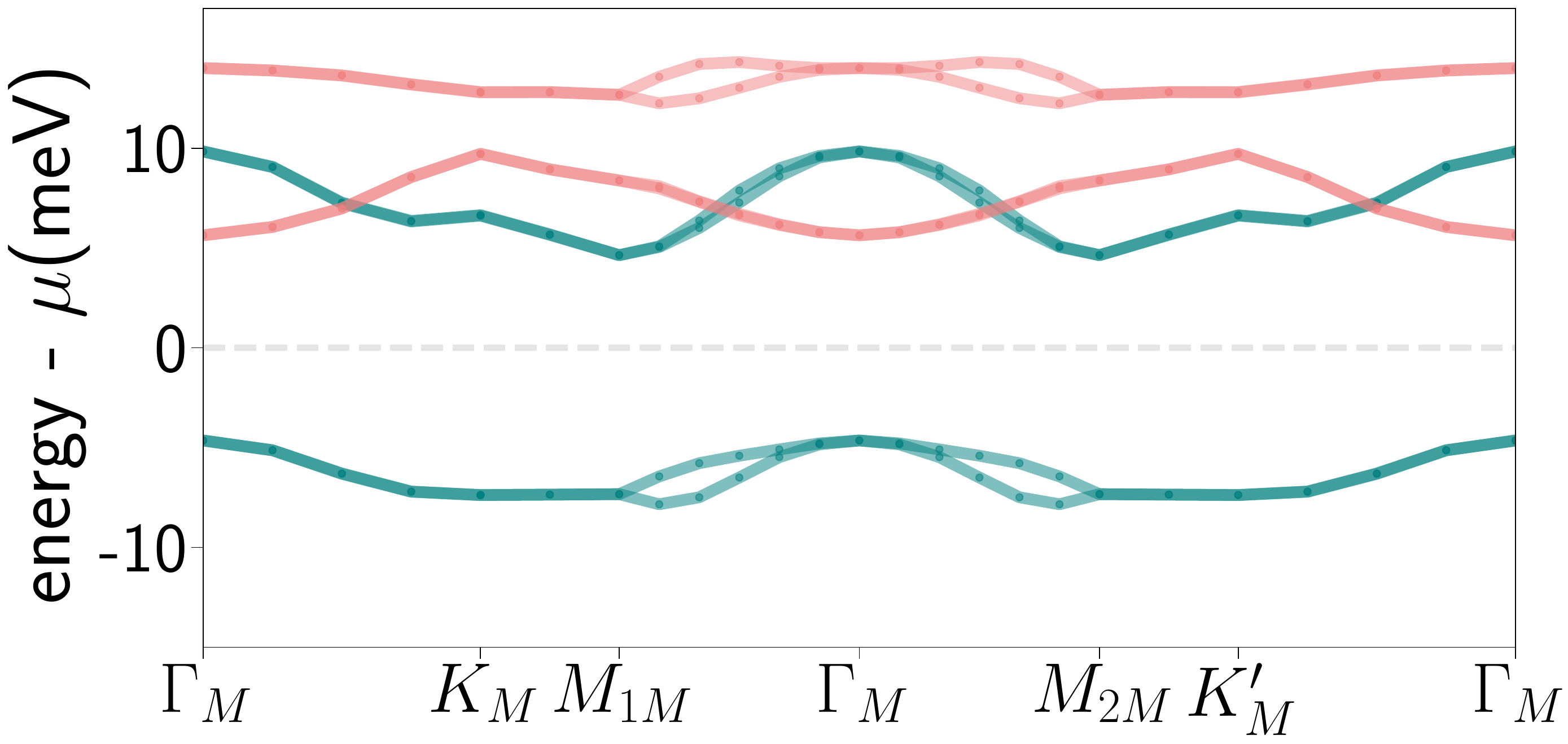}
    \includegraphics[width=.27\linewidth]{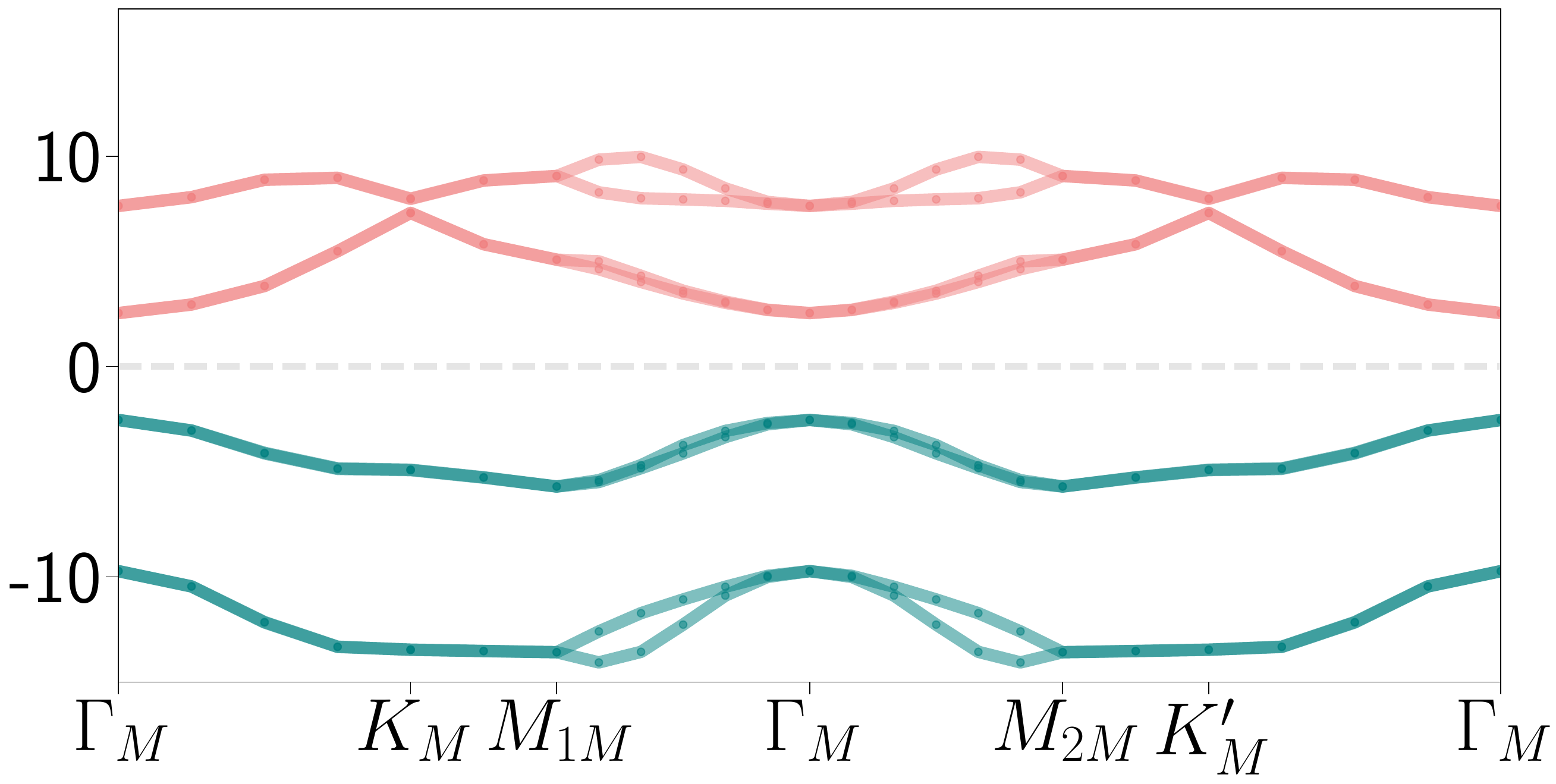}
    \includegraphics[width=.27\linewidth]{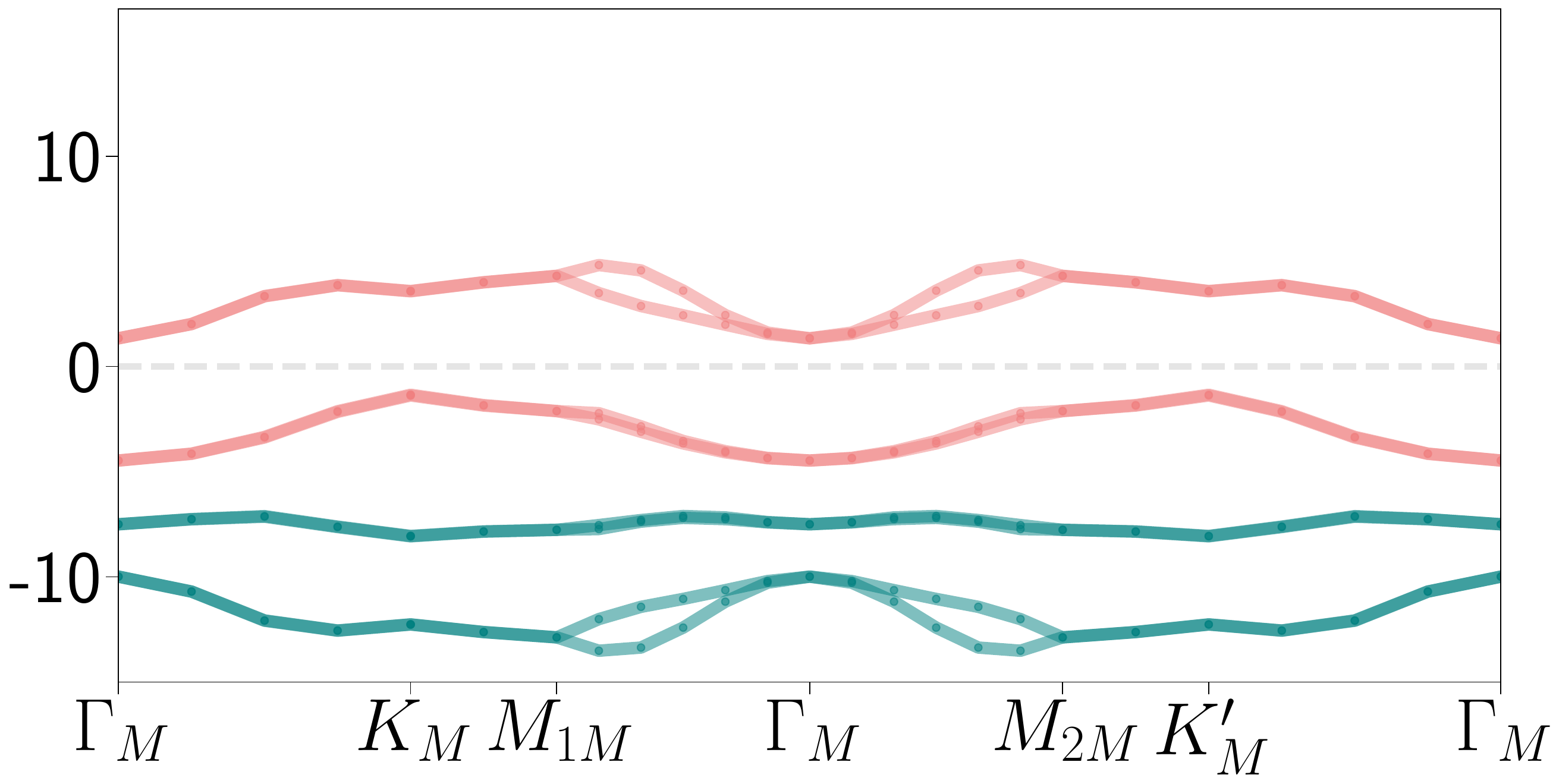} \\
    \normalsize \hspace{-1.8cm} \textbf{b)} \hspace{3.5cm} \textbf{c)} \hspace{3.2cm} \textbf{d)} \hspace{3.2cm} \textbf{e)} \hspace{3.1cm} \textbf{f)} \\
    \centering
    \includegraphics[width=.22\linewidth]{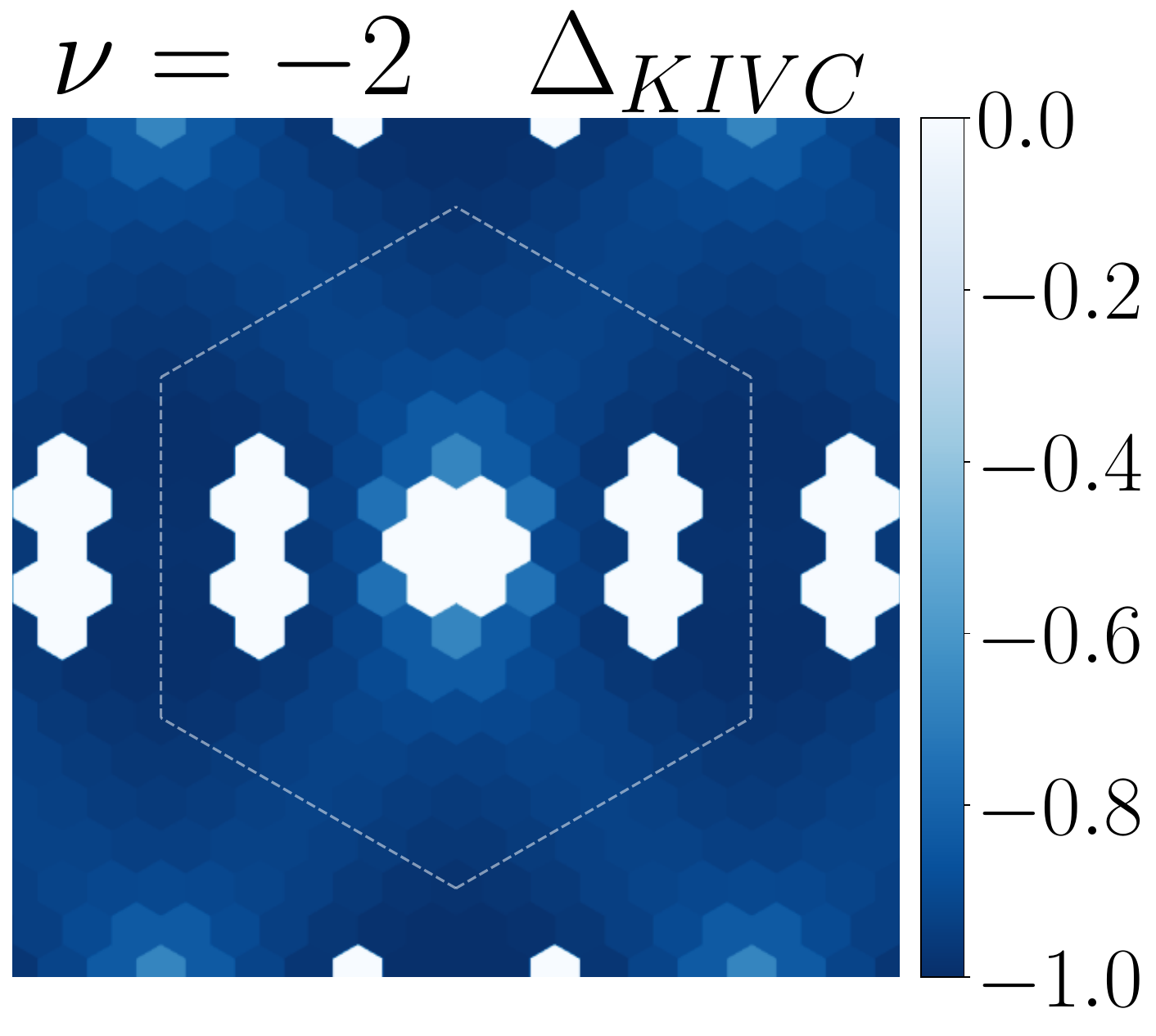}
    \includegraphics[width=.205\linewidth]{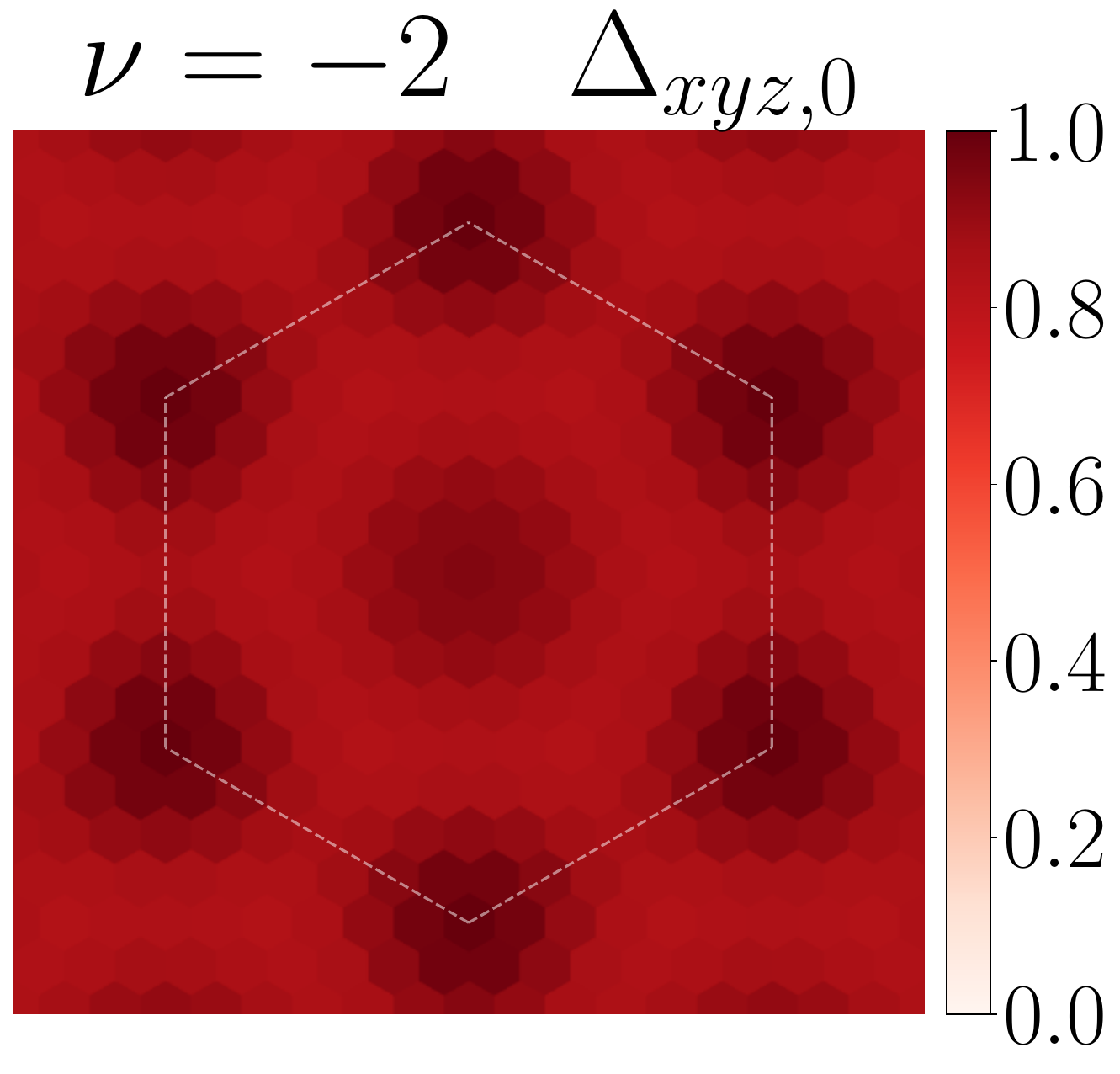}
    \includegraphics[width=.205\linewidth]{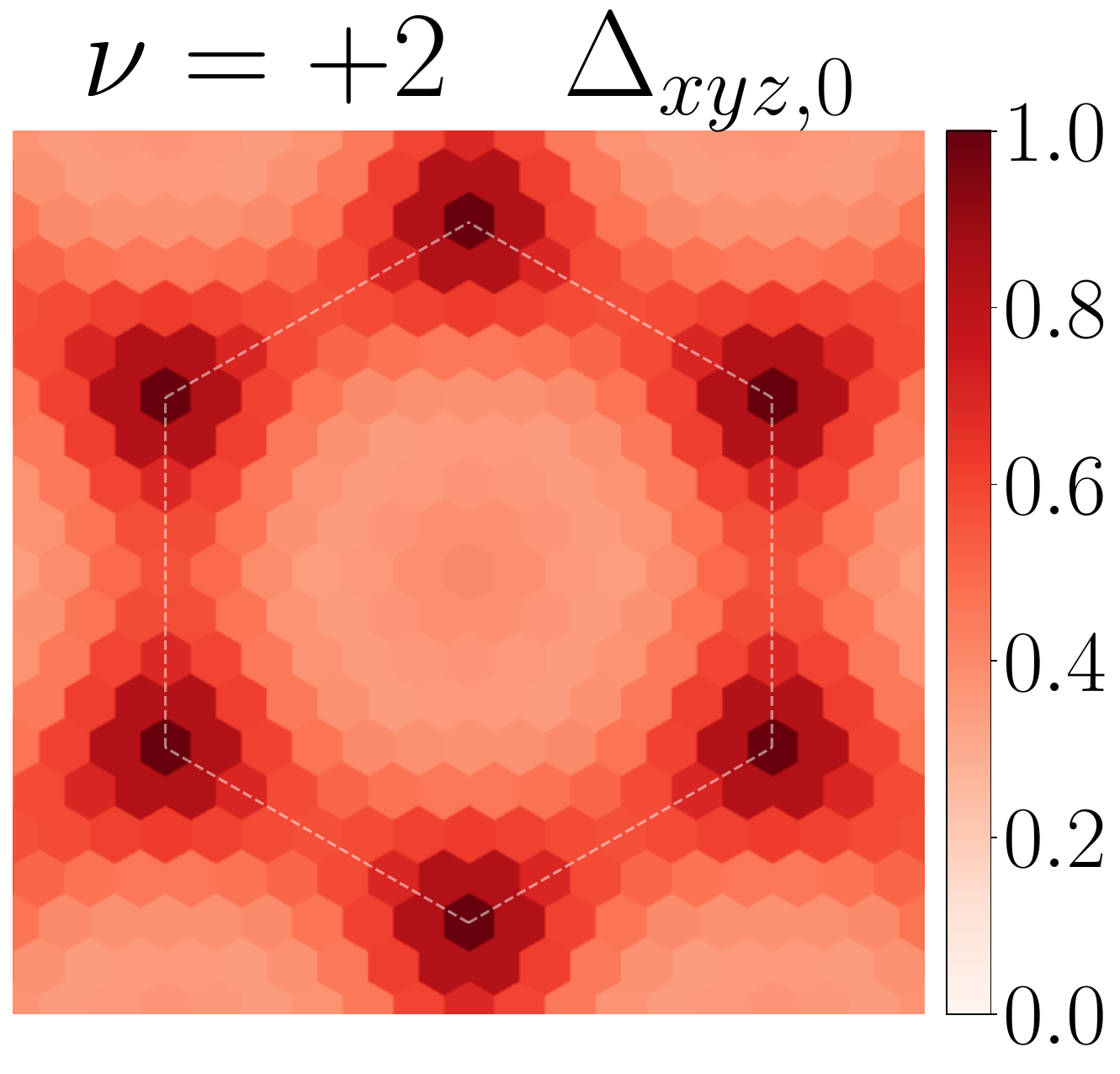}
    \includegraphics[width=.205\linewidth]{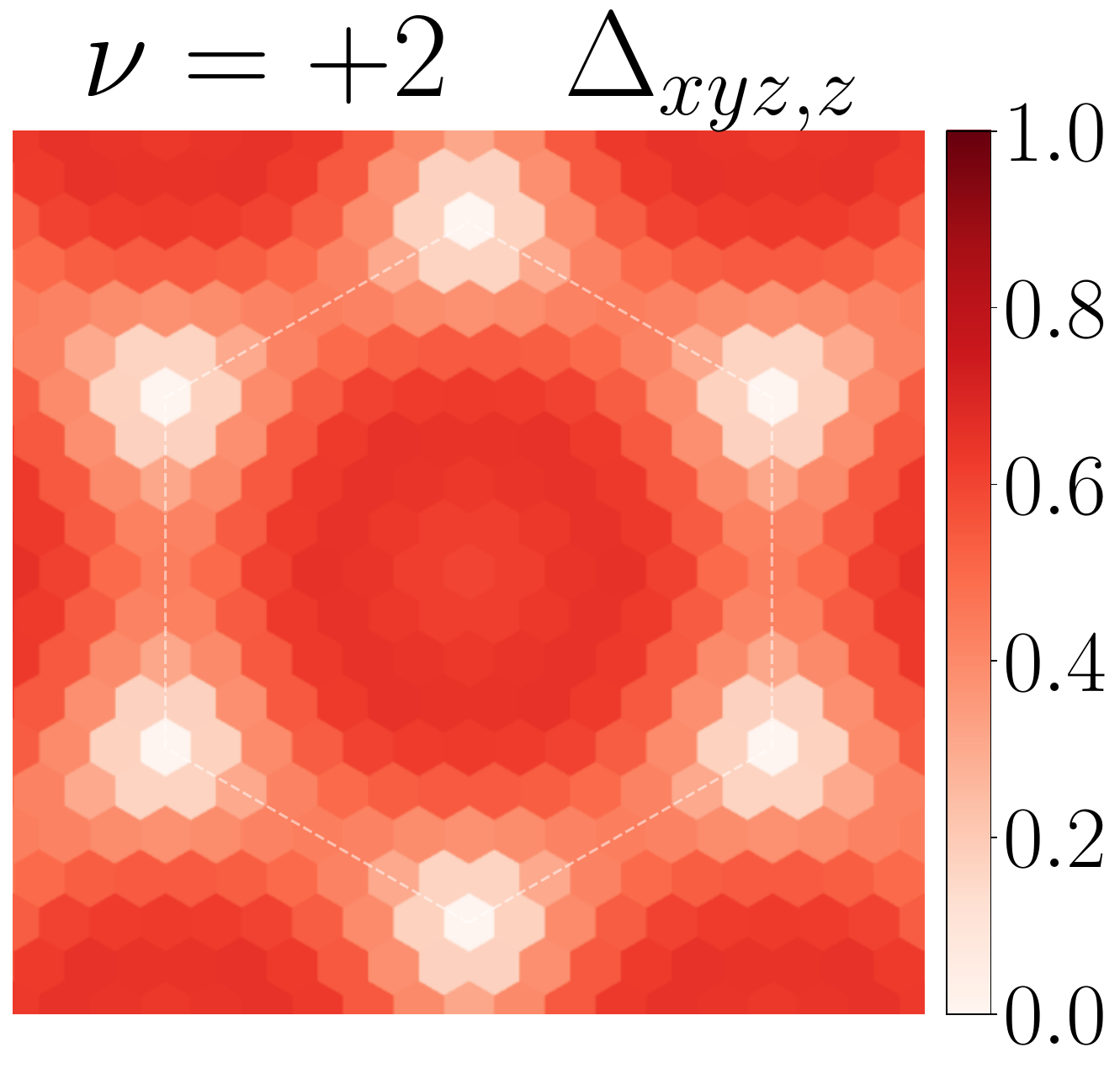}
    \includegraphics[width=.11\linewidth]{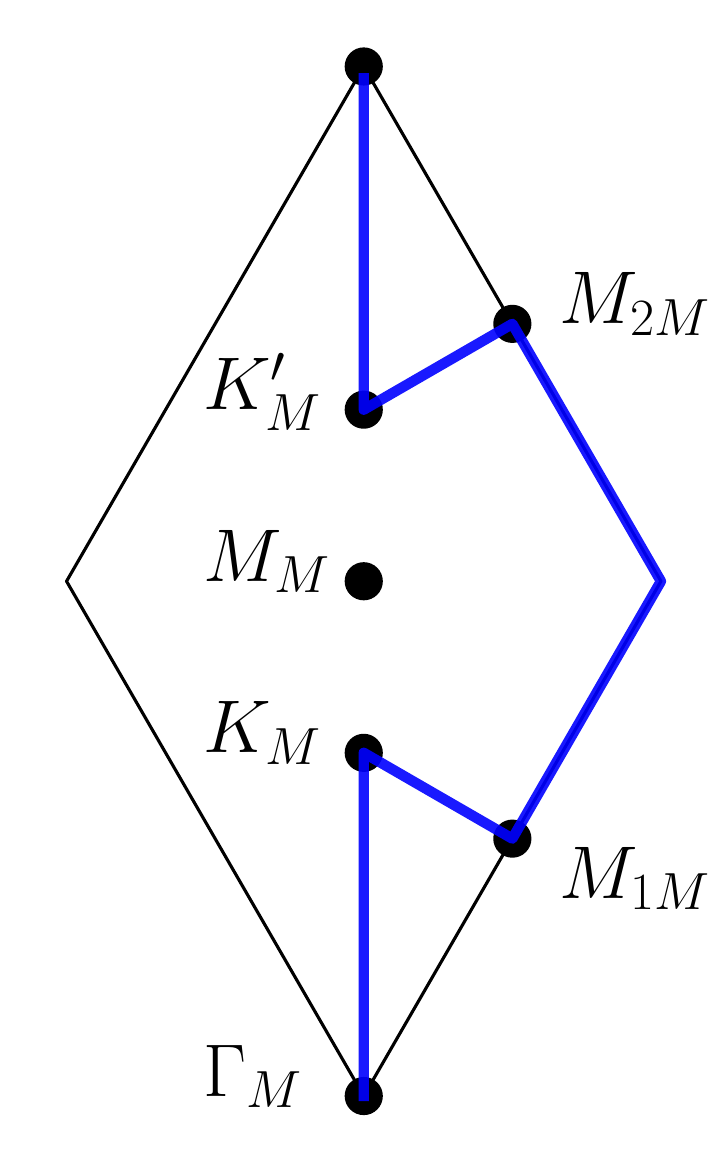}
    \caption{\textbf{Hartree-Fock results for $\boldsymbol{\epsilon=30, \ U=4}$ eV. a)} The KIVC band structures, for $\Phi = 0$, and the ground state bands, for $\Phi_0$. Spin up bands are in blue and spin down bands in red. From left to right the doping levels are $\nu=-2,0,+2$. The critical value for the SP-KIVC transition is found to be $U_c = 1.1$ eV. %Very similarly to the results reported in the main text, we observe gapped insulators for $\nu=0,+2$ at zero field and for all fillings at $\Phi_0$ flux. 
    In \textbf{b)} we show $\Delta_{\text{KIVC}}$ for $\nu=-2$ at zero flux.  There is a hole pocket around $\Gamma_M$ and two electron pockets at its sides, where the order parameter is ill-defined. In Fig. \ref{kivcorder}b) the extra electrons come from the opposite spin, so the electron pockets are not visible. In this phase $C_{3z}$ is broken but $C_{2z}$ and $C_{2x}$ are preserved. $C_{3z}$ breaking is not evident in the plot of a) because the $C_{3z}$ (or $C_{6z}$) related lines are related by $C_{2x}$ (or $C_{2y}=C_{2z}C_{2x}$) also. In \textbf{c)} we plot $\Delta_{xyz,0}$ of the hole doped state under one flux quantum, and in \textbf{d)} and \textbf{e)} the two dominant order parameters for the state at $\nu=+2$ and $\Phi_0$ flux. Unlike for $\epsilon=10$, $U=4$ eV, both parameters contribute equitably. The phase transition to the the IVC state takes place at some value of $\epsilon$ between $30$ and $50$. \textbf{f)} The path of the band structure plots.} 
\end{figure}

\begin{figure}[H]
    \textbf{a)} \hspace{8cm} \textbf{b)} \\
    \noindent \hspace*{3.7cm} \large{$\Phi=0$} \hspace{7cm} \large{$\Phi=\Phi_0$} \\
    \includegraphics[width=.23\linewidth]{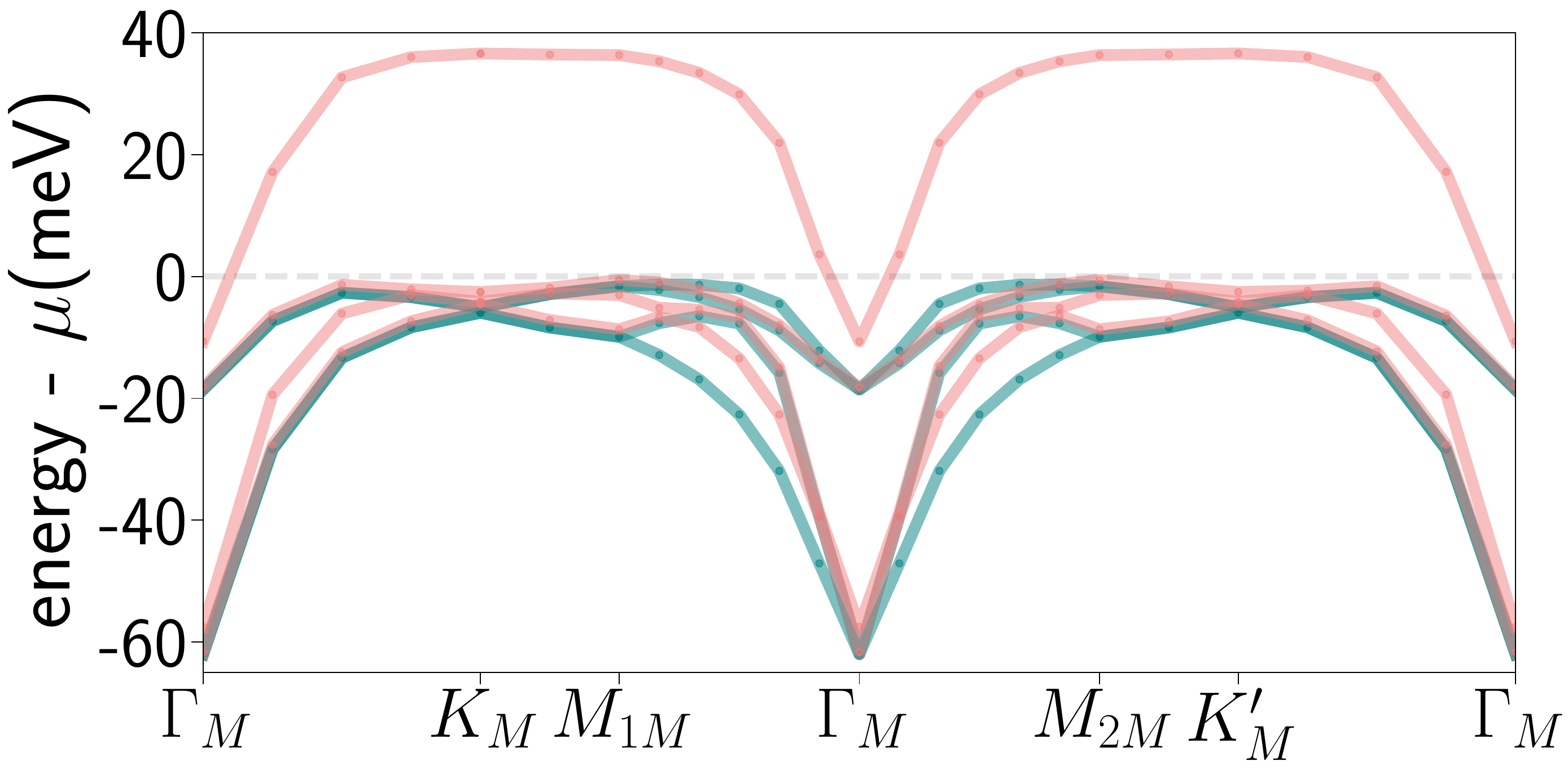}
    \includegraphics[width=.23\linewidth]{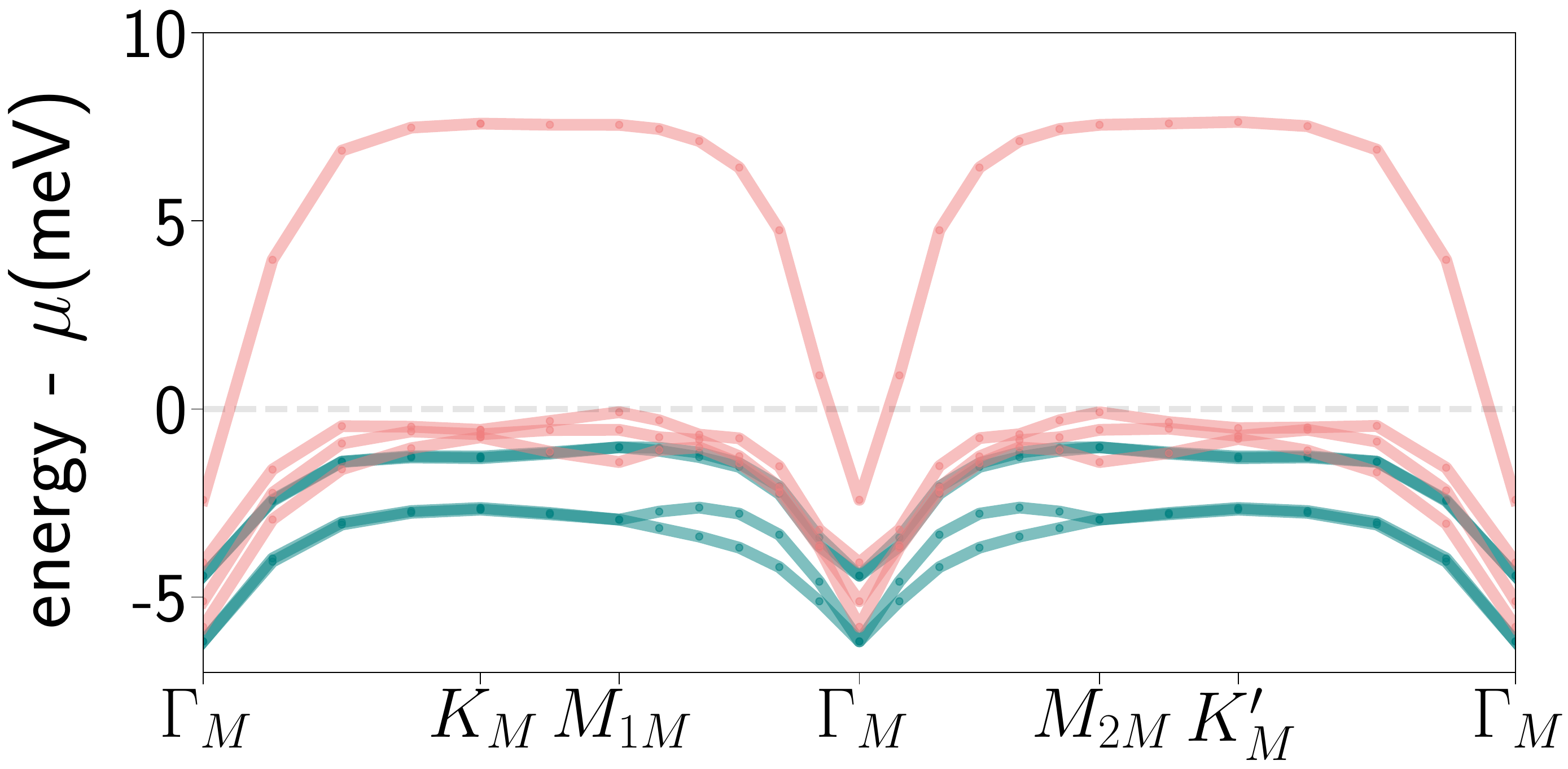}
    % \RaggedRight \hspace{3.3cm} \normalsize \textbf{b)} \\
    % \centering \large{$\Phi=\Phi_0$}\\
    \includegraphics[width=.23\linewidth]{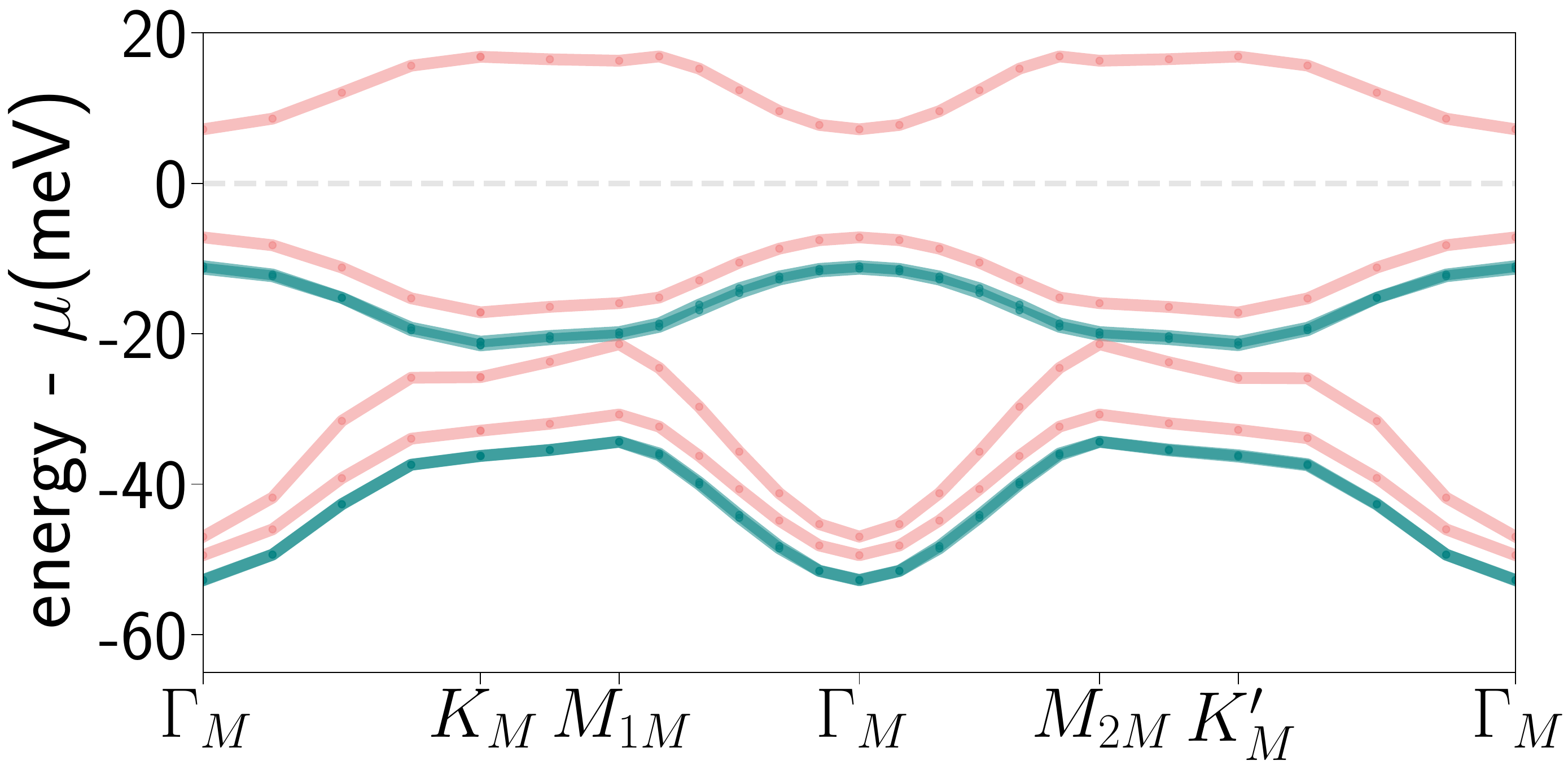}
    \includegraphics[width=.23\linewidth]{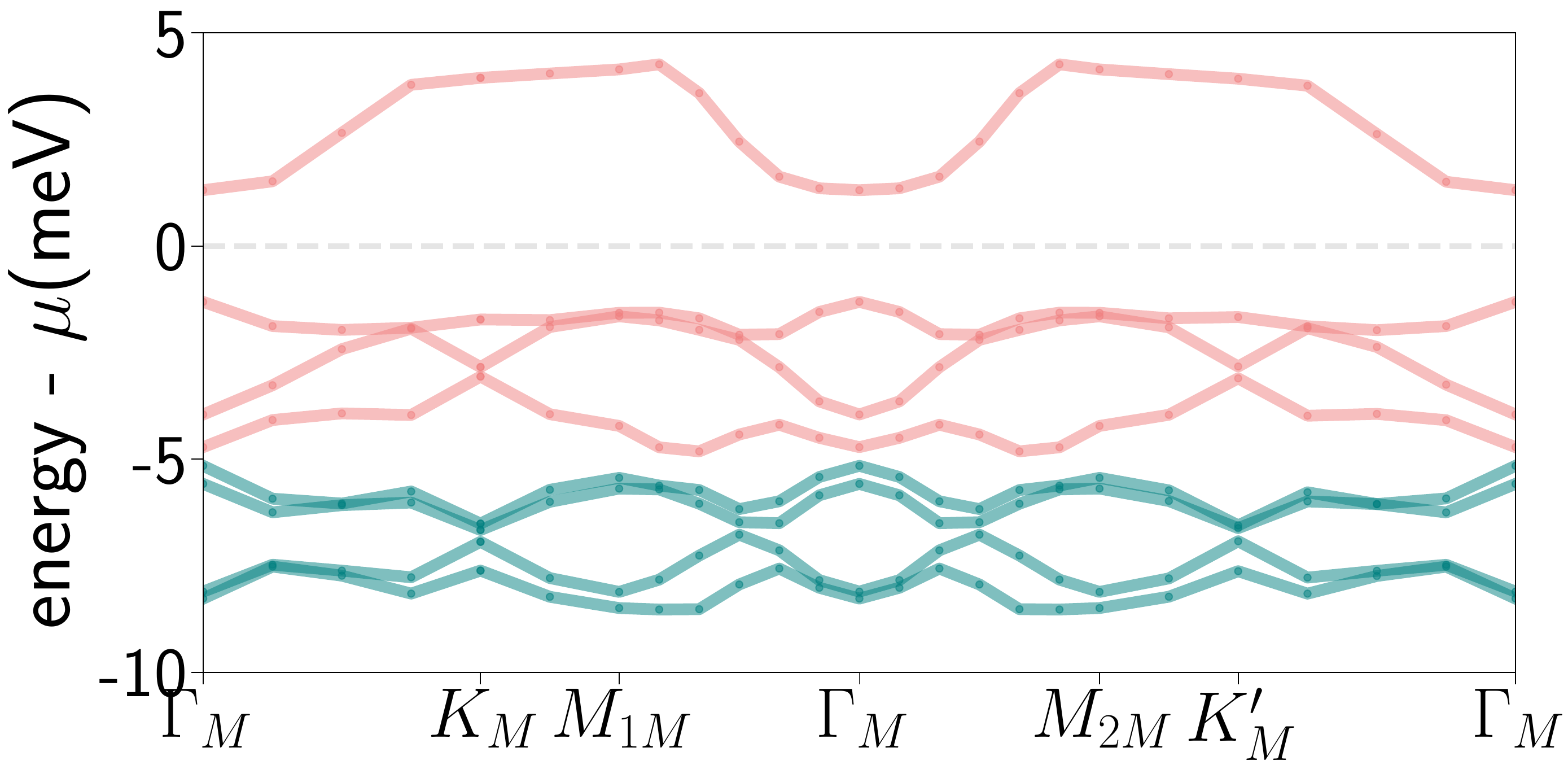}

    \caption{\textbf{Correlated insulators at $\boldsymbol{\nu=+3}$. a)} Chern insulators with Chern number $1$ that have been reported in samples with explicit $C_{2z}$ breaking due to the substrate\cite{Sharpe2019,Sharpe21}. We obtain incipient valley polarized anomalous Hall states without the need for $C_{2z}$ breaking at zero field, both for $\epsilon = 10$, $U=4$ eV, plotted to the left, and $\epsilon=50$, $U=4$ eV to the right. \textbf{b)} Gapped insulators at $B=26.5$ T. In both cases we get valley polarized states with Chern number $+1$. Interestingly, the band structures are very dissimilar for $\epsilon=10$, $U=4$ eV plotted to the left, and $\epsilon = 50$, $U=4$ eV to the right.}
\end{figure}

\begin{figure}[H]
    \centering \large{$\epsilon=10$, $U=4$ eV}\\
    \includegraphics[width=.32\linewidth]{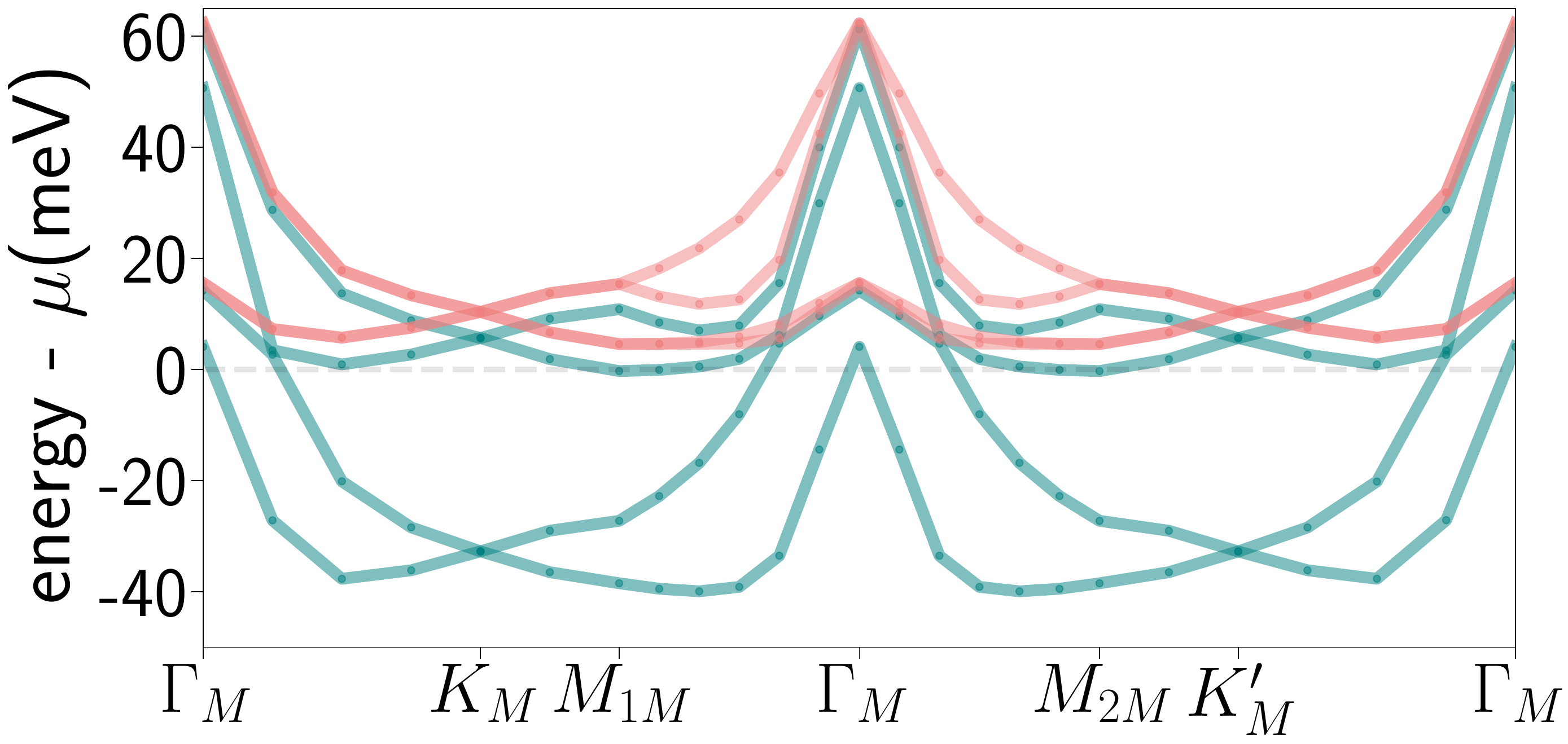}
    \includegraphics[width=.3\linewidth]{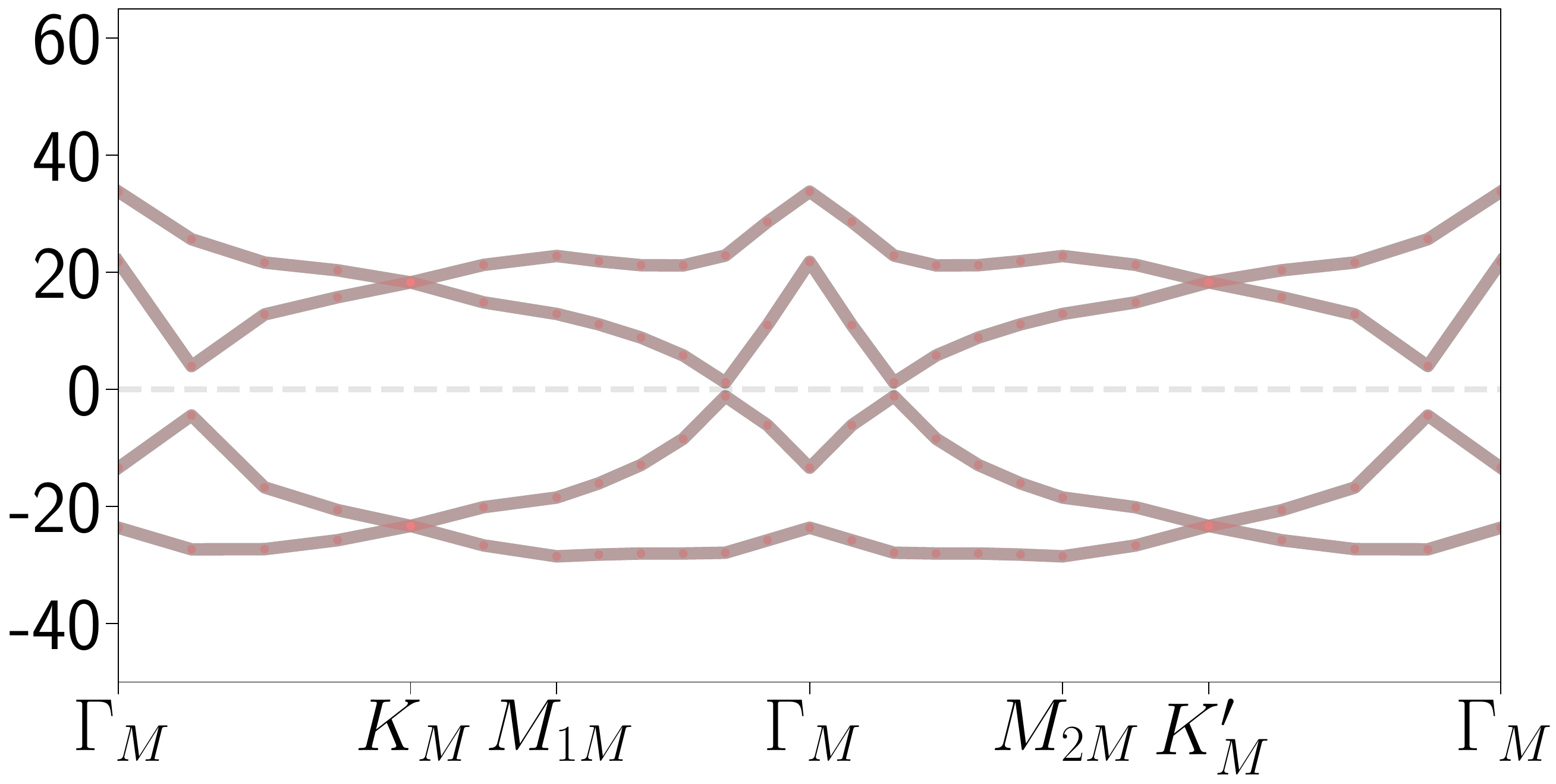}
    \includegraphics[width=.3\linewidth]{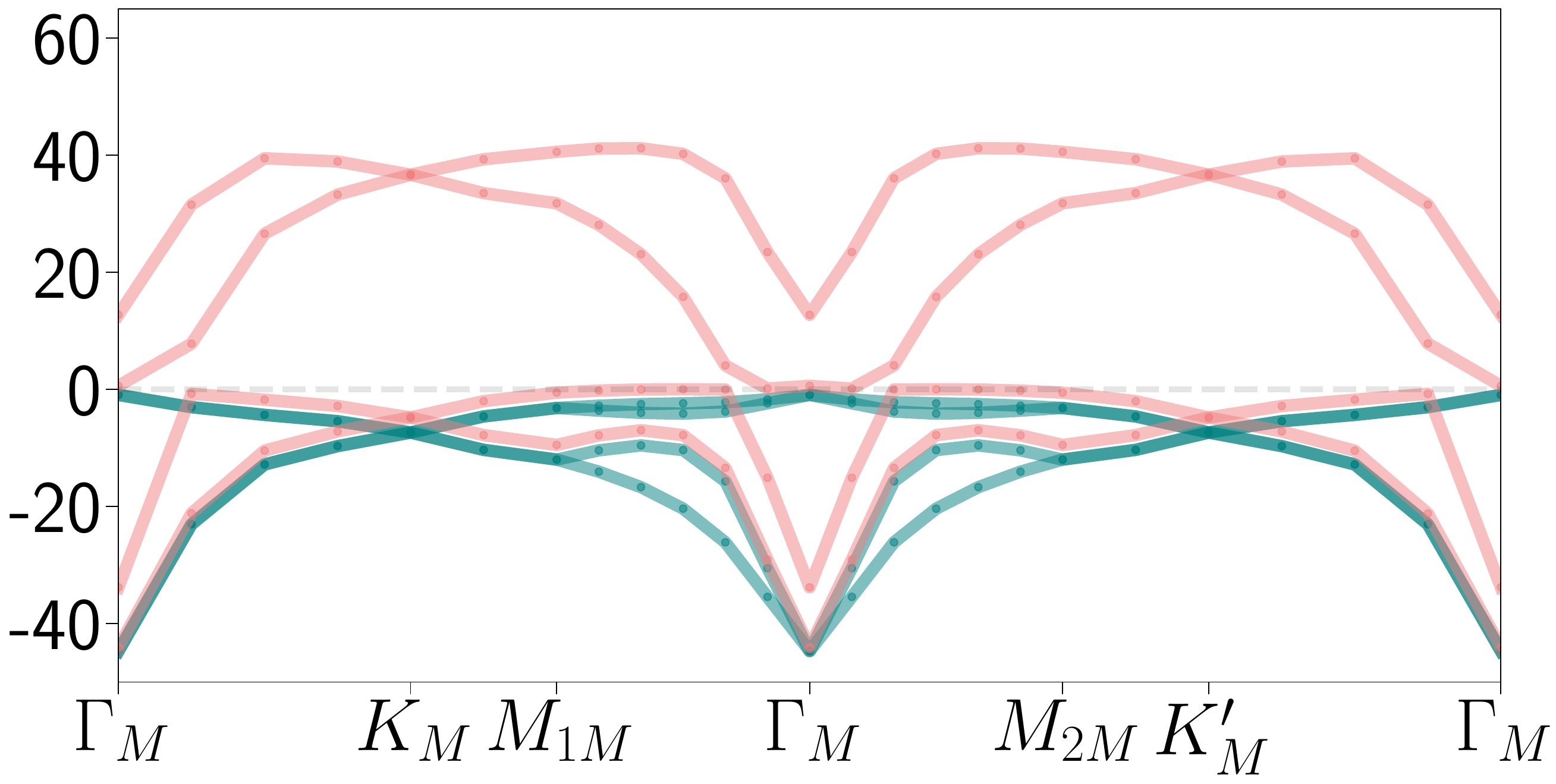}\\
    \centering \large{$\epsilon=50$, $U=0.5$ eV}\\ 
    \includegraphics[width=.32\linewidth]{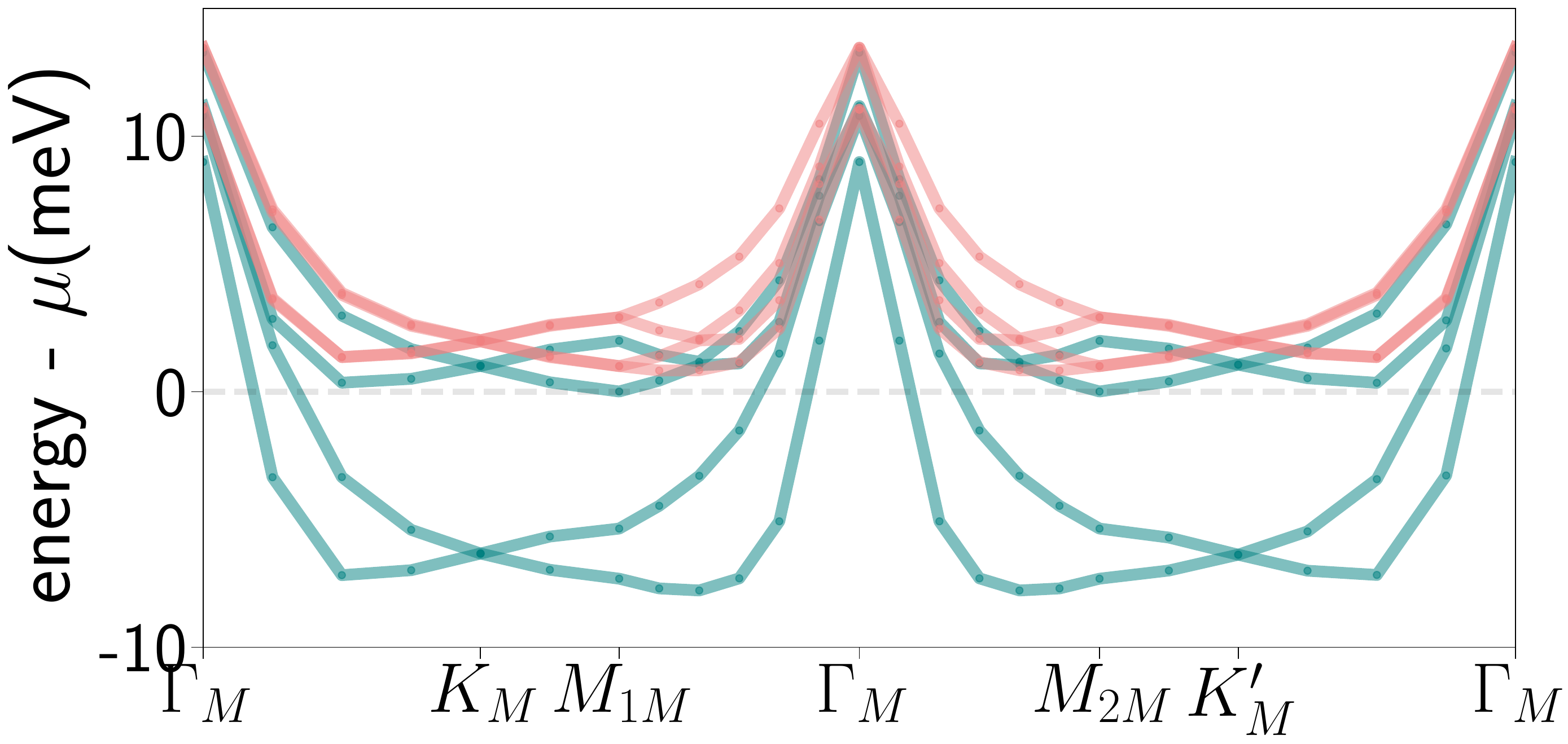}
    \includegraphics[width=.3\linewidth]{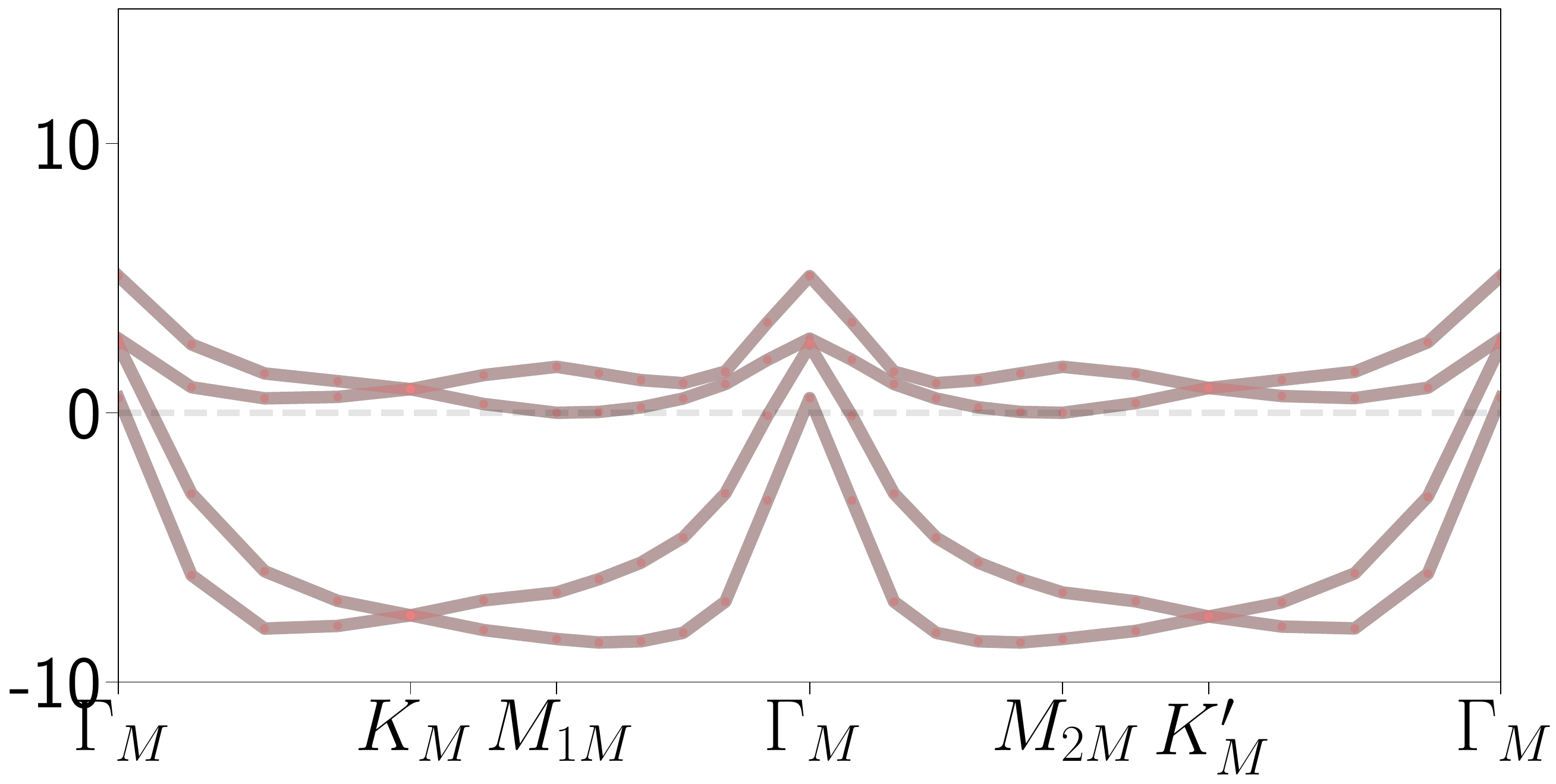}
    \includegraphics[width=.3\linewidth]{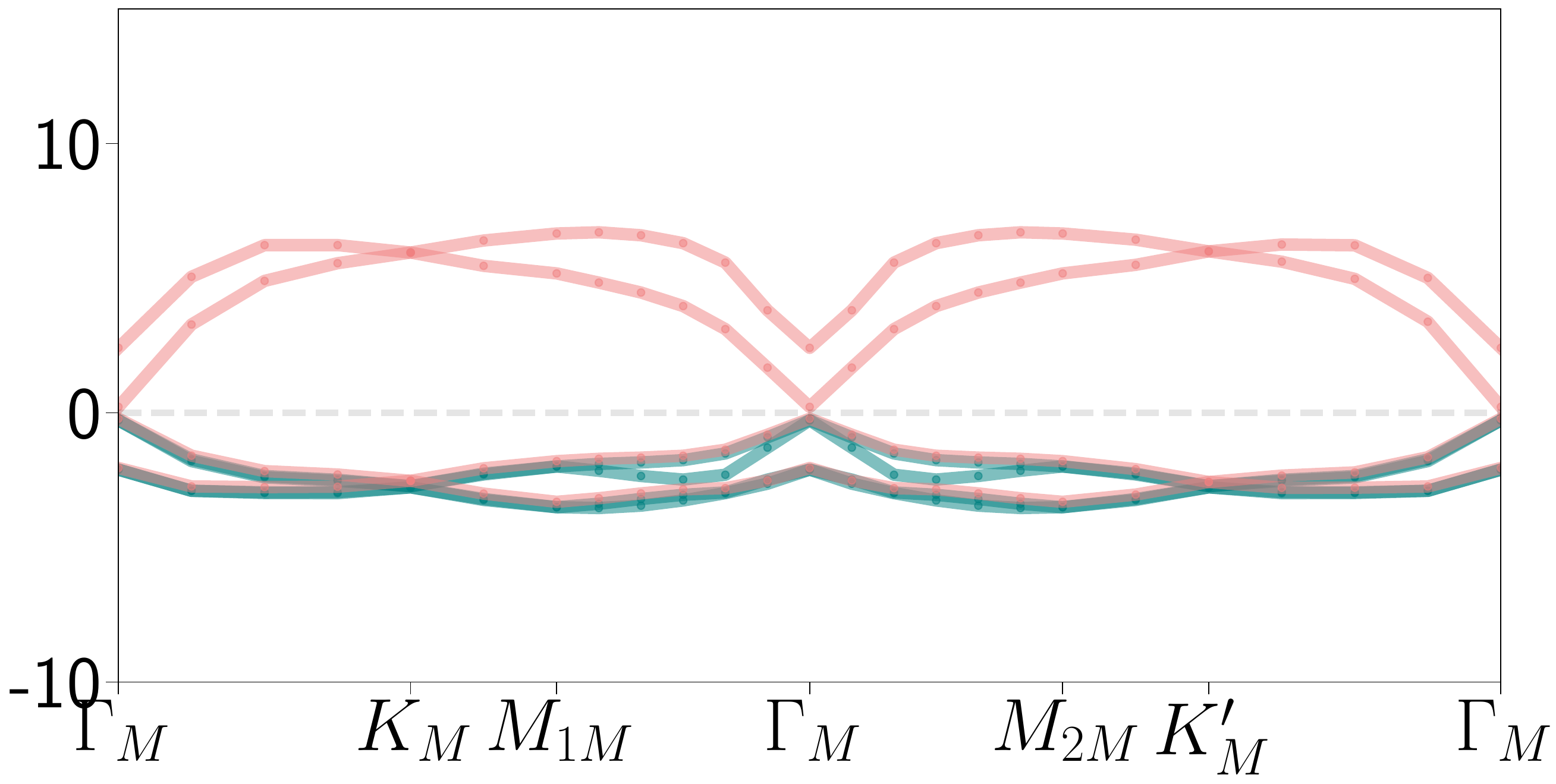} \\
    \caption{\textbf{The self-consistent valley polarized states.} Spin up bands are shown in blue and spin down bands in red. From left to right the doping levels are $\nu=-2,0,+2$. The order parameter is the valley polarization $\langle \sigma_0 \tau_z \rangle$ predominantly. Only for $\epsilon=10$, $U=4$ eV the inter-Chern parameter appears around $\Gamma_M$.} 
    \label{vplhf}
\end{figure}

\clearpage
\section{Tables with the energies of the candidate and self-consistent states}\label{appf}

\begin{table}[H]
\centering
\begin{tabular}{ |p{1.7cm}|p{6cm}|p{1.7cm}|p{1.7cm}|p{1.7cm}|p{1.7cm}|p{1.7cm}|}
 \hline
 \multicolumn{7}{|c|}{\Large{$\nu = -2$, $\xi = 10$ nm, $\Phi=\Phi_0$}} \\
 \hline
 state & wave function & kinetic & $\epsilon \times$Hartree & $\epsilon \times$Fock & $\epsilon \times$Coulomb & Hubbard$/U$   \\
 \hline
 IVC   & $\prod_{\boldsymbol{k}}\prod_\lambda \frac{1}{\sqrt{2}}(d^{\dagger}_{\boldsymbol{k}K \lambda \uparrow} + id^{\dagger}_{\boldsymbol{k}K'\lambda \uparrow})|0\rangle$ & $-48144.42$ & $328.32$ &  $-38862.09$ & $-38533.77$ &    $11.52$ \\
 VP & $\prod_{\boldsymbol{k}} d^{\dagger}_{\boldsymbol{k}K +1 \uparrow}d^{\dagger}_{\boldsymbol{k}K -1 \uparrow}|0\rangle$ & $-48144.42$ & $327.78$ & $-38861.38$ & $-38533.60$ & $11.52$ \\
  non int. GS & $-$ & $-48150.18$ & $225.32$ & $-38753.07$ & $-38527.76$ & $11.37$ \\
 \hline
\end{tabular}
% \caption{Expectation values of the energy of candidates states for the correlated insulators of MATBG, relative to the energy of the kinetic hamiltonian ground state.}
% \label{table4}
% \end{table}

% \begin{table}[h!]
% \centering
\begin{tabular}{ |p{1.7cm}|p{6cm}|p{1.7cm}|p{1.7cm}|p{1.7cm}|p{1.7cm}|p{1.7cm}|}
 \hline
 \multicolumn{7}{|c|}{\Large{$\nu = -2$, $\xi = 20$ nm, $\Phi=\Phi_0$}} \\
 \hline
 state & wave function & kinetic & $\epsilon \times$Hartree & $\epsilon \times$Fock & $\epsilon \times$Coulomb & Hubbard$/U$   \\
 \hline
 IVC   & $\prod_{\boldsymbol{k}}\prod_\lambda \frac{1}{\sqrt{2}}(d^{\dagger}_{\boldsymbol{k}K \lambda \uparrow} + id^{\dagger}_{\boldsymbol{k}K'\lambda \uparrow})|0\rangle$ & $-48144.42$ & $919.39$ &  $-40665.49$ & $-39746.10$ &    $11.52$ \\
 VP & $\prod_{\boldsymbol{k}} d^{\dagger}_{\boldsymbol{k}K +1 \uparrow}d^{\dagger}_{\boldsymbol{k}K -1 \uparrow}|0\rangle$ & $-48144.42$ & $918.85$ & $-40664.80$ & $-39745.95$ & $11.52$ \\
  non int. GS & $-$ & $-48150.18$ & $816.56$ & $-40547.45$ & $-39730.89$ & $11.37$ \\
 \hline
\end{tabular}
% \caption{Expectation values of the energy of candidates states for the correlated insulators of MATBG, relative to the energy of the kinetic hamiltonian ground state.}
% \label{table4}
% \end{table}

% \begin{table}[h!]
% \centering
\begin{tabular}{ |p{1.7cm}|p{6cm}|p{1.7cm}|p{1.7cm}|p{1.7cm}|p{1.7cm}|p{1.7cm}|}
 \hline
% \multicolumn{7}{|c|}{\Large{$\nu=0$, $\xi=10$ nm, $\Phi=\Phi_0$}} \\
\multicolumn{7}{|c|}{\Large{$\nu=0$, $\xi=10$ nm, $\Phi=\Phi_0$}} \\
 \hline
 state & wave function & kinetic &$\epsilon \times$Hartree & $\epsilon \times$Fock & $\epsilon \times$Coulomb & Hubbard$/U$    \\
 \hline
 IVC   & $\prod_{\boldsymbol{k}}\prod_{s \lambda} \frac{1}{\sqrt{2}}(d^{\dagger}_{\boldsymbol{k}K \lambda s} + id^{\dagger}_{\boldsymbol{k}K'\lambda s})|0\rangle$  & $-46554.33$ &  $-343.05$ & $-38855.98$  &  $-39199.03$   &  $11.45$ \\
 % IVC2   & $\prod_{\boldsymbol{k}}\prod_{s} \frac{1}{2}(d^{\dagger}_{\boldsymbol{k}K +1 s} + isd^{\dagger}_{\boldsymbol{k}K'+1 s}) (d^{\dagger}_{\boldsymbol{k}K -1 s} + is d^{\dagger}_{\boldsymbol{k}K'-1 s}) |0\rangle$   & $-351.497$ & $-218.037$ & $-569.534$  &  $ - 0.363$ \\
 VP   & $\prod_{\boldsymbol{k}}\prod_{s} d^{\dagger}_{\boldsymbol{k}K +1 s} d^{\dagger}_{\boldsymbol{k}K -1 s} |0\rangle$ & $-46554.33$ & $-345.21$ & $-38854.56$ & $-39199.77$  &  $11.52$ \\
 VSP   & $\prod_{\boldsymbol{k}} d^{\dagger}_{\boldsymbol{k}K +1 \uparrow} d^{\dagger}_{\boldsymbol{k}K -1 \uparrow} d^{\dagger}_{\boldsymbol{k}K' +1 \downarrow} d^{\dagger}_{\boldsymbol{k}K' -1 \downarrow} |0\rangle$  &  $-46554.33$ & $-342.64$ & $-38854.56$ & $-39197.20$ & $11.33$ \\
 SP & $\prod_{\boldsymbol{k}}\prod_{\eta \lambda }d^{\dagger}_{\boldsymbol{k}\eta \lambda \uparrow} |0\rangle$ & $-46554.33$ & $-342.64$ & $-38849.33$ & $-39191.97$  &  $10.75$ \\
  non int. GS & $-$ & $-46565.86$ & $8.86$ & $-38637.94$ & $-38629.08$  &  $11.77$ \\
 \hline
\end{tabular}

\begin{tabular}{ |p{1.7cm}|p{6cm}|p{1.7cm}|p{1.7cm}|p{1.7cm}|p{1.7cm}|p{1.7cm}|}
 \hline
 \multicolumn{7}{|c|}{\Large{$\nu = 0$, $\xi = 20$ nm, $\Phi=\Phi_0$}} \\
 \hline
 state & wave function & kinetic & $\epsilon \times$Hartree & $\epsilon \times$Fock & $\epsilon \times$Coulomb & Hubbard$/U$   \\
 \hline
 IVC   & $\prod_{\boldsymbol{k}}\prod_{s \lambda} \frac{1}{\sqrt{2}}(d^{\dagger}_{\boldsymbol{k}K \lambda s} + id^{\dagger}_{\boldsymbol{k}K'\lambda s})|0\rangle$ & $-46554.33$ &  $-346.62$ & $-40657.36$  &  $-41003.99$   &  $11.45$ \\
 % IVC2   & $\prod_{\boldsymbol{k}}\prod_{s} \frac{1}{2}(d^{\dagger}_{\boldsymbol{k}K +1 s} + isd^{\dagger}_{\boldsymbol{k}K'+1 s}) (d^{\dagger}_{\boldsymbol{k}K -1 s} + is d^{\dagger}_{\boldsymbol{k}K'-1 s}) |0\rangle$   & $-351.497$ & $-218.037$ & $-569.534$  &  $ - 0.363$ \\
 VP   & $\prod_{\boldsymbol{k}}\prod_{s} d^{\dagger}_{\boldsymbol{k}K +1 s} d^{\dagger}_{\boldsymbol{k}K -1 s} |0\rangle$  & $-46554.34$ & $-348.80$ & $-40655.99$ & $-41004.79$  &  $11.52$ \\
 VSP  & $\prod_{\boldsymbol{k}} d^{\dagger}_{\boldsymbol{k}K +1 \uparrow} d^{\dagger}_{\boldsymbol{k}K -1 \uparrow} d^{\dagger}_{\boldsymbol{k}K' +1 \downarrow} d^{\dagger}_{\boldsymbol{k}K' -1 \downarrow} |0\rangle$  &  $-46554.33$  & $-346.21$ & $-40655.99$ & $-41002.20$ & $11.33$ \\
 SP & $\prod_{\boldsymbol{k}}\prod_{\eta \lambda }d^{\dagger}_{\boldsymbol{k}\eta \lambda \uparrow} |0\rangle$ & $-46554.33$ &  $-346.21$ & $-40650.70$ & $-40996.92$  &  $10.75$ \\
  non int. GS & $-$ & $-46565.86$ &  $4.22$ & $-40421.29$ & $-40417.07$  &  $11.77$ \\
 \hline
\end{tabular}
% \caption{Expectation values of the energy of candidates states for the correlated insulators of MATBG, relative to the energy of the kinetic hamiltonian ground state.}
% \label{table4}
% \end{table}
\begin{tabular}{ |p{1.7cm}|p{6cm}|p{1.7cm}|p{1.7cm}|p{1.7cm}|p{1.7cm}|p{1.7cm}|}
 \hline
 \multicolumn{7}{|c|}{\Large{$\nu = +2$, $\xi = 10$ nm, $\Phi=\Phi_0$}} \\
 \hline
 state & wave function & kinetic & $\epsilon \times$Hartree & $\epsilon \times$Fock & $\epsilon \times$Coulomb & Hubbard$/U$   \\
 \hline
 IVC   & $\prod_{\boldsymbol{k}}\prod_\lambda \frac{1}{\sqrt{2}}(d_{\boldsymbol{k}K \lambda \downarrow} + id_{\boldsymbol{k}K'\lambda \downarrow})|+4\rangle$ & $-44964.25$ & $474.60$ & $-38843.22$ & $-38368.62$ &    $11.34$ \\
 VP & $\prod_{\boldsymbol{k}} d_{\boldsymbol{k}K +1 \downarrow}d_{\boldsymbol{k}K -1 \downarrow}|+4\rangle$ & $-44964.25$ & $474.05$ & $-38842.51$ & $-38368.45$ & $11.34$ \\
 non int. GS & $-$ & $-44971.37$ & $821.80$ & $-38664.38$ & $-37842.58$ & $11.73$ \\
 \hline
\end{tabular}

\begin{tabular}{ |p{1.7cm}|p{6cm}|p{1.7cm}|p{1.7cm}|p{1.7cm}|p{1.7cm}|p{1.7cm}|}
 \hline
 \multicolumn{7}{|c|}{\Large{$\nu = +2$, $\xi = 20$ nm, $\Phi=\Phi_0$}} \\
 \hline
 state & wave function & kinetic & $\epsilon \times$Hartree & $\epsilon \times$Fock & $\epsilon \times$Coulomb & Hubbard$/U$   \\
 \hline
 IVC   & $\prod_{\boldsymbol{k}}\prod_\lambda \frac{1}{\sqrt{2}}(d_{\boldsymbol{k}K \lambda \downarrow} + id_{\boldsymbol{k}K'\lambda \downarrow})|+4\rangle$ & $-44964.25$ & $1065.69$ & $-40642.57$ & $-39576.89$ &    $11.34$ \\
 VP & $\prod_{\boldsymbol{k}} d_{\boldsymbol{k}K +1 \downarrow}d_{\boldsymbol{k}K -1 \downarrow}|+4\rangle$ & $-44964.25$ & $1065.14$ & $-40641.89$ & $-39576.74$ & $11.34$ \\
 non int. GS & $-$ & $-44971.37$ & $1412.04$ & $-40439.88$ & $-39027.84$ & $11.73$ \\
 \hline
\end{tabular}
\caption{ 
\textbf{Expectation  values of the energy of candidates states for the correlated insulators of MATBG at one magnetic flux quantum.} We list the states for fillings $\nu=0,\pm 2$ setting $\xi$ to $10$ and $20$ nm. The values are in units of meV per unit cell, and $U$ is given in eV. The state non int. GS is the ground state of the non interacting system. Coincidentally the kinetic ground state at hole doping is given by the filled valence band of the spin $\uparrow$, and at charge neutrality by the valence bands of both spins, as can be seen in Fig. \ref{bands}. $|0\rangle$ denotes the state with filled remote bands and $|+4\rangle$ the $\nu=+4$ insulator. The Zeeman energy is $\pm 1.535$ meV per electron and is not explicitly included.}
\label{energiesp1}
\end{table}

\begin{table}[H]
\centering
\begin{tabular}{ |p{1.7cm}|p{6.5cm}|p{1.7cm}|p{1.7cm}|p{1.7cm}|p{1.7cm}|p{1.7cm}|}
 \hline
 \multicolumn{7}{|c|}{\Large{$\nu = -2$, $\xi = 10$ nm, $\Phi=0$}} \\
 \hline
 state & wave function & kinetic & $\epsilon \times$Hartree & $\epsilon \times$Fock & $\epsilon \times$Coulomb & Hubbard$/U$  \\
 \hline
 KIVC   & $\prod_{\boldsymbol{k}} \frac{1}{2}(d^{\dagger}_{\boldsymbol{k}K A \uparrow} + d^{\dagger}_{\boldsymbol{k}K'B \uparrow}) (d^{\dagger}_{\boldsymbol{k}K B \uparrow} - d^{\dagger}_{\boldsymbol{k}K'A \uparrow})|0\rangle$ & $-47981.11$ & $642.52$ & $-38976.71$ & $-38334.19$ &  $12.07$ \\
 VP & $\prod_{\boldsymbol{k}} d^{\dagger}_{\boldsymbol{k}K A \uparrow}d^{\dagger}_{\boldsymbol{k}KB \uparrow}|0\rangle$  & $-47981.11$& $642.52$ & $-38975.58$  & $-38333.06$  & $12.07$ \\
 TIVC & $\prod_{\boldsymbol{k}} \frac{1}{2}(d^{\dagger}_{\boldsymbol{k}K A \uparrow} + d^{\dagger}_{\boldsymbol{k}K'B \uparrow}) (d^{\dagger}_{\boldsymbol{k}K B \uparrow} + d^{\dagger}_{\boldsymbol{k}K'A \uparrow})|0\rangle$ & $-47981.11$  & $641.62$ & $-38966.78$ & $-38325.16$ & $12.07$\\
 QAH-IVC    & $\prod_{\boldsymbol{k}}\prod_s \frac{1}{\sqrt{2}}(d^{\dagger}_{\boldsymbol{k}K A s} + d^{\dagger}_{\boldsymbol{k}K'B s)}|0\rangle$ & $-47981.11$ & $641.62$&  $-38969.92$ &$-38328.30$ & $ 12.41$\\
 QAH-VP &  $\prod_{\boldsymbol{k}}\prod_s d^{\dagger}_{\boldsymbol{k}K A s} |0\rangle $ & $-47981.11$ &$637.34$ &$-38967.80$ &$-38330.46$ &  $12.53$\\
 VH &  $\prod_{\boldsymbol{k}} d^{\dagger}_{\boldsymbol{k}K A \uparrow} d^{\dagger}_{\boldsymbol{k}K' A \uparrow} |0\rangle $ & $-47981.11$ & $637.34$& $-38963.67$ &$-38326.34$ &  $ 12.07$\\
 non int. GS &  $-$ & $-47982.83$ & $581.99$ & $-38598.70$ &$-38016.71$ &  $ 12.34$\\
 \hline
\end{tabular}
% \caption{Expectation  values of the energy of candidates states for the correlated insulators of MATBG, relative to the energy of the kinetic hamiltonian ground state.}
% \label{table1}
% \end{table}

% \begin{table}[h!]
% \centering
\begin{tabular}{ |p{1.7cm}|p{6.5cm}|p{1.7cm}|p{1.7cm}|p{1.7cm}|p{1.7cm}|p{1.7cm}|}
 \hline
 \multicolumn{7}{|c|}{\Large{$\nu = -2$, $\xi = 20$ nm, $\Phi=0$}} \\
 \hline
 state & wave function & kinetic & $\epsilon \times$Hartree & $\epsilon \times$Fock & $\epsilon \times$Coulomb & Hubbard$/U$  \\
 \hline
 KIVC   & $\prod_{\boldsymbol{k}} \frac{1}{2}(d^{\dagger}_{\boldsymbol{k}K A \uparrow} + d^{\dagger}_{\boldsymbol{k}K'B \uparrow}) (d^{\dagger}_{\boldsymbol{k}K B \uparrow} - d^{\dagger}_{\boldsymbol{k}K'A \uparrow})|0\rangle$ & $-47981.11$ & $1232.78$ & $-40791.19$ & $-39558.42$ &  $12.07$ \\
 VP & $\prod_{\boldsymbol{k}} d^{\dagger}_{\boldsymbol{k}K A \uparrow}d^{\dagger}_{\boldsymbol{k}KB \uparrow}|0\rangle$  & $-47981.11$& $1232.78$ & $-40790.08$  & $-39557.30$  & $12.07$ \\
 TIVC & $\prod_{\boldsymbol{k}} \frac{1}{2}(d^{\dagger}_{\boldsymbol{k}K A \uparrow} + d^{\dagger}_{\boldsymbol{k}K'B \uparrow}) (d^{\dagger}_{\boldsymbol{k}K B \uparrow} + d^{\dagger}_{\boldsymbol{k}K'A \uparrow})|0\rangle$ & $-47981.11$  & $1231.88$ & $-40780.57$ & $-39548.70$ & $12.07$ \\
 QAH-IVC    & $\prod_{\boldsymbol{k}}\prod_s \frac{1}{\sqrt{2}}(d^{\dagger}_{\boldsymbol{k}K A s} + d^{\dagger}_{\boldsymbol{k}K'B s)}|0\rangle$ & $-47981.11$ & $1231.88$ &$-40783.73$ &$-39551.85$ &  $12.41$\\
 QAH-VP &  $\prod_{\boldsymbol{k}}\prod_s d^{\dagger}_{\boldsymbol{k}K A s} |0\rangle $ & $-47981.11$ & $1227.56$ &$-40781.60$ &$-39554.04$ &  $12.53$\\
 VH &  $\prod_{\boldsymbol{k}} d^{\dagger}_{\boldsymbol{k}K A \uparrow} d^{\dagger}_{\boldsymbol{k}K' A \uparrow} |0\rangle $ & $-47981.11$ & $1227.56$& $-40777.43$ &$-39549.87$ &  $ 12.07$\\
  non int. GS &  $- $ & $-47982.83$ & $1172.38$& $-40353.58$ &$-39181.20$ &  $ 12.34$\\
 \hline
\end{tabular}
\begin{tabular}{ |p{1.7cm}|p{6.5cm}|p{1.7cm}|p{1.7cm}|p{1.7cm}|p{1.7cm}|p{1.7cm}|}
 \hline
 % \multicolumn{7}{|c|}{\Large{$\nu = 0$, $\xi = 10$ nm, $\Phi=0$}} \\
 \multicolumn{7}{|c|}{\Large{$\nu=0$, $\xi=10$ nm, $\Phi=0$}} \\
 \hline
 state & wave function & kinetic & $\epsilon \times$Hartree & $\epsilon \times$Fock & $\epsilon \times$Coulomb & Hubbard$/U$   \\
 \hline
 KIVC   & $\prod_{\boldsymbol{k}}\prod_{s} \frac{1}{2}(d^{\dagger}_{\boldsymbol{k}K A s} + d^{\dagger}_{\boldsymbol{k}K'B s}) (d^{\dagger}_{\boldsymbol{k}K B s} - d^{\dagger}_{\boldsymbol{k}K'A s})|0\rangle$ & $-46395.22$    &  $-364.23$ & $-38971.96$  & $-39336.20$ & $11.98$ \\
 VP & $\prod_{\boldsymbol{k}}\prod_{s} d^{\dagger}_{\boldsymbol{k}K A s}d^{\dagger}_{\boldsymbol{k}KB s}|0\rangle$  & $-46395.22$  & $-364.23$& $-38969.71$ &  $-39333.94$ & $11.98$ \\
 TIVC & $\prod_{\boldsymbol{k}}\prod_{s} \frac{1}{2}(d^{\dagger}_{\boldsymbol{k}K A \uparrow} + d^{\dagger}_{\boldsymbol{k}K'B \uparrow}) (d^{\dagger}_{\boldsymbol{k}K B \uparrow} + d^{\dagger}_{\boldsymbol{k}K'A \uparrow})|0\rangle$& $-46395.22$ & $-367.82$ & $-38952.11$ & $-39319.93$ & $12.18$\\
 % VP'   & $\prod_{\boldsymbol{k}} d^{\dagger}_{\boldsymbol{k}K A \uparrow} d^{\dagger}_{\boldsymbol{k}KB \uparrow} d^{\dagger}_{\boldsymbol{k}K' A \downarrow} d^{\dagger}_{\boldsymbol{k}K'B \downarrow} |0\rangle$    & $2.977 - 529.138\epsilon^{-1} - 0.024U(\text{eV})$\\
 SP &  $\prod_{\boldsymbol{k}} \prod_{\eta \sigma} d^{\dagger}_{\boldsymbol{k}\eta \sigma \uparrow} |0\rangle$ & $-46395.22$  & $-364.23$ & $-38959.38$ & $-39323.62$ & $ 10.81$\\
 VH &  $\prod_{\boldsymbol{k}}\prod_{s} d^{\dagger}_{\boldsymbol{k}K A s} d^{\dagger}_{\boldsymbol{k}K' A s} |0\rangle$ & $-46395.22$ & $-384.95$ & $-38945.89$ & $-39330.85$ &  $12.67$\\
  non int. GS &  $-$ & $-46398.20$ & $-364.92$ & $-38439.73$ & $-38804.65$ &  $12.00$\\
 \hline
\end{tabular}
% \caption{Expectation  values of the energy of candidates states for the correlated insulators of MATBG, relative to the energy of the kinetic hamiltonian ground state.}
% \label{table1}
% \end{table}

% \begin{table}[h!]
% \centering
\begin{tabular}{ |p{1.7cm}|p{6.5cm}|p{1.7cm}|p{1.7cm}|p{1.7cm}|p{1.7cm}|p{1.7cm}|}
 \hline
 \multicolumn{7}{|c|}{\Large{$\nu = 0$, $\xi = 20$ nm, $\Phi=0$}} \\
 \hline
 state & wave function & kinetic & $\epsilon \times$Hartree & $\epsilon \times$Fock & $\epsilon \times$Coulomb & Hubbard$/U$   \\
 \hline
 KIVC   & $\prod_{\boldsymbol{k}}\prod_{s} \frac{1}{2}(d^{\dagger}_{\boldsymbol{k}K A s} + d^{\dagger}_{\boldsymbol{k}K'B s}) (d^{\dagger}_{\boldsymbol{k}K B s} - d^{\dagger}_{\boldsymbol{k}K'A s})|0\rangle$ & $-46395.22$    &  $-368.00$ & $-40784.72$  & $-41152.72$ & $11.98$ \\
 VP & $\prod_{\boldsymbol{k}}\prod_{s} d^{\dagger}_{\boldsymbol{k}K A s}d^{\dagger}_{\boldsymbol{k}KB s}|0\rangle$  & $-46395.22$  & $-368.00$& $-40782.49$ &  $-41150.49$ & $11.98$ \\
 TIVC & $\prod_{\boldsymbol{k}}\prod_{s} \frac{1}{2}(d^{\dagger}_{\boldsymbol{k}K A \uparrow} + d^{\dagger}_{\boldsymbol{k}K'B \uparrow}) (d^{\dagger}_{\boldsymbol{k}K B \uparrow} + d^{\dagger}_{\boldsymbol{k}K'A \uparrow})|0\rangle$& $-46395.22$ & $-371.62$ & $-40763.47$ & $-41135.10$ & $12.18$\\
 % VP'   & $\prod_{\boldsymbol{k}} d^{\dagger}_{\boldsymbol{k}K A \uparrow} d^{\dagger}_{\boldsymbol{k}KB \uparrow} d^{\dagger}_{\boldsymbol{k}K' A \downarrow} d^{\dagger}_{\boldsymbol{k}K'B \downarrow} |0\rangle$    & $2.977 - 529.138\epsilon^{-1} - 0.024U(\text{eV})$\\
 SP &  $\prod_{\boldsymbol{k}} \prod_{\eta \sigma} d^{\dagger}_{\boldsymbol{k}\eta \sigma \uparrow} |0\rangle$ & $-46395.22$  & $-368.00$ & $-40772.06$ & $-41140.16$ & $ 10.81$\\
 VH &  $\prod_{\boldsymbol{k}}\prod_{s} d^{\dagger}_{\boldsymbol{k}K A s} d^{\dagger}_{\boldsymbol{k}K' A s} |0\rangle$ & $-46395.22$ & $-388.88$ & $-40757.19$ & $-41146.07$ &  $12.67$\\
  non int. GS &  $-$ & $-46398.20$ & $-368.72$ & $-40176.25$ & $-40545.97$ &  $12.00$\\
 \hline
\end{tabular}
% \caption{\textbf{Energy of the ground sate candidates at charge neutrality and zero field.} The zero point of the energy is set such that the energy of the ground state of the non interacting system  is $0$ for each value of $\epsilon$ and $U$. The values are in the unit of meV per unit cell. The gate distance is set to $\xi = 10$ nm.}
% \label{table2}
% \end{table}

% \begin{table}[h!]
% \centering
\begin{tabular}{ |p{1.7cm}|p{6.5cm}|p{1.7cm}|p{1.7cm}|p{1.7cm}|p{1.7cm}|p{1.7cm}|}
 \hline
 \multicolumn{7}{|c|}{\Large{$\nu = +2$, $\xi = 10$ nm, $\Phi=0$}} \\
 \hline
 state & wave function & kinetic & $\epsilon \times$Hartree & $\epsilon \times$Fock & $\epsilon \times$Coulomb & Hubbard$/U$  \\
 \hline
 KIVC   & $\prod_{\boldsymbol{k}} \frac{1}{2}(d_{\boldsymbol{k}K A \uparrow} + d_{\boldsymbol{k}K'B \uparrow}) (d_{\boldsymbol{k}K B \uparrow} - d_{\boldsymbol{k}K'A \uparrow})|+4\rangle$ & $-44809.34$ & $543.00$ & $-38954.64$   &  $-38411.64$ & $11.88$ \\
 VP & $\prod_{\boldsymbol{k}} d_{\boldsymbol{k}K A \uparrow}d_{\boldsymbol{k}KB \uparrow}|+4\rangle$ & $-44809.34$ & $543.00$ &$-38953.51$ & $-38410.51$ &$11.88$ \\
 TIVC & $\prod_{\boldsymbol{k}} \frac{1}{2}(d_{\boldsymbol{k}K A \uparrow} + d_{\boldsymbol{k}K'B \uparrow}) (d_{\boldsymbol{k}K B \uparrow} + d_{\boldsymbol{k}K'A \uparrow})|+4\rangle$ & $-44809.34$ & $542.11$ & $-38944.71$ & $-38402.61$ & $11.88$ \\
 QAH-IVC & $\prod_{\boldsymbol{k}}\prod_s \frac{1}{\sqrt{2}}(d_{\boldsymbol{k}K A s} + d_{\boldsymbol{k}K'B s)}|+4\rangle$ & $-44809.34$  & $542.11$ & $-38947.85$ & $-38405.74$ & $12.22$\\
 QAH-VP &  $\prod_{\boldsymbol{k}}\prod_s d_{\boldsymbol{k}K A s} |+4\rangle $ & $-44809.34$ & $537.82$ & $-38945.73$ & $-38407.91$ &  $ 12.35$\\
  VH &  $\prod_{\boldsymbol{k}} d_{\boldsymbol{k}K A \uparrow} d_{\boldsymbol{k}K' A} |+4\rangle $ & $-44809.34$ & $537.82$ & $-38941.61$ & $-38403.78$  & $11.88$\\
   non int. GS &  $-$ & $-44812.85$ & $683.22$ & $-38576.60$ & $-37893.38$  & $12.29$\\
 \hline
\end{tabular}
% \caption{Expectation  values of the energy of candidates states for the correlated insulators of MATBG, relative to the energy of the kinetic hamiltonian ground state.}
% \label{table11}
% \end{table}

% \begin{table}[h!]
% \centering
\begin{tabular}{ |p{1.7cm}|p{6.5cm}|p{1.7cm}|p{1.7cm}|p{1.7cm}|p{1.7cm}|p{1.7cm}|}
 \hline
 \multicolumn{7}{|c|}{\Large{$\nu = +2$, $\xi = 20$ nm, $\Phi=0$}} \\
 \hline
 state & wave function & kinetic & $\epsilon \times$Hartree & $\epsilon \times$Fock & $\epsilon \times$Coulomb & Hubbard$/U$  \\
 \hline
 KIVC   & $\prod_{\boldsymbol{k}} \frac{1}{2}(d_{\boldsymbol{k}K A \uparrow} + d_{\boldsymbol{k}K'B \uparrow}) (d_{\boldsymbol{k}K B \uparrow} - d_{\boldsymbol{k}K'A \uparrow})|+4\rangle$ & $-44809.34$ & $1133.46$ & $-40765.58$   &  $-39632.12$ & $11.88$ \\
 VP & $\prod_{\boldsymbol{k}} d_{\boldsymbol{k}K A \uparrow}d_{\boldsymbol{k}KB \uparrow}|+4\rangle$ & $-44809.34$ & $1133.46$ &$-40764.47$ & $-39631.01$ &$11.88$ \\
 TIVC & $\prod_{\boldsymbol{k}} \frac{1}{2}(d_{\boldsymbol{k}K A \uparrow} + d_{\boldsymbol{k}K'B \uparrow}) (d_{\boldsymbol{k}K B \uparrow} + d_{\boldsymbol{k}K'A \uparrow})|+4\rangle$ & $-44809.34$ & $1132.55$ & $-40754.96$ & $-39622.41$ & $11.88$ \\
 QAH-IVC & $\prod_{\boldsymbol{k}}\prod_s \frac{1}{\sqrt{2}}(d_{\boldsymbol{k}K A s} + d_{\boldsymbol{k}K'B s)}|+4\rangle$ & $-44809.34$  & $1132.55$ & $-40758.11$ & $-39625.56$ & $12.22$\\
 QAH-VP &  $\prod_{\boldsymbol{k}}\prod_s d_{\boldsymbol{k}K A s} |+4\rangle $ & $-44809.34$ & $1128.23$ & $-40755.98$ & $-39627.75$ &  $ 12.35$\\
  VH &  $\prod_{\boldsymbol{k}} d_{\boldsymbol{k}K A \uparrow} d_{\boldsymbol{k}K' A} |+4\rangle $ & $-44809.34$ & $1128.23$ & $-40751.81$ & $-39623.58$  & $11.88$\\
   non int. GS &  $-$ & $-44812.85$ & $1273.27$ & $-4032.06$ & $-39058.79$  & $12.29$\\
 \hline
\end{tabular}
\caption{\textbf{Expectation values of the energy of candidates states for the correlated insulators of MATBG at zero magnetic field.} We list the states for fillings $\nu=-2,0,+2$ setting $\xi$ to $10$ and $20$ nm. The energies are in units of meV per unit cell, and the value of $U$ is given in eV. The state non int. GS is the ground state of the non interacting system. QAH (Quantum anomalous Hall) states have Chern number 2 and are stabilized by small out of plane magnetic fields\cite{Wu2021,Stepanov21}. $|0\rangle$ denotes the state with the filled remote bands and $| + 4\rangle$ the $\nu = +4$ insulator.}
\label{energiesp0}
\end{table}

\begin{table}[H]
\centering
\begin{tabular}{ |p{2.2cm}|p{1.7cm}|p{1.7cm}|p{1.7cm}|p{1.7cm}|p{1.7cm}|}
 \hline
 $\nu$, ($\epsilon$, $U$) & kinetic & $\epsilon \times$Hartree & $\epsilon \times$Fock & $\epsilon \times$Coulomb & Hubbard$/U$   \\
 \hline
 $-2$, $(10,4$ eV)   & $-48147.31$ & $228.05$ &  $-38903.47$ & $-38675.42$ &  $11.39$ \\
 $-2$, $(50,0.5$ eV) & $-48149.36$ & $222.52$ & $-38842.95$ & $-38620.43$ & $11.37$ \\
 $0$, $(10, 4$ eV) & $-46552.57$ & $-340.91$ & $-38890.94$ & $-39231.85$ & $10.90$ \\
 $0$, $(50, 0.5$ eV) & $-46554.48$ & $-341.91$ & $-38845.54$ & $-39187.45$ & $10.76$ \\
 $+2$, $(10, 4$ eV) & $-44959.04$ & $319.40$ & $-38846.62$ & $-38527.22$ & $11.08$ \\
 $+2$, $(50, 0.5$ eV) & $-44964.44$ & $476.25$ & $-38852.51$ & $-38376.26$ & $11.34$ \\
 \hline
\end{tabular}
\caption{ 
\textbf{Expectation  values of the energy of the ground states for $\boldsymbol{B=26.5}$ T.} The fillings are $\nu=0,\pm2$ and the interaction parameters $\epsilon=10$, $U=4$ eV and $\epsilon=10$, $U=0.5$ eV. The gate distance is set to $\xi=10$ nm. The values are in units of meV per unit cell, and $U$ is given in eV. The Zeeman energy is $\pm 1.535$ meV per electron and is not explicitly included.}
\label{energieshfp1}
\end{table}

\begin{table}[H]
\centering
\begin{tabular}{ |p{1.7cm}|p{1.7cm}|p{1.7cm}|p{1.7cm}|p{1.7cm}|p{1.7cm}|}
 \hline
 \multicolumn{6}{|c|}{\Large{$\nu = -2$. $\epsilon=10$, $U=4$ eV}} \\
 \hline
 order & kinetic & $\epsilon \times$Hartree & $\epsilon \times$Fock & $\epsilon \times$Coulomb & Hubbard$/U$   \\
 \hline
 KIVC   & $-47980.57$ & $631.44$ &  $-39002.21$ & $-38370.76$ &  $12.07$ \\
 VP & $-47981.24$ & $613.44$ & $-38958.61$ & $-38345.17$ & $12.05$ \\
 \hline
\end{tabular}

\begin{tabular}{ |p{1.7cm}|p{1.7cm}|p{1.7cm}|p{1.7cm}|p{1.7cm}|p{1.7cm}|}
 \hline
 \multicolumn{6}{|c|}{\Large{$\nu = -2$. $\epsilon=50$, $U=0.5$ eV}} \\
 \hline
 order & kinetic & $\epsilon \times$Hartree & $\epsilon \times$Fock & $\epsilon \times$Coulomb & Hubbard$/U$   \\
 \hline
 KIVC    & $-47981.42$ &  $602.60$ & $-38951.76$  &  $-38349.17$   &  $12.12$ \\
 VP   & $-47981.58$ & $600.96$ & $-38936.46$ & $-38335.50$  &  $12.03$ \\
 \hline
\end{tabular}

\begin{tabular}{ |p{1.7cm}|p{1.7cm}|p{1.7cm}|p{1.7cm}|p{1.7cm}|p{1.7cm}|}
 \hline
 \multicolumn{6}{|c|}{\Large{$\nu = 0$. $\epsilon=10$, $U=4$ eV}} \\
 \hline
 order & kinetic & $\epsilon \times$Hartree & $\epsilon \times$Fock & $\epsilon \times$Coulomb & Hubbard$/U$   \\
 \hline
 KIVC   & $-46393.94$ & $-364.72$ &  $-39036.56$ & $-39401.29$ &    $11.99$ \\
 SP & $-46394.61$ & $-364.36$ & $-38980.15$ & $-39344.50$ & $10.88$ \\
 VP  & $-46394.61$ & $-364.36$ & $-38989.93$ & $-39354.29$ & $11.98$ \\
 \hline
\end{tabular}

\begin{tabular}{ |p{1.7cm}|p{1.7cm}|p{1.7cm}|p{1.7cm}|p{1.7cm}|p{1.7cm}|}
 \hline
 \multicolumn{6}{|c|}{\Large{$\nu = 0$. $\epsilon=50$, $U=0.5$ eV}} \\
 \hline
 order & kinetic & $\epsilon \times$Hartree & $\epsilon \times$Fock & $\epsilon \times$Coulomb & Hubbard$/U$   \\
 \hline
 KIVC & $-46394.82$ & $-365.14$ & $-39003.10$ & $-39368.25$ &  $11.98$ \\
 SP & $-46395.38$ & $-364.90$ & $-38953.35$ & $-39318.25$ & $10.84$ \\
 VP & $-46395.38$ & $-364.90$ & $-38963.48$ & $-39328.38$ & $11.98$ \\
 \hline
\end{tabular}

\begin{tabular}{ |p{1.7cm}|p{1.7cm}|p{1.7cm}|p{1.7cm}|p{1.7cm}|p{1.7cm}|}
 \hline
% \multicolumn{7}{|c|}{\Large{$\nu=0$, $\xi=10$ nm, $\Phi=\Phi_0$}} \\
\multicolumn{6}{|c|}{\Large{$\nu=+2$. $\epsilon=10$, $U=4$ eV}} \\
 \hline
 order & kinetic &$\epsilon \times$Hartree & $\epsilon \times$Fock & $\epsilon \times$Coulomb & Hubbard$/U$    \\
 \hline
 KIVC    & $-44808.71$ &  $545.40$ & $-38986.29$  &  $-38440.89$   &  $11.89$ \\
 VP  & $-44809.02$ & $541.18$ & $-38961.14$ & $-38419.95$  &  $11.88$ \\
 \hline
\end{tabular}

\begin{tabular}{ |p{1.7cm}|p{1.7cm}|p{1.7cm}|p{1.7cm}|p{1.7cm}|p{1.7cm}|}
 \hline
 \multicolumn{6}{|c|}{\Large{$\nu = +2$. $\epsilon=50$, $U=0.5$ eV}} \\
 \hline
 order & kinetic & $\epsilon \times$Hartree & $\epsilon \times$Fock & $\epsilon \times$Coulomb & Hubbard$/U$   \\
 \hline
 KIVC  & $-44809.07$ & $544.39$ & $-38973.52$ & $-38429.13$ &    $11.89$ \\
 VP & $-44809.34$ & $543.00$ & $-38953.51$ & $-38410.51$ & $11.88$ \\
 \hline
\end{tabular}

\caption{ 
\textbf{Expectation  values of the energy of the self-consistent states for $\boldsymbol{B=0}$ T.} The fillings are $\nu=0,\pm2$ and the interaction parameters $\epsilon=10$, $U=4$ eV and $\epsilon=10$, $U=0.5$ eV. The gate distance is set to $\xi=10$ nm. The states are labeled by the dominant order parameter. The values are in units of meV per unit cell, and $U$ is given in eV.}
\label{energieshfp0}
\end{table}

\end{document}